\documentclass[brazil,12pt,a4paper,openany]{book}
\usepackage{anysize}
\usepackage[utf8]{inputenc}
\usepackage[english]{babel}
\usepackage[ruled,chapter]{algorithm}
\usepackage{makeidx}
\usepackage{subfig}
\usepackage{amsmath}
\usepackage{bm}
\usepackage{breqn}
\usepackage[subnum]{cases}
\usepackage{amssymb}
\usepackage{amsthm}
\usepackage{mathrsfs}
\usepackage{graphicx}
\usepackage{fancyhdr}
\usepackage{array}
\usepackage{geometry}
\usepackage{latexsym}
\usepackage{eufrak}
\usepackage{euscript}
\usepackage{enumerate}
\usepackage{dsfont}
\usepackage{slashed}
\usepackage[plainpages=false,pagebackref=false]{hyperref}
\usepackage{float}
\usepackage{pdfpages}
\usepackage{mathtools}
\usepackage{graphicx}
\usepackage{amsmath}
\usepackage{amssymb}
\usepackage{colortbl}

\usepackage[LGR,T1]{fontenc}

  \usepackage{lipsum}

\hypersetup{
pdftitle={QCD Matter in Compact Stars},
pdfsubject={PhD Thesis},
pdfauthor={José Carlos Jiménez Apaza},
pdfkeywords={QCD phase diagram, high densities, extreme conditions}
}

\title{QCD Matter in Compact Stars}
\author{José Carlos Jiménez Apaza}

\usepackage{multirow}

\makeindex

\marginsize{25mm}{25mm}{25mm}{25mm}

\hypersetup{pdfpagelabels,hyperindex,colorlinks=true,breaklinks=true,bookmarks=true,linkcolor=black,citecolor=black,
urlcolor=black}

\usepackage[sort&compress,numbers]{natbib}
\usepackage{doi}

\begin{document}

\setlength{\abovecaptionskip}{0.0cm}
\setlength{\belowcaptionskip}{0.0cm}
\setlength{\baselineskip}{24pt}

\pagestyle{fancy}
\lhead{}
\chead{}
\rhead{\thepage}
\lfoot{}
\cfoot{}
\rfoot{}

\fancypagestyle{plain}
{
	\fancyhf{}
	\lhead{}
	\chead{}
	\rhead{\thepage}
	\lfoot{}
	\cfoot{}
	\rfoot{}
}

\renewcommand{\headrulewidth}{0pt}


\frontmatter 

\thispagestyle{empty}

\begin{figure}[h]
	\includegraphics[scale=0.8]{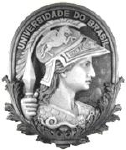}
\end{figure}

\vspace{15pt}

\begin{center}

\textbf{UNIVERSIDADE FEDERAL DO RIO DE JANEIRO}

\textbf{INSTITUTO DE FÍSICA}

\vspace{30pt}

{\uppercase{ \Large \bf Interacting Quark Matter Effects on the Structure of Compact Stars}}

\vspace{25pt}

{\large \bf José Carlos Jiménez Apaza}

\vspace{35pt}

\begin{flushright}

\parbox{10.3cm}{PhD thesis presented to the Graduate Program in Physics of the Institute of Physics of the Federal University of Rio de Janeiro, as part of the requirements for obtaining the title of Doctor in Sciences (Physics).}

\vspace{18pt}

{\large \bf Advisor: Eduardo Souza Fraga}

\vspace{12pt}

\end{flushright}

\vspace{90pt}

\textbf{Rio de Janeiro}

\textbf{March/2020}

\end{center}

\newpage
\noindent

\vspace*{15pt}
\begin{center}
{\LARGE\bf Abstract}\\
\vspace{15pt}
{\Large\bf Interacting Quark Matter Effects on the Structure of Compact Stars}\\
\vspace{6pt}
{\bf José Carlos Jiménez Apaza}\\
\vspace{12pt}
{\bf Advisor: Eduardo Souza Fraga}\\
\vspace{20pt}
\parbox{14cm}{\emph{Abstract} da Tese de Doutorado apresentada ao Programa de Pós-Graduação em Física do Instituto de Física da Universidade Federal do Rio de Janeiro - UFRJ, como parte dos requisitos necessários à obtenção do título de Doutor em Ciências (Física).}
\end{center}
\vspace*{35pt}

In this work, we study the effects that interacting quark matter has on the stellar structure of strange and charm quark stars. Additionally, their stability against radial pulsations is analyzed using a first-order formalism for adiabatic general relativistic oscillations. Besides, the early stage of stellar evolution of neutron stars after the supernovae explosion, i.e. protoneutron stars, is investigated by considering the possibility of a first-order phase transition to quark matter rich in leptons, where the dynamics of conversion between phases is studied within a thermal nucleation model.  For each kind of compact star mentioned we use for the quark phase the equation of state calculated within cold perturbative QCD (pQCD), parametrized only by its renormalization scale. We note that the original pQCD framework is manipulated appropriately to include neutrinos and extended to add heavy quarks to the original system composed only by up, down and strange quark flavors.

\vspace{15pt}


\newpage


\noindent

\vspace*{20pt}

\begin{center}

{\LARGE\bf Acknowledgements}

\end{center}

\vspace*{40pt}

Since I started as a physics student in 2008, I have benefited from many people who helped me and inspired me along the years to reach this point of my professional training as a physicist. I am very thankful to life for putting them along this formation process. In order to show no misleading preferences to some of them, and also due to space limitations, I am forced to mention only the ones that directly influenced the existence of this work through emotional, intellectual, motivational, and financial support.

I must start by thanking my parents, Felipe Jiménez and Angelina Apaza, for all the emotional support along these years, especially in the hard and difficult moments. Besides, it was very helpful what I learned about hard work from my dad, and about patience and imagination (much before knowing it was important to understand Nature's physical laws) from my mom. There are no words to express how much I love them. The rest of my family, my sisters Claudia and Alejandra, with my nephews Kiara and Felipe Ruben, represent a motivation to my life. In these years far from home, they supported me emotionally in the difficult times and encouraged me to follow my goals, always willing to help me at any time and situation. This thesis is dedicated to all of them.

Eduardo was a fundamental character along the six years I worked with him as his student in the MSc and PhD studies. I cannot consider him only as my thesis advisor but also as an inspiring and motivational friend since I met him in Arequipa-Peru where I got strongly attracted by his professional attitude and vision of physics research. While working with him I realized that, in a way, he behaves similarly to QCD through its asymptotic freedom (me being a quark), i.e. my highly-energetic work would represent more freedom to think, study and calculate. Eduardo's presence would be hardly felt if I was doing things ``right''. On the other hand, if I was working low energetically, he would immediately act as the strong force he is, compelling me to come back to the ``right'' path and stay confined to my duties as his student. This was what I needed and fortunately fitted exactly with my personality and way of doing research in physics. I will always appreciate and be indebted to Eduardo for all he has given to me in the form of knowledge and example as a physicist and, more importantly, as a human being. I hope our friendship lasts forever.

Patricia T. was my emotional support along these years. She was like the blood inside my veins, like the blow of my shining days, like the blue of the sky after my internal storms. In fact, whenever I think of her I remember what love is and what is its human name.

I thank the Institute of Physics of UFRJ for providing me with many good and excellent professors and colleagues, all of them interacting in a friendly environment. Besides, I thank all the friends I got here in these years. Especially, I would like to thank my dear friends Daniel Kroff, Elvis do Amaral Soares, Mauricio Hippert and Anderson Kendi. Their help and advise, about intellectual and maturity matters, at the beginning and at critical points of my studies, were important to finally achieve this thesis. I am very grateful to Brazil for giving me these good friends and also good physicists.

Thanks also to Carlos José Delgado and Igor de Souza Silva, for their efficiency and helpfulness with all kinds of bureaucratic problems, at the beginning and ending of my PhD studies, respectively.

I take the opportunity to congratulate UFRJ this year in which turns 100 years old.
 
Finally I thank Brazil through its agencies CNPq and FAPERJ for the financial support along all these four years of PhD studies. Without their support none of these research projects would be possible. Besides, this work is part of the project also supported by the project INCT-FNA Proc. No. 464898/2014-5. 



\newpage
\phantomsection
\addcontentsline{toc}{chapter}{Table of Contents}
\let\cleardoublepage\clearpage
\tableofcontents

\mainmatter
\begin{chapter}{Introduction}
	\label{chap:intro}	
	
\hspace{5 mm}


Quantum chromodynamics (QCD) is considered to be the fundamental theory of strong interactions. At high energies it exhibits $\textit{asymptotic freedom}$, where colored quarks and gluons are the relevant degrees of freedom. For low energies it offers a strongly-coupled behavior, hadrons being the relevant degrees of freedom. Therefore, there must exist some mechanism intrinsic to QCD which $\textit{confines}$ colored degrees of freedom at an intermediate energy scale. In a QCD $\textit{medium}$\footnote{This must be understood as the substance in which strong interactions propagate. For instance, since at high temperatures the QCD vacuum is melt, quarks and gluons are allowed to interact over larger volumes than only inside nucleons. The medium in this case is known as (hot) quark matter.} in equilibrium, strongly-interacting matter can be characterized by external parameters such as temperature $T$, baryon chemical potential $\mu_{B}$, magnetic field $B$, and so on. These systems allow us to get insights on the different thermodynamic phases of the strong interactions and build its phase diagram.

The study of these phases in the context of QCD for very high values of the external parameters, e.g. $T$ and $\mu_{B}$, can be done using perturbative techniques because the strong coupling, $\alpha_{s}$, is reasonably small in this limit. An important case to be analyzed along all this thesis occurs for large values of $\mu_{B}$ and low/moderate $T$ in compact star matter \cite{Glendenning:2000}.

Neutron stars\footnote{Usually observed as \textit{pulsars} (although recent evidence \cite{Buckley:2016fqu} proves that some pulsars can also be found as white dwarfs) but not necessarily \cite{Klochkov:2013voa}, they are the cool remnant of protoneutron stars produced in supernova explosions, having masses between 0.9 and 2 solar masses and radii around 15 km.} provide a unique laboratory for the investigation of the strong interaction under extreme conditions of density (compression), or equivalently high-$\mu_{B}$ \cite{Glendenning:2000}. Especially now, with observations entering a new era: NASA's Neutron star Interior Composition Explorer (NICER) mission \cite{NICER}, which will allow for measurements of neutron star masses and, especially, radii to unprecedented precision, the NICA/MPD \cite{Sissakian:2009zza} experiment that will probe the region of intermediate densities, and more importantly the first multimessenger observation of a binary neutron star merger that has been performed with great success \cite{TheLIGOScientific:2017qsa}, so that gravitational waves can now be used to probe the internal properties of neutron stars \cite{Andersson:2009yt}, giving us information about possible exotic phases in their cores, for instance, quark matter. Since the baryon densities inside these stars must be of the order of the nuclear saturation density, $n_{0}=0.16\rm fm^{-3}$, and above, one is allowed to use them as astrophysical laboratories to probe the high-baryon-density ($n\gtrsim{5}n_{0}$ or $\mu_{B}\sim{2}$ GeV) and cold regimes of the QCD phase diagram.

A crucial ingredient for the description of the structure and phases in the interior of neutron stars is the equation of state (EoS) for neutron star matter, which needs the understanding of the thermodynamics of strong interactions at densities of the order of the saturation density, $n_{0}$, and above. Unfortunately, such region in the parameter space of QCD is not accessible to a first-principle, nonperturbative (lattice) approach due to the stringent restrictions brought about by the sign problem \cite{deForcrand:2010ys}. The alternative that still provides controlled calculations in the fundamental theory of strong interactions would be cold and dense perturbative QCD \cite{Kapusta:2006pm,Laine:2016hma}. The state-of-the-art perturbative EoS for cold and dense QCD was obtained in Ref. \cite{Kurkela:2009gj}, and goes way beyond a simple description based on the MIT bag model as was shown in Ref. \cite{Fraga:2013qra}, where the equation of state was also cast into a simple pocket formula assuming local charge neutrality and beta-equilibrium\footnote{It is also possible to build an effective bag model from the two-loop massless cold QCD EoS \cite{Fraga:2001id} and use it to study hybrid stars \cite{Fraga:2001xc,Alford:2004pf}.}. 
Of course, due to asymptotic freedom, this approach is valid only at high enough densities and has to be matched either onto a phenomenological hadronic equation of state at lower densities or to the other controlled limit of the theory of strong interactions, chiral effective field theory (see, e.g. Refs. \cite{Hebeler:2013nza,Tews:2012fj}). One can also use both limits of the theory of strong interactions and a parametrization of the ignorance about the intermediate density region in terms of multiple polytropes to constrain the neutron star matter EoS down to $30 \%$ \cite{Kurkela:2014vha}. By implementing the astrophysical constraints on the maximum mass from measurements of the pulsars PSR J$1614-2230$, with M$=1.97\pm{0.04}$M$_{\odot}$ \cite{Demorest:2010bx}, and PSR J$0348+0432$, with M$=2.01\pm{0.04}$M$_{\odot}$ \cite{Antoniadis:2013pzd}, for instance, the previous approach sets limits on the properties of rotating neutron stars \cite{Gorda:2016uag} beyond the mass-radius diagram.

The aim of this thesis is to investigate the effects that the short-distance QCD interactions have on the cold quark matter equation of state when leptons, in particular neutrinos, and heavy quarks are introduced in the system, thus enabling us to probe the cold and dense sector of the QCD phase diagram. This is basically done by adopting as a fundalmental tool the perturbative QCD (pQCD) result of Kurkela \textit{et al.} \cite{Kurkela:2009gj} at intermediate and high densities. This approach is taken in this thesis because simple models only offer crude estimates of the physics at the core of neutron stars, where interactions are still very important. These effects are partially captured by the pQCD model, giving us further insights on astrophysical observables. 

In order to do that we proceed as follows. In Chapter \ref{chap:PNS} we  address the implementation of leptons in the pQCD equation of state, which is appropriate if one is interested in studying the early stage of the core-collapse supernovae explosions when a protoneutron star is formed. Besides, it is investigated how the mechanism of bubble nucleation is affected by the perturbative QCD corrections to the ideal gas limit. In Chapter \ref{chap:radialQS} the stability of quark stars (satifying or not the Bodmer-Witten hypothesis) against adiabatic radial oscillations is studied, where additionally the main aspects of the stellar structure of compact stars with spherical symmetry and the radial pulsation equations are reviewed. In Chapter \ref{chap:charm} a novel method is developed to include heavy quarks in a QCD system with only light quarks, i.e. up, down, and strange. This allows us to explore the heavy sector of the QCD phase diagram, in particular the possibility of charm quark stars existing in Nature. Finally, in Chapter \ref{chap:conclusion} we present our conclusions and future perspectives.

Along this thesis we use natural units, i.e. $G=c=k_{\rm B}=\hbar=1$, unless otherwise indicated. Besides, we use the signature for the metric tensor as being $g_{\mu\nu}={\rm diag}(+,-,-,-)$ which is standard when performing field theory calculations for compact star physics. All numerical calculations carried out along this thesis were done using \textit{Mathematica 10.3}. We note that apart from all the natural packages included in this version, when finding the radial-oscillation frequencies of compact stars for Chapters \ref{chap:radialQS} and \ref{chap:charm}, we have also installed ``The RootSearch Package'' (easily found in the Internet) which helps us to find numerically ordered values of roots of functions in a given domain, something which is not possible using the ``FindRoot'' for complicated functions. This package is crucial for our computations since they depend strongly on repeated routines (\textit{loops}) for a given equation of state.
  
\end{chapter}

\begin{chapter}{Quark Matter in Protoneutron Stars}
	\label{chap:PNS}

\hspace{5 mm}

\section{Introduction}

The possible occurrence of phase transitions in supernovae explosions was first proposed by Migdal et al. \cite{Migdal:1979je}. Then, some decades later, the general relativistic simulations of the Boltzmann neutrino transport equations were solved to analyse this hypothesis although still adopting bag-model type equations of state for a potential first-order transition to quark matter at high densities \cite{Nakazato:2008su,Sagert:2008ka,Fischer:2010wp}. In particular, in Ref. \cite{Sagert:2008ka} it was shown that the quark transition can occur during the early post-bounce phase of a core collapse supernova explosion\footnote{This happens at the end of the stellar evolution of massive stars (around $\sim 10 M_{\odot}$ and mildly above, otherwise it would collapse to a black hole) usually through a core-collapse supernovae explosion (also known as a Type IIb supernovae), in contrast to main-sequence stars \cite{Glendenning:2000,Carroll:2007}.}, then producing a second shock wave (the first being the usual shock wave after the bounce) which triggers a delayed supernova explosion\footnote{The produced quark matter emits a neutrino burst, typically a few hundred of milliseconds after the first neutronization burst, which could be detected by currently available neutrino detectors \cite{Benvenuto:1989qr,Aguilar-Arevalo:2016khx}.} for masses of the progenitor up to 15 $M_{\odot}$, being $M_{\odot}$ the mass of the Sun. 

Although all this process is summarized in Fig. \ref{fig:PNSevol}, it is better to give further insights on the timescales characterizing each evolving stage by means of the \textit{dynamical} timescale, $\tau_{\rm dyn}$, which depends on the balance between gravity and pressure. Especifically, in the case when the \textit{free-fall} (dynamical) timescale of gravity, $\tau_{\rm ff}=\sqrt{3\pi/32\bar{\epsilon}}$, equals or approximates the \textit{expansion} (dynamical) timescale of pressure\footnote{Being $\bar{R}$ an average stellar radius, $\bar{P}$ an average pressure and $\bar{\epsilon}=3\bar{M}/4\pi\bar{R}^{3}$ the average energy density of a constant-density sphere where $M$ is an average stellar mass. For our case of PNSs we consider $\bar{M}=1M_{\odot}$ and $\bar{R}=18~$km, which are standard values in the literature}, $\tau_{\rm exp}=\bar{R}(\bar{\epsilon}/\bar{P})^{1/2}$, one obtains the \textit{hydrodynamical} timescale, i.e. $\tau_{\rm ff}\simeq\tau_{\rm explo}\equiv\tau_{\rm hydro}$ \cite{Kippenhahn:2012ccc}. In particular, for our case of PNSs, we note that since their evolution takes $\tau^{\rm PNS}_{\rm evol}{~\sim~}15~$s (after deleptonization) and $\tau^{\rm PNS}_{\rm hydro}{~\sim~}0.233~$ms, i.e. $\tau^{\rm PNS}_{\rm evol}\gg\tau^{\rm PNS}_{\rm hydro}$, one can safely decouple the PNS evolution equations from the dynamics between spacetime and PNS matter, which represents an important simplification in detailed PNS simulations by only analyzing their relevant timescales \cite{Pons:1998mm}.

On the other hand, instead of the naive MIT bag model for the quark matter phase, cold and dense perturbative QCD (pQCD) can be used as a first-principle inspired model to describe the high-density sector of the EoS for supernova matter. However, as it is currently known 	\cite{Kurkela:2009gj}, it is not adequate to investigate the early stages of their lives as protoneutron stars (PNSs), since during the early post-bounce stage of core collapse, supernova matter is still hot and lepton rich. Additionally, in these conditions the neutrino mean free path is small compared to the size of the star. So, one has to include this trapped neutrinos in the framework\footnote{Reference  \cite{Fraga:2015xha} provides the first attempt of an extension including neutrinos and thermal effects in the case of massless quarks.}.   

\begin{figure}[h]
\begin{center}
\resizebox*{!}{7.8cm}{\includegraphics{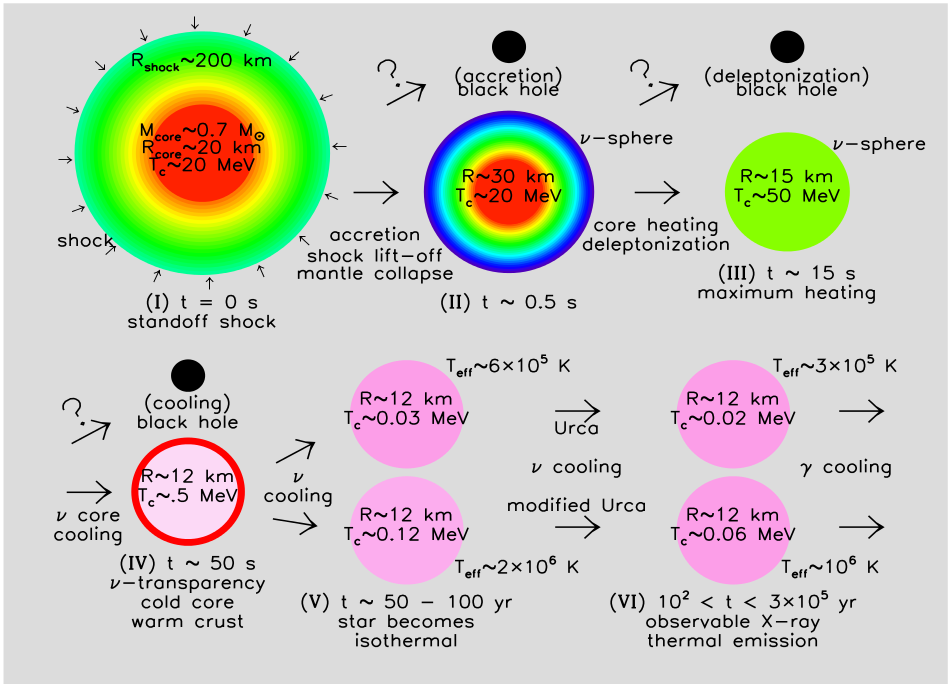}}
\end{center}
\caption{\label{fig:PNSevol} The main stages of evolution of a P(NS), for increasing values of the Roman numerals. The radius $R$ and central temperatures $T_{c}$ are indicated as it evolves in time $t$. Usually instead of the temperatures, a fixed net entropy per baryon (or entropy density per baryon density) $S_{B}=(s/k_{B})/n_{B}$ is used in the literature quantifying finite-temperature effects. When this $S_{B}$ is fixed (constant), it allows the temperature in stars to increase toward the center since in most cases $S_{B}\varpropto{T}/n^{1/3}_{B}$. In PNSs, it usually varies (obtained from numerical simulations) between $S_{B}\simeq{1}$ at around 1 s in stage (II) corresponding to $T\sim 20$ MeV, and $S_{B}\simeq2$ at around 15 s after deleptonization in stage (III) correspoding to $T\sim 50$ MeV, with densities around $\mu_{B}\sim{1500}$ MeV. For more details see Refs. \cite{Bethe:1979vh,Bethe:1988vn,Bethe:1990mw,Pons:1998mm} and Secs. 18.2 and 18.6 of Ref. \cite{Shapiro:1983}. Image taken from Ref. \cite{Lattimer:2006xb}.}
\end{figure}

Additionally, if we want to study the thermal nucleation process of a quark phase using a more adequate EoS at high densities in supernova matter, which was previously investigated using a simplified description of the EoS for quark matter in Refs. \cite{Mintz:2009ay,Mintz:2010mh}, we need to build a framework from pQCD satisfying the above mentioned core-collapse supernova conditions. In Fig. \ref{fig:PNSdiagram} it is shown the many phases that strongly-interacting matter can take and, more importantly, the supernova sector of this QCD phase diagram which can be probed by using lepton-rich equations of state for the nuclear and quark matter sectors. A phase transition from hadronic to quark matter may take place already during the early post-bounce stage of core collapse supernova. If the phase transition is of first order and exhibits a barrier, the formation of the new phase occurs via the nucleation of droplets.

\begin{figure}[h]
\begin{center}
\resizebox*{!}{7.0cm}{\includegraphics{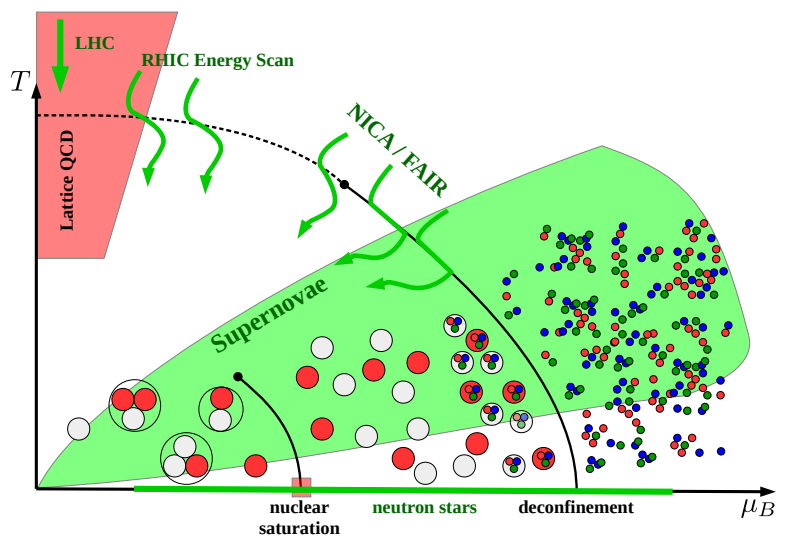}}
\end{center}
\caption{\label{fig:PNSdiagram} Schematic view of the phase diagram of strongly interacting matter. The solid lines represent the liquid-gas and hadronic-quark matter phase transitions. The green shaded area is the domain of the phase diagram accessible by supernova explosions. The green line on the chemical potential axis depicts the expected range of values accessible in compact star interiors. Image taken from Ref. \cite{Bastian:2018wfl}.}
\end{figure}

This chapter investigates the lepton-rich quark matter equation of state within pQCD. Besides, the effects that interacting QCD matter would have if a confinement-deconfinement transition occurs in the few milliseconds after the core-collapse supernovae explosion are studied in the metastable phase known as protoneutron star. This chapter follows the discussion found in Ref. \cite{Jimenez:2017fax}. So, we summarize in Sec. \ref{sec:2poorQM} the main properties of the state-of-the-art equation of state from perturbative QCD. In Sec. \ref{sec:2richQM} we include the presence of trapped electrons and neutrinos to build the lepton-rich pQCD equation of state parametrized by the renormalization scale parameter. We also analyze the allowed parameter space for stable strange quark matter. In Sec. \ref{sec:2deconfined} we discuss the framework for the description of nucleation of quark matter droplets in protoneutron stars. Furthermore, by computing the stellar structure that emerge form the Tolman-Oppenheimer-Volkov
(TOV) equations, we present our results for the mass-radius relations for which we analyse and calculate the nucleation times of unpaired quark matter in the core of protoneutron stars by matching the lepton-rich QCD pressure onto a hadronic equation of state, namely TM1 with trapped neutrinos. Using the inherent dependence of perturbative QCD on the renormalization scale parameter, we provide a measure of the uncertainty in the observables we compute. Associated with the phase conversion dynamics we compare them with the dynamical time scale of core collapse supernovae during which quark matter might be eventually formed. Finally, in Sec. \ref{sec:2conclusion} we present our concluding remarks.

\section{Lepton-poor unpaired quark matter}
	\label{sec:2poorQM}

In order to fix the thermodynamics and statistical mechanics terminology to be used along this thesis, in this section we summarize the main ideas needed to understand  observables calculated for a QCD system \textit{in medium}. In particular, we focus on the perturbative calculation of the equation of state for the quark-gluon plasma (QGP) in the cold limit, i.e., $T\rightarrow{0}$, the so-called cold \textit{quark matter}, which might exist in compact stars.

\subsection*{Statistical Mechanics of Relativistic Quantum Fields}

The principles of equilibrium statistical mechanics offer us a systematic way to calculate (macroscopic) thermodynamic observables in terms of the (microscopic) fundamental degrees of freedom of the system \cite{reichl1980modern}.  When these degrees of freedom become relativistic, the total number of particles is not kept constant and the appropriate thermodynamic potential to be used is the Landau thermodynamic potential $\Omega$ (also known as the grand canonical thermodynamic potential), being defined as \cite{Landau5}
\begin{equation}
\Omega(\left\lbrace\mu_{i}\right\rbrace,T,V)=
E-TS-\sum_{i}\mu_{i}N_{i}=-T\log{\mathcal{Z}},
\end{equation}
where $E$ is the (microcanonical) energy, $V$ is the volume of the system, $\mathcal{Z}$ is the grand canonical partition function, $\left\lbrace\mu_{i}\right\rbrace$ is a set of chemical potentials inherent to the system, $N_{i}$  is the liquid conserved charge of kind ``$i$'' and $T$ the absolute temperature. This potential additionally obeys \textit{the first law of thermodynamics}
\begin{equation}
d\Omega=-SdT-PdV-\sum_{i}N_{i}d\mu_{i},
\end{equation}
where $P$ is the total pressure, and $S$ is the entropy. 
From this law, one could obtain the pressure from 
\begin{equation}
P=-\left(\frac{\partial\Omega}{\partial{V}}\right)_{T,~\lbrace\mu_{i}\rbrace}.
\end{equation}
However, compact stars are macroscopic objects and in turn translational invariant, which makes $\Omega$ to be (to a very good approximation) proportional to the volume of the system. Therefore, $\partial\Omega/\partial{V}=\Omega/V$, which implies
\begin{equation}
\Omega=-PV.
\end{equation}
For a spatially uniform system, it is convenient to introduce the energy density $E/V$, number density $N/V$, and the entropy density $S/V$ as follows:
\begin{equation}
\epsilon=\frac{E}{V}=-P+Ts+\sum_{i}n_{i}d\mu_{i},
\end{equation}
and
\begin{equation}
n_{i}=\frac{N_{i}}{V}=-\frac{1}{V}\left(\frac{\partial\Omega}{\partial\mu_{i}}\right)_{T}=\left(\frac{\partial{P}}{\partial\mu_{i}}\right)_{T},
\end{equation}
and
\begin{equation}
s=\frac{S}{V}=-\frac{1}{V}\left(\frac{\partial\Omega}{\partial{T}}\right)_{\lbrace\mu_{i}\rbrace}=\left(\frac{\partial{P}}{\partial{T}}\right)_{\lbrace\mu_{i}\rbrace}.
\end{equation}
For this thesis we will consider the particle densities $n_{i}$ as the fundamental quantities. The reason for this will become clear later. Now we are just left with the calculation of the grand canonical partition function defined as \cite{Landau5}
\begin{equation}
\mathcal{Z}=\mathrm{Tr}\exp\left[-\beta(\hat{H}
-\sum_{i}\mu_{i}\hat{N}^{i})
\right],
\end{equation}
where the trace, $\mathrm{Tr}$, is taken over a complete set of quantum states, $\hat{H}$ is the total Hamiltonian of the system, $\beta=T^{-1}$, and $\hat{N}^{i}$ are conserved charge operators with the chemical potentials $\mu_{i}$'s acting as Lagrange multipliers, fixed by the eigenvalues of the $\hat{O}^{i}$'s. We note the micro to thermodynamic relations for $\langle{\hat{H}}\rangle=E$ and $\langle{\hat{N}_{i}}\rangle=N_{i}$.

For the physics of compact stars to be considered in the following sections, we use the known fact that the baryon number ($i=B$) is always conserved within the standard model and therefore it is always appropriate to be employed when writing the equation of state of neutron stars\footnote{Additionally, sometimes it is appropriate to also assume that the electric charge ($i=Q$), isospin ($i=I$), and/or the strangeness ($i=S$) might be globally/locally conserved. As we will see, only the electric charge case becomes relevant for our discussion.}.


\subsection*{The Partition Function for QCD at Finite $\mu$}

QCD is considered to be the fundamental theory for the strong interactions between quarks and gluons. If one is interested in studying a system placed within a medium (in equilibrium) described by QCD, it is known the system will behave as a strongly-interacting plasma (or more precisely, a liquid), the so-called QGP. The QGP thermodynamics can be obtained from Eqs. (2.1) and (2.8) using the QCD Hamiltonian. On the other hand, it turns out to be more appropriate to perform the calculation of this partition function through its functional representation given as follows \cite{Baluni:1977ms,Kapusta:2006pm,Vuorinen:2003yda}:
\begin{equation}
\mathcal{Z}_{\rm QCD}=\int\left[dA^{\mu}_{a}\right]_{\rm P}
\left[d\bar{\psi}\right]_{\rm A}\left[d\psi\right]_{\rm A}
{\Delta_{\rm FP}[A]\delta(G(A))}
\exp\left(-\int^{\beta}_{0}{d\tau}\int{d^{d}x~
\mathcal{L}^{\rm E}_{\rm QCD}}\right),
\end{equation}
where $\left\lbrace[d\phi_{i}]_{P,A}\right\rbrace$ are the functional measures for each corresponding field which can be periodic (P) or antiperiodic (A) in the compact dimension $\tau$ along the interval $[0,\beta]$, $G(A)$ is a gauge-fixing function, $\Delta_{\rm FP}[A]$ is the Faddeev-Popov (FP) Jacobian (determinant), and the QCD Lagrangian in Euclidean space (being the Lagrange multipliers $\lbrace\mu_{i}\rbrace$ added by convenience) is given by 
\begin{equation}
\mathcal{L}^{\rm E}_{\rm QCD}=
\frac{1}{4}F^{a}_{\mu\nu}F^{a}_{\mu\nu}+\bar{\psi}_{i}(\gamma_{\mu}D_{\mu}+m^{i}_{B}-
\mu_{i}\gamma_{0})\psi_{i},
\end{equation}
where 
\begin{equation}
F^{a}_{\mu\nu}=\partial_{\mu}A^{a}_{\nu}-
\partial_{\nu}A^{a}_{\mu}+
g_{B}f^{abc}A^{b}_{\mu}A^{c}_{\nu}, 
\end{equation}
and
\begin{equation}
D_{\mu}=\partial_{\mu}-ig_{B}A_{\mu},\hspace{1cm}
A_{\mu}=A^{a}_{\mu}T^{a},
\end{equation}
being the subscript $B$ for bare quantities and $\mu_{i}$ the finite quark chemical potentials\footnote{Notice that introducing $\mu_{i}$ in $\mathcal{L}^{\rm E}_{\rm QCD}$ is an abuse of language since it is a quantity that only makes sense in the thermodynamic (macroscopic) limit, i.e. the UV (microscopic) renormalization difficulties appearing in pQCD later will not affect the (pseudo) \textit{bare parameter} $\mu_{i}$. See Ref. \cite{Baluni:1977ms} for details.} with the flavor indices $i$. For the purposes of the next subsection, it will be appropriate for this $i$ to run for massless flavors between $1\leq{i}\leq{N_{l}}$ with $m_{i}=0$, and for the massive one being $i=N_{l}+1{~\equiv~}N_{f}$ having mass $m_{N_{f}}=m$. Besides,
we denote the chemical potential corresponding to this massive quark by $\mu_{N_{f}}=\mu$. 


\subsection{Cold and Dense Perturbative QCD}

It was proven by Gross and Wilczek \cite{Gross:1973id}, and Politzer \cite{Politzer:1973fx}, that at asymptotically high energies, the unique coupling between quark-quark, quark-gluon, and gluon-gluon degrees of freedom, $\alpha_{s}(=g^{2}/4\pi)$, tends to decrease logarithmically. This quantum effect is known as \textit{asymptotic freedom} and it is used widely to perform QCD perturbative calculations. When this strong coupling has small values, one can define the weak coupling expansion of physical observables (see Fig. \ref{fig:coupling}).

\begin{figure}[h]
\begin{center}
\resizebox*{!}{7cm}{\includegraphics{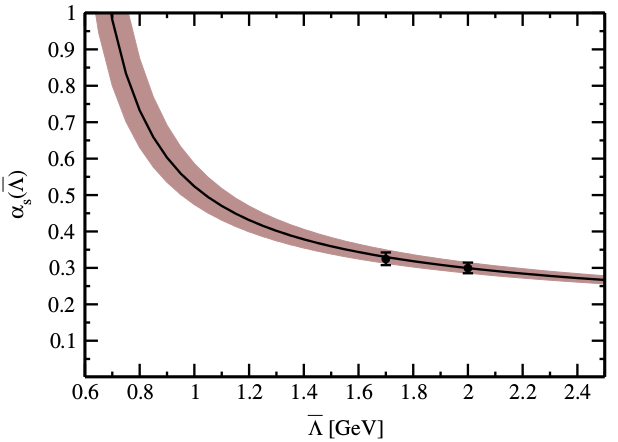}}
\end{center}
\caption{\label{fig:coupling} Dependence of the strong coupling $\alpha_{s}$ on the renormalization scale energy $\bar{\Lambda}$ in a range relevant for compact stars. The uncertainty band is generated through the uncertainty in the values obtained for $\Lambda_{\overline{\rm MS}}$. For more details, see Ref. \cite{Kurkela:2009gj}.}
\end{figure} 

For \textit{in-medium} QCD this weak coupling property at high energies is still valid and it allows us to calculate perturbatively the thermodynamical potential for quarks and gluons at finite temperature and finite chemical potentials. For this, one could start by applying the same formal techniques of quantum field theory in vacuum to Eq. (2.9) in order to obtain the finite temperature Feynman rules for QCD. For more details on these issues, we refer the reader to Ref. \cite{Kapusta:2006pm}.

A relevant remark on the equations of state to be used along this thesis is in order. We will consider the nuclear and quark matter phases in the limit of zero temperature. Strictly speaking, notice that this cold limit of temperature means that we are only dealing with the ground state of a gas with many fermionic components, which in our case are proton, neutrons, quarks, electrons, neutrinos, and even muons. 

The equation of state for quark matter at high densities and zero temperature was first obtained in pQCD by Freedman and McLerran \cite{Freedman:1976ub,Freedman:1977gz}, and Baluni \cite{Baluni:1977ms}, in a modified momentum subtraction scheme over four decades ago (cf. also Ref.  \cite{Farhi:1984qu,Toimela:1984xy}). Later, it was computed in the modern $\overline{\rm MS}$ renormalization scheme for massless quarks in Refs. \cite{Blaizot:2000fc,Fraga:2001id,Fraga:2001xc,Andersen:2002jz,Vuorinen:2003fs}. 
These results were then extended to include the role of a massive quark at two loops by Fraga and Romatschke \cite{Fraga:2004gz} and three loops by Kurkela et al. \cite{Kurkela:2009gj}.

For completeness, we summarize the main ideas behind this perturbative calculation of the thermodynamical potential, $\Omega$, first considering the massless limit and then adding a massive flavor,   together with the renormalization group equations\footnote{We note that any perturbative expansion for the thermodynamic potential is not analytic in the coupling $\alpha_{s}$, but has nonanalytic terms of type $\alpha^{n+1/2}$, and also logarithmic-type contributions $\alpha^{n/2}\log\alpha_{s}$.}.

\subsection*{Massless limit for $N_{f}=N_{l}=3$ and the KRV prescription}

In general it is difficult to obtain well-defined analytical expressions for the thermodynamic observables and one must to resort to numerical evaluations or to consider limiting cases. Besides, the obtained perturbative thermodynamic potential must be taken with care and the extraction of physical quantities (such as the EoS) requires to adopt particular strategies. This comes from the fact that perturbative observables must be truncated at some given order while still ensuring thermodynamic consistency, i.e., forcing the statistical-mechanics results to satisfy the laws of thermodynamics.	
	
 As a warm-up and since it is a system from which we are going to learn many techniques to be applied in later sections, we first consider the limiting case of
$N_{f}=3$ massless up, down and strange quarks, which in this limit ensures directly its (local) electric charge neutrality, i.e., there is no need to introduce electrons, since the $\beta-$equilibrium\footnote{This weak-interaction condition is fundamental for all the physics of compact stars \cite{Glendenning:2000}. For instance, one can have a non-interacting theory for nucleons (producing a Fermi pressure) which, without this condition, it is impossible to build its EoS. Strong interactions can be introduced (non-trivially) but the weak interaction cannot be disregarded from the beginning. In this sense, one can estimate the theoretical error when the strong force is not considered but that cannot be done with weak force.} condition produces $\mu_{u}=\mu_{d}=\mu_{s}\equiv~{\mu}$. In this limit, the total (perturbative) quark pressure would simply be obtained from $\Omega_{\rm pQCD}$ by using $P_{\rm pQCD}=-\Omega_{\rm pQCD}/V$, thus giving\footnote{Details on the renormalization process will be given after this warm-up accomplish its main purpose.} (up to second order in the strong coupling $\alpha_{s}$, to be defined later) \cite{Gorda:2018gpy,Kneur:2019tao}

\begin{equation}
\frac{P_{\rm pQCD}}{P_{\rm SB}}=1-0.636620\alpha_{s}(\bar{\Lambda})-\alpha^{2}_{s}(\bar{\Lambda})\left\lbrace{0.303964\log\alpha_{s}(\bar{\Lambda})+\left[0.874355
+0.911891\log\frac{\bar{\Lambda}}{\mu}\right]}\right\rbrace,
\end{equation}
where $\mu$ is the strange quark chemical potential (defined above) and $P_{\rm SB}=(3/4\pi^{2})\mu^{4}$ is the Stefan-Boltzmann pressure for $N_{f}=3$. 

Again, we note that a perturbative calculation of the thermodynamic potential (pressure) necessarily produces an unknown scale, $\bar{\Lambda}$, associated with the subtraction point for renormalization. One expects that at higher orders in pQCD this unphysical dependence will diminish. On the other hand, this feature offers  a quantitative way to estimate the contribution of the remaining, undetermined orders, i.e., it provides a measure of the inherent uncertainty in the result. It is usual to define the renormalization scale parameter $\bar{\Lambda}$ in terms of the natural scale at ultra-high densities (where quarks are massless) as being $\bar{\Lambda}=2\mu_{s}$ for any $N_{f}$, and consider variations by a factor of $2$, which produces an uncertainty band for the obtained observables. In general, e.g., when adding also massive flavors, it is convenient to write this parameter in the form $\bar{\Lambda}=X\sum_i{\mu_{i}}/N_{f}$, where the sum runs over all quark flavors that are present in the system, and the dimensionless parameter $X$ sits between $1$ and $4$ \cite{Kurkela:2009gj}. 

On the other hand, when obtaining other thermodynamic observables like, e.g. the total energy density, one would have to calculate first the total quark number density, $n$, as (using the fact that in this massless limit all the quark chemical potentials are equal)
\begin{equation}
	n_{\rm pQCD}=n_{u}(\mu,X)+n_{d}(\mu,X)+n_{s}(\mu,X) =-\frac{1}{V}\frac{\partial\Omega_{\rm pQCD}}{\partial\mu}=\frac{\partial{P}_{\rm pQCD}}{\partial\mu}.
\end{equation}
However, applying this derivative would give rise to perturbative corrections being of higher order than $\alpha^{2}_{s}$, as we pass to explain in some detail. Starting from Eq. (2.13) and using Eq. (2.14) one would obtain (being ``*'' a label to distinguish it from the actual technique to be used after these remarks were made)
\begin{equation*}
n^{*}_{\rm pQCD}=(...)\frac{\partial{P}_{\rm SB}}{\partial\mu}+P_{\rm SB}\frac{\partial}{\partial\mu}\left(1-0.636620\alpha_{s}(\bar{\Lambda})-\alpha^{2}_{s}(\bar{\Lambda})[{0.303964\log\alpha_{s}(\bar{\Lambda})+...}]\right),
\end{equation*}
where $(...)$ is the right-hand side of Eq. (2.13). Thus, it is found that
\begin{equation*}
n^{*}_{\rm pQCD}(\mu,\bar{\Lambda})=n_{\rm SB}\times(...)+P_{\rm SB}\left(-0.636620\frac{\partial\alpha_{s}}{\partial\mu}-
0.303964\frac{\partial(\alpha^{2}_{s}\log\alpha_{s})}{\partial\mu}-...\right),
\end{equation*}
where $n_{\rm SB}=3\mu^{3}/\pi^{2}$ is the Stefan-Boltzmann (massless) quark number density. From this last equation for $n^{*}_{\rm pQCD}$ it can be seen that, for instance, derivatives like $\frac{\partial}{\partial\mu}\alpha_{s}$ could produce complicated mixtures of logs and powers of $\mu$ (assuming a proportionality between $\bar{\Lambda}$ and $\mu$). In fact, by doing a simple calculation using $\alpha_{s}$ with $N_{f}=3$ (to be defined soon in Eq. (2.18)), it can be proved that (by fixing for simplicity $\bar{\Lambda}=\mu$ without losing generality)
\begin{equation*}
\frac{\partial\alpha_{s}}{\partial\mu}=-\frac{\alpha_{s}}{\mu\log(\mu/\Lambda_{\overline{\rm MS}})}-\frac{4\pi}{45\mu\log^{3}(\mu/\Lambda_{\overline{\rm MS}})}\left(\frac{1}{\log(\mu/\Lambda_{\overline{\rm MS}})-\log(2\log(\mu/\Lambda_{\overline{\rm MS}}))}\right).
\end{equation*}
From this we infer that the 2nd contribution on the right-hand side of $n^{*}_{\rm pQCD}$ (derived above) spoils any hope of interpreting it as an observable calculated perturbatively, which is beyond the accuracy of the calculation. Besides, these higher-order terms are typically ill behaved, but cannot be simply dropped because this would imply violations of thermodynamic consistency producing, e.g. $n^{*}_{\rm pQCD}d\mu\neq{dP_{\rm pQCD}}$, dictated by Eq. (2.14). Thus, this standard procedure ruins the thermodynamic consistency of the theory.

Instead of the above (somewhat inconsistent) approach, we begin by obtaining $n_{\rm pQCD}$ through Eq. (2.14) for which, and more importantly, the whole contribution depending on derivatives like $\partial\alpha_{s}/\partial\mu$ is neglected (in contrast to $n^{*}_{\rm pQCD}$), then only obtaining (for any $N_{f}$) \cite{Kurkela:2009gj}
\begin{equation}
\frac{n_{\rm pQCD}(\mu,\bar{\Lambda})}{n_{\rm SB}}=1-
2\frac{\alpha_{s}}{\pi}-\left(\frac{\alpha_{s}}{\pi}\right)^{2}\left(\frac{61}{4}-11\log{2}-0.369165N_{f}
+N_{f}\log\frac{N_{f}\alpha_{s}}{\pi}+\beta_{0}\log\frac{\bar{\Lambda}}{\mu}\right),
\end{equation}
where $\beta_{0}=11-2N_{f}/3$ is a constant obtained after solving the renormalization group equations \cite{Kurkela:2009gj} and $n_{\rm SB}=N_{f}\mu^{3}/\pi^{2}$ the Stefan-Boltzmann quark number density for any $N_{f}$. Now, from this point we consider the total quark number density, $n_{\rm pQCD}$, as the fundamental quantity keeping only terms up to (and including) $\mathcal{O}(\alpha^{2}_{s})$, and use it to determine the other parts of the EoS by requiring thermodynamic consistency. Furthermore, it can be easily seen that this $n_{\rm pQCD}$ vanishes for a particular chemical potential, $\mu_{0}(\bar{\Lambda})$. This $\mu_{0}(\bar{\Lambda})$ is below the regime where the computation is reliable, and the fact that the quark number densities become negative for $\mu<\mu_{0}$ can be seen to signal the breakdown of the calculation\footnote{Similar issues occur at low temperatures in effective models which aim to determine nonperturbative contributions to the equation of state of the QGP by comparing lattice results, then obtaining negative pressures  \cite{Guimaraes:2015vra,Canfora:2016xnc}.}. Note that this conclusion is made for the \textit{quark} chemical potential. Then, it is natural to set $n_{\rm pQCD}(\mu,\bar{\Lambda})=0$ for $\mu<\mu_{0}(\bar{\Lambda})$.

Finally, the total pressure is obtained from the integration of Maxwell relation connecting $n_{\rm pQCD}$ and $P_{\rm pQCD}$, then giving us
\begin{equation}
	P_{\rm pQCD}(\mu, \bar{\Lambda})=-B+\int^{\mu}_{\mu_{0}(\bar{\Lambda})}d\mu~{n}_{\rm pQCD}(\mu,\bar{\Lambda}),
	\end{equation}
where $B$ is an integration constant, equal to minus the pressure at $\mu=\mu_{0}(\bar{\Lambda})$. This $B$ represents the pressure difference between the physical (nonperturbative) and perturbative vacua, and it can be interpreted as the bag constant of the well-known MIT bag model. In a purely perturbative calculation, $B$ would usually be set to zero. On the other hand, in a nonperturbative study (i.e., supposing that this EoS is also reliable at low densities) this constant should be nonzero to ensure the energy density to be positive. Following this reasoning, the possible values of $B$ are, however, typically rather restricted, allowing us to make quantitative statements that are not possible in the original MIT bag model.   
	
If the perturbative EoS is to be used down to (not lower values of) $\mu=\mu_{0}(\bar{\Lambda})$, one must require $B\geq{0}$, otherwise the energy density (for the massless case $\mu_{u}=\mu_{d}=\mu_{s}=\mu$)
	\begin{dmath}
	\epsilon(\mu,\bar{\Lambda})=-P_{u}(\mu,\bar{\Lambda})-P_{d}(\mu,        					\bar{\Lambda})-P_{s}(\mu,\bar{\Lambda})+\mu_{u}n_{u}+\mu_{d}n_{d}+\mu_{s}n_{s}=-P(\mu,\bar{\Lambda})+\mu{n}(\mu,\bar{\Lambda})
	\end{dmath}
	would be negative at $\mu=\mu_{0}(\bar{\Lambda})$. This definition of the energy density is obtained from Eq. (2.5). It is only if one ceases to use the perturbative EoS at some $\mu\hspace{0.1cm}\geq\hspace{0.1cm}\mu_{0}$, for instance when matching to a hadronic EoS at smaller $\mu$, that $B$ can take negative values.
	
	Along all this thesis we will call this systematic procedure the KRV \textit{prescription} and, since we will consider a purely perturbative contribution to the QCD thermodynamic potential, we always put $B=0$. We note that by adopting this approach no physics is being dismissed as long as one uses properly the resulting pQCD EoS. For instance, when applied to quark stars, the unreliability of the EoS at low densities will be parametrized by a strong dependence of the calculated observables on the renormalization scale. This is consistent with standard approaches for dilute strongly-interacting matter since, by construction, it is expected that quark stars exist only at the core of neutron stars and in order to obtain a satisfactory and realistic treatment a mantle of nuclear matter would be required near the star's surface for which a matching procedure has to be performed. 
	
	Before continuing, a warning is in order. Although we will be talking about \textit{deconfined matter} and/or quark matter without further details, we will really be using a jargon to mean that quarks are liberated from their hadrons only in the special condition of high pressure. This will produce that the region of space of color confinement is extended up to a macroscopic scale. However, the totality of this matter is still color-neutral, i.e. strictly speaking the so-called deconfinement transition does not exist. This comes from the fact that asymptotic states do not exist in QCD \cite{Smilga:2001}, i.e. (hot/cold and dense) quark matter always confine color. Even in the case of mixed matter in 1st-order phase transitions for low values of the QCD surface tension, one is tempted to think that both kinds of matter are admixed \cite{Glendenning:2000}. Nevertheless, quarks are still confining their color quantum number.

\subsection*{(Perturbative) Renormalization Group Equations}
In the total quark pressure and number density shown above we considered the strong coupling as being small at high values of $\bar{\Lambda}$ but without giving a mathematical representation dependent on the number of flavors $N_{f}$ active in the system. This is done by means of the perturbative renormalization group equations. These equations can be solved in the $\overline{\rm MS}$ scheme\footnote{This scheme was used for the first time in Ref. \cite{Fraga:2001id} for pQCD thermodynamic calculations. It is more appropriate than, e.g. the old $\overline{\rm MOM}$ scheme, since in $\overline{\rm MS}$ one can fix straightforwardly the quark masses and strong coupling at known high experimental/lattice energies where their respective values are more reliably obtained. After this work, the generalizations to higher orders (and loops) were obtained more clearly than the past calculations. See Refs. \cite{Fraga:2001id, Fraga:2004gz} for further details.} for the strong coupling, for which one gets \cite{Fraga:2004gz}\footnote{Although this strong coupling is calculated up to two loops, in contrast to the thermodynamic potential of three loops, it will be used as defined here along this and the following chapters. We note that although a proper evaluation would imply to use the three-loop definition of $\alpha_{s}$ (available in Ref. \cite{Fraga:2001id}), it does not change our quantitative nor qualitative results.}
     \begin{equation}
     \alpha_{s}(\bar{\Lambda})=\frac{4\pi}{\beta_{0}L}\left(
     1-\frac{2\beta_{1}}{\beta^{2}_{0}}\frac{\ln{L}}{L}\right),
     \label{eq:alphas}
     \end{equation}
where $\beta_{0}=11-2N_{f}/3$, $\beta_{1}=51-19N_{f}/3$, $L=2\ln\left(\bar{\Lambda}/\Lambda_{\rm \overline{MS}}\right)$ with $\Lambda_{\overline{\rm MS}}$ being the $\overline{\rm MS}$ point (scale). 

Furthermore, the quark masses also run following this renormalization-scale dependence, but first a comment is in order. Since the physical quark masses $\left\lbrace{m}_{f}\right\rbrace$ are a measure of the chiral-symmetry breaking, they can be decomposed as $m_{f}=m_{D}+m_{B}$, where $m_{D}$ is generated dynamically from the chiral condensate $\left\langle\bar{\psi}\psi\right\rangle$, whereas $m_{B}$ is the (renormalization-group invariant) bare mass appearing in the QCD Lagrangian. We note that since the spontaneously broken nature of the chiral symmetry is responsible of the (nonperturbative) confining forces, our pure perturbative\footnote{Note that apart from our simplifications on the chiral condensate, when one goes to higher densities it becomes less important or, as usually said, it is melted. The same will happen for the bare masses for which at sufficient high densities their effect become negligible, i.e. the chiral symmetry is restored.} approach sets $m_{D}=0$ and $m_{f}=m_{B}$. From now on, this bare mass will be written as $\hat{m}_{f}$ for a given quark flavor.

Now for this chapter we consider in particular the running of the strange quark mass having the solution \cite{Vermaseren:1997fq}
	\begin{eqnarray}
	\begin{aligned}
	m_{s}(\bar{\Lambda})=\hat{m}_{s}\times\left(\frac{\alpha_{s}}{\pi}\right)^{4/9}\left(1+0.895062\left(\frac{\alpha_{s}}{\pi}\right) +1.37143\left(\frac{\alpha_{s}}{\pi}\right)^{2}\right) \;,
	\end{aligned}
	\label{eq:smass}
	\end{eqnarray}
where it must be noted that since $\alpha_{s}$ depends on $N_{f}$, fixing the massive quark at some energy scale to get a value of $\hat{m}_{s}$, will also be dependent on the number of flavors. 
\begin{figure}[h]
\begin{center}
\resizebox*{!}{7cm}{\includegraphics{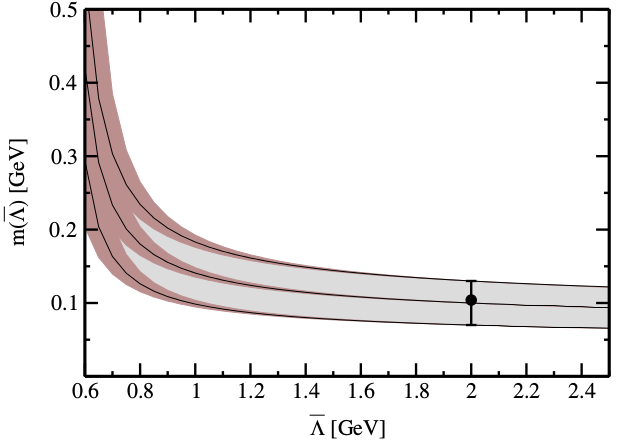}}
\end{center}
\caption{\label{fig:runstrange} Renormalized strange quark mass $m_{s}$ dependence on the renormalization scale $\bar{\Lambda}$ in a range relevant for compact stars. Due to the dependence of $m_{s}$ on $\alpha_{s}$, the uncertainties are two-fold, i.e. one from $\Lambda_{\overline{\rm MS}}$ and the other from the error in the values of $m_{s}$ at a given fiducial scale.  For more details on these matters, see Ref. \cite{Kurkela:2009gj}.}
\end{figure} 

A few comments are in order. As can be seen from Figs. \ref{fig:coupling} and \ref{fig:runstrange}, 
the values taken by $\alpha_{s}(\bar{\Lambda})$ and $m_{s}(\bar{\Lambda})$ below $\bar{\Lambda}{~\sim~}0.8$GeV become dominated by various uncertainties. This sets a lower limit for the values of $\bar{\Lambda}$ for which one can trust in any perturbative calculation. 

\subsection*{Cold pQCD+ Matter with $N_{f}$=2+1}

Now we discuss the quark mass effects on the QCD thermodynamic potential. As it was already mentioned, Fraga and Romatschke \cite{Fraga:2004gz} extended the above massless thermodynamic potential to include the role of a massive quark at two loops and also including the renormalization-group running of the strong coupling and mass parameters. The state-of-the-art framework was designed up to three loops by Kurkela et al. \cite{Kurkela:2009gj} and it deals with $N_{f}=N_{l}+1$ quark flavors, i.e. $N_{l}$ massless quarks plus $1$ massive quark. Originally, the massive flavor was chosen to be the strange quark in order to study its influence on the stellar structure of quark stars.

 This perturbative QCD thermodynamic potential up next-to-next-to-leading-order (NNLO) in the strong coupling $\alpha_{s}$, including the massless contribution plus a massive term, together with the mixed vacuum-matter (VM) diagrams and the corresponding ring terms, up to three loops\footnote{Keep in mind that higher-loop calculations do not bring higher-energy contributions into our compact-star context like, e.g. coming from the electroweak scale which occur at energies above 100 GeV. On the other hand, exotic objects governed by this very energetic scale are theorized to exist like the so-called \textit{electroweak stars} \cite{Dai:2009br}, which are not the aim of this work.} (shown in Figs. \ref{fig:2feyn} and \ref{fig:2plasmon}) can be written as
	\begin{equation}
	\Omega=\Omega^{m=0}(\vec{\mu})+\Omega^{m}(\tilde{\mu},m)+\Omega^{x}_{{\rm VM}}(\vec{\mu}, m)+\Omega_{{\rm ring}}(\vec{\mu}, \tilde{\mu}, m),
	\label{eq:potential}
	\end{equation}
  	where $\tilde{\mu}$ corresponds to the massive quark chemical potential, $\vec{\mu}~{\equiv}~(\mu_{1}, ...,\mu_{N_{l}})$ represents the vector chemical potential for the massless quarks, and $m$ is the renormalized (up to the same order in the strong coupling) mass\footnote{Strictly speaking, one must be careful when defining these quark masses (even in the UV regime where asymptotic freedom works), since color confinement forbids to considered it as an asymptotic state.} associated to the massive quark flavor. 	
\begin{figure}[h]
\begin{center}
\resizebox*{!}{2.7cm}{\includegraphics{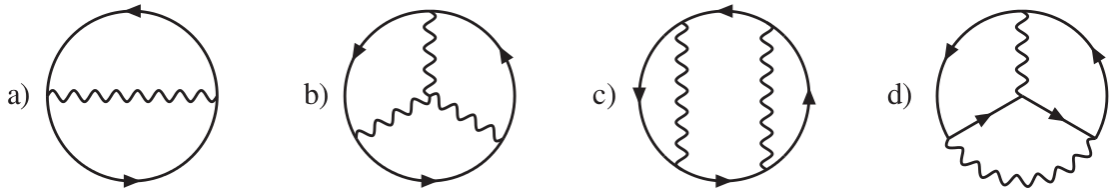}}
\end{center}
\caption{\label{fig:2feyn} The two- and three-loop two-gluon-irreducible (2GI) diagrams contributing to the thermodynamic potential of QCD. For more details, see Ref. \cite{Kurkela:2009gj}.}
\end{figure} 
\begin{figure*}[h]
\begin{center}
\resizebox*{!}{3.3cm}{\includegraphics{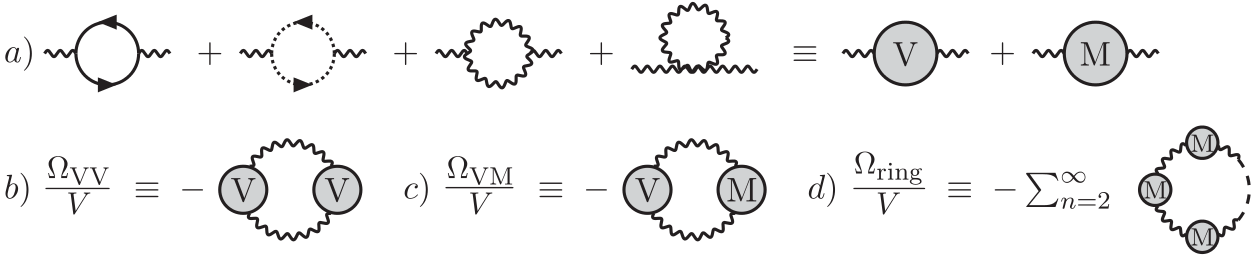}}
\end{center}
\caption{\label{fig:2plasmon} a) Sum of vacuum (V) and matter (M) terms contributing to the gluonic ring diagrams. b) and c) represent the vacuum-vacuum (VV) and vacuum-matter (VM) graphs, respectively. d) stands for the ``matter'' (vacuum subtracted) ring sum. Notice that only the matter terms (i.e., dependent on $\mu$) contribute to the thermodynamic potential of QCD. For more details, see Ref. \cite{Kurkela:2009gj}.}
\end{figure*} 

We note that although diagrams like the one shown in Fig. \ref{fig:2loop} might be included diagrammatically up to 2nd order in $\alpha_{s}$ and with three loops, they do not appear naturally in the perturbative expansion of the QCD thermodynamic potential \cite{Kapusta:1979fh,Toimela:1982hv,
Toimela:1984xy,Arnold:1994ps,Zhai:1995ac,
Braaten:1995jr,Vuorinen:2003yda} (bearing in mind that the Feynman diagrams are the same at zero and finite temperatures, where the former is only the limit of the latter), and it is the reason why they must not be considered when performing pQCD calculations. Additionally, and related to this issue, one can ask if diagrams like in Fig. \ref{fig:2loop} might enter into the ring diagrams, i.e. Case d) of Fig. \ref{fig:2plasmon}. However, from its very definition in Case a) of the same figure, it can be seen straightforwardly that these ring diagrams only consider one-loop diagrams but not a loop within another loop, which would be the case occurring in Fig. \ref{fig:2loop}.

\begin{figure*}[h]
\begin{center}
\resizebox*{!}{2.7cm}{\includegraphics{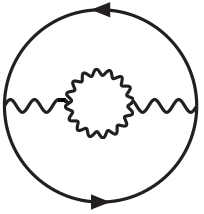}}
\end{center}
\caption{\label{fig:2loop} Prohibited NNLO Feynman diagram including an internal gluon loop in the diagram occurring in Case a) of Fig. \ref{fig:2feyn}.}
\end{figure*} 

Again, this perturbative thermodynamic potential depends on the renormalization scale parameter $\bar{\Lambda}$. For our case of a massive quark, the error band can be estimated by choosing a reasonable fiducial scale and varying $\bar{\Lambda}$ by a factor of $2$ (since larger factors do not affect significantly our results), for which it is usually adopted \cite{Kurkela:2009gj} the fiducial scale $\bar{\Lambda}=(2/3)\mu_{B}$, where $\mu_{B}$ is the baryon chemical potential\footnote{On phenomenological grounds, one can argue that reasonable values for $\bar{\Lambda}/\mu_q$, where $\mu_q$ is the quark chemical potential, lie between $2$ and $3$, if one takes perturbative QCD as a model for the equation of state for cold strongly interacting matter \cite{Fraga:2001id}.}. 

Now we can apply the KRV \textit{prescription} to this thermodynamic potential including a massive quark flavor. For the purposes of this chapter, we consider two massless quarks, i.e., up and down, and one massive one which will be the strange quark flavor. In the language of the above formalism, this means $N_{f}=2+1$. In order to obtain the thermodynamic observables we proceed as in the massless case summarized above. We begin by evaluating the up quark number density taking the derivative of pQCD thermodynamic potential with respect to its respective up chemical potential up to $\mathcal{O}(\alpha^{2}_{s})$, truncate terms of higher order, and use it to obtain expressions for the down and strange quark number densities that are thermodynamically consistent, i.e., satisfy relations such as $\partial{n}_{s}/\partial\mu_{u}=
\partial{n}_{u}/\partial\mu_{s}$. This leads to

		\begin{equation}
		n_{u}(\mu_{u},\mu_{d},\mu_{s}){\hspace{0.1cm}}{\equiv}{\hspace{0.1cm}}n^{(2)}_{u}(\mu_{u},\mu_{d},\mu_{s}),
		\end{equation}
		\begin{equation}
		n_{d}(\mu_{u},\mu_{d},\mu_{s})=\int^{\mu_{u}}_{u_{0}(\mu_{d},\mu_{s})}{d\mu{'}_{u}}\partial_{\mu_{d}}n^{(2)}_{u}(\mu{'}_{u},\mu_{d},\mu_{s})+n^{(2)}_{d}(\mu_{u}=u_{0},\mu_{d},\mu_{s}),
		\end{equation}
		\begin{equation}
		\begin{aligned}
n_{s}(\mu_{u},\mu_{d},\mu_{s})=\int^{\mu_{u}}_{u_{0}(\mu_{d},\mu_{s})}d\mu{'}_{u}\partial_{\mu_{s}}n^{(2)}_{u}(\mu{'}_{u},\mu_{d},\mu_{s})\\
		+\int^{\mu_{d}}_{d_{0}(\mu_{s})}d\mu{'}_{d}\partial_{\mu_{s}}n^{(2)}_{d}(u_{0}(\mu{'}_{d},\mu_{s}),\mu{'}_{d},\mu_{s})	+n^{(2)}_{s}({\mu_{u}=u_{0}},{\mu_{d}=d_{0}},\mu_{s}),
		\end{aligned}
		\end{equation}
		where the functions $n^{(2)}_{i}$ are defined so that they contain no terms beyond $\mathcal{O}(\alpha^{2}_{s})$, and $u_{0}$ and $d_{0}$ are free integration functions coming from the thermodynamic consistency. At this point everything would stay depending on each quark chemical potential and the integration functions being unknown. However, for the physics of (quasi-static, i.e. a system with steady-state microphysical processes occurring constantly) quark stars where neutrinos are neglected (since they escape quickly for millions of years leading to a thermal cooling), the $\beta$-equilibrium\footnote{This condition introduces electrons but not positrons to the system since, if present, they all have been already annihilated on stellar timescales.}

      \begin{equation}
	\mu_{u}=\mu_{d}-\mu_{e},
     \label{eqBeta1}
      \end{equation}
      \begin{equation}
	\mu_{d}=\mu_{s}\equiv{\mu} \, ,
	\label{eqBeta2}     
      \end{equation}
and (local) electric charge neutrality 
      \begin{equation}
	\frac{2}{3}n_{u}-\frac{1}{3}n_{d}-\frac{1}{3}n_{s}=n_{e} \, ,
     \label{neutralYl}
      \end{equation}
occurring between quark species enable us to simplify Eqs. (2.21)--(2.23) considerably\footnote{This conditions are valid only in the \textit{bulk} of the system. Surface effects are not considered.}. Here, $\mu_{u}$, $\mu_{d}$, $\mu_{s}$, and $\mu_{e}$ are the chemical potentials of the up, down and strange quarks together with the electron being introduced as a free Fermi gas. The baryon chemical potential and the baryon number density are defined as $\mu_{B}=\mu_{u}+\mu_{d}+\mu_{s}$ and $n_{B}=n/3$, respectively.		

Then, the integration functions $d_{0}$ and $u_{0}$ must satisfy this compact star conditions, i.e., $d_{0}(\mu_{s})=\mu_{s}$ and $u_{0}(\mu_{d},\mu_{s})=\mu_{s}-\mu_{e}(\mu_{d},\mu_{s})$. From this it can be seen that the integrals in Eqs. (2.22)-(2.23) vanish on the 1D curve in the $\mu_{i}$ space where beta equilibrium and electric charge neutrality are maintained. Additionally, within this physical subspace -- in particular, when calculating $\mu_{e}(\mu_{d},\mu_{s})$ from local charge neutrality -- we may use the truncated expressions $n^{(2)}_{i}$ for all quark flavors,
	\begin{equation}
	n_{u}(\mu,\bar{\Lambda})=n^{(2)}_{u}(\mu-\mu_{e}(\mu),\mu,\mu)+\mathcal{O}(\alpha^{3}_{s}),
	\end{equation}
	\begin{equation}
	n_{d}(\mu,\bar{\Lambda})=n^{(2)}_{d}(\mu-\mu_{e}(\mu),\mu,\mu)+\mathcal{O}(\alpha^{3}_{s}),
	\end{equation}
	\begin{equation}
	n_{s}(\mu,\bar{\Lambda})=n^{(2)}_{s}(\mu-\mu_{e}(\mu),\mu,\mu)+\mathcal{O}(\alpha^{3}_{s}).
	\end{equation}

We also note that while our starting point in Eq. (2.21) was chosen to be the up quark number density (rather than the other quark flavors) as the fundamental quantity, our final result Eq. (2.27)--(2.30) treats all flavors symmetrically. It is in these equations that one neglects $\mathcal{O}(\alpha^{3}_{s})$ and imposes local charge neutrality numerically, for which one sets $n_{i}=0$ whenever the number density in question becomes negative at $\mu_{0}(X)$.
		
Finally, the total (i.e. quarks and electrons) pressure and energy density can be obtained from thermodynamic consistency and they are given as follows:
\begin{eqnarray}
P(\mu,X)=\int^{\mu}_{\mu_{0}(X)}d\bar{\mu}
\left[ n_{u}\left(1-\frac{d\mu_{e}}{d\mu_{s}}\right) \right.\nonumber
\left. +~ n_{d}+n_{s}+
n_{e}\frac{d\mu_{e}}{d\mu_{s}}\right] \, ,
\end{eqnarray}
\begin{equation}
\epsilon(\mu,X)=-P+\mu(n_{u}+n_{d}+n_{s})
-\mu_{e}(n_{u}-n_{e}).
\label{pressdensPQCD}
\end{equation}
From these two last equations it can be proved (numerically) that the presence of electrons (leptons) is \textit{poor} compared to the other particles, quarks in this case\footnote{This can be understood intuitively using the fact that since for massless quarks the electron contribution is zero, when adding a massive quark with a light mass (as the strange quark) the electron contribution will increase slightly. In general, its values go around 50 MeV, decreasing at high $\mu_{s}$.}. Then, when fixing the free parameters of the pQCD thermodynamic potential we choose for the strange quark mass, $m_{s}(2\rm GeV,N _{f}=2+1)=92 \rm MeV$ \cite{Aoki:2016frl} and, for the strong coupling constant, $\alpha_{s}(1.5$GeV$,N_{f}=3)=0.336$, which allows us to fix the renormalization point in the $\overline{\rm MS}$ scheme to $\Lambda_{\overline{\rm MS}}=315^{+18}_{-12}$ MeV \cite{Bazavov:2014soa}. For convenience, we define the dimensionless renormalization scale $X \equiv 3\bar{\Lambda}/\mu_{B}$. Figure \ref{fig:KRVcold} illustrates the behavior of the pressure as a function of $\mu_{B}$ produced by a gas of massless up and down quarks plus massive strange quarks, where electrons are also added in order to ensure charge neutrality and $\beta$-equilibrium, obtained from Eq. (\ref{pressdensPQCD}), which we dub KRV from now on\footnote{We note that from the astrophysical point of view, low values of $X$, i.e. tending to 1, would produce low-mass quark stars but at the same time very dense in the sense that one could have almost all massless quarks at their surfaces. This is in contrast with large $X$, where one would find massive strange quarks at their cores and only massless up/down quarks at their surfaces}.
\begin{figure}[h]
\begin{center}
\resizebox*{!}{7.0cm}{\includegraphics{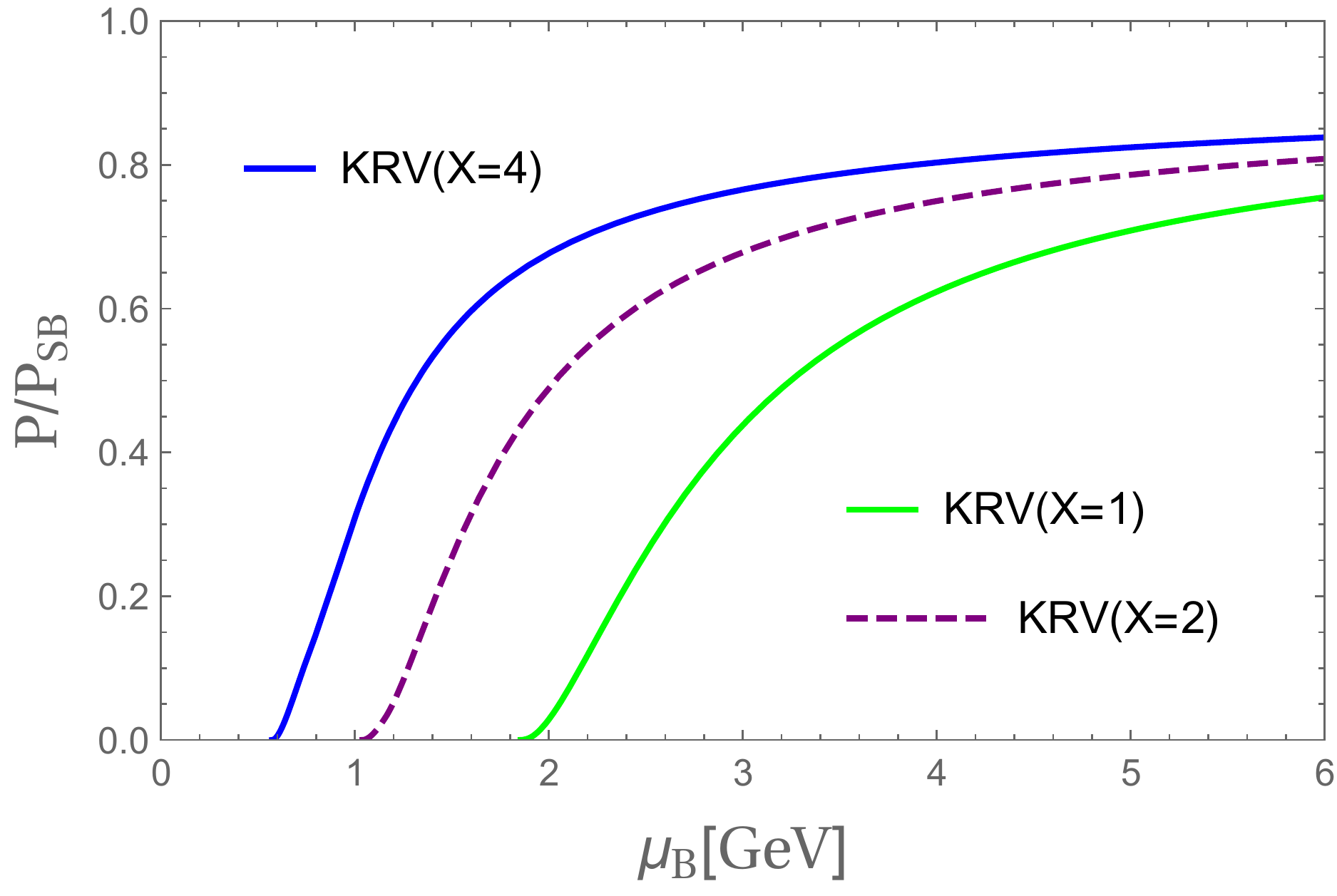}}
\end{center}
\caption{\label{fig:KRVcold}Total pressure of a gas of up, down and strange quarks plus electrons up to three-loops \cite{Kurkela:2009gj} normalized by the Stefan-Boltzmann free pressure for different values of $X$. Beta equilibrium and local charge neutrality were implemented. Taken from \cite{Jimenez:2017fax}.}
\end{figure} 

For clarity, in this thesis we differentiate between \textit{pure} quantum chromodynamics (QCD) matter, where only colored degrees of freedom play an important role entering into the pQCD thermodynamic potential, from the \textit{real}
theory of strongly interacting matter (QCD+) where other degrees of freedom and interactions contribute to the thermodynamic potential \cite{Kogut:2004su}. For example, up to now only the electrons entered in the calculations through the local electric charge neutrality condition representing the relevant (static) electromagnetic interaction, present to ensure vanishing electric repulsion between particles at each point of the compact star interior, which means that no quantum-eletrodynamic effects are included in this context. Of course, QCD+ is our ultimate goal of theoretical modelling and physical understanding\footnote{This distinction is also crucial when comparing results from lattice simulations being performed for QCD rather than QCD+ \cite{Kogut:2004su}.}.

Since the results from this section are obtained within a perturbative calculation, our system defined by Eq. (\ref{pressdensPQCD}) will be called perturbative QCD+ (pQCD+) matter.
The following section will consider additional leptons, i.e., electrons plus \textit{neutrinos}, considerably affecting the equation of state of supernova matter.

\section{Lepton-rich pQCD+ Matter}
 \label{sec:2richQM}

We now investigate the early stages of neutron stars lifes as protoneutron stars, i.e., during the early postbounce stage of core collapse supernova explosion. For this exotic matter, one has to consider a lepton-rich EoS. In particular, one has to include the trapped neutrinos\footnote{In this chapter, only electron neutrinos are considered. The other ones are dismissed since the weak equilibrium conditions do not demand their presence at the densities of interest, especially in this cold limit. Finite-temperature effects on the EoS would change this situation, as in supernova simulations.} in the framework. In what follows we consider a fixed lepton fraction $Y_{i}=n_{i}/n_{B}$, where $n_{B}$ is the baryon density and $n_{i}$ the density of lepton species $i$. So, we take $Y_{L}=(Y_{e}-Y_{e^{+}})+(Y_{\nu}-Y_{\bar{\nu}})=Y_{e}+2Y_{\nu}\equiv{0.4}$, where at zero-temperature it is known that positrons (antiparticles) do not contribute, i.e. $Y_{e^{+}}=0$ and $Y_{\bar{\nu}}=-Y_{\nu}$ since neutrinos contribute like $\mu_{\bar{\nu}}=-\mu_{\nu}$ and they are always left-handed, i.e. they have only one degree of degeneracy and electrons have two, at least within the Standard Model of particle physics. Although this value is chosen to be exact in this work, it sometimes changes by 0.01 compared to the values reached in simulations of the evolution of protoneutron stars \cite{Burrows:1986me,Pons:1998mm}.
Associated to this conserved quantity, we introduce an independent neutrino chemical potential $\mu_{\nu}$, from which we can also write the antineutrino chemical potential\footnote{One way to estimate (indirectly) a range of values for $\mu_{\nu}$ is through luminosity measurements of supernovae explosions which are related to the masses $M_{\rm PNS}$ and radii $R_{\rm PNS}$ of PNSs calculated using the TOV equations depending on the lepton-rich EoSs parametrized by $\mu_{B}=\mu_{B}(\mu_{\nu})$ \cite{Suwa:2020nee}.}.

For this lepton-rich pQCD+ phase, the free parameters entering into the renormalization group solutions of the running strange mass and strong coupling will be the same as the ones of the above section in order to make a comparison between lepton-poor and lepton-rich more clear. Then, using the quark densities obtained in Eqs. (2.28)--(2.30) of the preceding section, we can impose immediately (although non-trivially) $\textit{local}$ charge neutrality and $\textit{local}$ lepton fraction conservation, for which we have
      \begin{equation}
	\frac{2}{3}n_{u}-\frac{1}{3}n_{d}-\frac{1}{3}n_{s}=n^{Q}_{e} \, ,
     \label{neutralYl}
      \end{equation}
      \begin{equation}
	\frac{n^{Q}_{e}+2n^{Q}_{\nu}}{n^{Q}_{B}}=Y_{L}=0.4 \, ,
	\label{fixedYl}     
      \end{equation}
where $n_{e}=(1/3\pi^{3})\mu^{3}_{e}$ and $n_{\nu}=(1/6\pi^{3})\mu^{3}_{\nu}$, being the same definitions for this quark phase and hadronic phase, to be used later. The weak interaction equilibrium conditions imply
      \begin{equation}
	\mu_{d}+\mu^{Q}_{\nu}=\mu_{u}+\mu^{Q}_{e},
     \label{eqBetaYl}
      \end{equation}
      \begin{equation}
	\mu_{d}=\mu_{s}\equiv{\mu} \, .
	\label{strangeYl}     
      \end{equation}
Here, $\mu_{u}$, $\mu_{d}$, $\mu_{s}$, $\mu^{Q}_{e}$ and $\mu^{Q}_{\nu}$ are the chemical potentials of the up, down and strange quarks together with the electron and electron neutrino in the quark phase. The latter are introduced as free Fermi gas contributions. The definition of densities follow straightforwardly. Again, as in the lepton-poor case, the baryon number density is defined as $n^{Q}_{B}=n/3$. Notice that $\textit{antineutrinos}$, which can play a crucial role in the second neutrino burst signal of a possible QCD phase transition after the first bounce of the core-collapse supernova explosion \cite{Sagert:2008ka}, are also taken into account.

Given the constraints above, we can write all quark and lepton chemical potentials in terms of the strange quark chemical potential, $\mu_{s}\equiv{\mu}$, only. Then, again using the KRV \textit{prescription}, we use the quark and lepton number densities as the fundamental quantities from which one can construct the pressure, demanding thermodynamic consistency at each step of the calculation and preserving terms up to $\mathcal{O}(\alpha^{2}_{s})$. This procedure makes the implementations of the constraints above on charge neutrality, lepton fraction conservation and chemical equilibrium straightforward qualitatively although complicated numerically. 

\begin{figure}[h]
\begin{center}
\resizebox*{!}{7.0cm}{\includegraphics{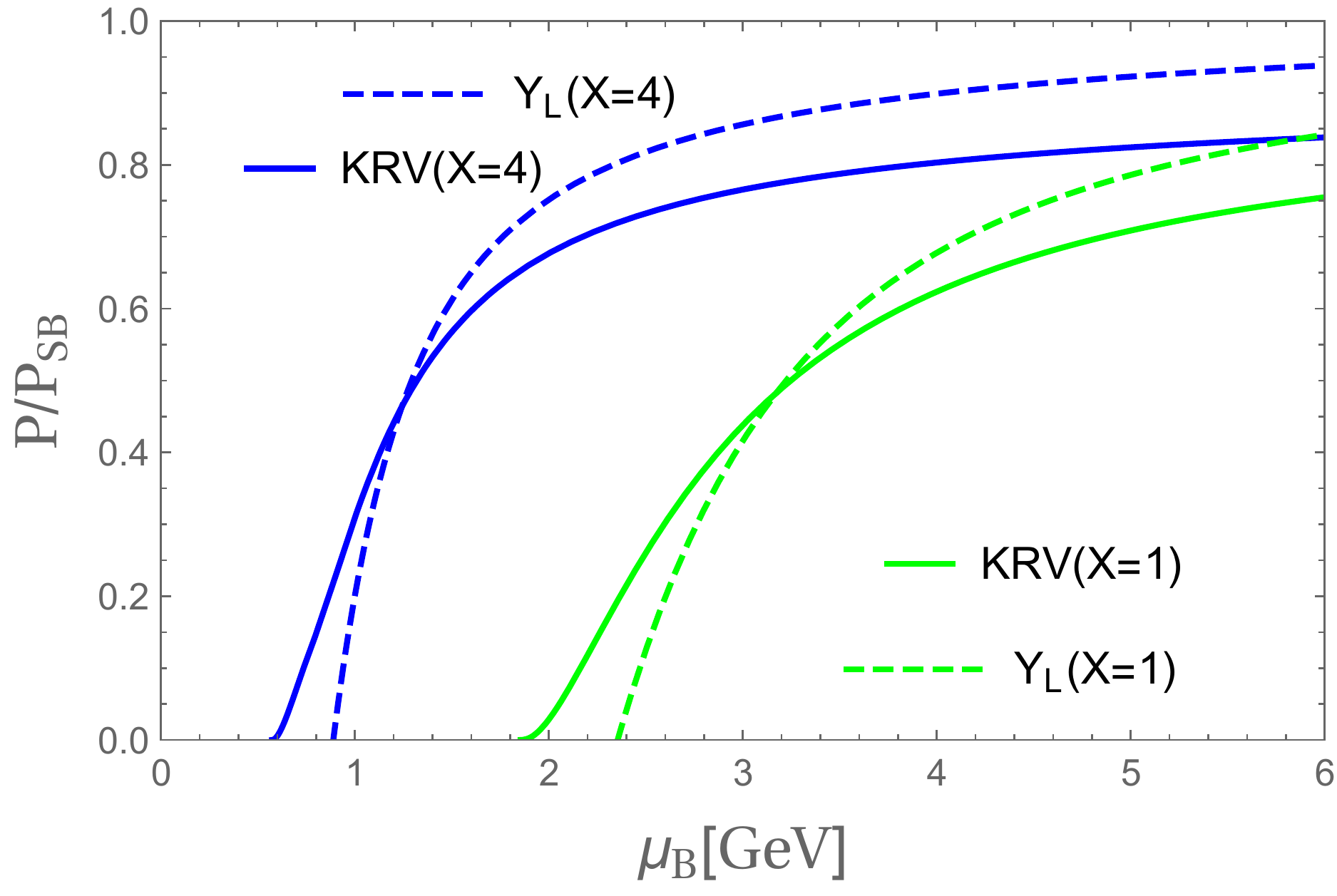}}
\end{center}
\caption{Total pressure of quarks and leptons for a fixed lepton fraction ($Y_{L}=0.4$) in dashed lines for the band between $X=1$ and $X=4$. In solid lines we show the lepton-poor case (KRV). Taken from Ref. \cite{Jimenez:2017fax}.}
\label{fig:KRVneutrino}
\end{figure}
	
Again, as in the massless case, in the cases where some quark density $n_{i}(\mu_{i},X)$ becomes negative below a given chemical potential, $\mu_{i}<\mu_i^{0}(X)$, we set it to $n_{i}\equiv{0}$ to obtain a consistent model. Integrating the number densities from their minimal value $\mu_i^{0}(X)$ to some arbitrary strange quark chemical potential $\mu$ and taking into account Eqs. (\ref{neutralYl}) --(\ref{strangeYl}), we obtain the total pressure for lepton-rich quark matter as follows:
\begin{equation}
P(\mu,X)=\int^{\mu}_{\mu_{0}(X)}d\bar{\mu}
\left[ n_{u}\left(1+\frac{d\mu^{Q}_{\nu}}{d\mu_{s}}-\frac{d\mu^{Q}_{e}}{d\mu_{s}}\right)+~ n_{d}+n_{s}+
n^{Q}_{e}\frac{d\mu^{Q}_{e}}{d\mu_{s}}+2n^{Q}_{\nu}\frac{d\mu^{Q}_{\nu}}{d\mu_{s}} \right].
\label{pressurePNS}
\end{equation}
Similarly, the associated lepton-rich energy density would have to be obtained from $\epsilon(\mu,X)=-P(\mu,X)+\left(\sum_{f=u,d,s}n_{f}\mu_{f}\right)
+n^{Q}_{e}\mu^{Q}_{e}+2n^{Q}_{\nu}\mu^{Q}_{\nu}$. From both, we can also express the pressure in Eq. (\ref{pressurePNS}) and energy density as functions of the baryon chemical potential, $P=P(\mu_{B},X)$ and $\epsilon=\epsilon(\mu_{B},X)$, respectively. Notice that this baryochemical potential carries contribution also from (anti)neutrinos. In Fig. \ref{fig:KRVneutrino} one can see how the cold quark matter EoS (KRV) is modified by the presence of trapped neutrinos ($Y_{L}$) for different values of the renomalization scale $X$. As expected, at high densities the total pressure becomes rapidly similar to the quark Stefan-Boltzmann limit due to the additive degeneracy of leptons plus quarks since the latter are quasi-free due to asymptotic freedom.

\begin{figure}[h]
\begin{center}
\resizebox*{!}{7.0cm}{\includegraphics{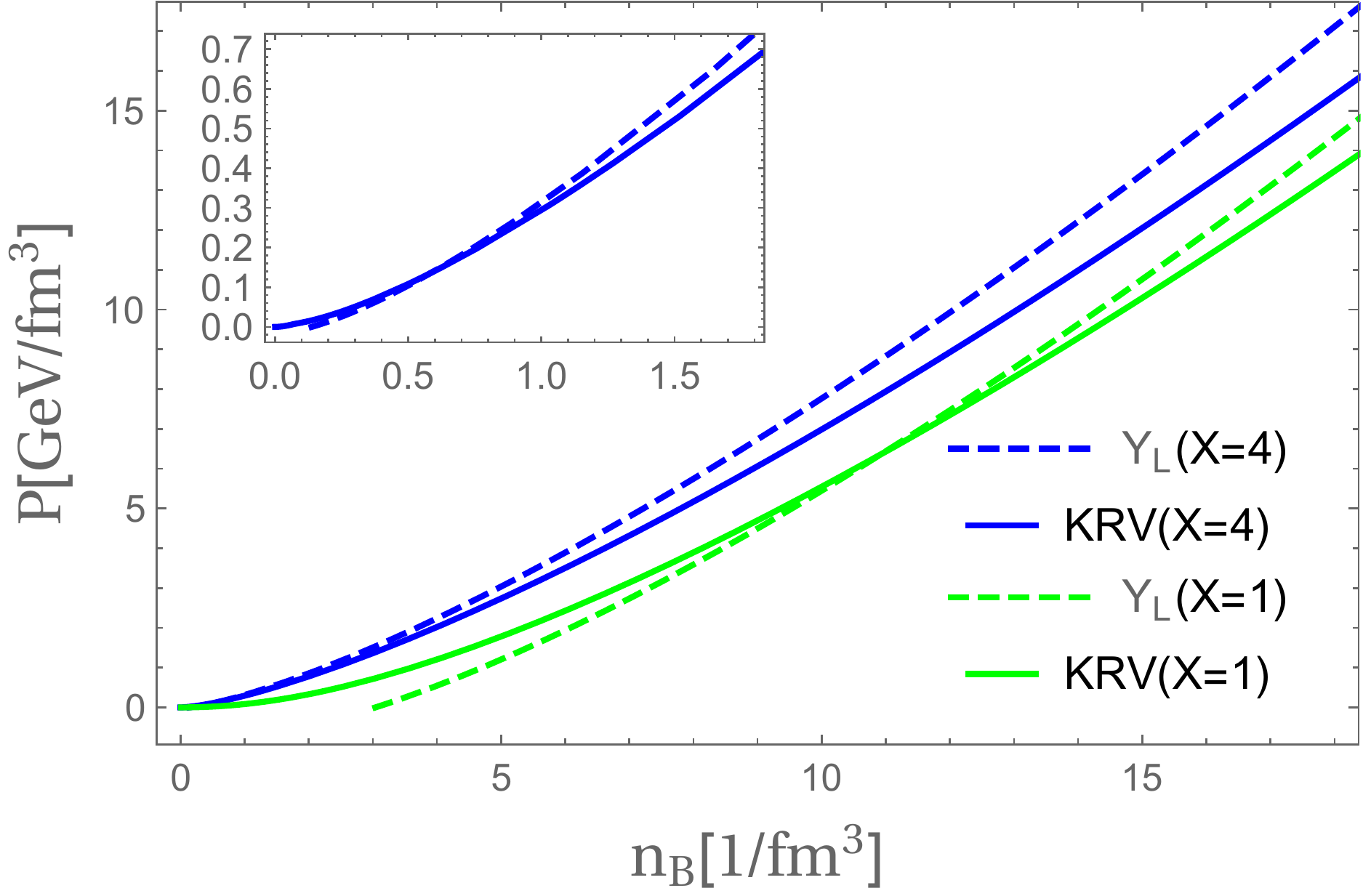}}
\end{center}
\caption{Total pressure of quarks and leptons for a fixed lepton fraction ($Y_{L}(X)$) and in the lepton-poor case, KRV(X) as a functions of the baryon number density $n_{B}$. Taken from Ref. \cite{Jimenez:2017fax}.}
\label{fig:PnBYl}
\end{figure} 

Notice that, at high $\mu_{B}$, since $m_{s}(X)$ tends to be constant \cite{Fraga:2004gz,Kurkela:2009gj} and $\alpha_{s}(X)$ is nonzero (unless we are at asymptotically high densities \cite{Kurkela:2009gj}, which are not relevant for the physics of compact stars), the total pressure of quarks and leptons will increase faster, in contrast to the lepton-poor case. However, at low $\mu_{B}$ the lepton-poor total pressure appears to be higher. This occurs due to the respective runnings of $m_{s}$ and $\alpha_{s}$. To clarify this issue, we show, in Fig. \ref{fig:PnBYl},  the total pressure of quarks and leptons as a function of the baryonic number density $n_{B}$. It turns out, as one can see from this figure, that the lepton presence makes our EoS stiffer\footnote{By \textit{stiffer} we meant that at a given energy density the pressure is higher compared to a \textit{softer} EoS.} at high densities. We note that by introducing temperature effects, one would soften the EoS due to the presence of additional degrees of freedom. Only with temperature one would get the qualitative agreement with the softening of the hadronic EoS having neutrinos, as it will be the case with the TM1-PNS EoS to be discussed later.

\begin{figure}[h]
\begin{center}
\resizebox*{!}{7.0cm}{\includegraphics{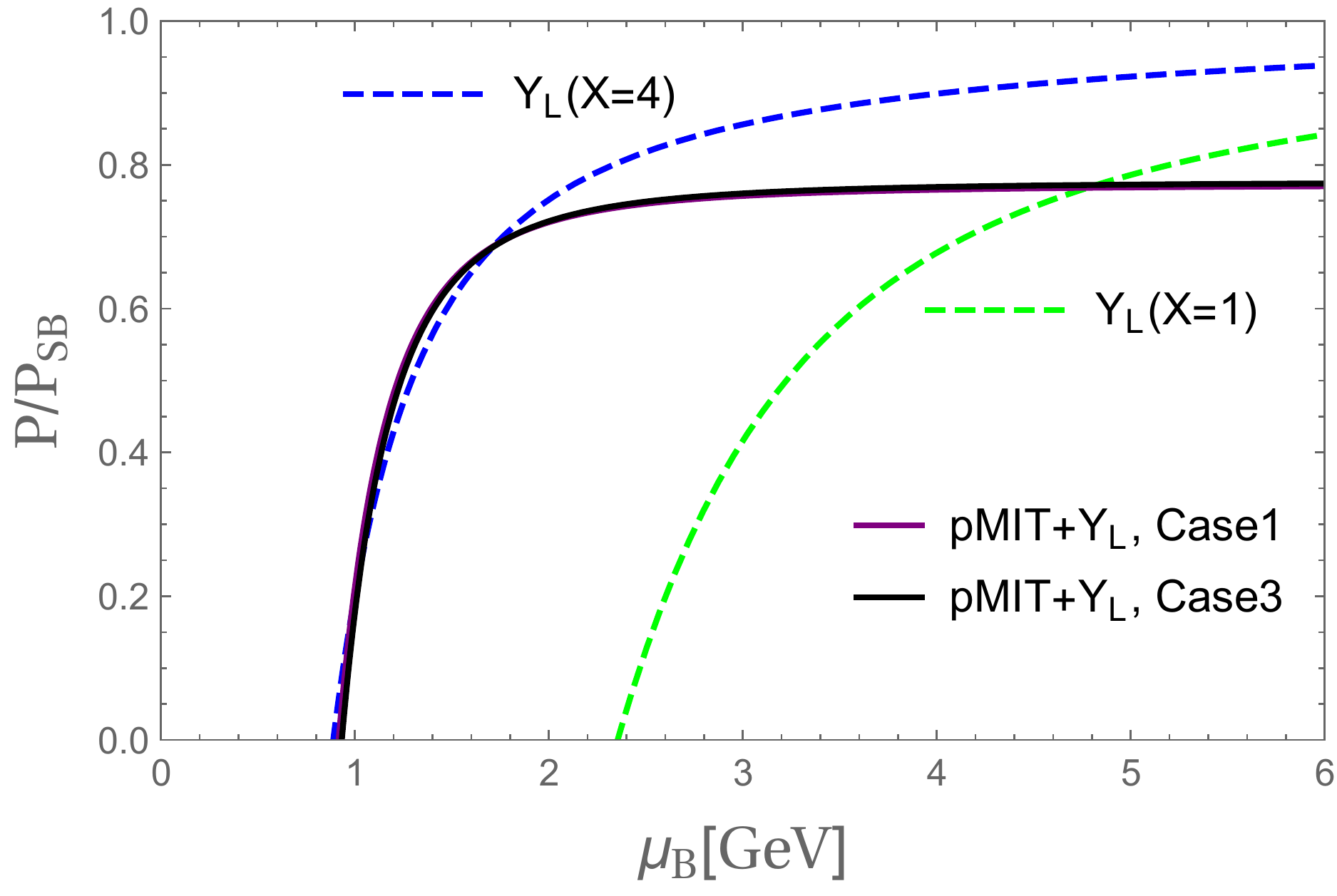}}
\end{center}
\caption{\label{fig:KRVcomparison} Total pressure for a gas of quarks and leptons in beta equilibrium with a fixed lepton fraction $Y_{L}=0.4$ for $X\in[1, 4]$ (dashed lines) compared to the effective MIT bag model \cite{Alford:2004pf,Mintz:2009ay} at the same conditions ($\rm pMIT+Y_{L}$, in solid lines). We show two lepton-rich bag models with effective bag constants, $B_{1}=(144.65\rm ~ MeV)^{4}$ (purple) and $B_{3}=(147.56\rm ~MeV)^{4}$ (black), but with the same effective perturbative correction $c=0.3$ and strange quark mass, $m_{s}=100 \rm MeV$. Taken from Ref. \cite{Jimenez:2017fax}.}
\end{figure} 

In Fig. \ref{fig:KRVcomparison} we compare our results to the lepton-rich modified version of the effective MIT bag model which takes into account effective corrections from pQCD \cite{Alford:2004pf} (pMIT+$Y_{L}$) for a set of particular values of the effective bag constant $B$ and strange quark mass $m_{s}$ considered in Ref. \cite{Mintz:2009ay}
\begin{equation}
	P_{\rm pMIT+Y_{L}}(\left\lbrace{\mu}\right\rbrace)=(1-c)\left[\sum_{i=u,d}\frac{\mu^{4}_{i}}{4\pi^{2}}\right]+P_{s}+\frac{\mu^{4}_{e}}{12\pi^{2}}+\frac{\mu^{4}_{\nu}}{12\pi^{2}}-B,
\end{equation}
	and
\begin{equation}
	P_{s}=(1-c)\frac{\mu^{4}_{s}}{4\pi^{2}}-\frac{3}{4\pi^{2}}m^{2}_{s}\mu^{2}_{s}.
\end{equation}
where the effective perturbative corrections are included in the constant $c~{\approx}~0.3$, expressing the nonideality of the gas\footnote{In order to obtain astrophysical observables, the associated energy density would have to be obtained from $\epsilon_{\rm pMIT+Y_{L}}(\left\lbrace{\mu}\right\rbrace)=3P_{\rm pMIT+Y_{L}}(\left\lbrace{\mu}\right\rbrace)+
(3/2\pi^{2})m^{2}_{s}\mu^{2}_{2}+4B$, which allow us to build the EoS}. However, in this reference, the parameters are set to obtain maximum masses of lepton-poor hybrid stars in agreement with observations available then, which were in the order of $1.67$ solar masses, which implies that lower values would required for the critical baryon density if a deconfinement transition were needed in lepton-rich matter. Besides, from this figure one can realize that the pMIT+$Y_{L}$ pressure becomes flat from some $\mu_{B}$ which comes from the constant factors $``1-c"$ and $m^{2}_{s}$ in $P_{\rm pMIT+Y_{L}}$, which explicitly not account for asymptotic-freedom and renormalization-group effects on the strange quark mass, something which is automatic in our lepton-rich pQCD results.

\subsection{Bodmer-Witten Hypothesis in Supernovae Matter} 
    
Long ago, Bodmer \cite{Bodmer:1971we} and later (independently) Witten \cite{Witten:1984rs} investigated a system formed by massless up, down and strange quarks that, if at zero pressure could have energy per baryon
  \begin{equation}
{E}/{A}\leq{~930}\rm MeV \, ,
\label{EoverA}
\end{equation}
i.e., lower than the most stable nuclei $\rm Fe^{56}$ (and  $\rm Ni^{62}$), one would find configurations of absolutely $\textit{stable strange quark matter}$ (SQM) as the true ground state of strong interactions, instead of nuclear matter \footnote{Notice that although this hypothesis is more easily satisfied in the cold limit $T=0$ (as we assume along all this thesis), it can also be analyzed at finite temperatures, e.g. in the primordial universe \cite{Witten:1984rs}.}.

Since then, there have been many improvements trying to add non-perturbative aspects of the strong interactions and take into account the mass of the strange quark (for reviews, see Refs. \cite{Madsen:1998uh,Weber:2004kj}). Nevertheless, the interactions among quarks are usually described in a very simplified fashion and the presence of neutrinos is overlooked \footnote{Obviously this statement is true only at the level of the EoS but not when studying dynamical processes like the compact star thermal evolution where neutrinos are important when calculating the corresponding transport coefficients like, e.g. neutrino emissivities.}. On the other hand, if one is interested in the very first seconds after the formation of stable strange quark matter (either in a cosmological QCD phase transition or in strange quark stars\footnote{For example, in the past it was believed that their neutrino cooling  would be faster than that of neutron star matter \cite{Alcock:1986hz}. However, later it was found that the neutrino cooling of nuclear matter could be of the same order as in strange matter \cite{Lattimer:1991ib,Pethick:1991mk,Schaab:1997nt}. So, this raised many questions upon the neutrino importance in different astrophysical and cosmological scenarios.}), one would have to include the presence of neutrinos thermodynamically.

 Thus, we pass to focus on the influence of these leptons when interacting pQCD+ matter is considered and analyse if its parameter space favors or disfavors its appearance. This can be reached by investigating the likelihood of satisfying the criterion given by Eq. (\ref{EoverA}) using the lepton-rich pQCD+ thermodynamic potential we have at hand. As before, it is more convenient to use the quark densities as fundamental quantities and analyze if the parameter space of our theory with strange quark matter is modified in the presence of neutrinos for the usual values of the renormalization scale $X\in[1,4]$. To do this, we use the Hugenholtz-Van Hove theorem \cite{Hugenholtz:1958zz} generalized to a system with many components \cite{R.C.Nayak:2011hyj}. It requires only the quark and lepton densities and chemical potentials as input, giving the following energy per baryon:
	\begin{equation}
	\frac{E(\mu_{s},X)}{A}=\frac{n_{u}}{n^{Q}_{B}}(\mu_{\nu}-\mu_{e})+3\mu_{s} \, ,
	\end{equation}
where we implicitly assumed that all the quantities on the rhs of the equation above are functions of the strange chemical potential $\mu_{s}$.

Constraining the values of $X$ such that $\mu_{s}$ and $n_{s}$ are not zero $\textit{and}$ satisfy Eq. (\ref{EoverA}) we obtain, for the cold case, $X\in[2.95, 4]$, and for the lepton-rich case $X\in[3.45, 4]$, as can be seen in Fig. \ref{fig:SQMcomparison}. Even if the parameter space of $X$ is not radically modified when trapped neutrinos are included, one can notice from Fig. \ref{fig:SQMcomparison} that the band for $X$ tends to shrink to $\mu_{B}\in[0.86, 0.88]\rm GeV$ for vanishing pressure (as compared to $\mu_{B}\in[0.803, 0.93]\rm GeV$ in the cold case). Figure \ref{fig:SQMcomparison} also indicates that lepton-rich strange quark matter becomes essentially $\textit{independent}$ of the renormalization scale $X$.

\begin{figure}[h]
\begin{center}
\resizebox*{!}{7.0cm}{\includegraphics{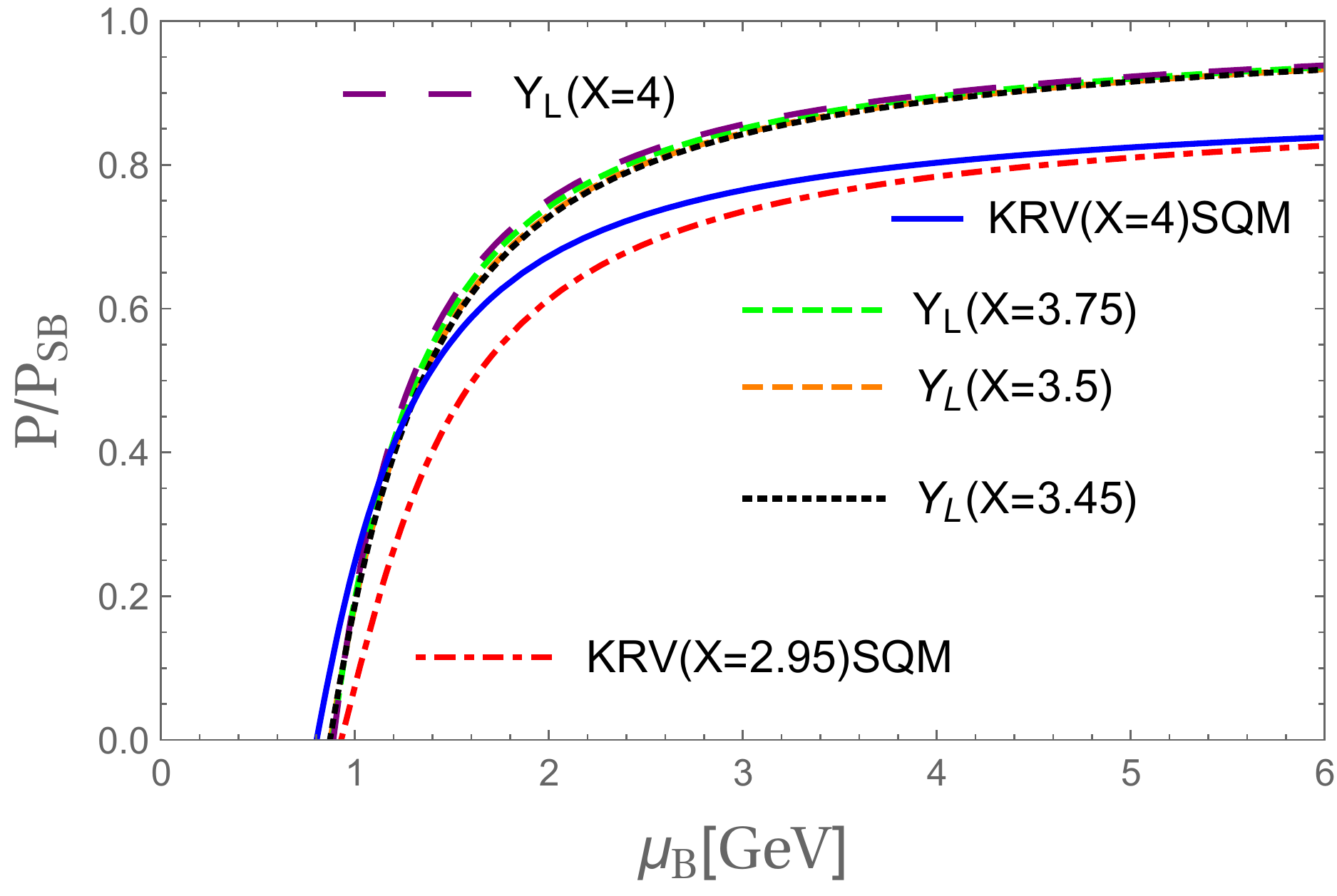}}
\end{center}
\caption{\label{fig:SQMcomparison} Total normalized pressure for quarks and leptons with $Y_{L}=0.4$ allowing for the SQM hypothesis ($Y_{L}$, in dashed lines). For comparison, we show also the pressure for lepton-poor strange matter ($\rm KRV[X]SQM$, in solid and dot-dashed lines). Taken from Ref. \cite{Jimenez:2017fax}.}
\end{figure} 

In this way one realizes that the presence of neutrinos makes the SQM hypothesis $\textit{less}$ favorable, i.e., the stability window of critical densities with vanishing pressure is narrower. A similar behavior was observed in Ref. \cite{Dexheimer:2013czv}, where the authors also included finite-temperature contributions in different quark models. This leaves us with $X\in[1, 3.44]$ for unpaired quark matter having as ground state hadronic matter in vacuum. 

\section{Deconfined Matter in Supernova Explosions}
	\label{sec:2deconfined}
Now we consider the nucleation of quark matter droplets\footnote{For this we assume strange quark matter not to be the true ground state of strong interactions} in the core of a protoneutron star (PNS) which, after a brief period of deleptonization, will produce a regular neutron star. In particular, we investigate whether unpaired quark matter can be nucleated as the density increases in a medium of hadronic matter rich in trapped neutrinos.

If this happens by means of a first-order phase transition, in which we still have the presence of a barrier to fluctuations, it is well known that there are two main mechanisms for the nucleation of the true ground-state phase: thermal activation and quantum tunneling \cite{Gunton:1983}. However, a few years ago, it has been shown that, in PNS conditions, thermal activation will dominate over quantum nucleation\footnote{Quantum nucleation and the spinodal decomposition processes can compete with thermal nucleation in driving the phase transition. However, effective potentials are needed in order to account for the spinodal instability appropriately \cite{Bessa:2008nw}. On the other hand, for quantum nucleation, it was proven in Refs. \cite{Iida:1997ay,Mintz:2009ay} that only for temperatures below 5 MeV this process would be comparable to thermal activation and then for supernova matter it becomes unimportant.} \cite{Mintz:2009ay}. Thus, our analysis will focus on the thermal nucleation of quark matter droplets in a medium of hadronic matter. 

In order to do that, we have to match the lepton-rich pQCD EoS discussed in the previous section onto an EoS for the hadronic phase. Given the difference in scale between the temperatures involved in PNS matter ($T\sim 10~$MeV) and the typical baryon chemical potential in the medium ($\mu_{B}\sim 1500~$MeV), we can keep cold ($T=0$) approximations for the EoSs at each phase. One can easily see this in the quark phase (in the free and massless limit) where $P{~\simeq~}P_{\rm SB}(1+{18\pi^{2}}(T/\mu_{B})^{2}+...)$, so thermal corrections would be of the order $\mathcal{O}(T^{2}/\mu^{2}_{B})\sim{1}\%$, for the above-mentioned scales. Therefore, temperature effects will matter only in the calculation of the nucleation rate, as implemented, e.g. in Refs. \cite{Mintz:2009ay,Palhares:2010be}.

\subsection*{Hadron-Quark Phase Transition}

There are many equations of state that can be used to describe the properties of lepton-poor cold nuclear matter at densities around the nuclear saturation density, $n_{0}$. Some of them describe correctly the many phases that can exist inside a (cold) neutron star. However, for PNS matter the most appropriate (and usual) choice of EoS is the one which comes from relativistic mean field theory using the so-called TM1 parametrization of Shen $\textit{et al.}$ \cite{Shen:1998gq}. Here, we have to generalize their lepton-poor result to the case where neutrinos are trapped\footnote{See Ref. \cite{ShenTables1998} for more details on the tabulated version of this EoS, where the pressure, neutron and proton chemical potentials with their associated number densities are written as 3D functions of $(Y_{P},n_{B},T)$, i.e. proton fraction, baryon density and temperature, for which we choose the cold limit.}. Notice that this stiff EoS softens when we add a fixed fraction of leptons, in contrast to our lepton rich QCD EoS. Besides, after solving the supernovae conditions for this lepton-rich quark matter, one obtains (for the specific case of $X=3.75$) relatively high values of the electron chemical potential ($\simeq$ 1.2 GeV at $\mu_{s}=1.5$ GeV) and neutrino chemical potential ($\simeq$ 1.3 GeV at $\mu_{s}=1.5$ GeV) which are vanishingly small and zero in the lepton-poor case, respectively. This can be seen in Fig. \ref{fig:LeptonsChemicalX375}.

\begin{figure*}[h]
\begin{center}
\hbox{\includegraphics[width=0.5\textwidth]{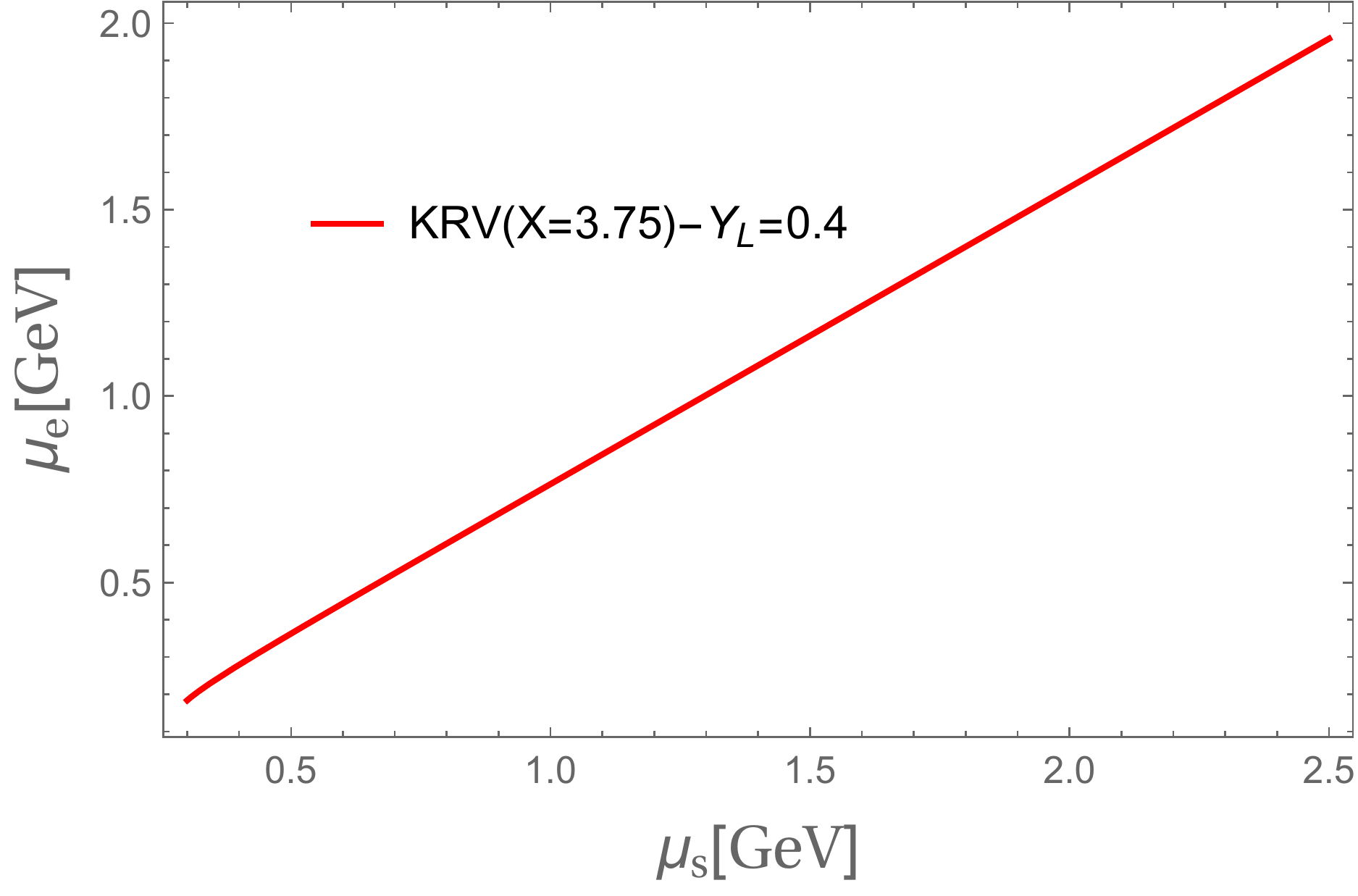}
	  \includegraphics[width=0.5\textwidth]{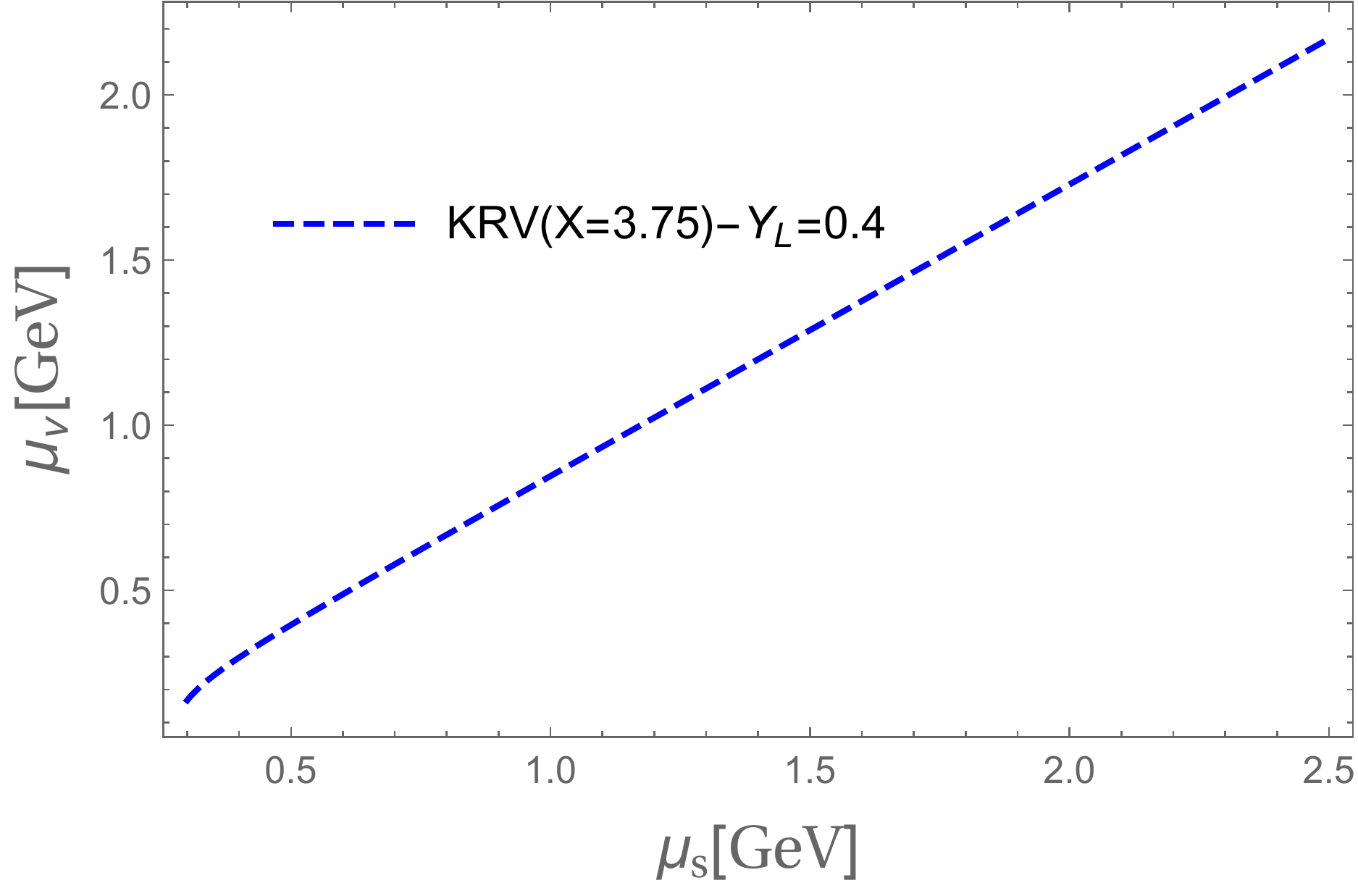}}
\vspace{5mm}
\caption{Electron and electron neutrino chemical potentials in the quark matter phase for $X=3.75$, as functions of the strange quark chemical potential, obtained after solving the beta-equilibrium, electric-charge-neutrality, and fixed-lepton-fraction conditions for supernova matter.}
\label{fig:LeptonsChemicalX375}
\end{center}
\end{figure*}

Thus, for the nuclear phase using the TM1 table without leptons, we impose again local charge neutrality, $n_{p}=n^{H}_{e}$ (implying also $Y_{p}=Y^{H}_{e}$), local lepton fraction conservation (cf. Ref.\cite{Pagliara:2009dg} for a global version of this constraint in PNS matter)

      \begin{equation}
	\label{Beta}
	\frac{n^{H}_{e}+2n^{H}_{\nu}}{n^{H}_{B}}=Y_{p}+Y^{H}_{\nu}=Y_{L}=0.4,
      \end{equation}
and the weak equilibrium condition $\mu_{n}+\mu^{H}_{\nu}=\mu_{p}+\mu^{H}_{e}$, where $\mu_{n}$, $\mu_{p}$, $\mu^{H}_{e}$ and $\mu^{H}_{\nu}$ are the chemical potentials of neutrons, protons, electrons and electron neutrinos in the hadronic phase, respectively. Also, $n_{p}$, $n_{n}$, $n^{H}_{e}$ and $n^{H}_{\nu}$ are the respective particle densities with $n^{H}_{B}=n_{n}+n_{p}$, being the baryon number density. Once more, lepton densities are introduced as free Fermi gases. When we refer to this lepton-rich hadronic matter EoS we will use the abbreviation TM1-PNS. See Fig. \ref{fig:2normPlepton} for a comparison of both cases. However, before going any further, a comment on the behavior of the neutrino density in the hadronic phase, $n^{H}_{\nu}$, near the surface of the PNS is in order. Since it is known \cite{Gondek:1997fd} that the very outer layer of a PNS becomes transparent to neutrinos by deleptonization and cooling on a very short timescale via electron-positron pair annihilation and plasmon decay, reaching temperatures of the order $T<1$ MeV, it seems natural to model the thermal structure of the PNS interior by a hot ($\sim$ 20 MeV) core limited by a neutrinosphere, and a much cooler ($\sim$ 10 MeV), neutrino transparent outer envelope. The transition through the neutrinosphere is accompanied by a temperature drop, which takes place over some interval of density just above the ``edge'' of the hot neutrino-opaque core, situated at some $n_{\nu}$. Detailed calculations \cite{Gondek:1997fd,Burrows:1986ajk} show that these values lie around $n^{\rm edge}_{\nu}=2\times10^{-3}{\rm fm^{-3}}$. However, the fact that our calculations are performed at $T=0$ allow us to disregard these transition temperatures, in other words neutrino cooling near the PNS surface affects more the temperature values than the range of neutrino density values. In this sense, one can choose $n^{\rm edge}_{\nu}$ only as an estimate of lowest values taken by this neutrino density approaching the surface. Notice that this approximation is consistent with all the above supernova equilibrium conditions since, on the PNS life timescale, the neutrinos are almost in complete thermal and chemical equilibrium at the surface of the PNS, and other emission (microscopic) effects can be safely neglected.

Now, we continue the building of the scenario we have in mind, i.e. the core-collapse of a supernova, similar to the one studied in Refs. \cite{Mintz:2009ay,Palhares:2010be}. Not taking into account the strange quark matter hypothesis, we know that for some density region there could be a deconfinement phase transition between hadronic and quark matter phases in a very dense and neutrino-rich environment found in PNS matter \cite{Glendenning:2000}. Since initially the hadronic phase does not contain strangeness, one would expect weak interactions could trigger a phase transition to unpaired quark matter. However, this would be too slow to produce strange quarks compared to the fast deconfinement transition driven by strong interactions \cite{Norsen:2002qw,Mintz:2009ay}.

Another scenario would be to consider a fast production of strange quarks due to the environment conditions of temperature and density of PNS matter, where a small amount of strangeness may appear through the presence of hyperons\footnote{Hyperons could then convert two-flavor quark matter into unpaired quark matter. Although it is well known that the presence of neutrinos inhibits the presence of hyperons at high densities, statistical fluctuations can be important \cite{Ishizuka:2008gr,Pons:1998mm}.} \cite{Bhattacharyya:2006vy}. The scenario adopted in this work is somewhat more inclusive since our EoS naturally adds a strange massive component to the two-flavor quark matter EoS as one goes from low to high densities, so that this scenario unifies the ones above for the formation of unpaired quark matter.

From the neutrino-rich pQCD EoS, the phase transition could be of first-order, depending on the chosen value for the renormalization scale $X$. In that case, we can use the modified Maxwell construction of Ref. \cite{Hempel:2009vp} for PNS matter, which mimics the out-of-equilibrium conditions. Then, our conditions for phase coexistence are the equality of the total pressures of the two phases, $P^{H}=P^{Q}$, and the condition of chemical equilibrium
  \begin{equation}
  \mu_{n}+Y_{L}\mu^{H}_{\nu}=\mu_{u}+2\mu_{d}+Y_{L}\mu^{Q}_{\nu}=\mu_{B}\equiv
  \mu_{\rm eff} \, .
  \label{mu_eff}
  \end{equation}
Indeed, although $P^{H}=P^{Q}$ is valid only at the transition point, Eq. (\ref{mu_eff}) is the chemical potential associated to the $\textit{global}$ conservation of baryon number along all the PNS life. In any case, the $\mu^{\rm H,Q}_{\nu}$s will be written in terms of an independent chemical potential, e.g. the $\mu_{s}$, after using the weak equilibrium conditions which in turn allow to write $\mu_{B}$ as an independent (global) control parameter. We note that this baryon chemical potential accounts for neutrinos non-trivially in order to implement more realistically the fact that we build the nuclear and quark EoSs locally independent of each other without enforcing explicitly global Gibbs constructions at the phase transition. To avoid confusion with the cold lepton-poor case, where $\mu_{n}=\mu_{B}$ for the hadron phase and $\mu_{u}+2\mu_{d}=\mu_{B}$ for the quark phase, we define it as an effective chemical potential, $\mu_{\rm eff}$, valid through all the phases of the PNS and useful for our matching procedure. In Fig. \ref{fig:2normPlepton} we show the lepton-rich quark and nuclear EoSs from which one can see that both match at some values of $\mu_{B}=\mu_{\rm eff}$, in contrast to the $\mu_{B}$'s appearing in Figs. \ref{fig:KRVneutrino}, \ref{fig:KRVcomparison}, and
\ref{fig:SQMcomparison}, where neutrinos are included through $\mu_{B}=\mu_{u}+\mu_{d}+\mu_{s}$ but not as Eq. (\ref{mu_eff}). From it can also be easily seen that the values chosen for $X$ require first-order phase transitions, being soft or strong.

\begin{figure*}[h]
\begin{center}
\hbox{\includegraphics[width=0.5\textwidth]{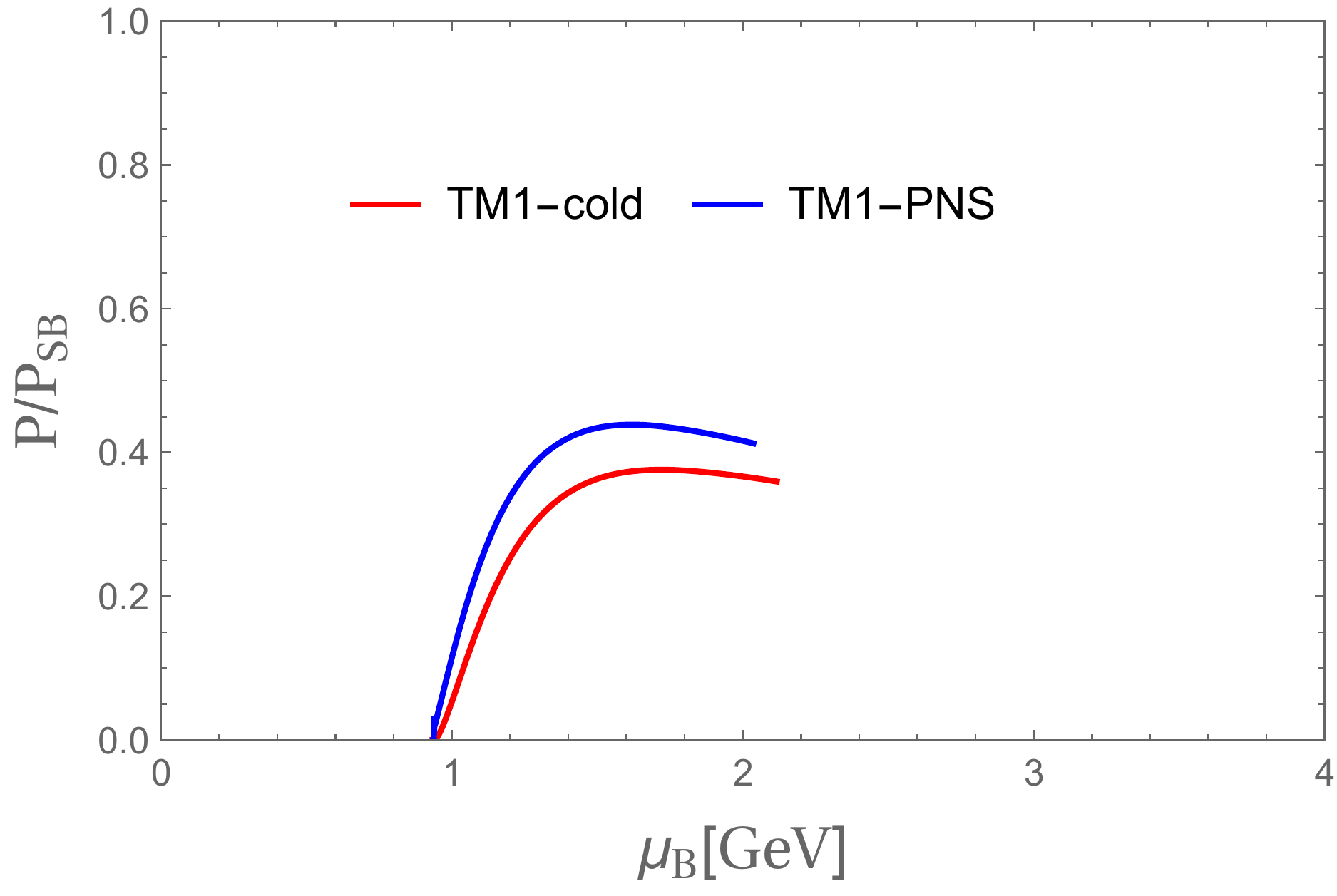}
	  \includegraphics[width=0.5\textwidth]{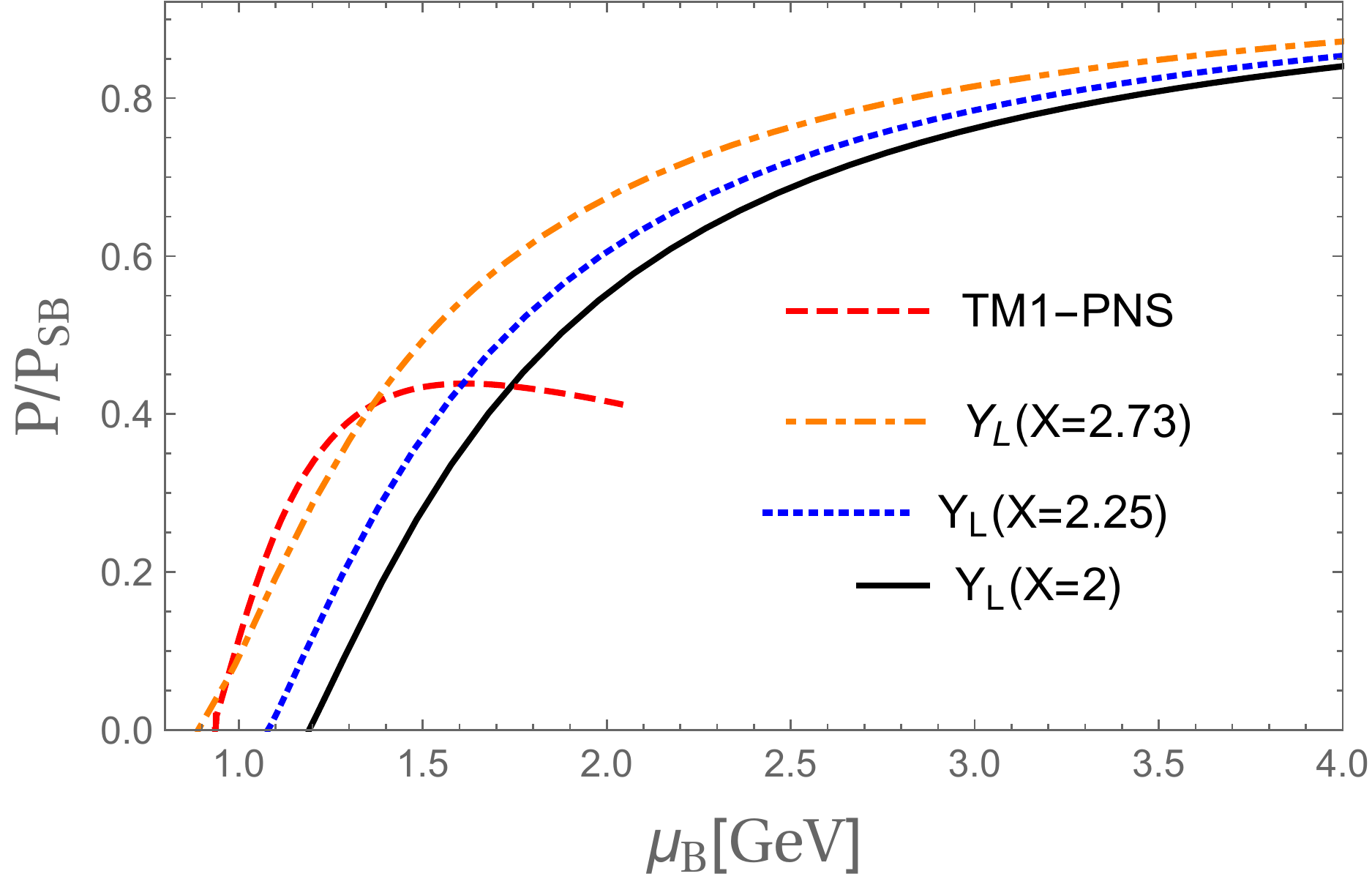}}
\vspace{5mm}
\caption{\label{fig:2normPlepton} Left panel: Nuclear pressure, normalized by the Stefan-Boltzmann gas, for the TM1 model without neutrinos (TM1-cold) and with trapped neutrinos (TM1-PNS).  Right panel: Total normalized pressure for lepton-rich quark matter matched onto lepton-rich hadronic EoS TM1-PNS (dashed line) for different values of $X$ that allow for nucleation of unpaired quark matter still consistent with measurements of two-solar mass pulsars. Notice that at low densities for $X=2.73$, one has a critical baryon chemical potential which would mean a hybrid star made up mainly of quark matter with a thin nucleonic crust. This situation was not analyzed in this work since the quark pressure is higher than the nuclear one at low densities. This signals a thermodynamic non-favourable situation. Notice that phase transitions at high densities signal small quark-matter cores. Taken from Ref. \cite{Jimenez:2017fax}.}
\end{center}
\end{figure*}
Since we only consider first-order phase transitions, it is appropriate to use a thermodynamic quantity which tells us how $\textit{strong}$ is the phase transition. This quantity will be the latent heat, defined as (and using Eq. (2.5)) $\Delta{Q}{~\equiv~}\epsilon_{\rm quark}(\mu^{c}_{\rm eff})-\epsilon_{\rm nuclear}(\mu^{c}_{\rm eff})=\mu^{c}_{\rm eff}\Delta{n_{B}}$ \cite{Kurkela:2014vha}, where $\mu^{c}_{\rm eff}$ is the critical effective chemical potential from which the quark phase starts and $\Delta{n_{B}}$ the baryonic number difference between phases at the critical point\footnote{In principle, a correct matching for both phases of lepton-rich matter would imply the existence of a mixed phase with possible nontrivial geometrical structures \cite{Glendenning:2000}. However, modifying our conditions of charge neutrality from local to global would not affect considerably our results in comparison to the intrinsic uncertainty brought about by the renormalization scale dependence of our results.}. This definition can be considered as a latent heat since it account for the liberated energy (density) when going discontinuously from a dilute to a dense phase. Notice that this $\Delta{Q}$ replaces the cold limit of the standard definition of latent heat at finite temperature $\Delta{Q}=TdS$, being $S$ is the entropy for a quasi-static transformation.

\subsection{Homogeneous Thermal Nucleation and Stellar Structure}

In order to estimate the time scales appropriate for the nucleation of deconfined lepton-rich quark matter with a fixed $Y_{L}$, we first need to elaborate the formalism required to compute nucleation rates $\Gamma$, i.e., the Langer formalism \cite{Langer:1969bc} of homogeneous nucleation\footnote{We note that general-relativistic effects can be neglected when studying this out-of-equilibrium processes since the spacing between the nucleated bubble of quark matter and surrounding nucleons ($R_{\rm bubble-N}\sim 5$fm) is too small compared to the PNS's size, $R_{\rm PNS}\sim 20$Km, i.e. $R_{\rm bubble-N}/R_{\rm PNS}\ll{1}$. In other words, gravitational effects are not needed when performing microscopic calculations of PNSs, as is usually done in cold NSs \cite{Glendenning:2000}. However, this is not necessarily true for similar processes in NS mergers.} via thermal activation \cite{Gunton:1983}.
  
In first-order phase transitions, the conversion from one phase to the other usually occurs slowly and very close to the thermodynamic equilibrium, as in the Maxwell construction. However, when some relevant control parameter, e.g., the density, changes abruptly when a system is near the transition, the system finds itself in an unstable situation. The system at hand could be initially homogeneous in a low-density phase (e.g., nuclear), and close to the transition line to a high-density phase (e.g., quark), and then suffering a sudden compression. Although the system was prepared at the dilute phase, the free energy at the new, higher density disfavors the dilute phase and the high-density phase now becomes the stable one. This marks the beginning of phase conversion \cite{Mintz:2011yda}. 

The omnipresent fluctuations (thermal and quantum) will not be suppressed (as expected in equilibrium) due to the instability of the system. Such fluctuations will drive the system to another point of stability \cite{Mintz:2010mh,Mintz:2011yda}. Our main goal is to study the evolution of those fluctuations in time. If a homogeneous system is brought into instability close enough to the coexistence line of the phase diagram, its dynamics will be dominated by the creation of bubbles (or droplets) called \textit{nucleation}. We focus on thermal nucleation of quark matter as nuclear matter is compressed in a stellar collapse. 

\subsection*{Langer's theory of thermal nucleation}

A usual field-theoretical approach for thermal nucleation in one-component metastable systems was developed by Langer in 1969 \cite{Langer:1969bc}. This microscopic model was able to calculate the rate of creation of critical bubbles per unit volume. This assumes that the system under study is in contact with a thermal reservoir. In this formalism, a key quantity for the calculation of the rate of nucleation is the coarse-grained free energy functional \cite{Mintz:2010mh,Mintz:2011yda}
\begin{equation}
F[\phi]=\int{d^{3}r}\left\lbrace{\frac{1}{2}[\nabla\phi(\vec{r})]^{2}+V[\phi(\vec{r})]}\right\rbrace,
\end{equation}
where $\phi(\vec{r})$ is the order parameter of the phase transition at a given point $\vec{r}$ of space. By assumption, the ``potential'' $V(\phi)$ has a global (true) minimum at $\phi_{t}$ and a local (false) at $\phi_{f}$. At a given baryon chemical potential $\mu$ of the metastable phase, the difference $\Delta{V}=V(\phi_{t})-V(\phi_{f})$ is identified with the pressure difference between the stable and metastable phases, with opposite sign: $\Delta{V}=-\Delta{p}(\mu)=p_{t}-p_{f}$, where $p_{t}(p_{f})$ is the pressure for the true (false) phase at baryon chemical potential $\mu$ \cite{Mintz:2010mh,Mintz:2011yda}.

The field equation for $\phi(\vec{r})$ is given by a minimum of the functional F. Two of them are the trivial ones given by homogeneous field configurations with $\phi(\vec{r})=\phi_{t}$ or $\phi(\vec{r})=\phi_{f}$. The third is a spherically symmetric bubblelike solution that has as boundary conditions \cite{Mintz:2010mh,Mintz:2011yda}
\begin{eqnarray}
\phi(r=0)=\phi_{t},\\
\phi(r\rightarrow{\infty})=\phi_{f}.
\end{eqnarray}
Roughly speaking, this means that the stable phase ($\phi_{t}$) is found deep in the bubble and the metastable one ($\phi_{f}$) is found away from it. Somewhere in-between, the order parameter must change from its central value $\phi_{t}$ to $\phi_{f}$ at $r\rightarrow{\infty}$. The relatively thin region which marks the border between ``inside'' ($\phi=\phi_{t}$) and ``outside'' ($\phi=\phi_{f}$) the bubble is called the bubble wall \cite{Mintz:2010mh,Mintz:2011yda}.

Exactly at the coexistence line, one can prepare one (infinite) system with the two homogeneous phases in equal proportions divided by a plane wall with a small width. This configuration is static, once no phase is  favored. Further, each phase occupies a semi-infinite volume. If the system is slightly pushed into metastability, the static solution for $\phi(\vec{r})$ is a bubble with a very large radius and still a small wall width. This is the starting point for the \textit{thin-wall approximation} \cite{Mintz:2010mh,Mintz:2011yda}: the free energy Eq. (2.42) of the system of volume $({4\pi}/{3})L^{3}(L\rightarrow\infty)$ is determined by the outcome of a competition between a surface energy term, which is positive and comes from $|\nabla\phi|^{2}$ in Eq. (2.42), and a bulk term, which is negative and corresponds to the potential V, or to the pressure difference between the phases. Notice that, within this approximation, $\phi(r)$ is constant, except over the (thin) wall of the bubble, and so $V(\phi)$ is also essentially constant both inside and outside the bubble. This means that the free energy for the bubble configuration of radius R in the thin-wall approximation of Eq. (2.42) is given by
\begin{equation}
F_{\rm bubble}(R)=4\pi{R}^{2}\sigma-\frac{4\pi}{3}(L^{3}-R^{3})p_{f}-\frac{4\pi}{3}R^{3}p_{t},
\end{equation}
whereas the homogeneous metastable configuration has $|\nabla\phi|^{2}=0$ and 
\begin{equation}
F_{\rm metastable}=-\frac{4\pi}{3}L^{3}p_{f}.
\end{equation}
In Eq. (2.45), we introduced the surface tension $\sigma$, which is merely the energy per unit area of the bubble wall. For this Langer theory \cite{Langer:1969bc}, the nucleation rate has as its main ingredient the free energy shift when a bubble is created from fluctuations in the homogeneous metastable phase. From Eqs. (2.45) and (2.46) we have
\begin{equation}
\Delta{F}(R)=F_{\rm bubble}(R)-F_{\rm metastable}=4{\pi}R^{2}\sigma-\frac{4\pi}{3}R^{3}(\Delta{p}),
\end{equation}
where $\Delta{p}=p_{t}-p_{f}>0$. Here, the pressures in each of the phases are calculated for the same value of $\mu_{\rm eff}$. Notice that this implies different baryon chemical potentials and densities for each phase, due to the supernova conditions \cite{Mintz:2011yda}. 

\subsection*{Nucleation rates and Surface Tensions}

Bubble configurations of given radii R arise from the homogeneous metastable phase due to thermal fluctuations, and each of those has an associated value of $\Delta{F}(R)$ \cite{Mintz:2010mh,Mintz:2011yda}. From Eq. (2.47), we can see that $\Delta{F}(R)$ has a maximum at the critical radius $R_{c}={2\sigma}/{\Delta{p}}$. The equations of motion show that any bubble with $R<R_{c}$ will shrink and disappear whereas any bubble with $R>R_{c}$ will grow, as a consequence of the competition between the positive surface energy and the negative bulk energy. Hence, the critical bubbles are the smallest bubbles that can start to drive the phase conversion dynamics \cite{Mintz:2010mh,Mintz:2011yda}. To give a quantitative meaning to the process of nucleation, one can calculate the rate $\Gamma$ of critical bubbles created by fluctuations per unit volume, per unit time. In Langer's formalism \cite{Mintz:2010mh,Mintz:2011yda}:
\begin{equation}
\Gamma=\frac{\mathcal{P}_{0}}{2\pi}\exp\left[-
\frac{\Delta{F}(R_{c})}{T}\right],
\end{equation}
where the prefactor $\mathcal{P}_{0}$ is usually factorized into two parts: a statistical prefactor, which measures the rate of successful creation of a critical bubble by thermal fluctuations, and a dynamical prefactor, which measures the early growth rate of the bubble. An exact calculation \cite{Csernai:1992tj,Scavenius:2000bb} of this $\mathcal{P}_{0}$ reveals that it is strongly dependent on the transport coefficients of shear $\eta_{i}$ and bulk $\zeta_{i}$ viscosities, where ``$i$'' depends if we are studying the nucleation of a hadronic bubble in a QCD plasma ($i=q$, and not our case) or a quark bubble in a hadronic gas ($i=n$, which is our case). The calculation gives (see Refs. \cite{Csernai:1992tj,Scavenius:2000bb} for more details)
\begin{equation*}
\frac{\mathcal{P}_{0}}{2\pi}=\frac{4}{\pi}\left(\frac{\sigma}{3T}\right)^{3/2}\frac{\sigma(\zeta_{n}+4\eta_{n}/3)R_{c}}{\xi^{4}_{n}(\Delta{w})^{2}},
\end{equation*}
where $\Delta{w}=w_{q}-w_{n}$ is the enthalpy difference between phases being $w=\epsilon+P$, and $\xi_{n}$ the correlation length in the nuclear phase.
However, and as usually done, here we adopt the simple estimate ${\mathcal{P}_{0}}/{2\pi}=T^{4}$, obtained on dimensional grounds corresponding to an overestimate of the above calculation of this prefactor since, as explicitly proven in Ref. \cite{Csernai:1992tj}, this exact prefactor is in the order and larger than $T^{4}$ only near the critical transition temperature $T_{c}\gg{T}$, being $T$ the expected temperatures occurring at the center of the PNS. This approximation is in agreement with the thin-wall approximation, which also gives an overestimate to $\Gamma$. Although this overestimate can lead to an overall factor of approx. $10^{2}$ or even higher, the qualitative aspects of the obtained results are hardly changed. Additionally, as we will soon see, this produces \textit{underestimates} for thermal nucleation time scales occurring in core collapse supernovae typical conditions, for which the details are not longer relevant. 

Our final formula for the nucleation rate, $\Gamma$, can be rewritten, after some straightforward algebra, as \cite{Mintz:2010mh,Mintz:2011yda}
\begin{equation}
\Gamma=T^{4}\exp\left[-\frac{16\pi}{3}\frac{\sigma^{3}}{(\Delta{p})^{2}T}\right], 
\label{eq:gamma}
\end{equation}
where $\Delta{F}(R_{c})$ is the difference in free energy between the metastable (nuclear) phase and the true stable (quark) phase, which can be written in terms of the radius of the critical bubble, $R_{c}$, which corresponds to a saddle point in functional space. Notice that the influence of the equation of state is present through $\Delta{p}$, which is the difference between the matched pressures\footnote{It may seem that the very strong dependence of $\Gamma$ on $\Delta{P}$ could produce quite different scenarios if some ingredients of the model are modified slightly, e.g. the values defining $m_{s}$. However, all these dependencies are vanishingly small compared to the dependency on $\bar{\Lambda}$. For this work we only explore a subset of the allowed values of $\bar{\Lambda}$ which in turn do not display all of its dependency as a band of results.}. It is clear that the surface tension, $\sigma$, plays a crucial role \cite{Mintz:2009ay,Palhares:2010be,Mintz:2011yda,Pinto:2012aq,Mintz:2012mz,Lugones:2013ema}.
This strong dependence of $\Gamma$ on $\sigma$ is determinant for the nucleation time. 

\subsection*{Nucleation time scales}

The nucleation time, $\tau$, to create the first single {\it critical} droplet of lepton-rich unpaired quark matter inside a volume of $1 \rm km^{3}$, which is the typical size of the core of a PNS, is given by \cite{Mintz:2009ay,Mintz:2010mh,Mintz:2011yda}
\begin{equation}
\tau_{\rm nucl}\equiv\left({\frac{1}{1 \rm km^{3}}}\right)\frac{1}{\Gamma} \, ,
\label{eq:tau_nucl}
\end{equation}
where we assume homogeneity of density and temperature in the core, a good approximation since the density profile in this region of the PNS is approximately flat \cite{Glendenning:2000}. This is the time scale to be compared with the duration of the early post-bounce phase of a supernovae event, few hundreds of milliseconds, during which it has been shown that quark matter formation could trigger the explosion \cite{Sagert:2008ka}. With this definition, we assume that the temperature and density are \textit{constant} within this central volume of $1{\rm km}^{3}$. Of course, this also goes in the direction of underestimating this time scale. Realistically, one should first compute the pressure and density profiles using the equations of hydrostatic equilibrium, then calculate the local value of $\Gamma$ as a function of the radial coordinate, and finally integrate over the region containing metastable matter. This calculation would lead us to include the contribution of a very small gradient pressure in the region near the phase transition is taken place. However, the density profiles are almost flat within the central kilometers of the star which, in turn, allow us to neglect the gradient pressure to a very good approximation, thus making these assumptions quite reasonable \cite{Mintz:2010mh,Mintz:2011yda}. 

Finally, we note that we calculate the time of production of one single critical bubble, which has a typical size of some fermi. Besides, when comparing $\tau$ with the bounce time scale as a criterion for the formation of a quark core,  we tacitly assume that the quark matter bubble becomes macroscopic almost instantaneously \cite{Mintz:2010mh,Mintz:2011yda}. Besides, the volume, V, in which we consider fluctuations corresponds to the volume of a drop of the new phase which critical radius $R_{c}$. Again, the surface tension is the crucial quantity that determines whether the phase transition occurs via the intermediate phase or directly. Taking into account the uncertainties on the value of $\sigma$ we estimate the critical radii to be of the order\footnote{Notice that these radii are also implicitly dependent on the pQCD renormalization parameter $\bar{\Lambda}$.} of $R_{c}\sim6$ fm. It can be proven that statistical fluctuations are efficient for radii of the order of 2 to 4 fm, and thus of the same order of magnitude as the critical radii. So, the first quark bubble nucleated that marks the beginning of the transition (within the hadronic phase) will be the critical one \cite{Mintz:2009ay,Mintz:2010mh,Mintz:2011yda}.
\begin{figure*}[h]
\begin{center}
\hbox{\includegraphics[width=0.5\textwidth]{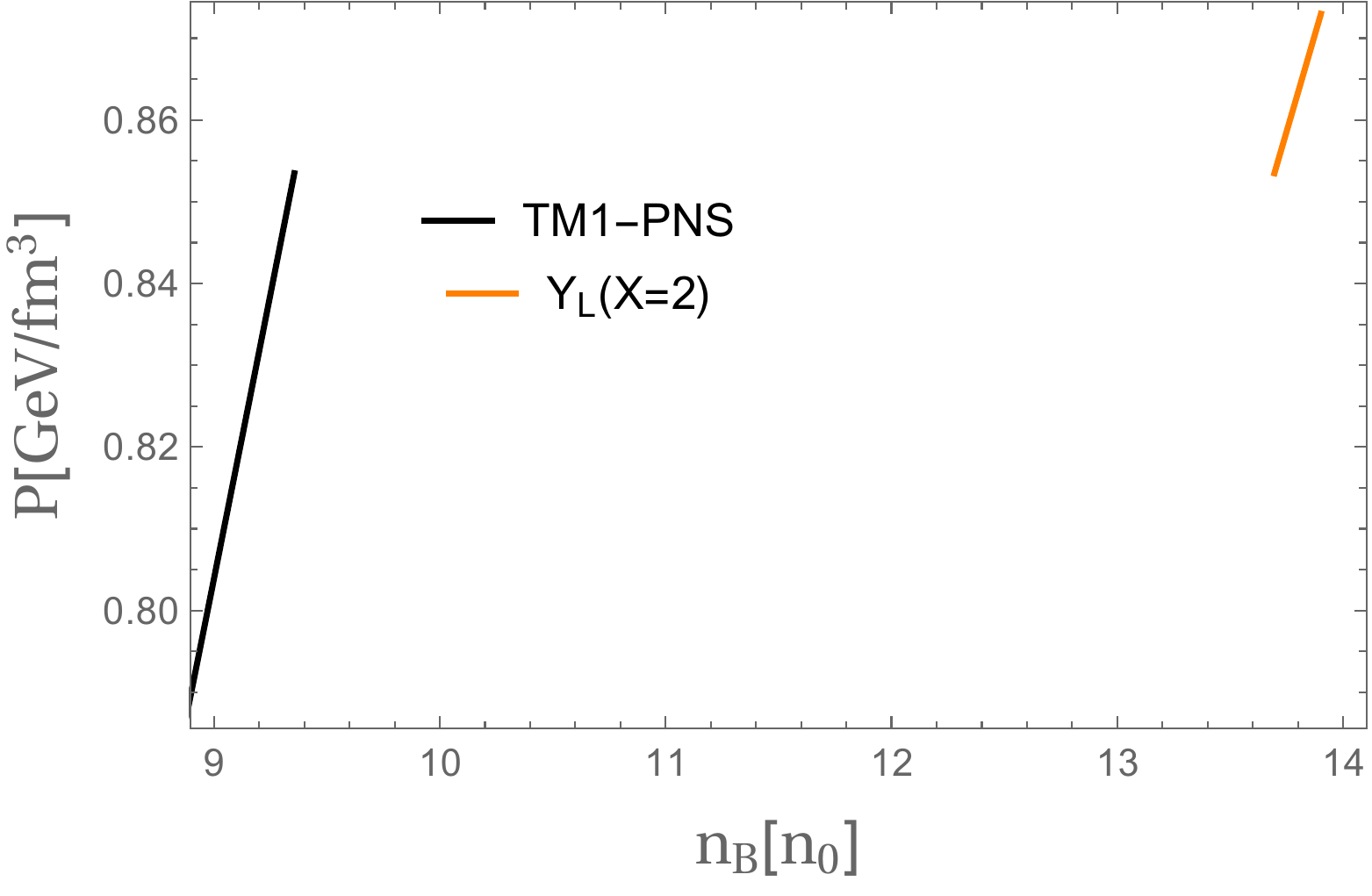}
	  \includegraphics[width=0.5\textwidth]{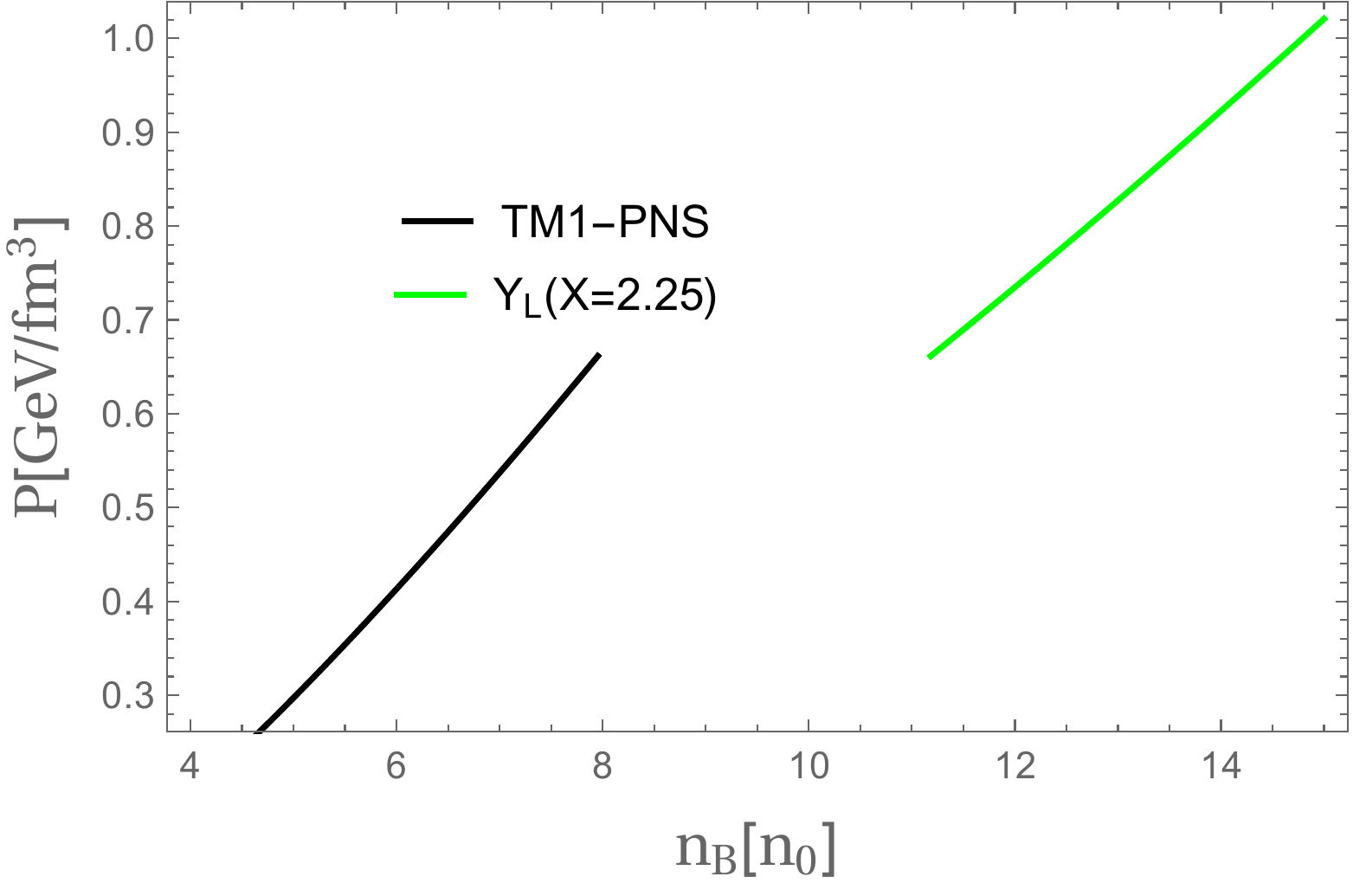}}
\vspace{5mm}
\caption{Left panel: Pressures for the lepton-rich nuclear, TM1-PNS, and lepton-rich quark matter for $Y_{L}(X=2)$ phases versus the baryon number density where a Maxwell construction between phases must be carried out in the jump density at constant pressure. Right panel: The same but for $Y_{L}(X=2.25)$.}
\label{fig:PnBX2X225}
\end{center}
\end{figure*}
\begin{figure*}[h]
\begin{center}
\hbox{\includegraphics[width=0.5\textwidth]{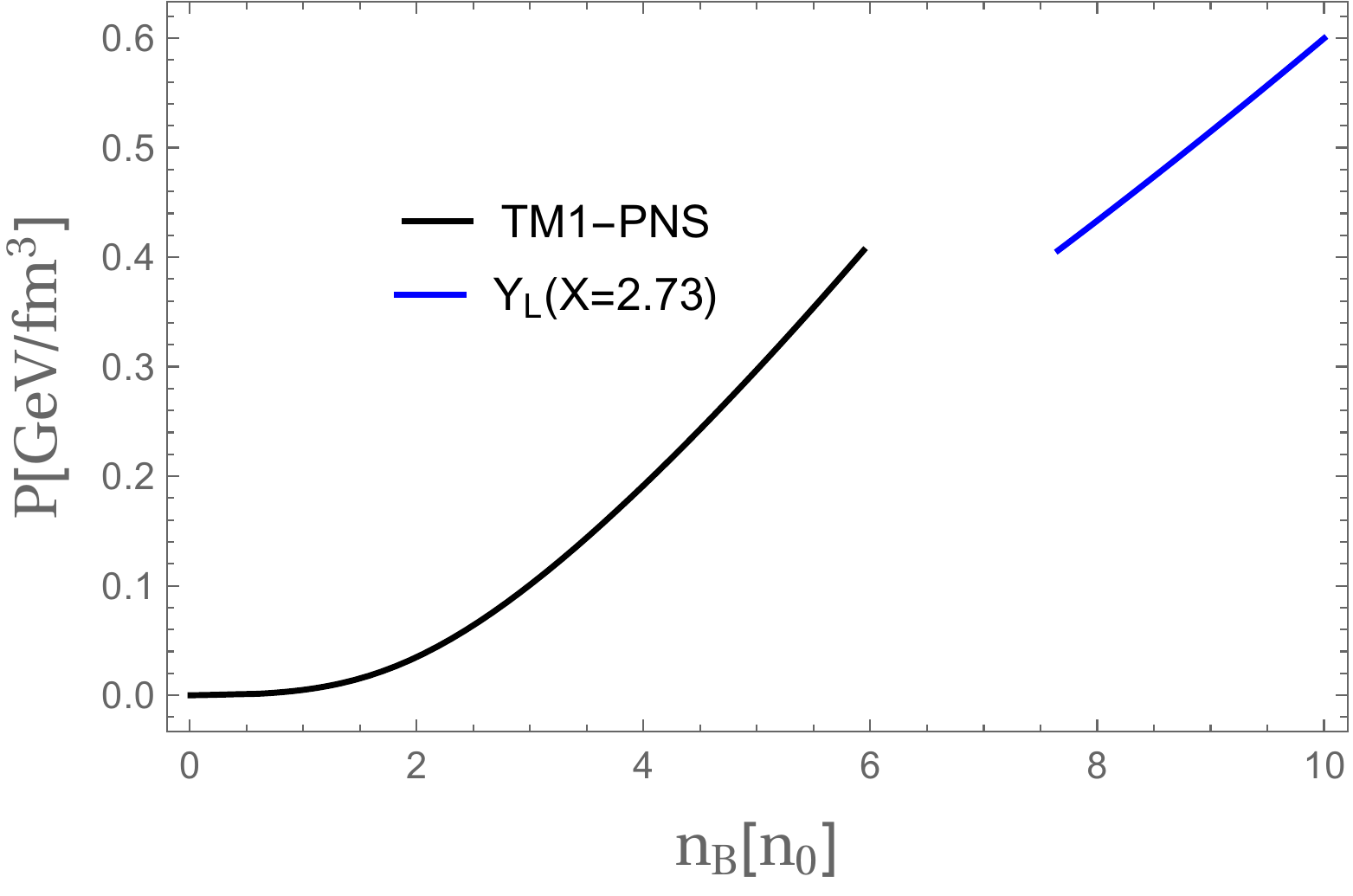}
	  \includegraphics[width=0.5\textwidth]{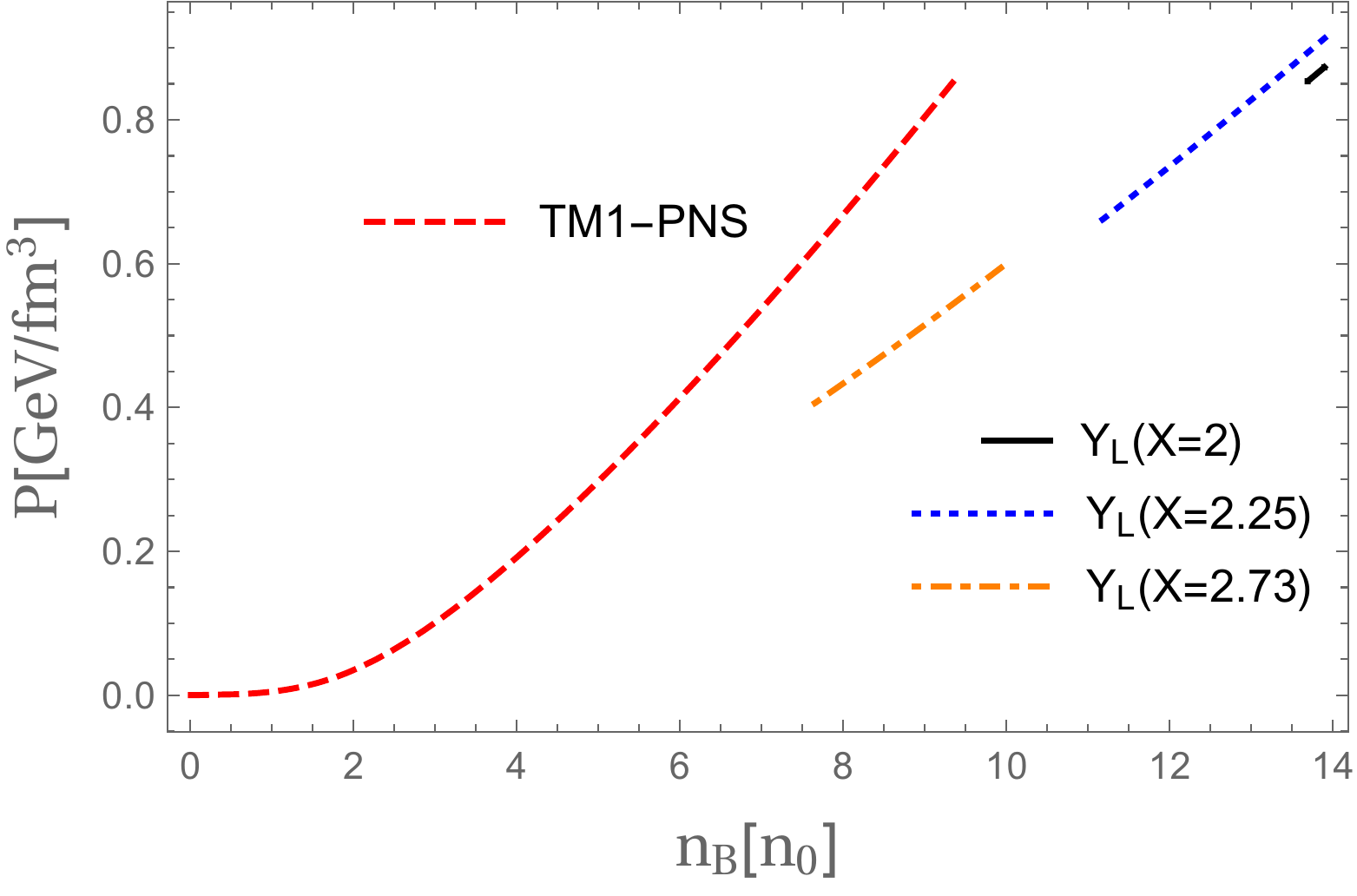}}
\vspace{5mm}
\caption{Left panel: Pressures for the lepton-rich nuclear, TM1-PNS, and lepton-rich quark matter for $Y_{L}(X=2.73)$ phases versus the baryon number density where a Maxwell construction between phases must be carried out in the jump density at constant pressure. Right panel: We plot together all the analysed cases treated in this work for different values of the renormalization scale $Y_{L}(X)$ in the lepton-rich quark phase. Notice that they begin at the critical baryon densities for each case.}
\label{fig:PnBX273Xall}
\end{center}
\end{figure*}
%
\subsection*{The \textit{usual} Tolman-Oppenheimer-Volkov Hydrostatic Equations}

To be consistent with current astrophysical observations of neutron star masses allowing for a core of quark matter \cite{Alford:2006vz}, one has to choose values for $X$ in the cold lepton-poor pQCD equation of state, KRV-EoS, that match the cold lepton-poor TM1-EoS\footnote{For the low-density region, we include the Baym-Pethick-Sutherland EoS \cite{Baym:1971pw} that is necessary for an adequate treatment of the crust.} at a given critical baryon number density (chemical potential) and generate at least two-solar mass stars as maximum masses. Thus, in order to do this we must resort to find the stellar configurations in hydrostatic equilibrium by means of the so-called Tolman--Oppenheimer-Volkov (TOV) equations, which we pass to explain systematically\footnote{We note that although these equations will be written in the $\hbar=G=c=1$ units along all this thesis (which implies $[P]=[\epsilon]=[\rm GeV]^{4}$), for clearness we show the behavior of the EoSs to be solved in their correct physical units, i.e. $[P]=[\epsilon]=[\rm GeV/fm^{3}]$, where the respective values of the $\hbar$, $G$, and $c$ were put back, in particular using the identity $\hbar{c}=197.33~{\rm MeV\times{fm}}$ \cite{Glendenning:2000}.}.

First of all, we note that along all this thesis we assume that the relativistic (or compact) stars are static and spherically symmetric, so that one can use the Schwarzschild-like line element, 
\begin{equation}
ds^{2}=e^{\nu}dt^{2}-e^{\lambda}dr^{2}-r^{2}(d\theta^{2}+\sin^{2}\theta{d\phi^{2}}),
\end{equation}
where $\nu$ and $\lambda$ are functions of $r$.
Note that it can be proven that (see, e.g., Ref. \cite{Misner:1974qy}), due to the intrinsic nature of general relativity, other set of coordinates can be used but the Schwarzschild-like are particularly useful when interpreting observations of different stellar parameters, e.g. delay times related to the gravitational redshift parameter $Z$.

In order to describe matter inside compact stars we can consider it to be made up of a perfect fluid with high precision (since shear stresses are negligible and energy transport is vanishing small on a hydrodynamic time scale, $\tau_{\rm hydro}$), and it is represented by
\begin{equation}
T^{\mu\nu}=(P+\epsilon)u^{\mu}u^{\nu}-Pg^{\mu\nu},
\end{equation}
where $P$ is the pressure, $\epsilon$ is the energy density, and $u^{\mu}=dx^{\mu}/d\tau$ (being $\tau$ the proper time) is the local fluid four-velocity which satisfies the normalization condition
\begin{equation}
g_{\mu\nu}u^{\mu}u^{\nu}=1.
\end{equation}
Then, introducing the metric functions and the perfect-fluid energy momentum tensor into the Einstein's equations, one obtains the \textit{usual} pair of TOV equations \cite{Glendenning:2000,Shapiro:1983} for stellar configurations in hydrostatic equilibrium,
  \begin{equation}
  \frac{dP}{dr}=-\frac{\epsilon\mathcal{M}}{r^{2}}\left(1+\frac{P}{\epsilon}\right)\left(1+\frac{4\pi{r^3}{P}}{\mathcal{M}}\right)
  \left(1-\frac{2\mathcal{M}}{r}\right)^{-1} \; ,
  \end{equation}
  \begin{equation}
  \frac{d\mathcal{M}}{dr}=4\pi{r}^{2}\epsilon \; ,
  \end{equation}
where $\mathcal{M}$ is the gravitational mass inside the radius $r$. To solve these equations, one needs the EoS, written as $P=P(\epsilon)$ for pure stars or $\epsilon=\epsilon(P)$ for hybrid stars (which will be our case for this chapter), as an input\footnote{Strictly, this distinction between the writing of the EoS is necessary especially when it involves a 1st-order phase transition at intermediate densities and it is manifested as a jump in the energy density at constant pressure. This is the main feature of the Maxwell construction.}. Then, one imposes that at the center of the star $\mathcal{M}(r=0)=0$ and $P(r=0)=P_{c}$, and the integration must end when $P(r=R)=0$, i.e. at the surface of the star, its total mass being $\mathcal{M}(r=R)=M$. 

For our case of the matched (hybrid) EoSs in Figs. \ref{fig:PnBX2X225} and \ref{fig:PnBX273Xall} in the lepton-rich limit, we find that their lepton-poor counterparts share their $\textit{soft}$ and $\textit{strong}$ first-order phase transitions\footnote{\textit{Soft} for small jumps in the density and \textit{strong} for large ones, depending on the values of $X$.} being able to accommodate NS masses above the latest measurements for pulsars in binary systems. We note that the lepton-rich EoSs would be obtained by building numerically a table for $P(\mu_{\rm eff})$ depending on $\epsilon({\mu_{\rm eff}})$ for each $\mu_{\rm eff}$, whereas for the lepton-poor case one only takes $\mu_{B}$ without any neutrino term. Nevertheless, our nucleation problem only requires a calculation of $\Delta{P}(\mu_{\rm eff})$, or more appropriately $\Delta{P}(n_{B})$, by building tables binding $\Delta{P}(\mu_{\rm eff})$ to $n_{B}(\mu_{\rm eff})$ for each $\mu_{\rm eff}$, where $n_{B}(\mu_{\rm eff})$ comes from a same procedure for $n_{B}(\mu_{s})$ and $\mu_{\rm eff}(\mu_{s})$. 

Then, after solving the TOV equations with our lepton-poor EoSs, in Fig. \ref{fig:MR} we show the mass-radius diagram for a few values of the renormalization scale $X$ featuring first-order phase transitions with different critical densities and intensities. We restrict the values of the renormalization scale to the interval $[2,2.73]$--- if one keeps higher values, one would produce hybrid stars that do not satisfy the observational constraint of two solar masses; if one keeps lower values of $X$, one would find purely nucleonic stars.

\begin{figure*}[h]
\begin{center}
\hbox{\includegraphics[width=0.5\textwidth]{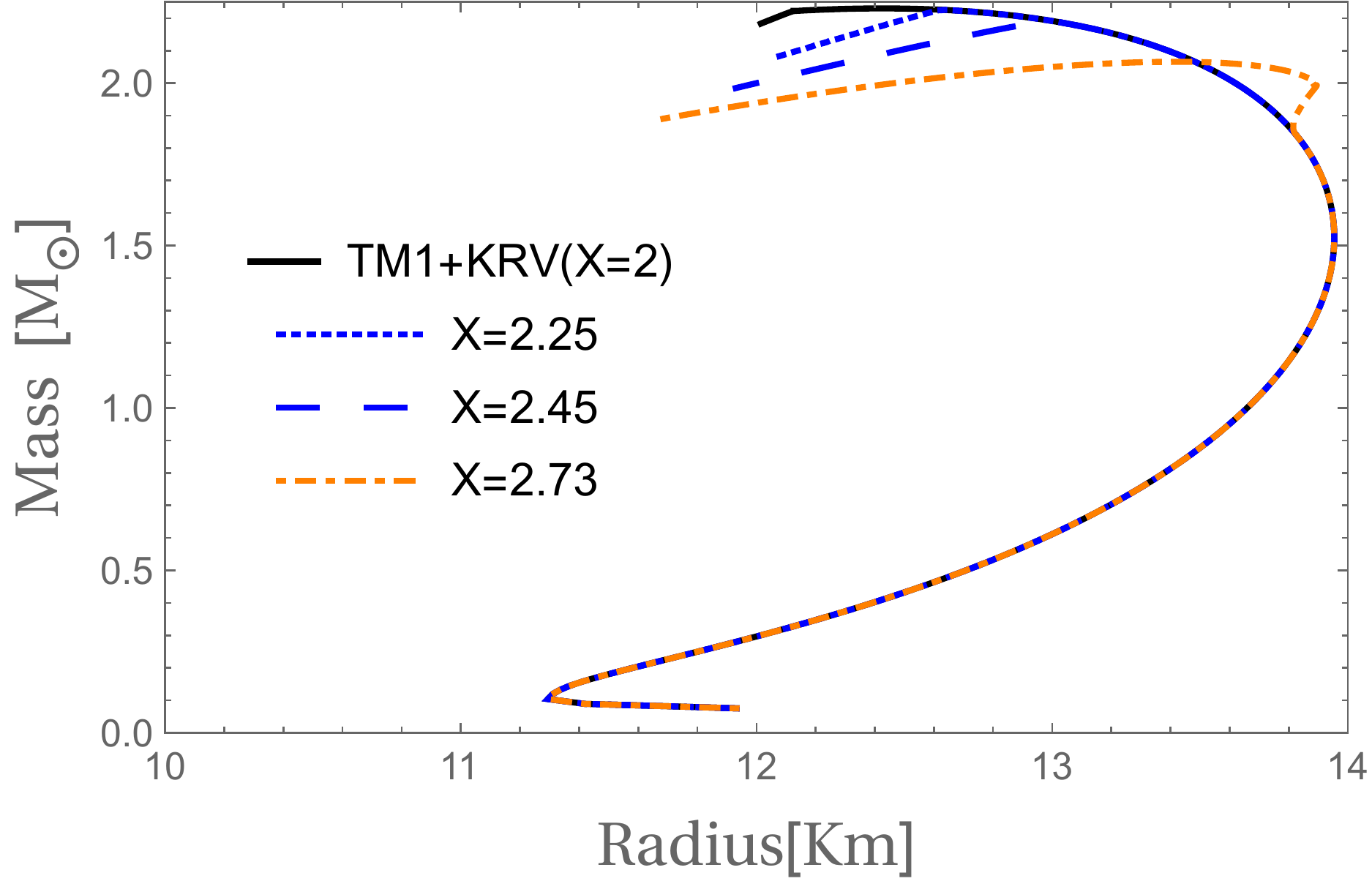}
	  \includegraphics[width=0.5\textwidth]{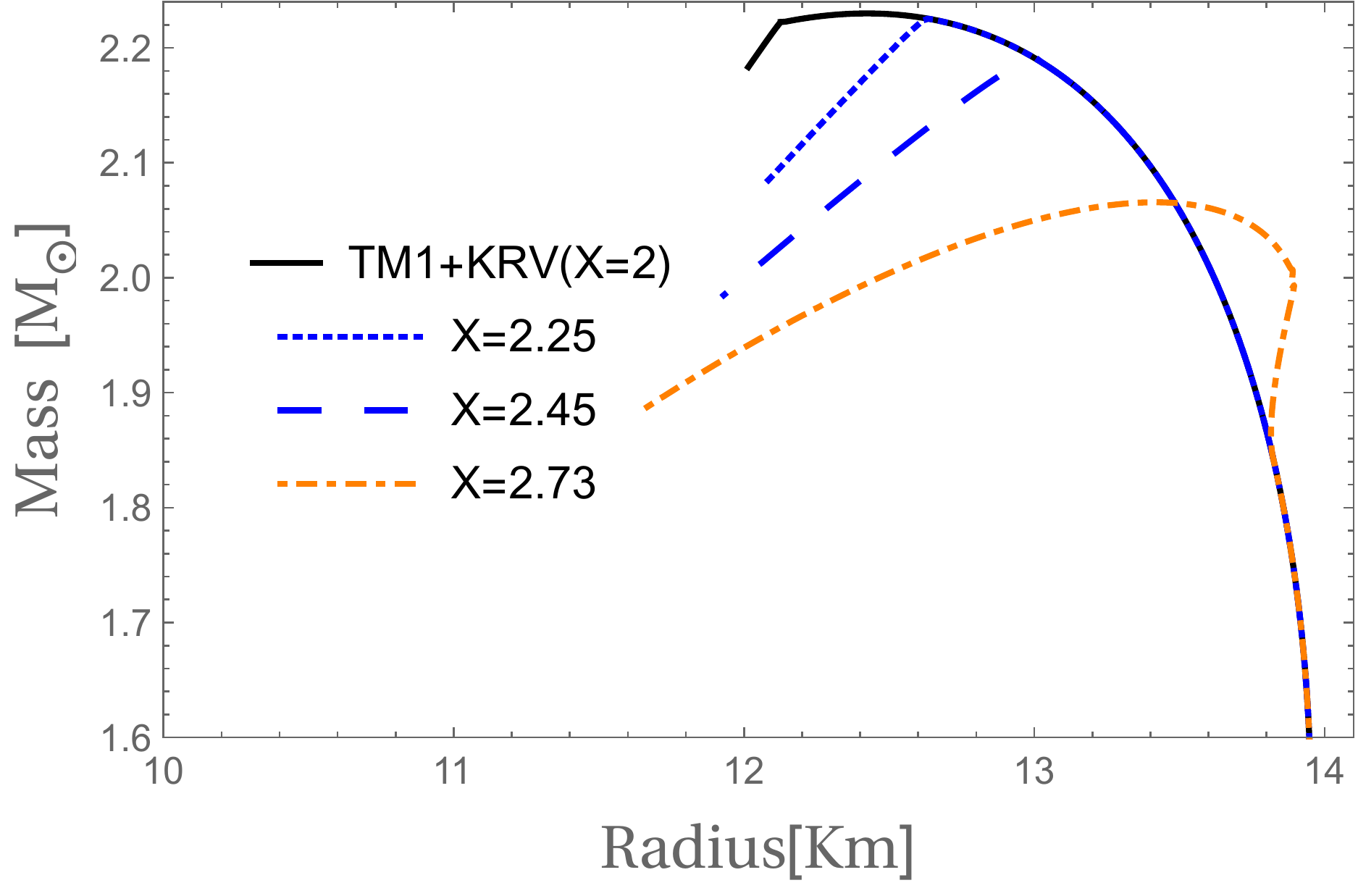}}
\vspace{5mm}
\caption{\label{fig:MR}Left panel: Mass-radius diagram for hybrid stars that masquerade (pure or mixed phase) quark matter cores. The matching is performed using the TM1-EoS for the hadronic phase and the KRV-EoS for the quark phase for different values of the renormalization scale $X$. Right panel: Zoom of this diagram for the maximal mass stellar configurations where 1st-order phase transitions occur. Both figures are taken from \cite{Jimenez:2017fax}.}
\end{center}
\end{figure*}

Notice that the softest possible matching corresponding to $X=2.73$ still exhibits a first-order phase transition to a quark mixed phase at a critical baryon density of $n_{\rm crit}=2.82n_{0}$, with a pure quark core with central density $n_{c}=4.2n_{0}$. The maximum mass in this case can reach $M=2.08M_{\odot}$, and the latent heat is given by $\Delta{Q}(\rm X=2.73)=(129.3\rm MeV)^{4}$, the lowest value of latent heat obtained in the matching procedure. Following the arguments of Ref. \cite{Kurkela:2014vha}, one expects to have soft first-order phase transitions if the latent heat $\Delta{Q}$ is smaller than $(\Lambda_{\rm QCD})^{4}{~=~}(\Lambda_{\overline{\rm MS}})^{4}{~\sim~}(300{\rm MeV})^{4}$, so that one has a large parameter space to surpass the two-solar mass limit with a quark content in hybrid stars. As we decrease the value of $X$, the first-order phase transition becomes stronger and happens at very high densities, as can be seen in Table \ref{tab:2.1}, making the hybrid star more nucleonic with a small mixed quark core.

\begin{table}[h!]
  \begin{center}
    \label{tab:table1}
    \begin{tabular}{c|c|c} 
      $X$ & $n_{\rm crit}$ & $\Delta{Q}$\\
      \hline
      $2$ & $9n_{0}$ & $(286.5\rm ~MeV)^{4}$\\
      $2.25$ & $6.85n_{0}$ & $(251.9\rm ~MeV)^{4}$\\
      $2.45$ & $5.35n_{0}$ & $(221.4\rm ~MeV)^{4}$\\
    \end{tabular}
    \vspace{5mm}
        \caption{Critical baryon densities and latent heats for different values of $X$. Taken from Ref. \cite{Jimenez:2017fax}.}
        \label{tab:2.1}
  \end{center}
\end{table}

The curves in the left panel of Fig. \ref{fig:MR} show the usual behavior for hybrid stars built using the Maxwell construction\footnote{This is justified since we are assuming implicitly two independent, homogeneous, and electrically neutral local phases. This is in contrast to the Gibbs/Glendenning construction \cite{Glendenning:2000}, where these conditions are imposed globally in a non-trivial way. Additionally, low enough values for the (unknown) surface tension are assumed in order to allow the formation of a mixed phase \cite{Palhares:2010be}.}. Besides, a closer look at this diagram near the maximum mass sector is shown in the right panel of Fig. \ref{fig:MR} allowing us to qualitatively analyze the \textit{dynamical stability} of these stellar configurations. From this figure one can clearly see the kinks (placed before or after the $M_{\rm max}$, depending on $X$) which mark the onset to instability with respect to gravitational collapse, being followed by a decreasing-mass/radius line indicating the set of unstable hybrid stars. This happens because the mixed (Maxwell) phase allow us to deal with stars of higher and higher central energy densities, i.e. more gravity, but while keeping a constant pressure, i.e. no corresponding (repulsive) pressure to balance the hydrostatic equilibrium, thus producing instability against the collapse \cite{Schramm:2015hba}. This embodies the fact that physically it is not completely acceptable to have the hadronic and quark phases sharply separated by an energy density barrier. Thus, we note that using the Gibbs/Glendenning construction allow us to have increasing values of energy density and pressure within the mixed phase, then the star having a mixture of quarks and baryons present in the core of the star, which could be seen in the mass-radius diagram as a softening of the kink but with a mild reduction of the maximum mass hybrid star \cite{Wei:2018mxy}. The case $X=2.73$ is characteristic of stars having a very small mixed phase, being stable only before the first turning point but not reaching the maximal mass neutron star (see Chapter 3 for more details on these matters for pure stars and, e.g. Ref. \cite{Alford:2013aca} for hybrid stars). One should be aware that these stability and thermodynamic construction issues are, of course, model dependent.

\begin{figure*}[h]
\begin{center}
\hbox{\includegraphics[width=0.5\textwidth]{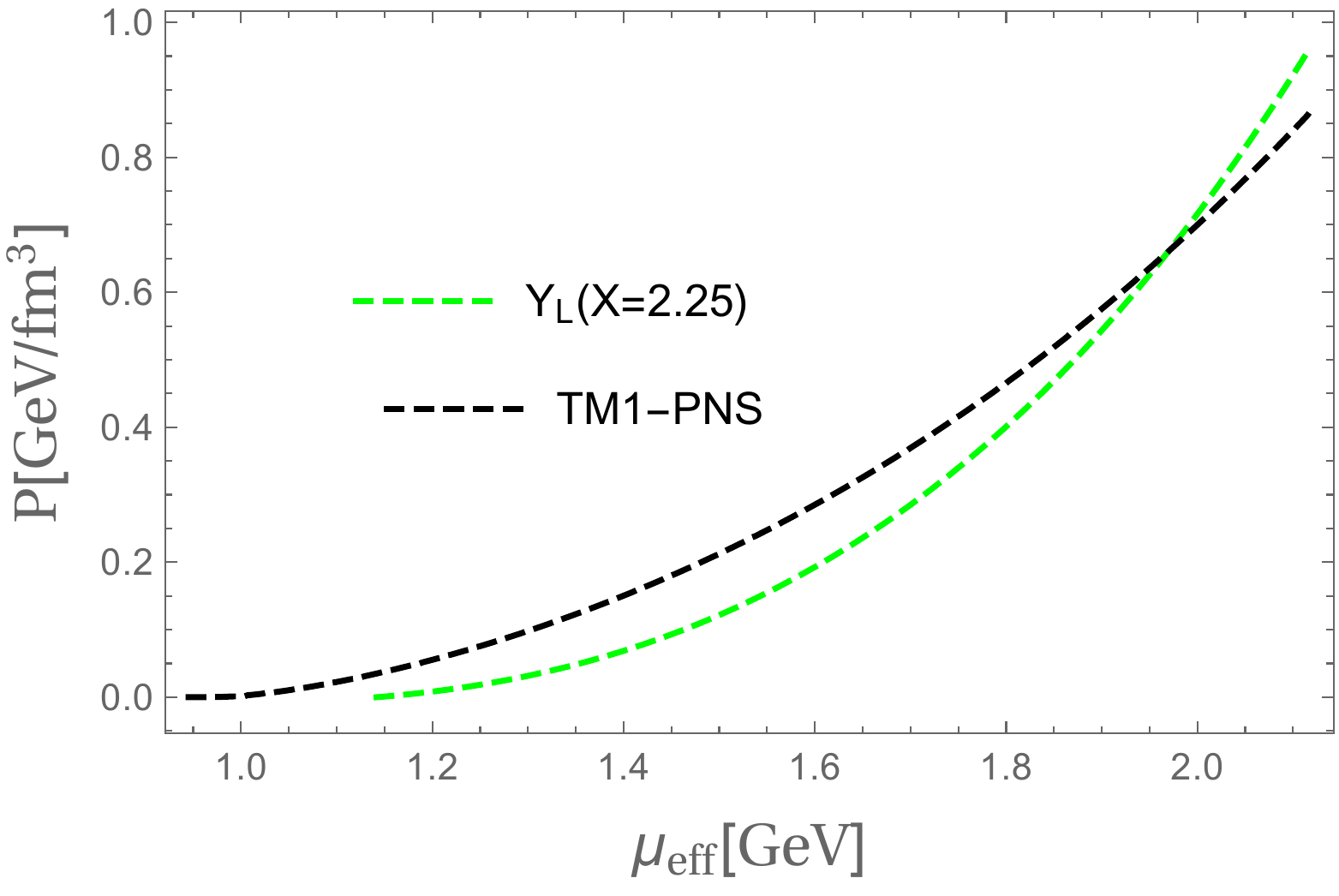}
	  \includegraphics[width=0.5\textwidth]{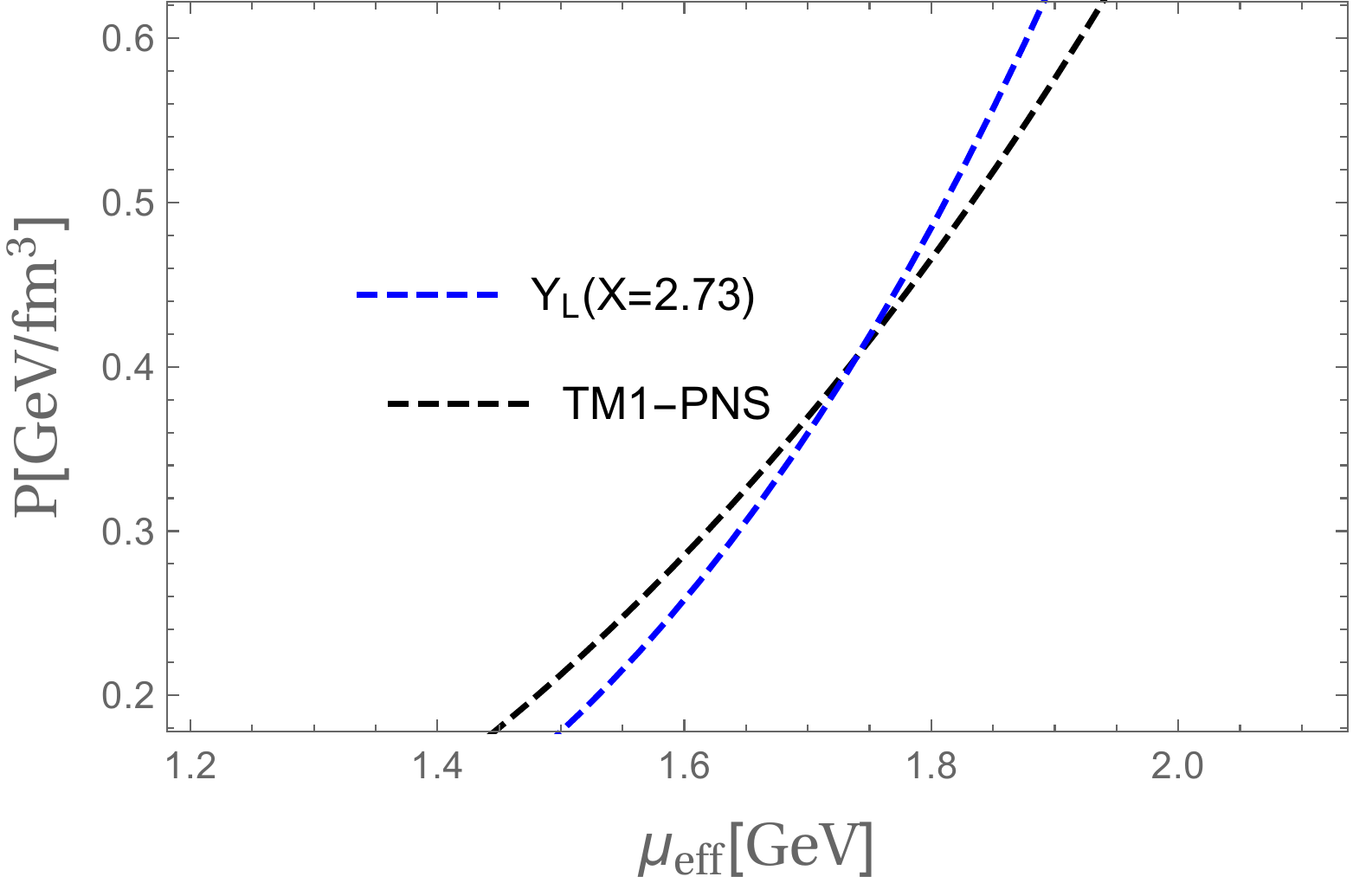}}
\vspace{5mm}
\caption{\label{fig:PmuEff} Left panel: Total pressure for lepton-rich quark matter and lepton-rich hadronic EoS TM1-PNS that match at some critical $\mu_{\rm eff}$ and from it allows the nucleation of unpaired quark matter that is still consistent with measurements of two-solar mass pulsars. Right panel: The same but for a different value of $X$. It can be seen that changing X changes the critical effective chemical potential considerably.}
\end{center}
\end{figure*}
%

\subsection{Nucleation to unpaired quark matter}

We turn our attention back to the early post-bounce state of core-collapse supernovae producing  protoneutron star (PNS) matter and the question of timescales for the nucleation of quark matter in the core of a PNS. In this situation, neutrinos are explicitly trapped, and one should use the complete lepton-rich pQCD EoS satisfying the constraints mentioned above.
\begin{figure*}[h]
\begin{center}
\hbox{\includegraphics[width=0.5\textwidth]{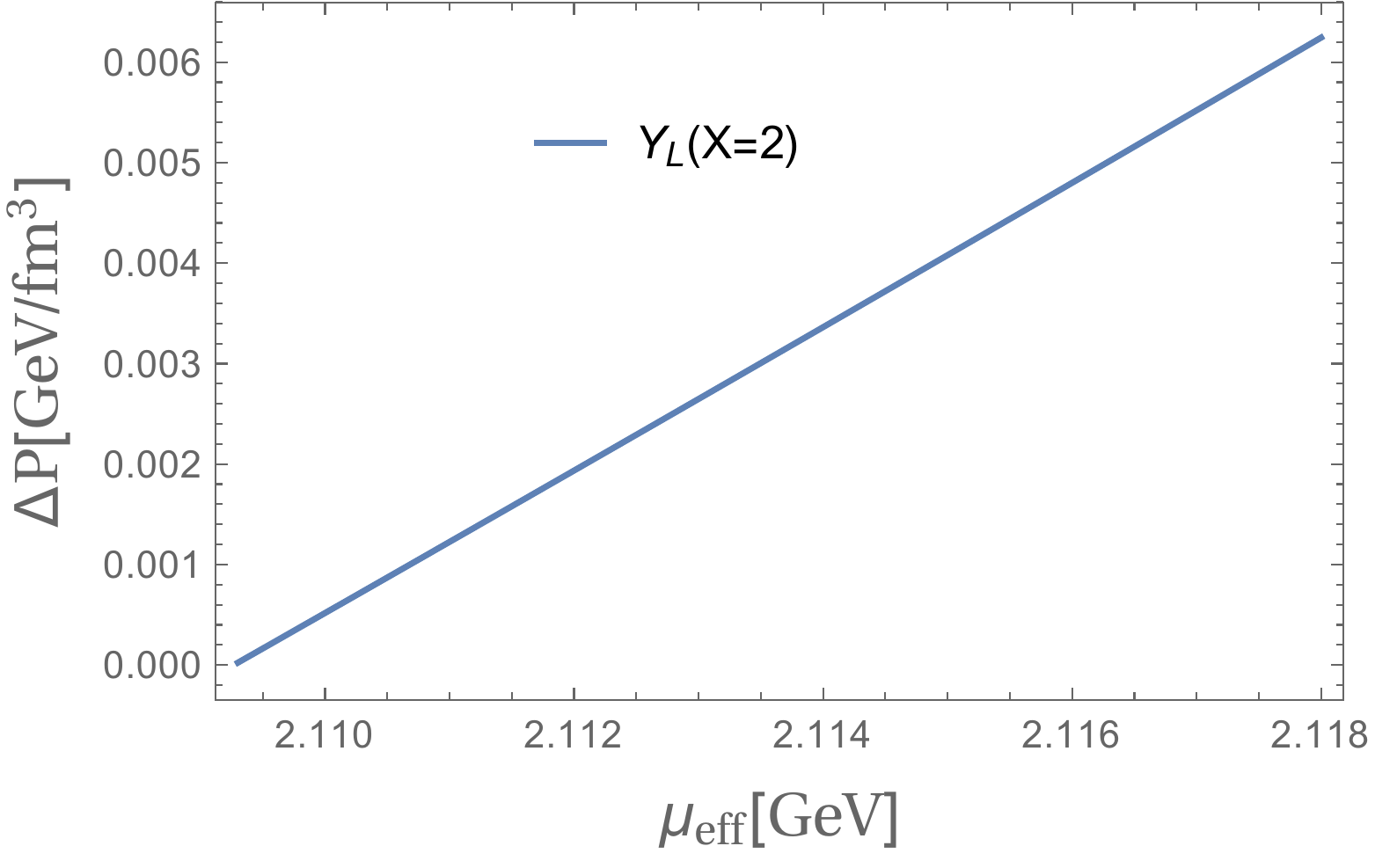}
	  \includegraphics[width=0.5\textwidth]{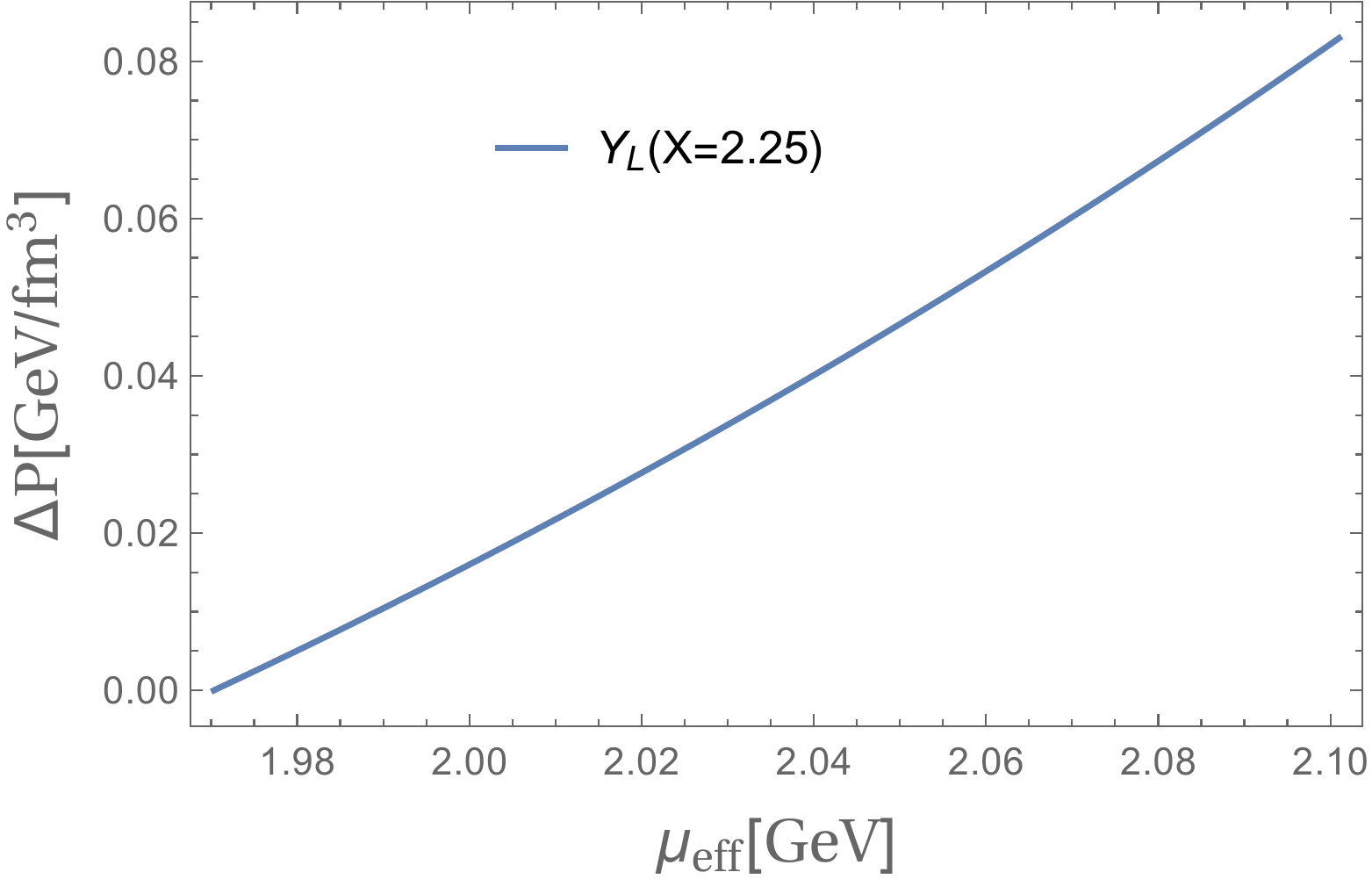}}
\vspace{5mm}
\caption{Pressure differences between phases as function of the effective chemical potential for $X=2$ and $X=2.25$. These will enter into the nucleation rates and then will define the allowed values of the surface tensions required for the hadron-quark transition.}
\label{fig:dPmuX2X225}
\end{center}
\end{figure*}
In Figs. \ref{fig:2normPlepton} and \ref{fig:PmuEff} we illustrate the matching of a few cases of the lepton-rich pQCD EoS with $X\in[2, 2.73]$ onto the TM1-PNS EoS. One can see that, for $X=2.73$, the phase transition is $\textit{not}$ soft anymore, in contrast to the lepton-poor case. Nevertheless, it occurs at a critical density which is still not very high. Something analogous happens for the other values of the renormalization scale $X$ displayed, so that one can conclude  that the presence of neutrinos shifts the critical densities towards larger values, turning $\textit{weak}$ first-order transitions (in deleptonized dense matter) into $\textit{strong}$ first-order transition in the lepton-rich case.
\begin{figure*}[h]
\begin{center}
\hbox{\includegraphics[width=0.5\textwidth]{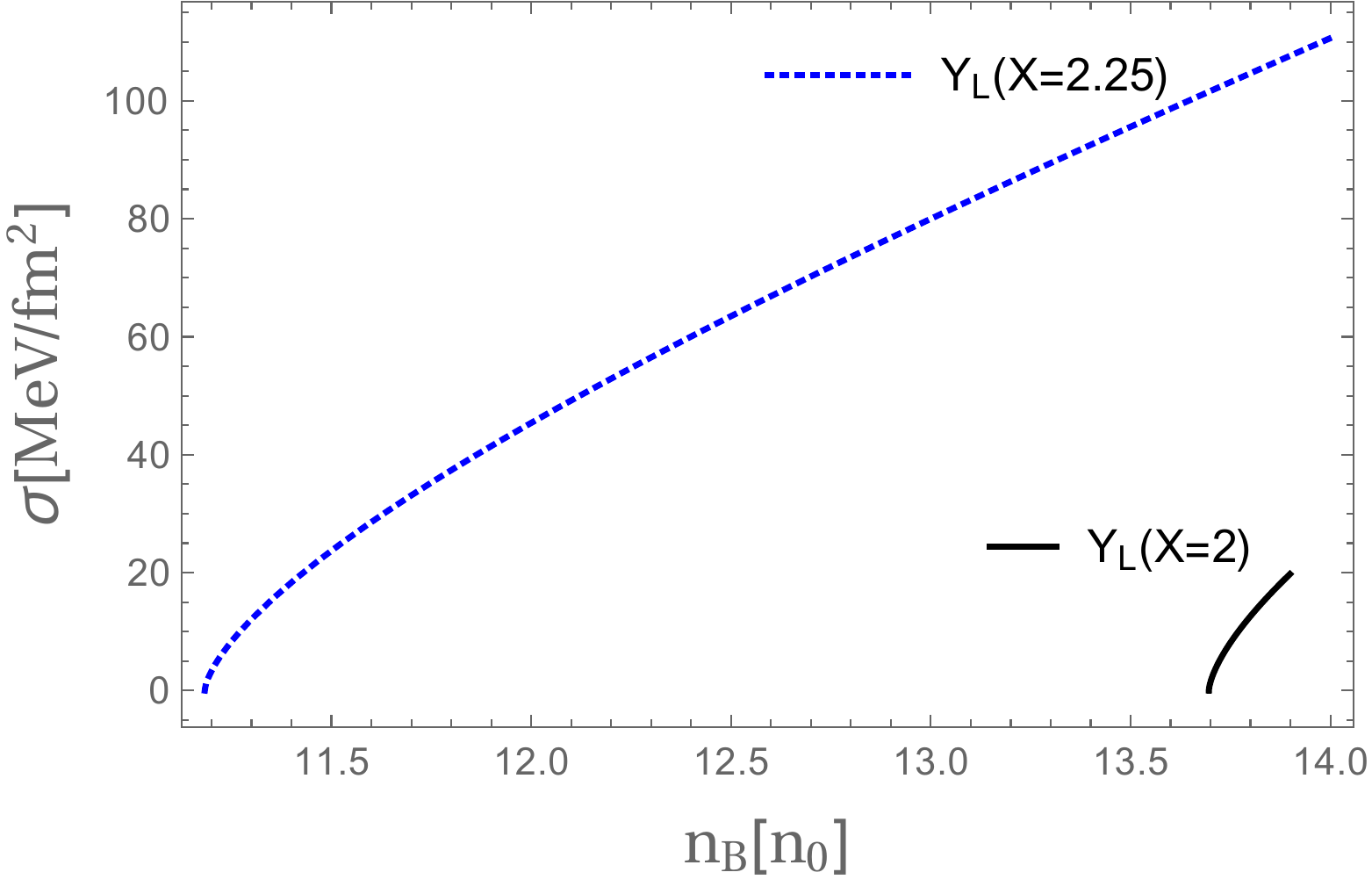}
	  \includegraphics[width=0.5\textwidth]{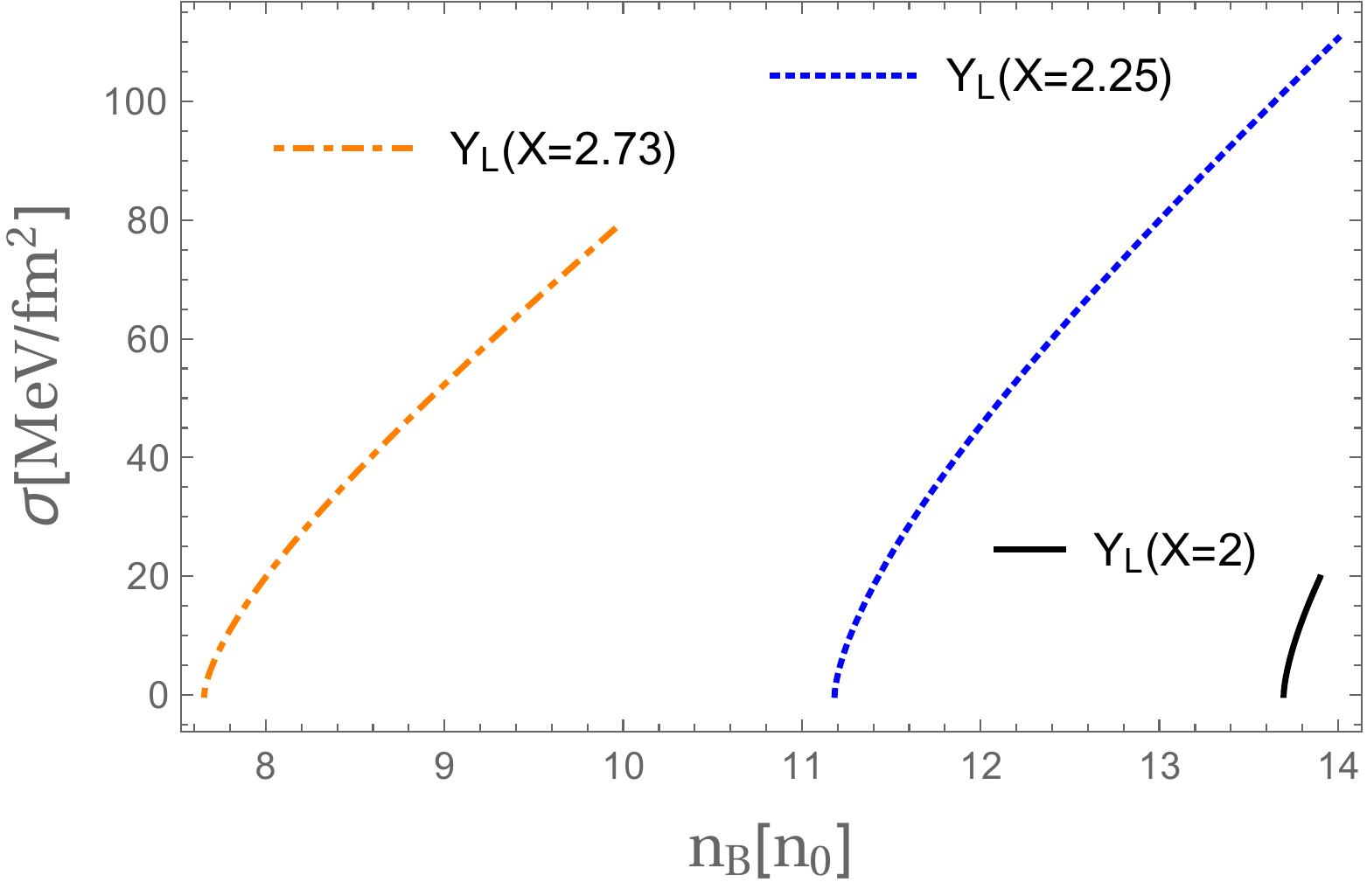}}
\vspace{5mm}
\caption{\label{fig:surfacetension} Lines of constant nucleation time $\tau_{\rm nucl}=\tau_{\rm PNS}=100 \rm ~ms$ for a PNS in terms of the surface tension and the baryon density for different values of $X$. Right panel taken from Ref. \cite{Jimenez:2017fax}.}
\end{center}
\end{figure*}
For nucleation to be effective, we consider that its typical time scale is (approximately) equal to the lifetime of the PNS matter, i.e., $\tau_{\rm nucl}=\tau_{\rm PNS}=100 \rm ~ms$, being taken as an underestimate. Then, following the procedure of Ref. \cite{Mintz:2009ay}, we can make a contour plot for different values of surface tension\footnote{For comparison, the droplets of liquid iron in vacuum have $\sigma_{\rm Fe}\sim10^{-19}\rm MeV/fm^{2}$ at $T\sim2000$K \cite{Fraser:1971def}. As expected, temperature and density effects change this situation by many orders of magnitude.}, $\sigma$, and baryon density, using Eqs. (\ref{eq:gamma}) and (\ref{eq:tau_nucl}). These results are shown in Figs. \ref{fig:dPmuX2X225}, \ref{fig:surfacetension} and \ref{fig:NuclTimeXall} for different values for the renormalization scale, namely  $X=2$, $2.25$ and $2.73$, which gives us qualitatively similar behaviors for the rising of surface tension with baryon density for different values of the critical baryon density.
\begin{figure*}[h!]
\begin{center}
\hbox{\includegraphics[width=0.5\textwidth]{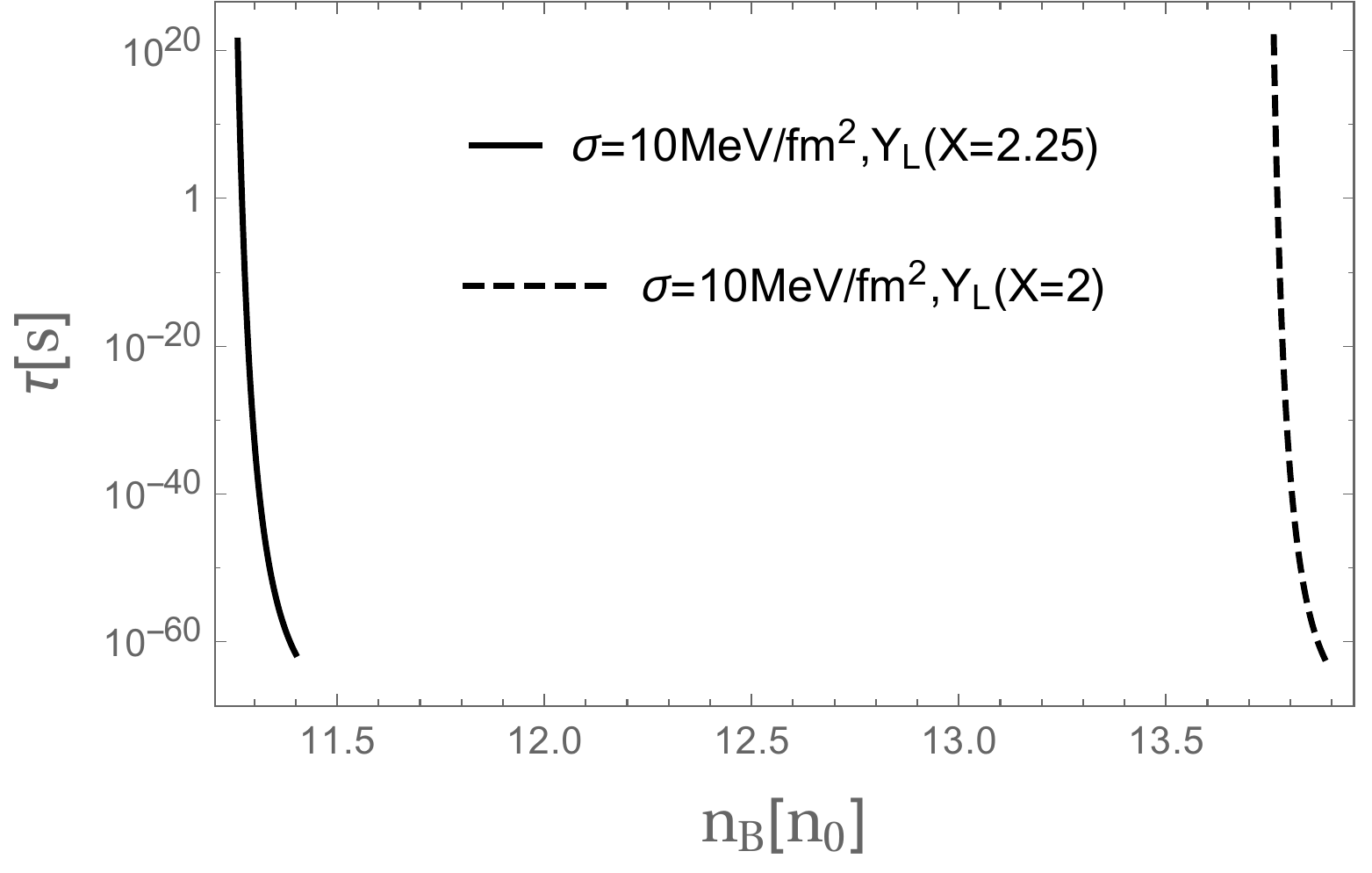}
	  \includegraphics[width=0.5\textwidth]{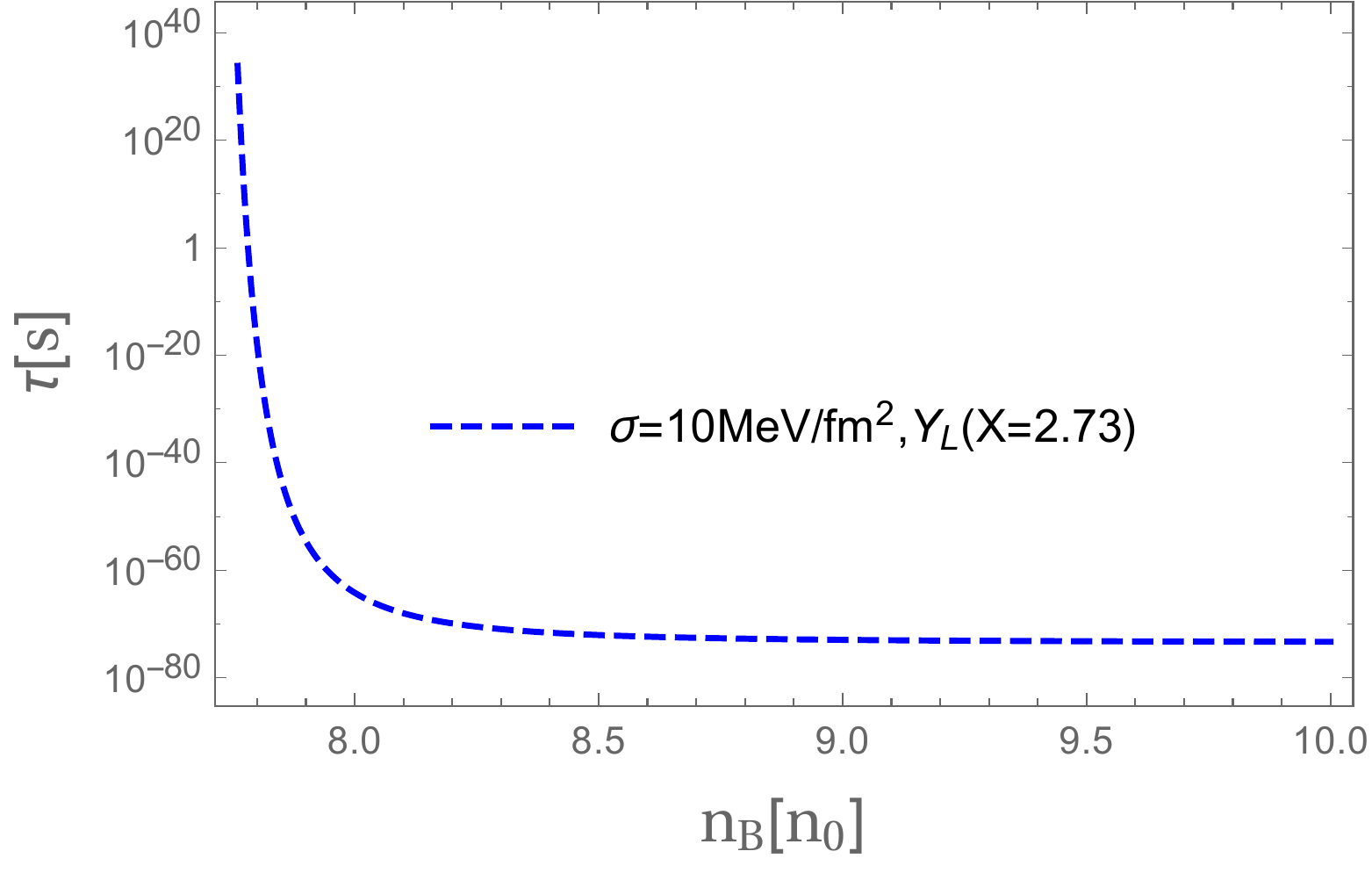}}
\vspace{5mm}
\caption{Nucleation times for different values of the renormalization scale X in lepton-rich quark matter $Y_{L}$ with a fixed typical (low) value of the surface tension between hadronic and phases. Notice that the age of the Universe is $\sim~4.4\times{10}^{17}$ seconds.}
\label{fig:NuclTimeXall}
\end{center}
\end{figure*}
As we go to higher values of $n_{B}$, there is a smaller window in the parameter space for nucleation of unpaired quark matter droplets, requiring lower and lower values of the surface tension. The corresponding values for the critical baryon density to form droplets of quark matter and latent heat released at the transition are shown in Table \ref{tab:2.2}. Notice that the presence of trapped neutrinos makes the critical densities noticeably higher (cf. Table {tab:2.1}), which is in line with previous results of Ref. \cite{Lugones:1997gg}. Besides, we note that from all these results one can deduce that although the case $X=2.73$ is interesting in the lepton-rich case having the more-or-less low value of critical baryon density ($\approx{~8~}n_{0}$), it is unlikely that it will survive as a hybrid neutron star with a quark core (satisfying the two-solar mass constraint) after deleptonization since we have proven that it is mostly unstable against collapse in the neighbourhood of the maximum mass and having to surpass a large barrier of energy density within the PNS when the phase conversion dynamics occurs which introduce additional instabilities. Thus, great care must be taken when using the Maxwell construction (even its modified version used here) in PNSs and it also should serve us to further restrict the values of $X$ by comparing their phase conversion stability in the lepton-rich case against their gravitational stability in the lepton-poor case.

Furthermore, the latent heat $\Delta{Q^{\nu}}$ is related to the second neutrino burst to be found in supernovae explosions in the case of a QCD transition, as proposed in Ref. \cite{Sagert:2008ka}. Moreover, the difference between latent heats in the lepton-rich and deleptonized cases, $\Delta{Q^{\nu}}-\Delta{Q}$, can be interpreted as the energy taken away in the form of neutrino emission during the deleptonization phase.

\begin{table}[h!]
  \begin{center}
    \label{tab:table2}
    \begin{tabular}{c|c|c} 
      $X$ & $n_{\rm crit}$ & $\Delta{Q^{\nu}}$\\
      \hline
      $2$ & $13.7 n_{0}$ & $(325.7\rm ~MeV)^{4}$\\
      $2.25$ & $11.2 n_{0}$ & $(297.4\rm ~MeV)^{4}$\\
      $2.73$ & $7.66 n_{0}$ & $(246.1\rm ~MeV)^{4}$\\
    \end{tabular}
    \vspace{5mm}
        \caption{Critical baryon densities and latent heats for different values of $X$ in the lepton-rich case. Taken from Ref. \cite{Jimenez:2017fax}.}
        \label{tab:2.2}
  \end{center}
\end{table}

\section{Concluding remarks} 
	\label{sec:2conclusion}

In this chapter we have investigated protoneutron star matter using the state-of-the-art perturbative equation of state for cold and dense QCD in the presence of a fixed lepton fraction in which both electrons and neutrinos are included. Finite-temperature effects in the EoS can be neglected since they have a minor effect in the PNS scenario at hand. Even if the presence of neutrinos does not modify appreciably the EoS at low densities, their presence significantly increases the pressure as one goes to higher densities, within the region that is relevant for the physics of PNS. We note that (as in Ref. \cite{Fraga:2004gz}), this perturbative lepton-rich pQCD+ EoS can be considered reliable at high densities and must be matched onto a hadronic EoS at lower densities to build a realistic framework for dense matter. 

	Besides computing the modifications in the equation of state due to the presence of trapped neutrinos, we have shown that stable strange quark matter is less favorable in this environment, i.e. the parameter space for the formation of strange quark matter with neutrinos decreases.

	In order to estimate the odds of nucleating unpaired quark matter in the core of protoneutron stars, we had to match the lepton-rich QCD pressure onto a hadronic equation of state, namely TM1 with trapped neutrinos. In doing so, we found that neutrinos make the deconfinement transition from nuclear matter to quark matter more $\textit{difficult}$, in line with previous results that use simplified models for the quark matter sector \cite{Lugones:1997gg,Benvenuto:1999uk,Menezes:2003pa}. However, whereas the previous descriptions require very high values of the  critical density, $n_{\rm crit} \sim 10 n_{0}$, so that the maximum masses of PNS surpass the current limit of two solar masses, the description using lepton-rich pQCD needs a critical density $n_{\rm crit} \sim 3 n_{0}$ to produce deleptonized stable hybrid stars satisfying the observational constraints.
	
	Additionally, the high dependency and sensitiveness of our results, in particular, the nucleation rates\footnote{This $\Gamma$ is also very sensitive to a slight changes in the free parameters of the nucleation theory due to their highly complicated functional form, so the conclusions extracted from it must be taken with care.} and contour-line surface tensions, to changes in the renormalization scale parameter $\bar{\Lambda}$ (inherited from perturbative QCD) can be seen as a measure of the uncertainty in the observables computed at low densities where pQCD is no longer reliable and the error band increases significantly. On the other hand, another results, like the strange quark matter hypothesis in lepton-rich conditions, display a robust behavior almost independent of the chosen $\bar{\Lambda}$'s.
	    
\end{chapter}
\begin{chapter}{Adiabatic Pulsations of Quark Stars}
	\label{chap:radialQS}

\hspace{5 mm}

\section{Introduction}

As every macroscopic object in the Universe, compact stars can oscillate in many different ways mainly as a whole e.g. radially, or in localized regions e.g. starquakes. Humans are familiarized with the latter in the form of earthquakes produced by localized perturbations of the Earth's interior crust, a phenomenon well known in geology. In particular, along the years many pulsar observations have shown that neutron stars (NS) must suffer different kinds of mechanical deformations\footnote{Neutron stars are good resonators, where different oscillation modes can be excited, e.g. the fundamental ($f$), pressure ($p$), gravity (g), and Rossby ($r$) modes. In particular, the fundamental period of their radial oscillations usually vary between $\sim(0.5-1)~$ milliseconds, depending on the EoS \cite{Haensel:2007yy}.}, e.g. radial and non-radial oscillations, rotation and glitches, along their lifetime in order to reproduce radio, X-ray, gamma-ray and other electromagnetic signatures \cite{Shapiro:1983,Glendenning:2000,Haensel:2007yy,Benhar:2004xg,Doneva:2015jba,Haskell:2015jra,Harding:2006qn,Watts:2006hk}. Recently, the LIGO and Virgo observatories measured gravitational waves coming from the merger of neutron stars, the GW170817 event, opening a new window to probe NS responses to some of the disturbances produced by tidal deformations in the inspiral phase \cite{TheLIGOScientific:2017qsa}. Besides the usual constraints -- e.g.  the existence of $\sim 2 M_\odot $ neutron stars \cite{Demorest:2010bx,Antoniadis:2013pzd,Kurkela:2014vha,Buballa:2014jta} and tidal deformabillities \cite{Annala:2017llu,Rezzolla:2017aly,Most:2018hfd,Abbott:2018exr} --, mechanical responses could potentially provide indications of the presence of quark matter in their cores \cite{Most:2018eaw,Wei:2018tts,Tanimoto:2019tsl,Annala:2019puf} or indicate the presence of strange quark stars \cite{Bauswein:2008gx,Bauswein:2009im}. 
	
In particular, the radial oscillation modes give information about the stability of the stellar model under consideration, as first investigated by S. Chandrasekhar in 1964 \cite{Chandrasekhar:1964zza}, being applied initially only to polytropic equations of state (EoSs), and only much later to more realistic nuclear EoSs \cite{Glass:1983}. This kind of oscillation is also the simplest of these categories since they are relatively easy to solve numerically its associated eigenvalue problem. Many radial oscillation modes were calculated for modern sets of EoSs for cold NS by Kokkotas and Ruoff \cite{Kokkotas:2000up}. Besides, in the absence of dissipative processes, the oscillation spectrum of a stable stellar model forms a complete set. Thus, it is possible to describe any arbitrary periodic radial motion of a neutron star as a superposition of its various eigenfunctions \cite{Kokkotas:2000up}.	
	
On the other hand, the radial pulsations of quark stars were analyzed mostly using the MIT bag model to build the equation of state for cold quark matter \cite{Benvenuto:1989kc,Benvenuto:1991,Vath:1992,Lugones:1995vg,Benvenuto:1998tx,VasquezFlores:2010eq}, in some cases adding constant corrections to the strange quark mass and interactions which behave effectively only as being of long range. Results suggested that the periods of the fundamental mode were very low to be detected \cite{Benvenuto:1991jz,Horvath:1991pv}, which motivated the search for the so-called non-radial oscillations that would have higher periods which could be measured through gravitational wave observations \cite{Flores:2013yqa,Flores:2018pnn}. We note also that since neutron star atmospheres also experience radial oscillations, which in turn are related to the luminosity measurements of pulsars at different frequencies of emission, they can also be used to obtain indirectly the values of the masses and radii, thus allowing us to draw diagrams mixing all these theoretical and observational stellar structure properties\footnote{However, for these purposes one would have to use non-adiabatic coefficients to model oscillating stellar matter consistently since the system is loosing energy by emitting electromagnetic waves, which is marked contrast with the formalism developed in this chapter.}.

It seems more appropriate to study the radial pulsations of quark stars by employing an equation of state for cold quark matter extracted from perturbative quantum chromodynamics (pQCD). It has a long story \cite{Freedman:1976ub,Freedman:1977gz,Baluni:1977ms,Toimela:1984xy,
Blaizot:2000fc,Fraga:2001id,Fraga:2001xc,Fraga:2004gz,Kurkela:2009gj,Fraga:2013qra} and, although its realm of validity corresponds to much higher densities, it is relevant in modelling the equation of state of compact stars, since QCD short-range interactions become important at intermediate densities, reachable in the interior of NS \cite{Fraga:2001id,Fraga:2001xc,Fraga:2004gz,Kurkela:2009gj,Fraga:2013qra}.   
This approach consistently incorporates the effects of interactions and includes a built-in estimate of the inherent systematic uncertainties in the evaluation of the equation of state, so that in several cases we present bands instead of lines. This goes beyond the misleading precision of the MIT bag model description, providing a more realistic range of possibilities, in a framework that can be systematically improved.

In this chapter, we perform the general relativistic stability analysis against adiabatic radial oscillations of unpaired quark stars, then obtaining their pulsation frequencies and periods coming from possible radial perturbations occurring at different stages of the pulsar's lifetime\footnote{Compact star oscillations can happen due to several reasons, e.g. accretion from a partner in a binary system or due to impact of interstellar objects as asteroids or comets \cite{Geng:2015vza,Dai:2016qem}.}. This chapter follows the discussion found in Ref. \cite{Jimenez:2019iuc}. So, in Sec. \ref{sec:3stability} we set up the framework of Ref. \cite{Gondek:1997fd}, where the original Sturm-Liouville form is turned into a pair of first-order coupled differential equations, and we review the pocket FKV formula cast in Ref. \cite{Fraga:2013qra} for the equation of state for cold quark matter from perturbative QCD of Ref. \cite{Kurkela:2009gj}. In Section \ref{sec:3radQCD} we present our results for the stability of quark stars for which we also take into account the constraints imposed by the recent gravitational wave event GW170817 to compact star masses and radii \cite{Rezzolla:2017aly,Most:2018hfd} to compute the fundamental and first excited mode frequencies ($n=0,1$) of unpaired quark stars, and restrict their vibrational spectrum. Section \ref{sec:3conclusion} presents our conclusions.

\section{Stellar Structure and Stability Analysis}
	\label{sec:3stability}

It is a well-known fact in physics that \textit{mechanical} equilibrium does not assure its \textit{dynamical} counterpart. In the preceding chapter we have seen that the TOV equations (which allow us to find the stellar structure of compact stars, e.g. mass, radius and inertia moment) only produce compact stellar configurations satisfying the former condition, i.e. in \textit{hydrostatic} equilibrium. In other words, the hydrostatic equilibrium does not ensure \textit{stability}. This happens because equilibrium configurations correspond to a maximum or to a minimum  in the energy with respect to radial compression or dilation, respectively \cite{Shapiro:1983}. This lead us to look for a systematic method to find the set of hydrostatic-equilibrated stellar configurations that are at the same time stable ones \cite{Glendenning:2000}.

In this section we summarize the main concepts behind the stability analysis of stellar configurations in hydrostatic equilibrium by performing linear radial perturbations in compact stars\footnote{We note that this problem can also be formulated in several other ways \cite{Shapiro:1983du}.}. Notice that non-linear radial oscillations could be important only for stellar configurations around the maximum mass, producing unstable modes. However, these fall out of the scope of this chapter. For more details, see Ref. \cite{Gourgoulhon:1995jjj}.

\subsection{The \textit{unusual} Tolman-Oppenheimer-Volkov Equation}

In the preceding chapter we have obtained the \textit{usual} pair of Tolman-Oppenheimer-Volkov (TOV) equations for stellar configurations in hydrostatic equilibrium,
  \begin{equation}
  \frac{dP}{dr}=-\frac{\epsilon\mathcal{M}}{r^{2}}\left(1+\frac{P}{\epsilon}\right)\left(1+\frac{4\pi{r^3}{P}}{\mathcal{M}}\right)
  \left(1-\frac{2\mathcal{M}}{r}\right)^{-1} \; ,
   \label{TOV1}
  \end{equation}
  \begin{equation}
  \frac{d\mathcal{M}}{dr}=4\pi{r}^{2}\epsilon \; ,
  \label{TOV2}
  \end{equation}
being $\mathcal{M}$ is the gravitational mass inside the radius $r$, and where boundary conditions must be imposed to solve these equations consistently. However, for the purposes of this chapter we should introduce the \textit{unusual} TOV equation given by
  \begin{equation}
  \frac{d\nu}{dr}=-\frac{2}{P+\epsilon}\frac{dP}{dr} \; ,
  \label{TOV3}
  \end{equation}
which is needed later for the pulsation equations. In order to solve Eq. (\ref{TOV3}) for $\nu$, we use the boundary condition\footnote{If the three TOV equations are solved simultaneously, one only needs boundary conditions at the origin and in this case one should impose $\nu(r=0)\equiv{\nu_{0}}$, where $\nu_{0}$ is any constant, e.g. 0 or 1. Then, $\nu(r)$ has to be adjusted by adding an overall constant that satisfied Eq. (\ref{TOV3bc}) at the surface.} when Eq. (\ref{TOV3}) is solved independently from the \textit{usual} TOV equations (i.e. having at hand $P(r)$ and $\epsilon(r)$)
\begin{equation}
 \nu(r=R)=\ln\left(1-\frac{2M}{R}\right),
 \label{TOV3bc}
\end{equation}  
which ensures that this metric function $\nu(r)$ will match continuously the Schwarzschild metric outside the star ($r>R$)
\begin{equation}
ds^{2}=\left(1-\frac{2M}{R}\right)dt^{2}-\left(1-\frac{2M}{R}\right)^{-1}dr^{2}-r^{2}(d\theta^{2}+\sin^{2}\theta{d\phi^{2}}),
\end{equation}
in agreement with Birkhoff's theorem \cite{Shapiro:1983}. Second, the solution for the metric function $\lambda(r)$ can be written immediately in terms of Eq. (\ref{TOV2}) giving
\begin{equation}
\lambda(r)=-\ln\left(1-\frac{2\mathcal{M}(r)}{r}\right),
\end{equation}
and which will enter as a fundamental ingredient as follows.

\subsection{Radial Pulsations of Relativistic Stars}

We now set up the framework of stellar pulsations. The system to be analyzed is a sphere (the star) made up of a perfect fluid (nuclear, quark matter, or a mixed phase made up of both) pulsating radially with very small amplitude. For these pulsations to be studied, one begins by using the equations governing the equilibrium configuration about which the star pulsates, i.e., the TOV equations.
Then, one must use an appropriate coordinate system for the vibrating sphere that reduces to the Schwarzschild static metric functions $\nu_{0}(r)$ and $\lambda_{0}(r)$ of the sphere at equilibrium (where the subscript ``0'' indicates the unperturbed configuration) for the zero pulsation amplitude. This is reached by using to the Schwarzschild dynamical metric functions $\nu(t,r)$ and $\lambda(t,r)$ where the time dependences are included. Physically, all this process preserves the spherical symmetry of the unperturbed star before and after the perturbation.

Then, one must perform perturbative expansions (up to first order, i.e. keeping only linear terms) in the pulsation functions (since they are assumed to be very small by construction), i.e. the metric functions become \cite{Misner:1974qy}
\begin{equation}
\nu(t,r)=\nu_{0}(r)+\delta\nu(t,r),\hspace{0.5cm}\lambda(t,r)=\lambda_{0}(r)+\delta\lambda(t,r),
\end{equation}
and the perfect-fluid components, i.e. pressure, energy density and number density, respectively, turn out to be \cite{Misner:1974qy}
\begin{equation}
P(t,r)=P_{0}(r)+\delta{P}(t,r),\hspace{0.4cm}\epsilon(t,r)=\epsilon_{0}(r)+\delta\epsilon(t,r),\hspace{0.4cm}
n(t,r)=n_{0}(r)+\delta{n}(t,r).
\end{equation}
Notice that all these metric and thermodynamic functions of the perturbed system ($P$, $\epsilon$, and $n$) are slightly shifted from their unperturbed values ($P_{0}$, $\epsilon_{0}$, and $n_{0}$) as measured in the fluid's rest frame so that these perturbations are performed at fixed coordinate locations, i.e. as \textit{Eulerian perturbations}. Notice that perturbations like $\epsilon=\epsilon_{0}+(\delta\epsilon)
e^{i\omega{t}+i\vec{k}\dot\vec{x}}$ also appear when studying the stability and causality of relativistic hydrodynamic equations where, for instance, it is proved that the viscous Navier-Stokes equations are non-causal nor stable in the relativistic limit \cite{Denicol:2008ha,Pu:2009fj}.

Finally, the radial \textit{Lagrangian} displacement\footnote{In general, small Lagrangian displacements $\xi^{i}(r,\theta,\phi,t)$ ($i=1,2,3$) of fluid elements in a perturbed star induces small changes $h_{ij}(r,\theta,\phi,t)$ in the metric. Usually one factors out the angular dependence on $\theta$, $\phi$ via spherical harmonics $Y_{lm}(\theta,\phi)$. For instance, small radial pulsations represent the case $l=m=0$, $\xi^{\theta}=\xi^{\phi}=0$, $\xi^{r}=\xi(r,t){~\equiv~}\Delta{r}(r,t)$; having the same complications in general relativity as in Newtonian theory \cite{Haensel:2007yy}. Besides, they do not affect the gravitational field in vacuum outside the star, i.e. $h_{ij}=0$. On the contrary, non-radial pulsations with $l>0$ are more complicated for relativistic stars \cite{Thorne:1967cam}.} of a fluid element from its equilibrium position located at the coordinate radius $r$ is displaced to the coordinate radius $r+\Delta{r}(r,t)$ at coordinate time $t$ in the vibrating configuration, where
\begin{equation}
\Delta{r}(t,r)=\chi(r)\exp(i\omega{t}),
\end{equation}
being $\chi(r)$ the purely radial part of the Lagrangian perturbation and the harmonic time dependence is assumed with ``$\omega$'' the oscillation angular frequency. Notice that this harmonic dependence is also assumed in the other perturbations of the metric functions and thermodynamic functions even if we do not write it explicitly. 

The equation of motion for these perturbations are obtained after introducing $\nu(t,r)$, $\lambda(t,r)$ into Einstein's equations, that must be rearranged with the help of the conservation laws for the perfect-fluid energy momentum tensor ($P(t,r)$, $\epsilon(t,r)$) and baryon number density current ($nu^{\mu}$) \cite{Misner:1974qy},
\begin{equation}
\nabla_{\mu}T^{\mu\nu}=0,\hspace{0.5cm}\nabla_{\mu}(nu^{\mu})=0,
\end{equation}
where $\nabla_{\mu}$ is the covariant derivative. After many algebraic manipulations and neglecting nonlinear terms, one is able to get a set of equations governing the evolution of the \textit{Lagrangian perturbation functions} $\Delta{r}$ and $\Delta{P}$, where their relation with the Eulerian (right-hand side) is given by displacements \cite{Misner:1974qy}
\begin{equation}
\nonumber \Delta{P}(t,r){~\equiv~}\delta{P}+
\frac{dP_{0}}{dr}\Delta{r},
\end{equation}
\begin{equation}
\Delta{\epsilon}(t,r){~\equiv~}\delta{\epsilon}+
\frac{d\epsilon_{0}}{dr}\Delta{r},
\end{equation}
\begin{equation}
\nonumber \Delta{n}(t,r){~\equiv~}\delta{n}+\frac{dn_{0}}{dr}\Delta{r},
\end{equation}
must be used along the process. 

\subsection*{The Chandrasekhar viewpoint: Second-order equation}

 Historically, it was S. Chandrasekhar \cite{Chandrasekhar:1964zza} who first derived the equation governing the dynamics of radial pulsations having the generic form

\begin{equation*}
\frac{\partial^{2}}{\partial{t}^{2}}\Delta{r}(r,t)=-\hat{\mathcal{H}}\Delta{r}(r,t),
\end{equation*}

where $\hat{\mathcal{H}}$ is a self-adjoint (Hermitian) linear second-order operator, independent of $t,\theta,\phi$, and determined by the equilibrium stellar model. Using the harmonic dependence of Eq. (3.12), we find a Sturm-Liouville problem for the (squared) eigenfrequencies $\lambda{~\equiv~}\omega^{2}$ and eigenvectors $\chi(r)$ of radial oscillations\footnote{This is analogous as in stationary quantum mechanics for the discrete eigenstates in a quantum system with Hamiltonian operator $\hat{\mathcal{H}}$.}

\begin{equation*}
\lambda\chi=\hat{\mathcal{H}}\chi,\hspace{0.5cm}\lambda{~\equiv~}\omega^{2}.
\end{equation*}

After expanding this $\hat{\mathcal{H}}$ operator and some algebra, one obtains the second-order pulsating differential equation (again in the form of a Sturm-Liouville problem) given by
\begin{equation}
\frac{d}{dr}\left(\Pi\frac{d\zeta}{dr}\right)+(Q+\omega^{2}_{n}W)\zeta=0
\end{equation}
where $\zeta=r^{2}\chi\exp(-{\nu}/{2})$ is the \textit{renormalized displacement function}\cite{Misner:1974qy}, and the radial function $\Pi$, $Q$, and $W$ are defined by

\begin{equation}
\nonumber \Pi=\Gamma\frac{P}{r^{2}}\exp\left(\frac{\lambda+3\nu}{2}\right),
\end{equation}
\begin{equation}
Q=\frac{1}{r^{2}}\left[\frac{1}{P+\epsilon}\left(\frac{dP}{dr}\right)^{2}-\frac{4}{r}\frac{dP}{dr}-8\pi(P+\epsilon)Pe^{\lambda}\right]
e^{{(\lambda+3\nu)}/{2}},
\end{equation}
\begin{equation}
\nonumber W=\frac{P+\epsilon}{r^{2}}\exp\left(\frac{3\lambda+\nu}{2}\right),
\end{equation}
being $\Gamma$ the adiabatic index related to the EoS, and the boundary conditions to solve this equation are
\begin{equation}
\zeta(r=0)=0,\hspace{0.5cm}\delta{P}(r=R)=0,
\end{equation}
where the Eulerian perturbation of the pressure is given by
\begin{equation}
\delta{P}=-\frac{dP}{dr}\frac{e^{\nu/2}\zeta}{r^{2}}-\frac{\Gamma{P}}{r^{2}}e^{\nu/2}\frac{d\zeta}{dr}.
\end{equation}
Then, all that remains is to solve this equation which yields the eigenvalues $\omega^{2}_{n}$ and the eigenfunctions $\zeta_{n}$ for the radial perturbations labeled by the index $n=0,1,2,...$ which enumerates the nodes of the radial function $\zeta(r)$ within the star. For instance, we will see that the fundamental mode is very simple since $\zeta(r)$ is nearly linear in $r$ and describes homologous contractions and rarefactions of the star.

An examination of these (squared) frequencies $\omega^{2}_{n}$ shows that they must obey the ordering $\omega^{2}_{0}<\omega^{2}_{1}<\omega^{2}_{2}<\cdot\cdot\cdot$. Besides, if $\omega^{2}_{n}>0$, the frequency is real, the mode is \textit{stable} and the star undergoes small-amplitude harmonic oscillations. On the other hand, if $\omega^{2}_{n}<0$, then the frequency is purely imaginary and the mode is \textit{unstable}, i.e. small displacements make the amplitude to depend upon terms like $Ae^{+|\omega|t}+Be^{-|\omega|t}$ (for some constants $A$ and $B$) which produces an indefinite increasing in the amplitude with time. For the global stability of the star, it is sufficient to look only at the fundamental (lowest) eigenvalue\footnote{Dimensional considerations allow us to estimate its values as $\omega_{0}{~\sim~}\sqrt{G\bar{\rho}}$, where $\bar{\rho}$ is the mean density of the star and it is valid for types of stars. In turn, the associated pulsation period $2\pi/\omega_{0}$ can serve as an estimate for the hydrodynamical timescale for a given star, e.g. neutron stars have $\omega_{0}{~\sim~}10$ kHz.}, $\omega^{2}_{0}$ (without nodes), since it suffers the highest instability increment, i.e. the most rapid exponential growth. Additionally, for this mode the star has \textit{neutral} stability, in other words it is neither stable nor unstable. If $\omega^{2}_{0}>0$, then all $\omega^{2}_{n}>0$ and the star is stable. If $\omega^{2}_{0}<0$, then there is at least one unstable mode and the star becomes unstable.
 
\subsection*{Adiabatic Index}

Since we already have introduced the last coefficient needed to solve the above pulsation equations, i.e. the adiabatic index $\Gamma$, now we define it quantitatively. But before doing so, we first note two of its main aspects being required to make physical sense. Firstly, this $\Gamma$-index must quantify the Lagrangian changes of pressure associated with variations of the particle number density when \textit{adiabatic} perturbations are being performed on the star, i.e. perturbations where the heat transfer between neighbouring fluid elements is negligible. Secondly, the preceding point immediately raises the question about to what extent the (static) stellar structure is affected by these pressure perturbations. In order to solve this issue we must verify if the pulsation timescale, $\tau_{\rm pul}$, is at least\footnote{If this would not be the case, the pulsations would begin to modify the stellar structure and dynamical effects could arise as, for instance, the tidal deformations producing gravitational waves \cite{Misner:1974qy}.} in the order or higher than the (hydro)dynamic timescale, $\tau_{\rm dyn}$. So, by applying the definitions given at the beginning of Chapter 2 for these timescales, it is obtained that for a quark (or neutron) star with a mass of 1.4 $M_{\odot}$ and radius of 15 km, the (hydro)dynamical timescale is $\approx{0.15}$ms, which (as we will see later in this chapter) is in the order of quark star (pulsation) periods, i.e. $\tau_{\rm pul}\sim\tau_{\rm dyn}$. Thus, the hydrostatic equilibrium is still a very good approximation when small (radial) pulsations are being applied. Another related timescale which we must compare is the so-called \textit{reaction} timescale,  $\tau_{\rm reac}$, which is the time needed to reach chemical equilibrium by means of microscopic reactions\footnote{In compact star physics, essentially the weak interaction determines this equilibrium since it lasts longer than the electromagnetic and strong interactions \cite{Haensel:2007yy}.} between the particle components, which after some time lead to thermal equilibrium\footnote{Strictly, full equilibrium is only reached partially since the chemical reactions are completed too slowly even compared with stellar scales, and, additionally, the star is still not at thermal equilibrium with the cosmic microwave background radiation surrounding it, which is manifested as the emission of photons and neutrinos from the star's crust \cite{Glendenning:2000}. Thus, we are actually in a \textit{limited} equilibrium \cite{Shapiro:1983du}.} at about 1 MeV, i.e. the cold limit on the nuclear scale \cite{Gondek:1997fd}. This issue is easily solved by remembering that in cold matter it is true that $\tau_{\rm reac}\gg\tau_{\rm dyn}(\approx{0.15}$ms), since by looking at the neutron star evolution (the PNS of Chapter 2), the beta-equilibrium is reached after the trapped neutrinos diffuse out of the star, which happens at around 15 s. This means that after performing the perturbations, the particle compositions are not modified and remain fixed\footnote{This is not so obvious since a change in density could disturb beta equilibrium. This in turn initiate weak interaction processes which would tend to move the matter toward the equilibrium on a characteristic relaxation timescale $\tau_{\rm reac}$ \cite{Haensel:2007yy}.} (or frozen, as it is usually known), i.e. constant values for the electron, $Y_{e}$, and other particle fractions, $Y_{i}$. In contrast, the case $\tau_{\rm reac}\approx\tau_{\rm dyn}$ would produce nuclear transformations (reactions) that could affect strongly the EoS (and $\Gamma$), thus complicating considerably the dynamical analysis of stability \cite{Meltzer:1966kip,Chanmugan:1971gab,Haensel:2007yy}. Then, after pointing out the above considerations, one is allowed to write the $\Gamma$-index for cold, adiabatic and isentropic (in this case zero entropy) compact star matter as (only in this equation we write explicitly constant particle fractions ``$Y_{i}$'')

\begin{equation}
\left(\frac{\partial\log{P}}{\partial\log{n}}\right)_{s=0, ~Y_{i}}{~\equiv~}\Gamma=\frac{n}{P}\frac{\Delta{P}}{\Delta{n}}.
\end{equation}		 

Besides, if one uses the EoS existence hypothesis $n=n(\epsilon,P)$, one is able to get the well-known formula (for the zero entropy and temperature limit)

\begin{equation}
\Gamma{~\equiv~}\left[\left(1+\frac{\epsilon}{P}\right)\frac{d{P}}{d{\epsilon}}\right]_{s=T=0}   \; . 
\end{equation}

For instance, for the case of the simple bag model for quark matter (to be studied later), it is easy to find the following analytic expression for this index:
\begin{equation}
\Gamma_{\rm MIT}=\frac{4}{3} \left( 1+ \frac{B}{P} \right)  \; . 
\end{equation}
However, for the FKV formula (to be also discussed later) this adiabatic index takes the form $\Gamma_{\rm FKV}=\Gamma(X_{\rm FKV})$ which turns out to be very involved analytically and must be evaluated numerically. 
 
\subsection*{The Gondek \textit{et al.} viewpoint: First-order equations} 
 
Decades after the seminal work of Chandrasekhar, it was realized by Vath and Chanmugan \cite{Vath:1992} that this 2nd order pulsation equation could be transformed into a set of two 1st order differential equations by choosing as appropriate variables $\Delta{r}/r$ and $\Delta{P}/P$. Some years later, Gondek \textit{et al.} \cite{Gondek:1997fd} found, after careful and non-trivial manipulations of the preceding equations, a convenient way to rewrite these oscillation equations for the relative radial displacement $\Delta{r}/r$ and the Lagrangian perturbation of the pressure $\Delta{P}$. This last set of equations is well adjusted to numerical techniques since one directly imposes the boundary condition at the star's surface. Moreover, these equations do not involve any derivatives of the relativistic adiabatic index, $\Gamma$, which is sensitive to the EoS being used. This is highly useful since for our later analysis we use mainly tabulated inputs, e.g. the adiabatic index and the equation of state.
 
Defining $\Delta{r}/r\equiv\xi$ and $\Delta{P}$ as the independent variables (again omitting their harmonic time dependence, ${e}^{i\omega{t}}$) for the pulsation problem, one obtains the following system of equations \cite{Gondek:1997fd}:
	\begin{equation}
	\frac{d\xi}{dr}=-\frac{1}{r}\left(3\xi+\frac{\Delta{P}}{\Gamma{P}}\right)-\frac{dP}{dr}\frac{\xi}{(P+\epsilon)} \; ,
	\label{3Rad1}
	\end{equation}
	\begin{multline}
	\frac{d\Delta{P}}{dr}=\xi\left\lbrace{\omega^{2}e^{\lambda-\nu}(P+\epsilon)r-4\frac{dP}{dr}}\right\rbrace+ \\
	\xi\left\lbrace\left(\frac{dP}{dr}\right)^{2}\frac{r}{(P+\epsilon)}-8\pi{e^{\lambda}}(P+\epsilon)Pr\right\rbrace + \\
	\Delta{P}\left\lbrace{\frac{dP}{dr}\frac{1}{P+\epsilon}-4\pi(P+\epsilon)r{e}^{\lambda}}\right\rbrace \; ,
	\label{3Rad2}
	\end{multline}	
where $\omega$ is the oscillation frequency.
	
The boundary conditions are given as follows:
	\begin{itemize}	
	 \item Physical smoothness at the center of the star requires that, when $r \to 0$, the coefficient of the $1/r$ term in Eq. (\ref{3Rad1}) must vanish. Thus, we impose that
	\begin{equation}
	(\Delta{P})_{\rm center}=-3(\xi\Gamma{P})_{\rm center}.
	\label{3BC1}
	\end{equation}

	\item Normalizing the eigenfunctions to $\xi(0)=1$ and knowing that $P(r \to R) \to 0$, we see that the Lagrangian perturbation in the pressure at the surface vanishes. Thus
	\begin{equation}
	(\Delta{P})_{\rm surface}=0.
	\label{3BC2}
	\end{equation}
	\end{itemize}
	
	In order to solve simultaneously Eqs. (\ref{3Rad1})-(\ref{3BC2}) numerically, we use the following recipe: 
	\begin{itemize}	
	\item Solve the TOV equations for the EoS to be used in the analysis to calculate the coefficients of Eqs. (\ref{3Rad1})-(\ref{3Rad2}), i.e. combinations of $\Gamma(r)$, $P(r)$, $\epsilon(r)$, $\lambda(r)$, and $\nu(r)$ for a given central pressure.
	
	 \item After solving (numerically) the desired equations with their boundary conditions and a set of trial values for $\omega^{2}$ by means of \textit{The Shooting Method}\footnote{The implementation of this numerical technique for the Chandrasekhar equations is highly non-trivial due to the complicate nature of its boundary conditions. In our case, the 1st-order formalism allows us to use the shooting method in a simple fashion on a range of values for trial $\omega$'s.}, obtain an oscillating behavior of $\Delta{P}$ and $\xi$ as functions of $\omega$.
	 
	 \item Only the discrete values of the frequency that satisfy $\Delta{P}(\omega^{2}_{i})=0$ are considered eigenfrequencies of the system.
	\end{itemize}	
	
Although this procedure is different from the Sturm-Liouville eigenvalue problem briefly review above, it also examines the transition between stable and unstable stellar configurations which occurs when the fundamental frequency vanishes, i.e. $\omega_{0}=2\pi{f}_{0}\rightarrow{0}$ \cite{Shapiro:1983,Bardeen:1966aa}. Additionally, within this first-order formalism of Gondek \textit{et al.}, we obtain the maximal stable mass configuration for a given equation of state (i.e. before gravitational collapse) when the fundamental mode becomes zero with $\Delta{P}(\omega^{2}_{0}=0)=0$ simultaneously.	

Finally, in order to be confident on the results of our numerical calculations, we have verified that our code reproduces the pulsation frequencies for the equations of state listed in Kokkotas and Ruoff \cite{Kokkotas:2000up}, which is a standard reference considering modern EoSs and with more reliable values of frequencies. Notice that although these authors use a different set of pulsation equations than Gondek \textit{et al.}, they are aware that their results can differ from other methods mainly due to the interpolation schemes, such as linear logarithmic or spline interpolations, which show an error of 3 percent in the values given for their frequencies. Thus, strictly speaking, our frequencies are within this error when using our code with their (mainly nuclear) equations of state.

\section{Stable Unpaired Quark Star Configurations}
	\label{sec:3radQCD}

Usually the physics of quark stars is only analyzed within the crude MIT bag model, where a (phenomenological) confining constant $B$ is added to the energy of the system producing a negative pressure that puts quarks inside a ``bag''. More precisely, it corresponds to $\Omega_{\rm pQCD}-\Omega_{\rm QCD}$, i.e. the thermodynamic potential difference of the pQCD and non-perturbative chiral symmetry breaking vacua.  However, no information is gained of the short-distance sector of the strong interactions in this model when calculating the EoS for cold quark matter, i.e., satisfying $\beta$-equilibrium and electric charge neutrality. Instead, it is more appropriate to use cold and dense perturbative QCD (pQCD) as a quark model, where its failing at low densities is parametrized by the renormalization scale parameter rather than the artificial precision of the bag constant value.	The bag model was used along the decades to study the radial oscillations of strange stars. Instead, in this section we analyze the radial pulsations of pQCD stars as quark stars existing in the core of neutron stars or as strange stars.

\subsection{NNLO pQCD Pocket Formula} 

As already mentioned in Chapter 2, the cold pQCD equation of state was computed up to next-to-next-to-leading-order (NNLO) in the strong coupling $\alpha_{s}$ by Kurkela et al. \cite{Kurkela:2009gj}, including the effects of the renormalization group on $\alpha_{s}(\bar{\Lambda})$ and the strange quark mass $m_{s}(\bar{\Lambda})$. Again, this perturbative calculation brings about an additional scale, the renormalization scale, $\bar{\Lambda}$, which provides an estimate of the inherent systematic uncertainties in the evaluation of the equation of state. This $\bar{\Lambda}$-parameter has to be varied within some range, and can be constrained by the phenomenology \cite{Fraga:2001id}.

Some years after this pQCD result appeared (and motivated by the attractive mathematical simplicity of the bag model), it was cast into the simple and easy-to-use pocket formula \cite{Fraga:2013qra}, which we call FKV, given by
	\begin{equation}
	P_{\rm QCD}=P_{\rm SB}(\mu_{B})\left(c_{1}-\frac{a(X_{\rm FKV})}{(\mu_{B}/{\rm GeV})-b(X_{\rm FKV})}\right)  \; ,
	\label{pressureFKV}
	\end{equation}	
where $P_{\rm SB}$ represents the Stefan-Boltzmann gas. This formula includes the contributions from massless up and down quarks, a strange quark with running mass, and massless electrons. It is in $\beta$-equilibrium and electrically neutral. Here, $\mu_{B}$ is the baryon chemical potential and we use the dimensionless version of the renormalization scale, $X_{\rm FKV}=3\bar{\Lambda}/\mu_{B}$, which can vary between $1$ and $4$, as discussed in Ref. \cite{Kurkela:2009gj}. The auxiliary functions that enter this pressure are defined as
	\begin{equation}
	a(X_{\rm FKV})=d_{1}X^{-\nu_{1}}_{\rm FKV},\hspace{0.5cm}b(X_{\rm FKV})=d_{2}X^{-\nu_{2}}_{\rm FKV},
	\end{equation}		
	with the following fit values (for more details about the procedure followed to obtain these numbers, see Ref. \cite{Fraga:2013qra})
	\begin{equation}
	c_{1}=0.9008,\hspace{0.2cm}d_{1}=0.5034,\hspace{0.2cm}d_{2}=1.452,
	\end{equation}
	\begin{equation}
	\nu_{1}=0.3553, \hspace{0.3cm}\nu_{2}=0.9101.
	\end{equation}

\subsection*{Discussion}	
	
For completeness and better understating of this formula, some comments are in order. First of all, unlike in Kurkela \textit{et al.} \cite{Kurkela:2009gj} (see also Chapter 2), Fraga \textit{et al.} \cite{Fraga:2013qra} decided to use for simplicity (which is allowed in the calculation) the thermodynamical potential $\Omega_{\rm pQCD}=-P_{\rm QCD}(\left\lbrace\mu_{i}\right\rbrace)$ directly  and truncate it at second order in $\alpha_{s}$. Then, electric charge neutrality and beta equilibrium allow to write all the quark and electron chemical potentials in terms of $\mu_{s}$ only, which in turn allows one to write $\Omega_{\rm pQCD}$ as depending only on this strange chemical potential. Thus, a table (later using an interpolating function) of values relating $\Omega_{\rm pQCD}(\mu_{s})$ for each $\mu_{B}(\mu_{s})$ can be built. The quark number densities (entering the electric charge neutrality) are obtained as usual by deriving the thermodynamic potential but not used as a fundamental ingredient nor imposing thermodynamic consistency, as all in Chapter 2. 

Following this, a technical detail one has to keep in mind when using the FKV pocket formula for the pressure and in obtaining other thermodynamic quantities is that, since $P_{\rm QCD}$ is a function of $\mu_{B}$ and $X_{\rm FKV}$, when obtaining the energy density $\epsilon_{\rm QCD}$ one has the freedom to choose a) a fixed value of $X_{\rm FKV}=X_{\rm FKV}(\mu_{B})$ (e.g., $X_{\rm FKV}=2$, i.e., $\bar{\Lambda}=(2/3)\mu_{B}$) in $P_{\rm QCD}$ and then build the energy density using the thermodynamic relation (being e.g. $n_{B}(X_{\rm FKV}=2)=dP_{\rm QCD}(X_{\rm FKV}=2)/d\mu_{B}$, and notice the total derivative)
\begin{equation}
\epsilon_{\rm QCD}=-P_{\rm QCD}+n_{B}\mu_{B},
\end{equation}
or b) consider $X_{\rm FKV}$ an independent constant and then build $\epsilon_{\rm QCD}$ by taking derivatives in $\mu_{B}$ keeping $X_{\rm FKV}$ constant when computing the baryon number density being now defined as $n_{B}=(dP_{\rm QCD}/d\mu_{B})_{X_{\rm FKV}}$, where one chooses the value of $X_{\rm FKV}$ only after obtaining this generic result for $n_{B}$ (see, e.g., Ref. \cite{Haque:2014rua} for a similar discussion at high temperatures). We note that both approaches only differ by a few percent in the values of $\mu_{B}$ near and at zero pressure, $\mu^{0}_{B}$, thus varying between $\mu^{0}_{B}$ and $\mu^{0}_{B}+\delta\mu_{B}$, which changes the pressure in this region by $(\delta\mu_{B}/\mu_{B})^{4}$, being vanishingly small. In turn, this $\delta\mu_{B}$ does not affect the minimum value taken by the energy density $\epsilon_{\rm QCD}(P_{\rm QCD}=0)=\epsilon_{\rm min}$ since the correction would be again of the order $(\delta\mu_{B}/\mu_{B})^{4}$, which is negligible. Additionally, we are allowed to argue from these considerations that this difference $\delta\mu_{B}$ will not affect our results for the oscillation frequencies and periods (to be calculated later) since they only require the knowledge of the equation of state of stellar matter. In this work, we have chosen the approach a) mentioned above from which we build the EoS, $P_{\rm QCD}=P_{\rm QCD} (\epsilon_{\rm QCD})$, as shown in Fig. \ref{fig:3EoSs}, together with other EoSs.

\begin{figure}[h]
\begin{center}
\resizebox*{!}{7.0cm}{\includegraphics{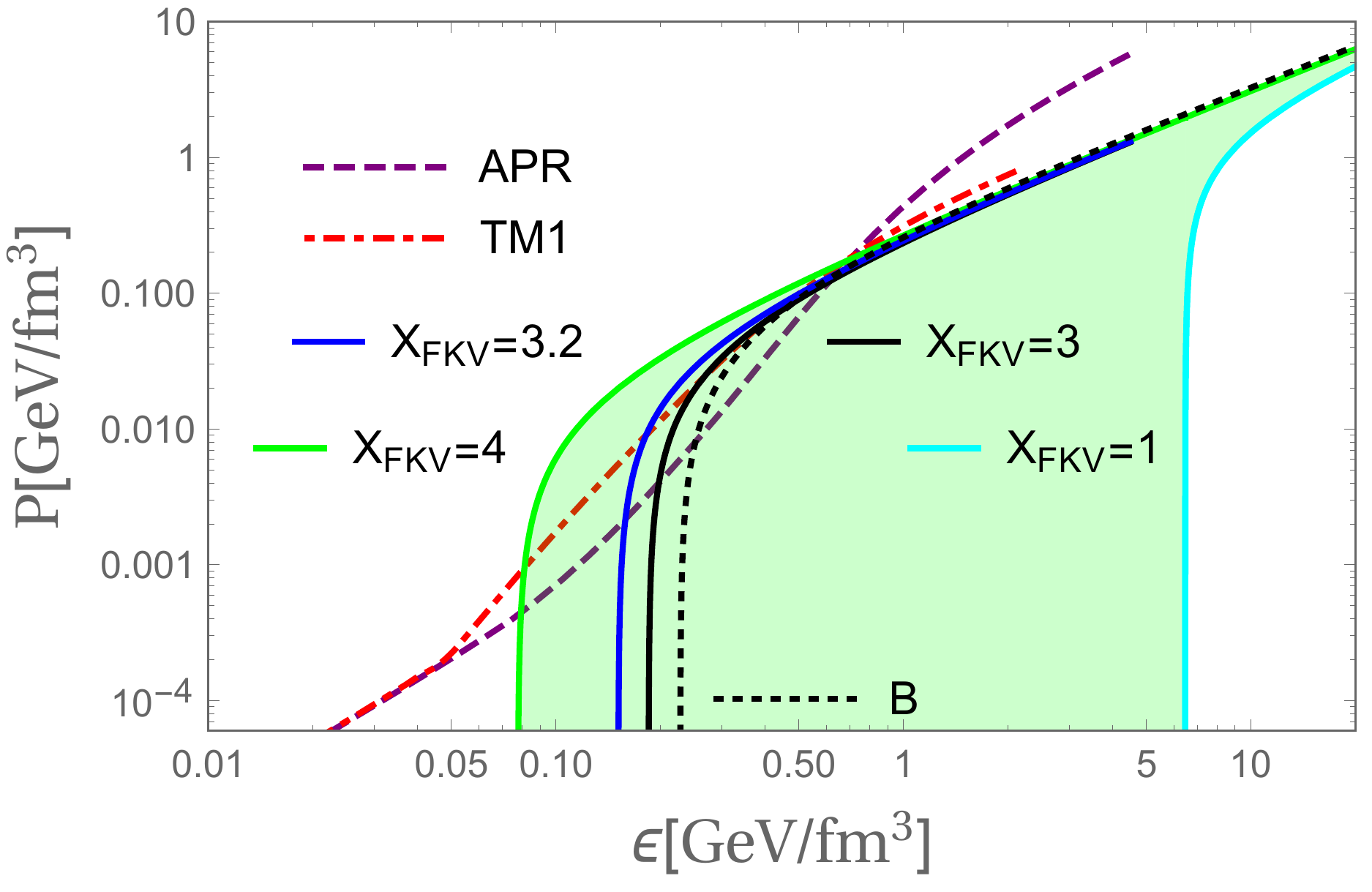}}
\end{center}
\vspace{5mm}
\caption{Equations of state, $P=P(\epsilon)$, for a few values of $X_{\rm FKV}$ and the bag model $B=(145\rm MeV)^4$ used along this chapter, to be discussed in the next section. We also show the nuclear matter APR \cite{Akmal:1998cf} and TM1\cite{Shen:1998gq} equations of state. Note the large dependency of the pQCD EoS on $X_{\rm FKV}$ at low densities and quite $X_{\rm FKV}$-independent behavior at high densities. Taken from Ref. \cite{Jimenez:2019iuc}.}
\label{fig:3EoSs}
\end{figure}

Although it was mentioned that this FKV pocket formula takes into account massless up, down and massive strange quarks, the appearance of the last ones begins even at zero pressure, unlike in the pQCD framework of Chapter 2. It must be noted that this is a simplification since it is expected that this EoS must be matched at low densities onto a nuclear matter EoS. Thus, their formula is perturbative and its practical purpose would be to replace the MIT bag model. Indeed, one is free to include other non-perturbative effects, like the color superconducting gap or a bag constant. Currently, this idea is widely used to match this high density EoS with the low density chiral perturbation theory result using polytropes at intermediate densities \cite{Kurkela:2014vha,Annala:2019puf}. We note that for this work we do not include any nuclear crust to the FKV EoS at low densities.
	
Finally, a crucial remark must be included on certain similarities with the bag model. Perturbative expansions of the cold QCD equation of state, such as the the one of Ref. \cite{Kurkela:2009gj}, not only include contributions in powers of $\alpha_{s}$, but also terms of the form $\alpha^{n+1}_{s}\log^{n}(\alpha_{s})$, where ``$n$'' is the order of the expansion, being $n\geq{1}$, which in our case is $n=1$. This logarithm of $\alpha_{s}$ arises from the ring (plasmon) diagrams which represent an infrared contribution to the QCD pressure. It is interesting that this effect produces numerically similar results with respect to an effective bag model (which includes a physical nonperturbative contribution explicitly) for some values of $X_{\rm FKV}$ (see Fig. \ref{fig:3Anom}). See Ref. \cite{Fraga:2001id} for similar findings at lower order in $\alpha_{s}$. Then, this is manifested astrophysically for quarks stars producing similar mass-radius diagrams qualitatively (see Fig. \ref{fig:3MassRad}).

One can also easily compute the perturbative trace anomaly of QCD  (normalized by the Stefan-Boltzmann gas) from Eq. (\ref{pressureFKV}), obtaining
   \begin{equation}
   t^{\mu}_{\mu}(\mu_{B}, X_{\rm FKV})=\frac{\epsilon_{\rm QCD}-3P_{\rm QCD}}{P_{\rm SB}}=\frac{\mu_{B}}{\rm GeV}\frac{a(X_{\rm FKV})}{\left[(\mu_{B}/{\rm GeV})-b(X_{\rm FKV})\right]^{2}} \; ,
   \end{equation}
which gives a measure of the role of interactions encoded in the breaking of conformal symmetry \cite{Andersen:2011ug}. In Fig. \ref{fig:3Anom} we show the pQCD normalized trace anomaly for different values of $X_{\rm FKV}$, and compare it to the result obtained from the bag model for $B=(145\rm MeV)^4$, which vanishes very quickly with $\mu_{B}$. For comparison we also plot in this figure the trace anomalies of standard nuclear matter APR and TM1 equations of state (to be discussed). For the perturbative case, we show a band that represents a measure of the actual uncertainties below, to be contrasted to the apparent, misleading precision of the bag model line.

\begin{figure}[h!]
\begin{center}
\resizebox*{!}{7.0cm}{\includegraphics{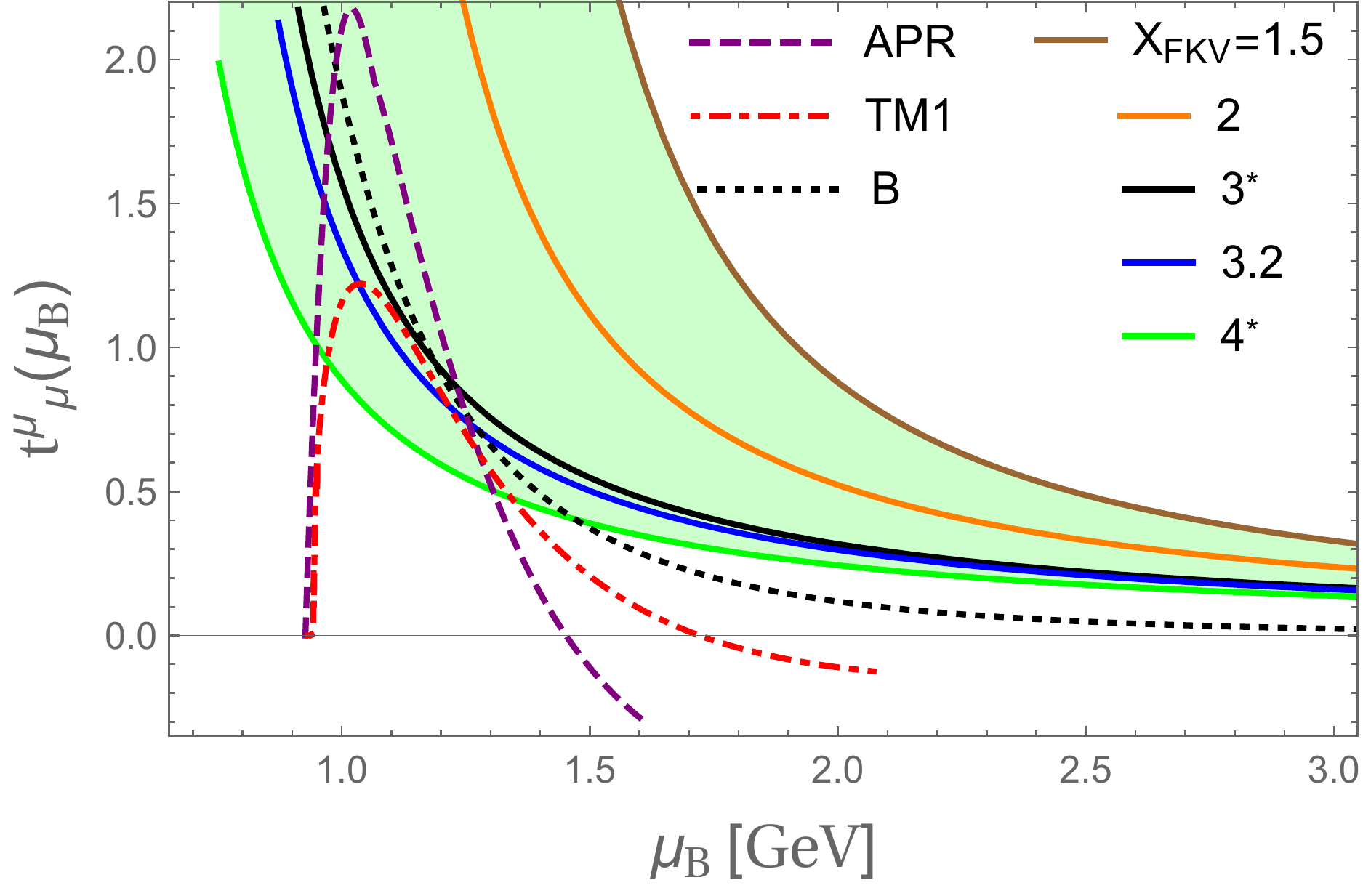}}
\end{center}
\vspace{5mm}
\caption{Trace anomaly for the cold pQCD result of Refs. \cite{Kurkela:2009gj,Fraga:2013qra} normalized by the Stefan-Boltzmann pressure as a function of the baryon chemical potential for different values of $X_{\rm FKV}$ (continuous) and for the bag model with $B=(145\rm MeV)^4$ (dotted). We mark with asterisks the cases of $X_{\rm FKV}$ between 3 and 4 since only within this range one obtains at least two-solar mass stars. For comparison, we also show the behavior of the trace anomaly for two well-known nuclear matter equations of state, APR and TM1 (see text). Taken from Ref. \cite{Jimenez:2019iuc}.}
\label{fig:3Anom}
\end{figure}

\subsection*{Strange quark matter parameter space}

Strange stars are quark stars that are self bound by QCD interactions and satisfy the Bodmer-Witten hypothesis of strange quark matter being the true ground state of nuclear matter. Strange matter configurations at zero pressure (and also at $T=0$ in this work) would have $E/A=\epsilon_{\rm QCD}/n_{B}<930$MeV, i.e. energy per baryon lower than iron-56 \cite{Glendenning:2000}.   
   
The FKV formula has a broad parameter space which allows for the existence of configurations of self-bound matter. It was shown in Ref. \cite{Kurkela:2009gj} that this condition is satisfied for values of $X_{\rm FKV} \sim 3-4$. A similar conclusion was obtained within the Nambu--Jona-Lasinio (NJL) model\footnote{This is an effective, i.e. non-renormalizable model for quark matter which includes the chiral condensate effects through spontaneous breaking of the chiral symmetry but without accounting for confinement.} for quark matter for the most common parametrizations of the EoS \cite{Buballa:2003qv}. In what follows, we consider only bare quark stars, i.e. stars without a nuclear crust, which depending on the value of $X_{\rm FKV}$ will be self-bound stars or ordinary quark stars possibly found in neutron star interiors. Notice that in this case we are not mentioning any cosmological QCD situation (where possible strangelets of strange quark matter might be formed) since high temperatures should be included which, in general, reduce the likelihood of satisfying the Bodmer-Witten hypothesis.

\subsection{Warm-up: Maximal-mass pQCD star for $X_{\rm FKV}=3$}

For a better understanding of the process of solving the oscillation equations, we summarize the path that must be followed in order to obtain stable stellar configurations being dynamically equilibrated under radial perturbations. For this, we take the particular case of $X_{\rm FKV}=3$ (producing a $2M_{\odot}$ quark star) in the FKV formula for pQCD with the central value of $\epsilon_{c}$ producing a maximal-mass star configuration:

\begin{itemize}

	\item One should start by solving the TOV equations for this $X_{\rm FKV}=3$ equation of state. First of all, the stellar radius $R$ and total gravitational mass $\mathcal{M}(R)=M$ must be determined, as shown in Fig \ref{fig:3MassRad} below (for more details on this, see Table \ref{tab:3table1}). Thus, one obtains the functions $\nu(r)$, $\lambda(r)$ through $\mathcal{M}(r)$, the energy density profile $\epsilon(r)$, and $P=P(r)$ by using the relation $P=P(\epsilon(r))$; all up to the radius $R$. All this is shown in Figs. \ref{fig:3MuNuX3} and \ref{fig:PressureEnergyProfilesX3}.
	
	\begin{figure*}[h!]
\begin{center}
\hbox{\includegraphics[width=0.5\textwidth]{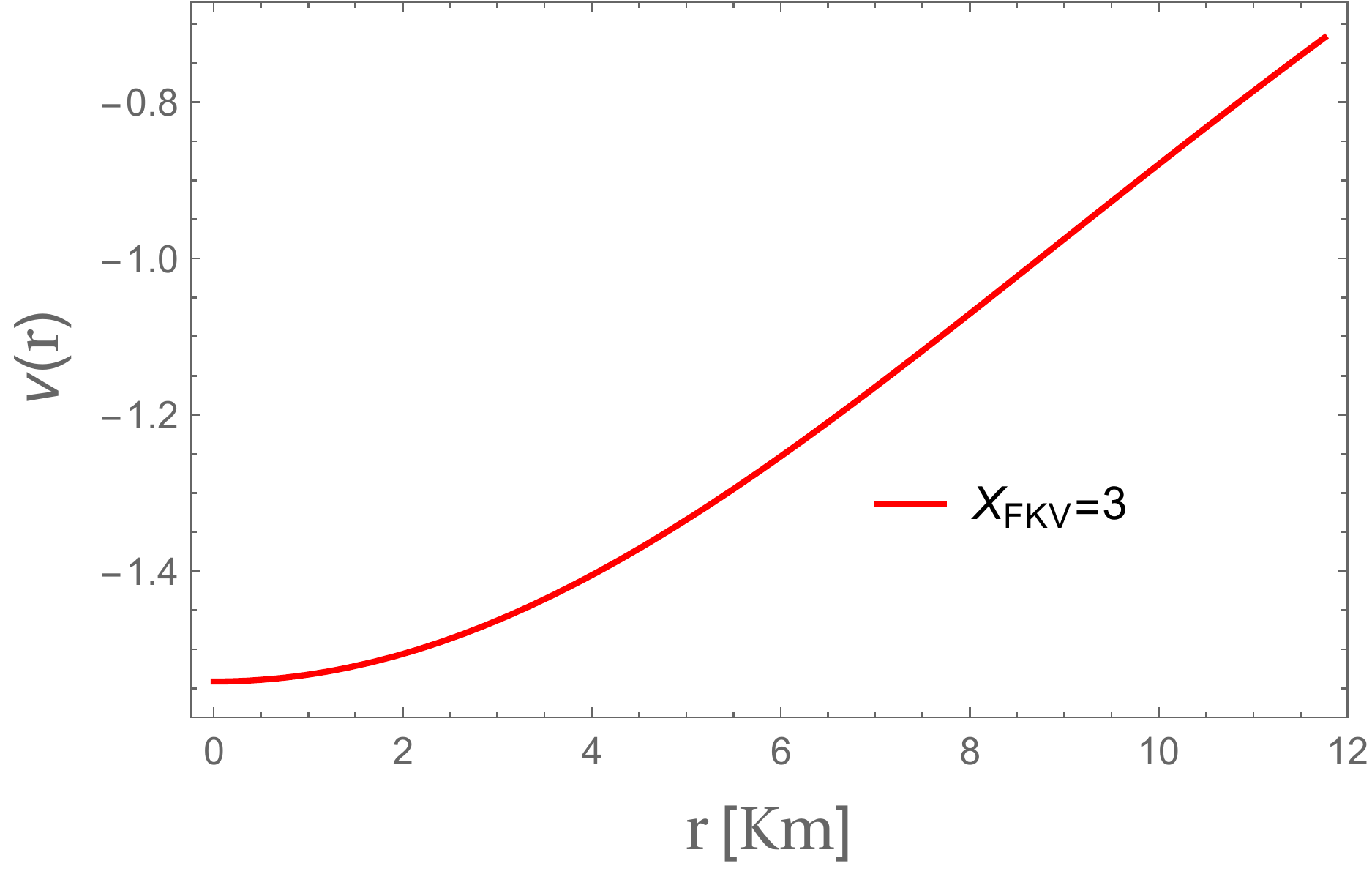}
	  \includegraphics[width=0.5\textwidth]{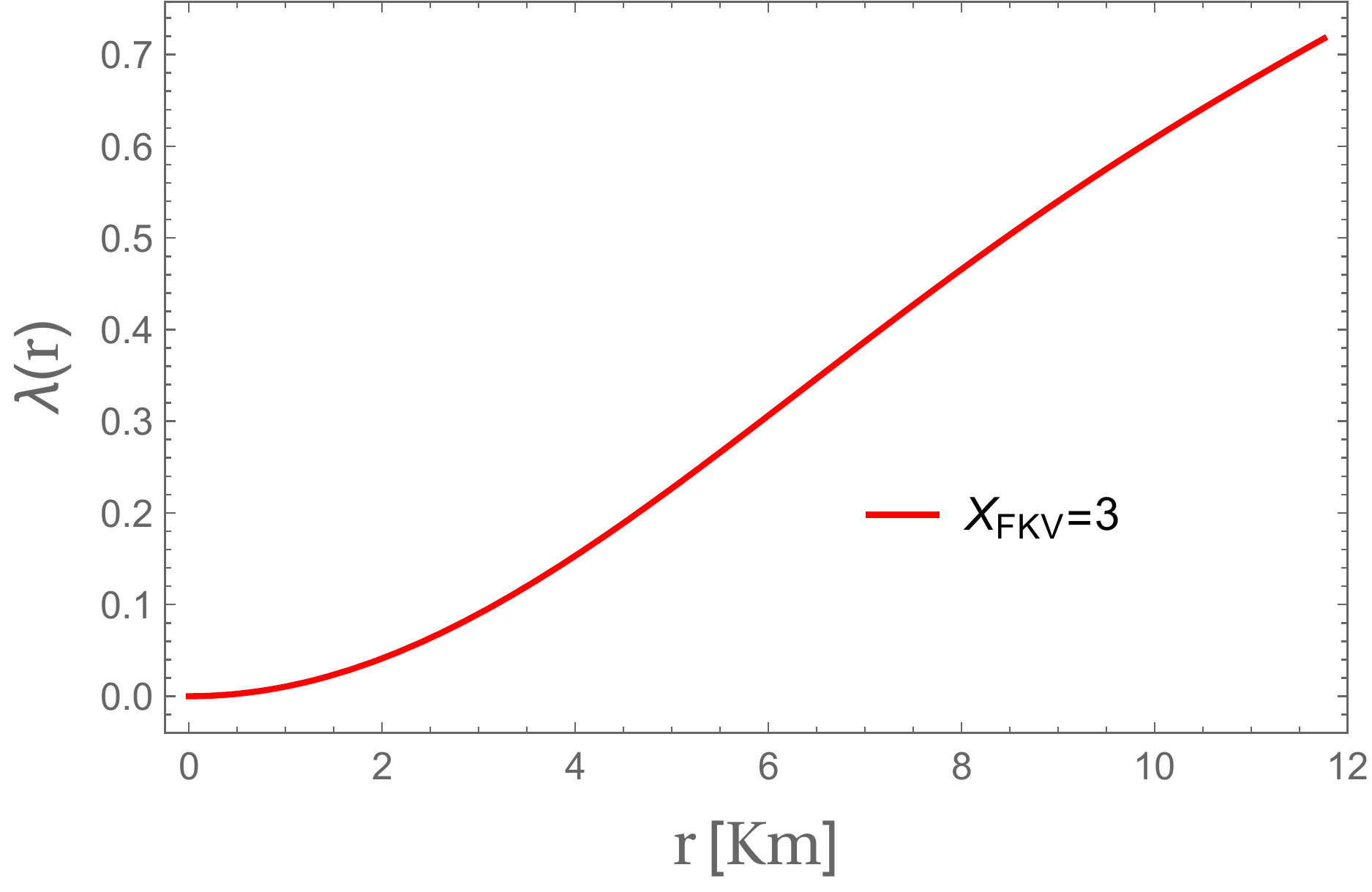}}
\vspace{5mm}
\caption{Left panel: Temporal metric function $\nu(r)$ for the interior of this maximal mass quark star obtained from FKV formula for $X_{\rm FKV}=3$. Right panel: The same but for the radial metric function $\lambda(r)$.}
\label{fig:3MuNuX3}
\end{center}
\end{figure*}

\begin{figure*}[h!]
\begin{center}
\hbox{\includegraphics[width=0.5\textwidth]{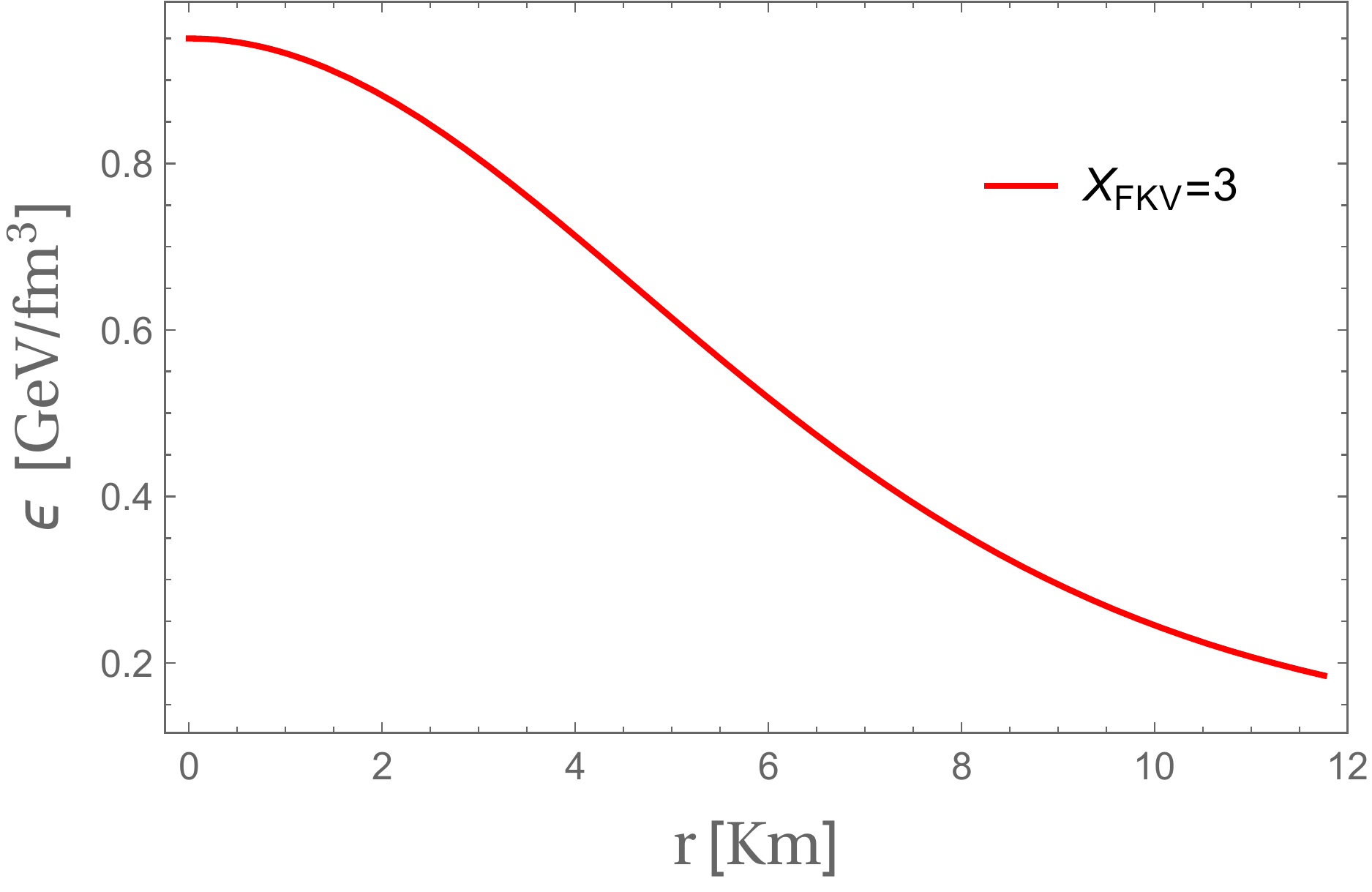}
	  \includegraphics[width=0.5\textwidth]{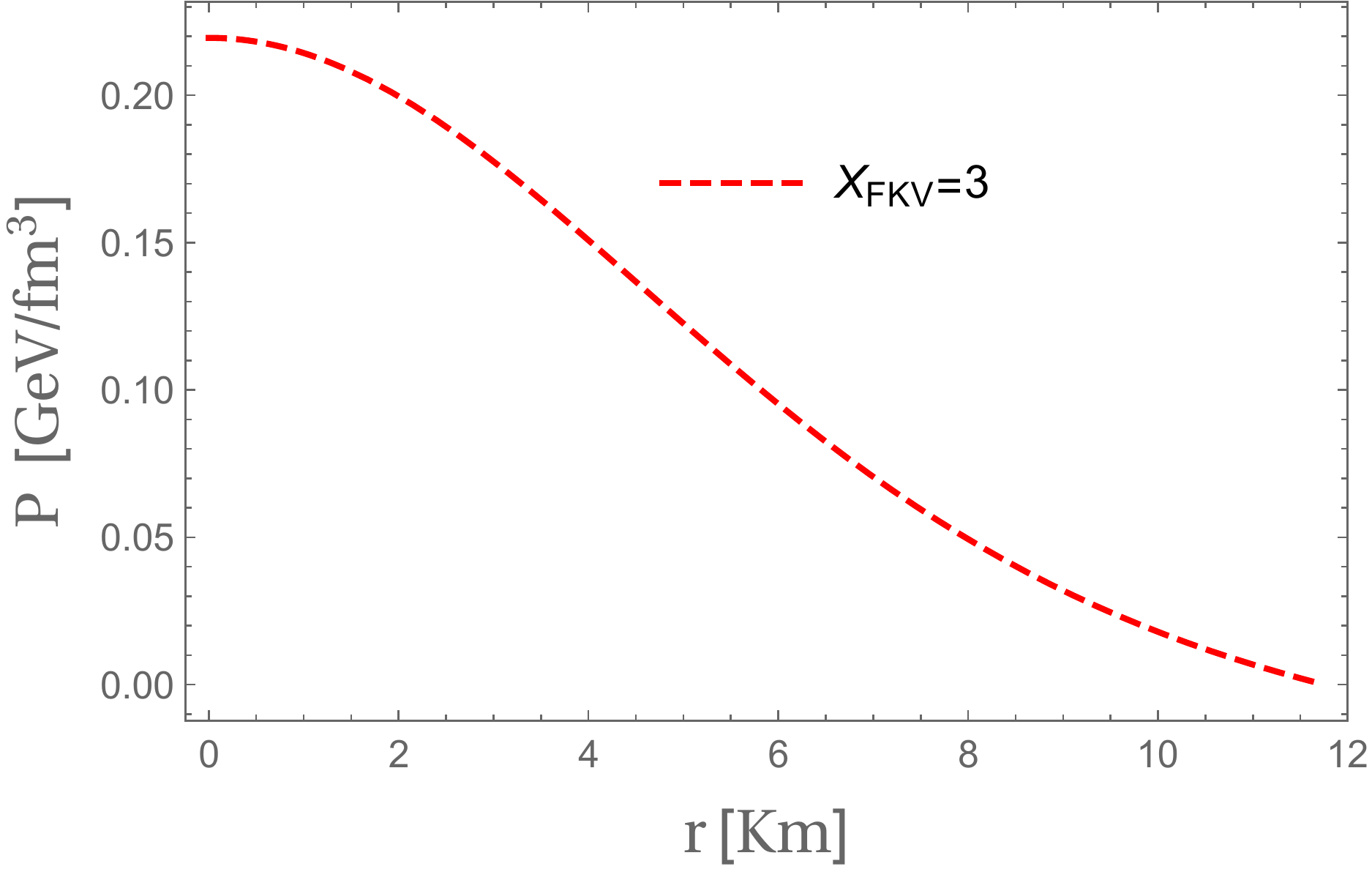}}
\vspace{5mm}
\caption{Left panel: Energy density profile for a particular stellar configuration chosen from the FKV formula for $X_{\rm FKV}=3$. Right panel: The same but for the total pressure.}
\label{fig:PressureEnergyProfilesX3}
\end{center}
\end{figure*}

	\item Then, the adiabatic index of the EoS needs to be calculated. For the case of $X_{\rm FKV}=3$, the result is shown in Fig. \ref{fig:3AdiabaticIX3}, depending only on the energy density $\epsilon$, which in turn depends on the radial coordinate $r$, i.e. it becomes suitable to be introduced into the first-order oscillation equations. All these dependencies can be seen in Fig. \ref{fig:3AdiabaticIX3}.

\begin{figure*}[h!]
\begin{center}
\hbox{\includegraphics[width=0.5\textwidth]{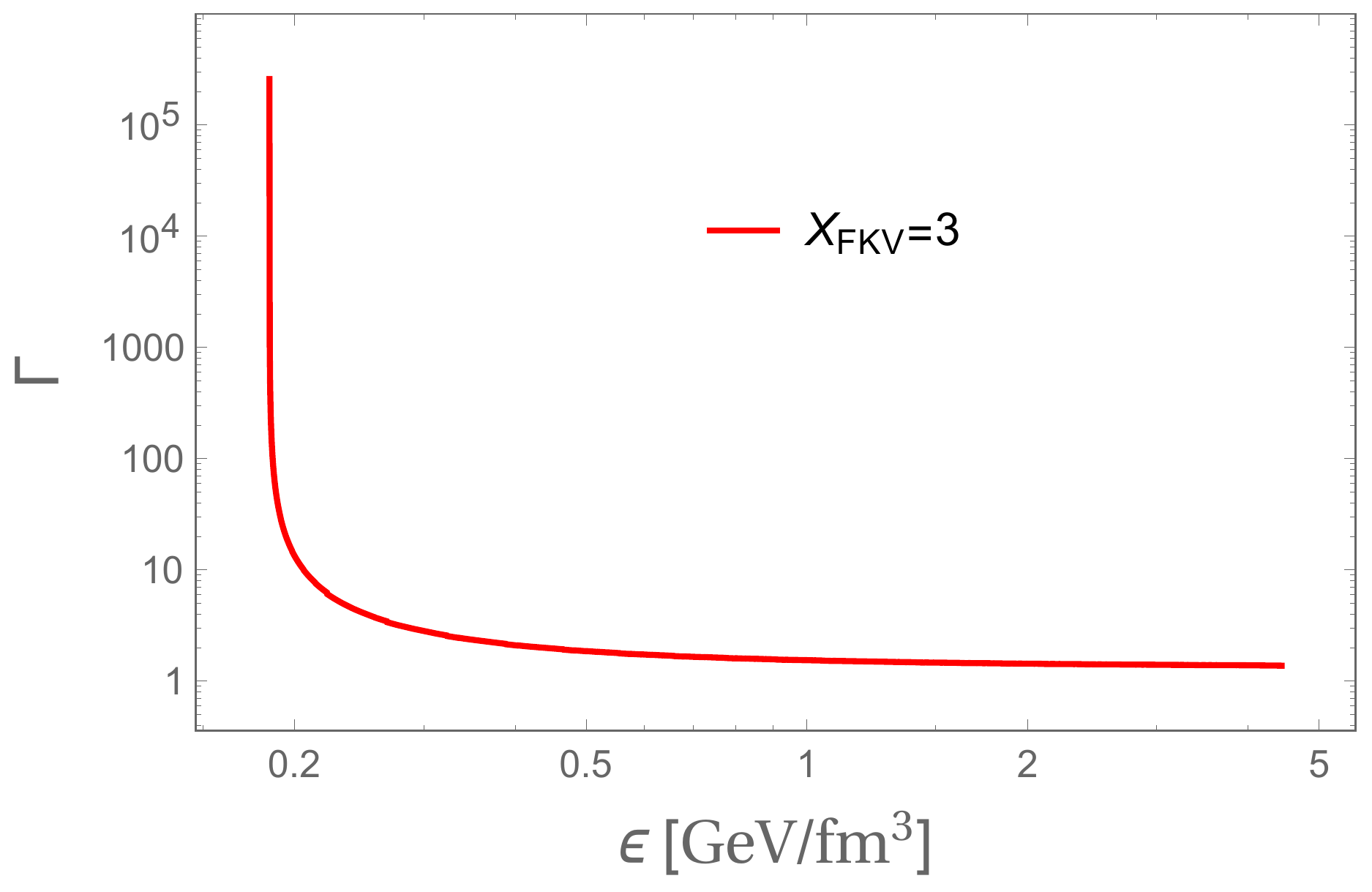}
	  \includegraphics[width=0.5\textwidth]{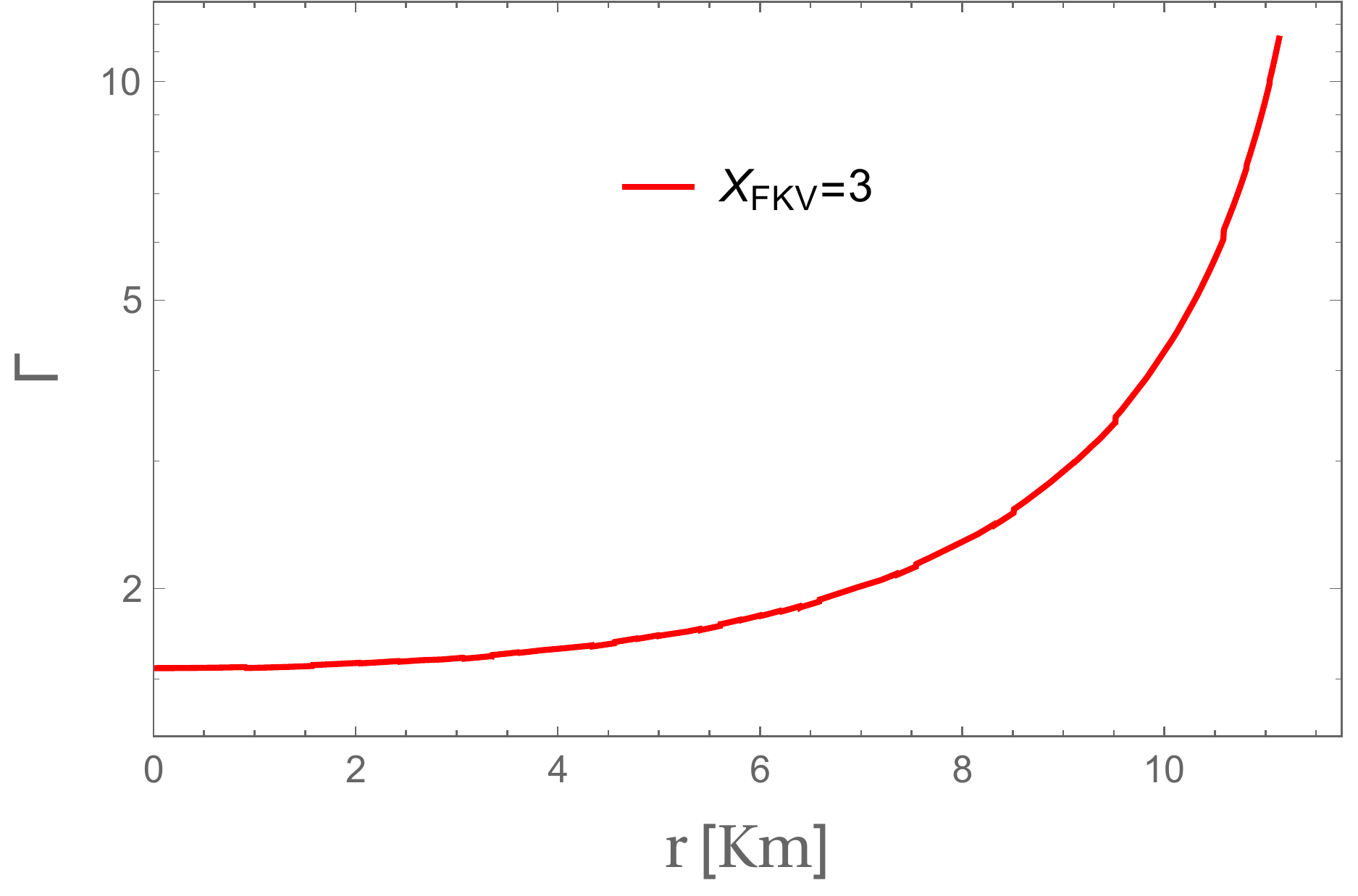}}
\vspace{5mm}
\caption{Left panel: Logarithmic plot for the adiabatic index versus (central) energy densities for the pQCD matter equation of state, FKV, in the specific case of $X_{\rm FKV}=3$. Right panel: Adiabatic index profile in the interior of the maximal mass quark star for $X_{\rm FKV}=3$ obtained at $\epsilon=0.982~{\rm GeV/fm^{3}}$ at the star's center ($r=0$). Notice that here $\Gamma$ varies in agreement with its values taken in the Left panel at the center ($\Gamma\sim{1}$) and near the surface ($\Gamma\sim{10}$), where at the exact value of the surface R, the $\Gamma$ index diverges. This does not represent a problem since in Eq. (\ref{3Rad1}) the divergence raised by the product ``$\Gamma{P}$'' is controled by the vanishing value of $\Delta{P}$ at the surface.}
\label{fig:3AdiabaticIX3}
\end{center}
\end{figure*}

	\item Finally, finding the maximal-mass quark star configuration, i.e. the critical point before reaching unstable stellar configurations, means that when imposing the boundary condition $\Delta{P}\rightarrow{0}$ at the stellar radius $r=R$ one must to look for the vanishing value of the linear frequency $f_{n}(=\omega_{n}/2\pi)$ and consider it as the turning point between stability and instability. The other zeroes should be considered as the frequencies of the excited modes $n=1,2,...$. This is illustrated in Fig. \ref{fig:3DeltaPfreqX3}, where the six first oscillation modes were found. Additionally in this figure, and for the purposes of the next chapter, we show the result of the exchanging $\omega^{2}_{n}\rightarrow{-\omega^{2}_{n}}$ in the Gondek \textit{et al.} equations to look for \textit{unstable} configurations, i.e. by finding complex frequencies in the neighbourhood of the maximal-mass star. A careful study (for this particular $X_{\rm FKV}=3$ and any other $X_{\rm FKV}$) indicates that there does not exist solutions satisfying the boundary condition on $\Delta{P}(R)$ for complex values of frequency for any mode $n$, i.e. only real frequencies (stable) stellar configurations are found up to the maximal-mass star.

\begin{figure*}[h!]
\begin{center}
\hbox{\includegraphics[width=0.5\textwidth]{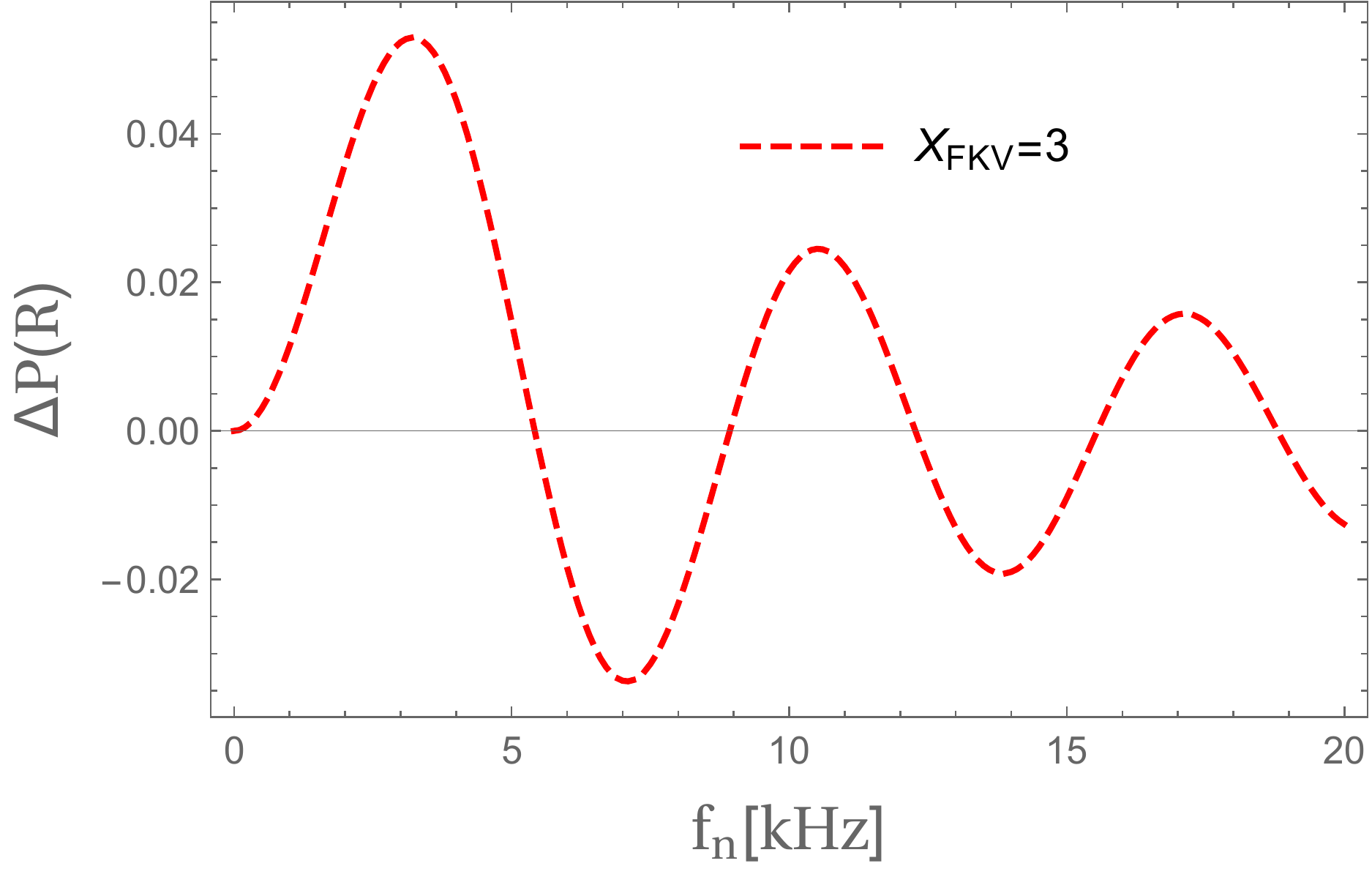}
	  \includegraphics[width=0.5\textwidth]{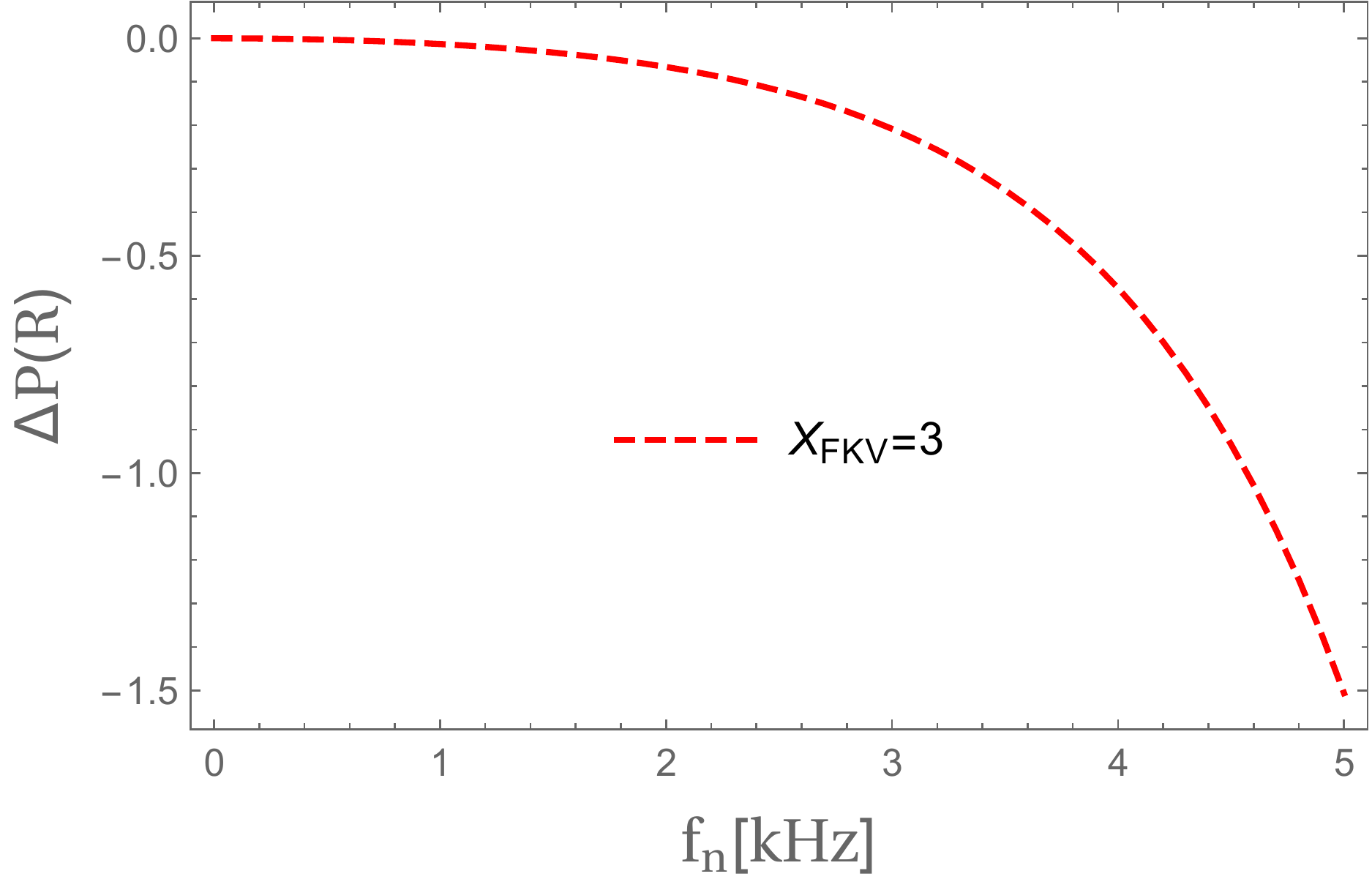}}
\vspace{5mm}
\caption{The Lagrangian displacement of the pressure, $\Delta{P}$ in $\rm GeV/fm^{3}$, at the surface of the star, $R$ in Km's, satisfying its respective boundary condition. Each of these figures represent the situation of stable (left panel) $\omega^{2}_{n}\geq{0}$ and unstable $\omega^{2}_{n}<{0}$ (right panel) quark star configurations. See text for more details.}
\label{fig:3DeltaPfreqX3}
\end{center}
\end{figure*}

	\item Introducing all the pulsating functions and the values obtained above of the frequencies for $n=0,1,2$ into the radial perturbation equations of Gondek \textit{et al.}, one could get the behavior for both Lagrangian variables $\Delta{P}$ and $\xi$, at each point of the star from its center up to its surface. Besides, they are different for each oscillation mode. This can be seen in Fig. \ref{fig:3LagranVarX3} where each mode has the same number of nodes as if we were dealing with the fundamental or excited states. 

\begin{figure*}[h!]
\begin{center}
\hbox{\includegraphics[width=0.5\textwidth]{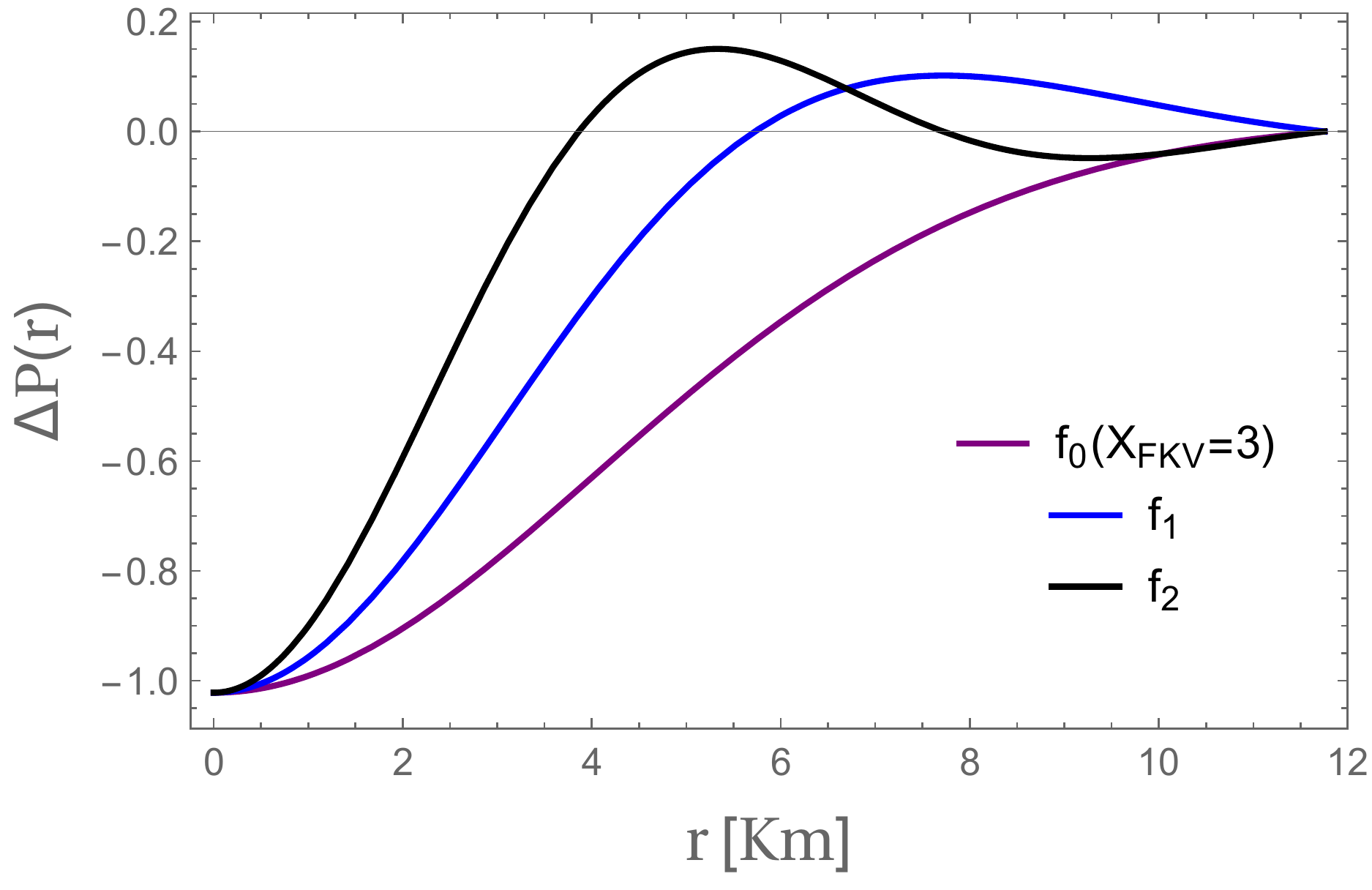}
	  \includegraphics[width=0.5\textwidth]{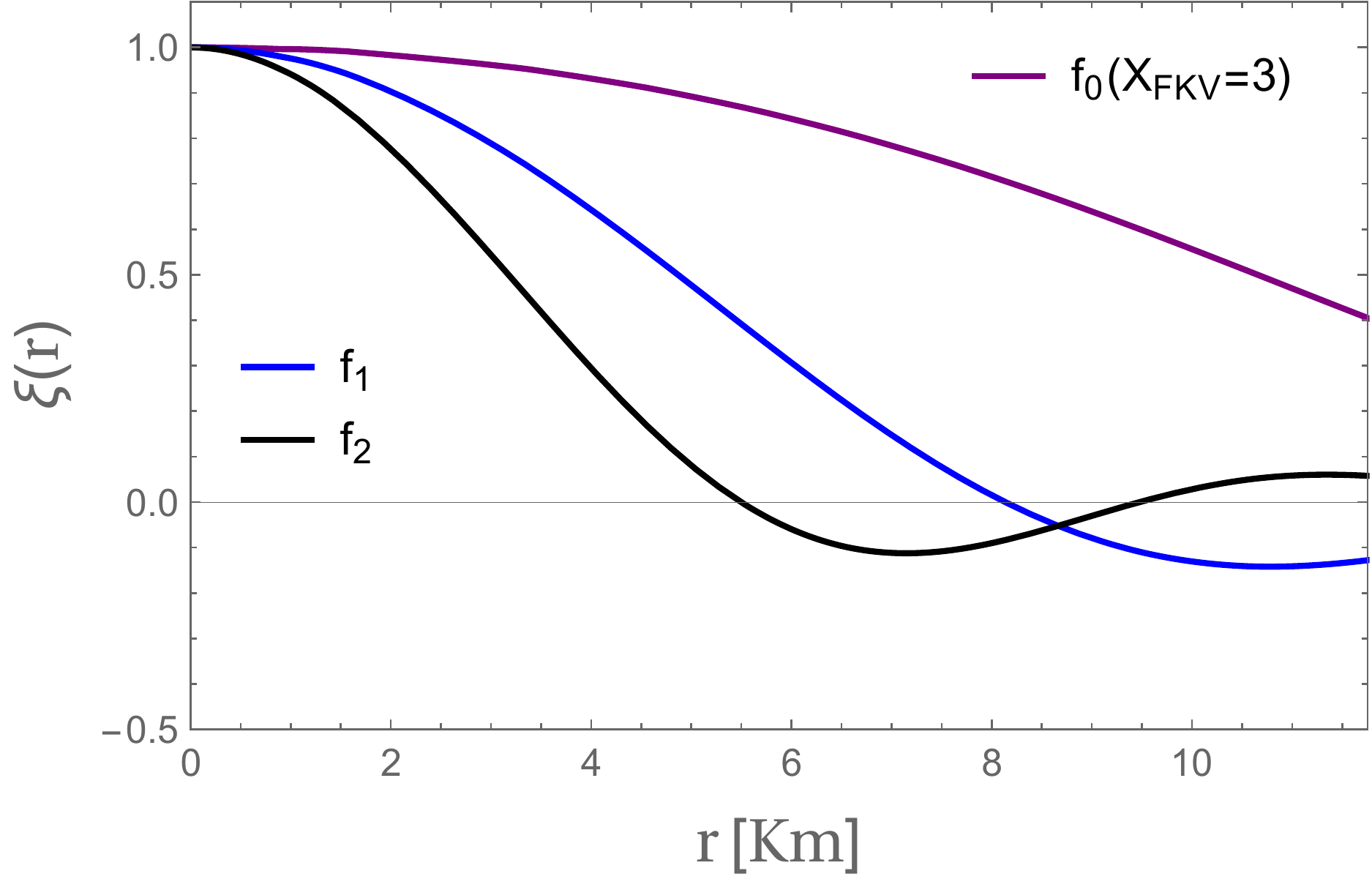}}
\vspace{5mm}
\caption{Behavior of the Lagrangian variables ($\Delta{P}$ in $\rm GeV/fm^{3}$) of the oscillating perturbations in the interior of a quark star for $X_{\rm FKV}=3$ at the maximal mass configuration.}
\label{fig:3LagranVarX3}
\end{center}
\end{figure*}

\end{itemize}

\subsection{Vibrational Spectrum of pQCD stars for any $X_{\rm FKV}$}

In order to explore all the stellar configurations obtainable by the pQCD FKV formula, one must perform the same process as above but also implementing \textit{loops} in the numerical code for each value of the central energy density and looking for values of frequency satisfying the boundary conditions of the pulsation equations. We present our global and particular results below.

Before proceeding with our study of quark stars, we present in Figs. \ref{fig:3modesNuclear1} and \ref{fig:3modesNuclear2} the oscillation frequencies of two standard equations of state for cold nuclear matter, the relativistic mean field theory calculation of Shen \textit{et al.} \cite{Shen:1998gq} in the TM1 parametrization (dubbed TM1) and the result of Akmal \textit{et al.} \cite{Akmal:1998cf} (dubbed\footnote{This EoS is also known as APR4 or A$18+\delta{v}+{\rm UIX}^{*}$.} APR) obtained from nucleon-nucleon potentials after performing Dirac-Brueckner-Hartree-Fock calculations. These particular oscillation frequencies are shown to highlight the difference in the qualitative behavior of nuclear and quark stars. For instance, in Fig. \ref{fig:3modesNuclear1} an expected effect for nuclear EoSs happens, the so-called \textit{avoided crossing} (a continuous knee at some value of energy density) occurs marking a transition between standing localized waves in the outer layer of the star to predominantly localized waves when going to the star's core. See Ref. \cite{Gondek:1999ad} for more details on this issue. Notice also that although they have their frequency modes of the same order of magnitude, they are numerically different due to the stiffness of each EoS and which is realized in the behavior their corresponding $\Gamma$'s.

\begin{figure*}[h!]
\begin{center}
\hbox{\includegraphics[width=0.5\textwidth]{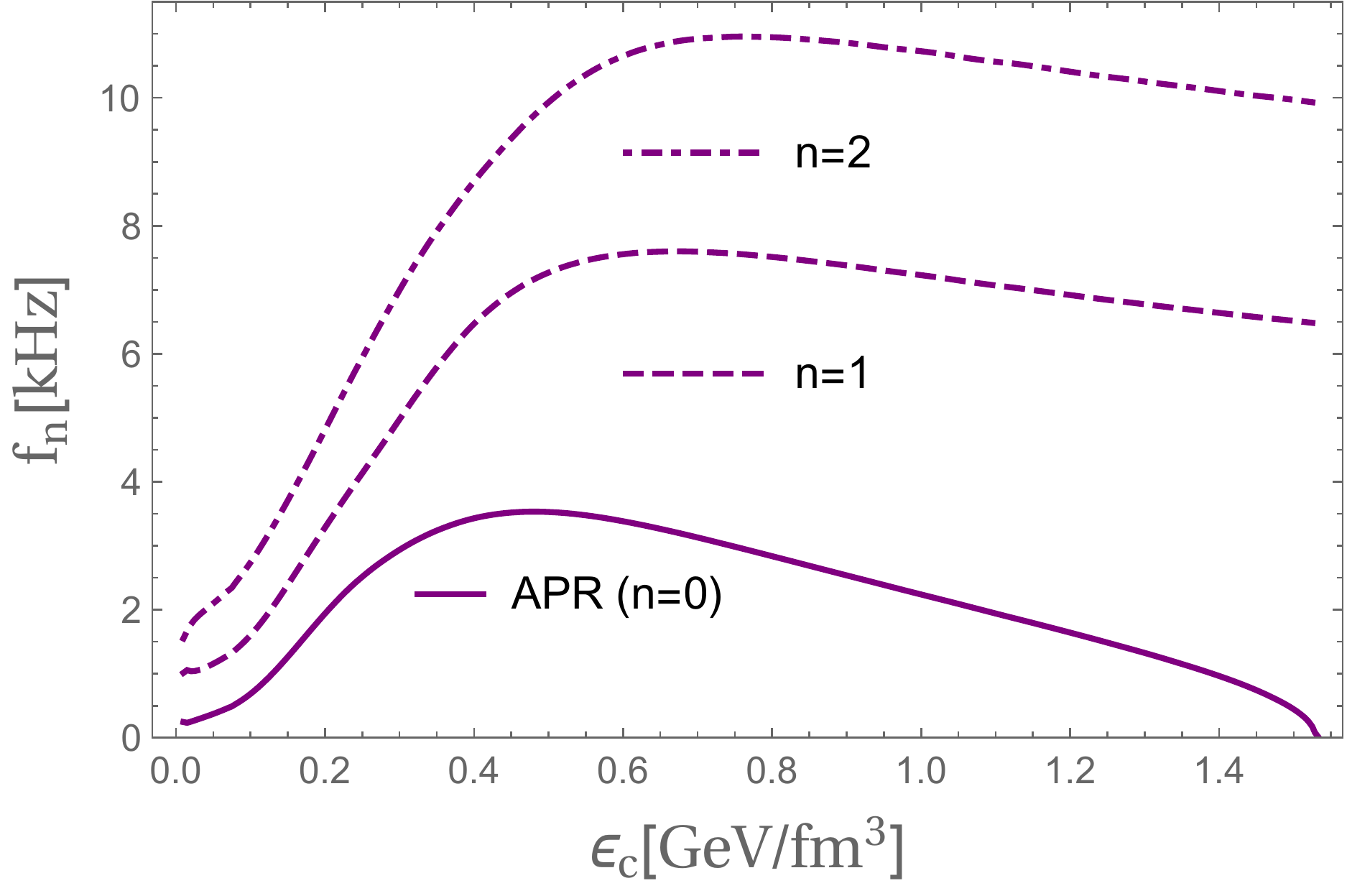}
	  \includegraphics[width=0.5\textwidth]{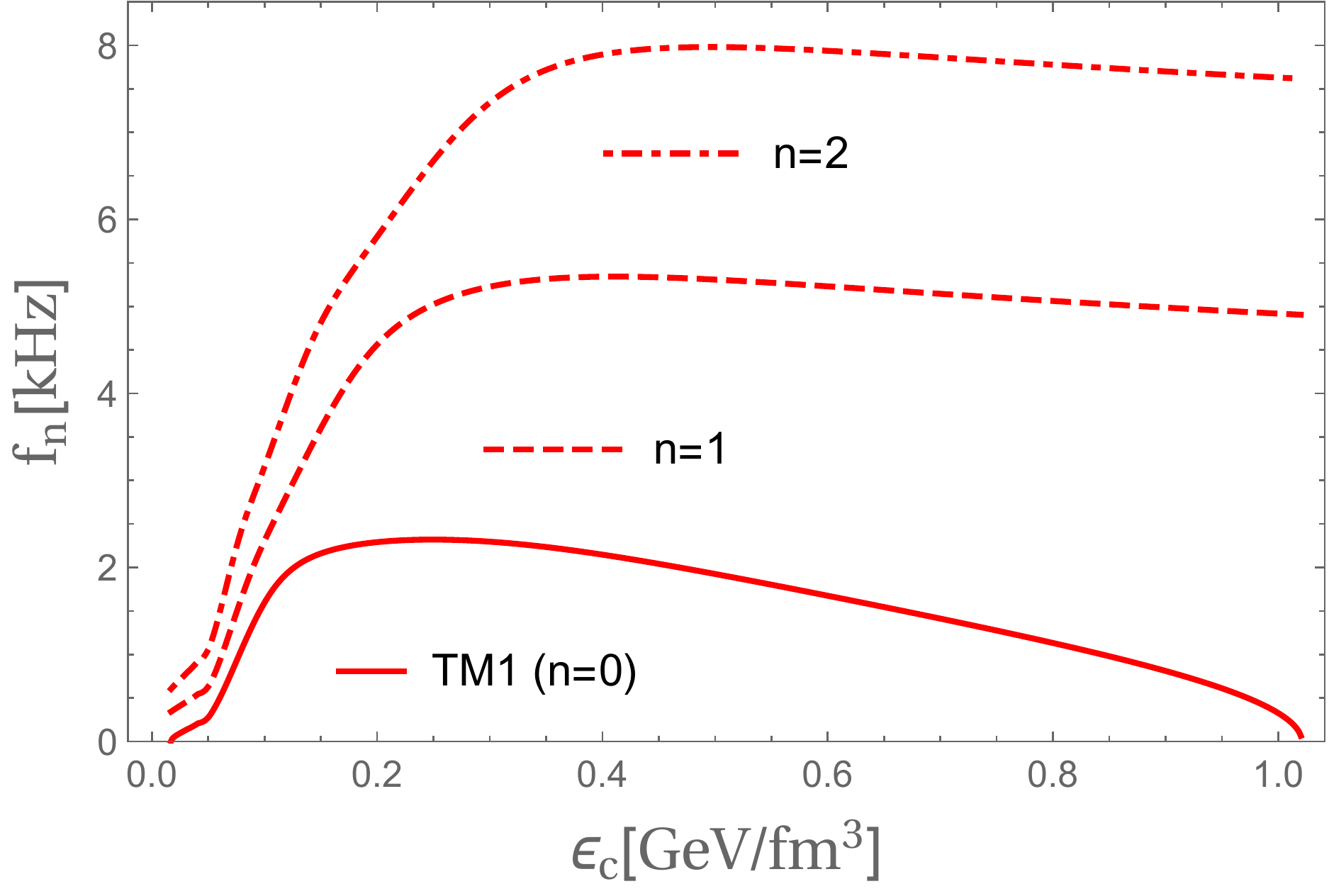}}
\vspace{5mm}
\caption{Left panel: The fundamental, $n=0$, and the first two excited modes, $n=1,2$, as function of the central energy density for the nuclear APR equation of state. Right panel: The same for the TM1 relativistic mean field theory.}
\label{fig:3modesNuclear1}
\end{center}
\end{figure*}
\begin{figure*}[h!]
\begin{center}
\hbox{\includegraphics[width=0.5\textwidth]{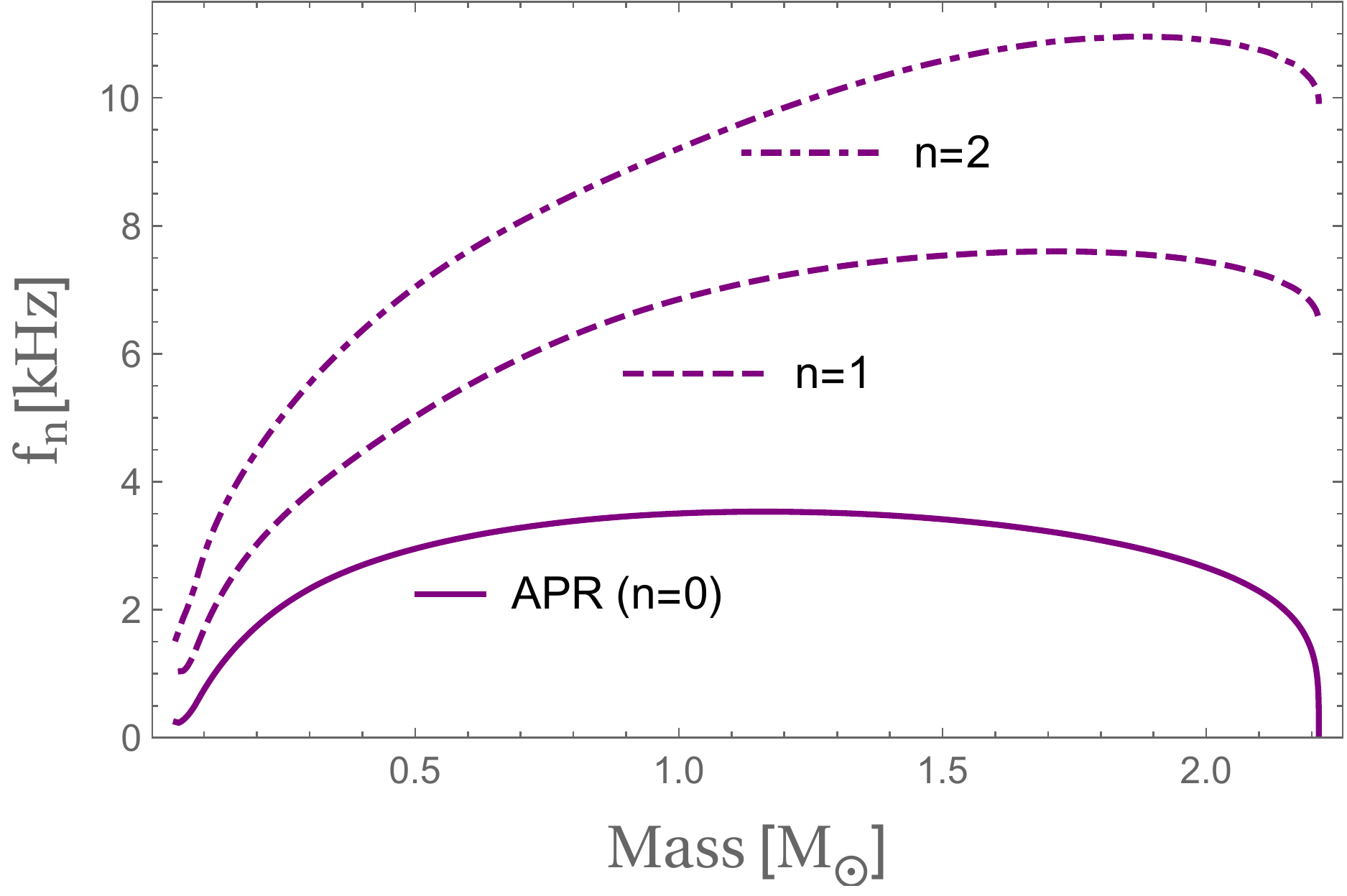}
	  \includegraphics[width=0.5\textwidth]{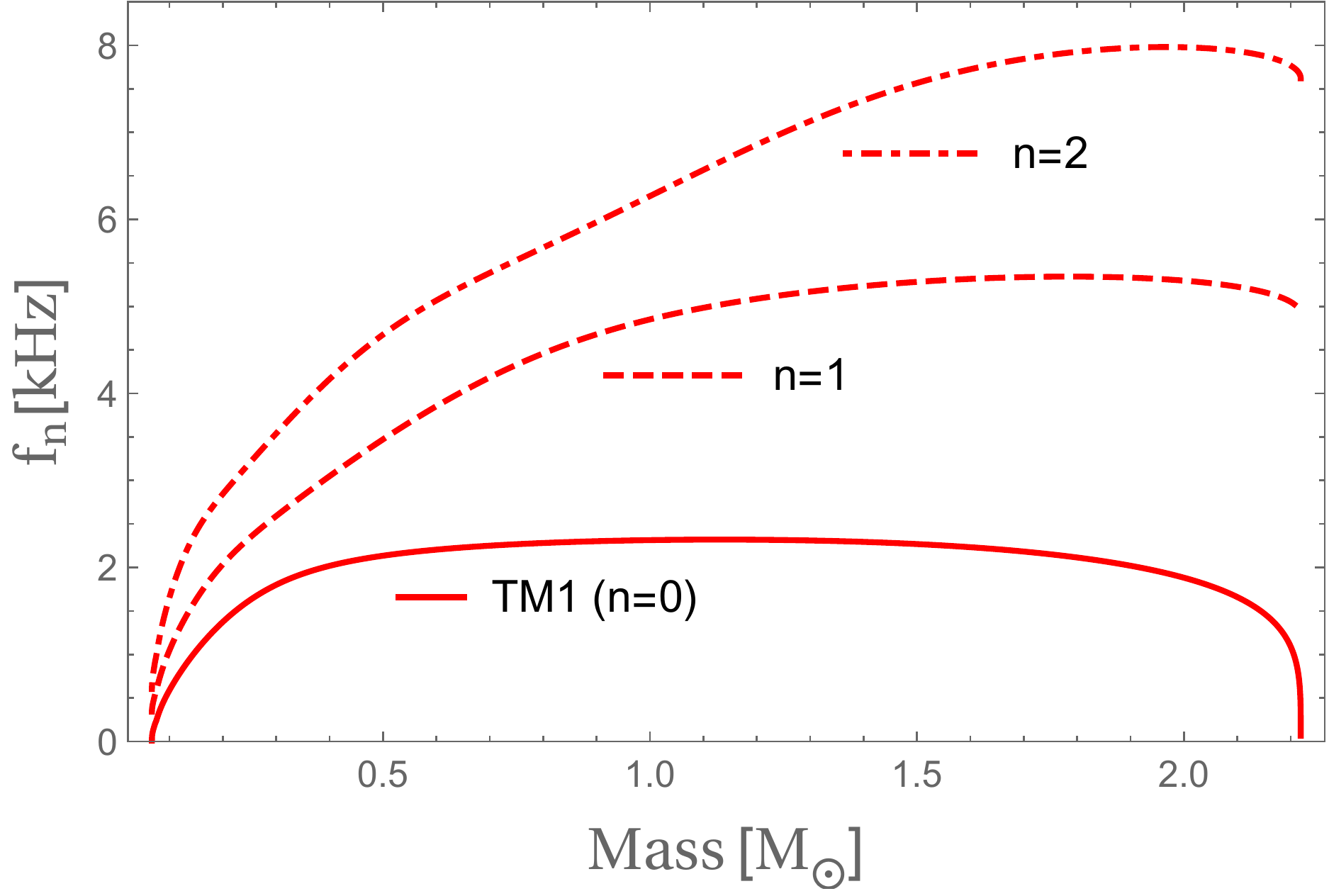}}
\vspace{5mm}
\caption{Left panel: The fundamental, $n=0$, and the first two excited modes, $n=1,2$, as function of the total gravitational mass for the nuclear APR equation of state. Right panel: The same but for the TM1 relativistic mean field theory.}
\label{fig:3modesNuclear2}
\end{center}
\end{figure*}	

\subsection*{Hydrostatically equilibrated quark stars}

After having grasped the main ideas needed to solve the radial oscillation equations, we now investigate the behavior of different eigenfrequencies, $\omega_{n}$, of the fundamental and first excited modes produced by a radial perturbation in a quark (or strange) star for any value of $X_{\rm FKV}$, finding the transition point before gravitational collapse.
	
Again, we first solve the TOV equations for the perturbative QCD EoS using the FKV formula for some arbitrary values of the renormalization scale $X_{\rm FKV}$. For the sake of comparison, we also display results obtained for the nuclear matter EoSs.

\begin{figure}[h!]
\begin{center}
\resizebox*{!}{7.0cm}{\includegraphics{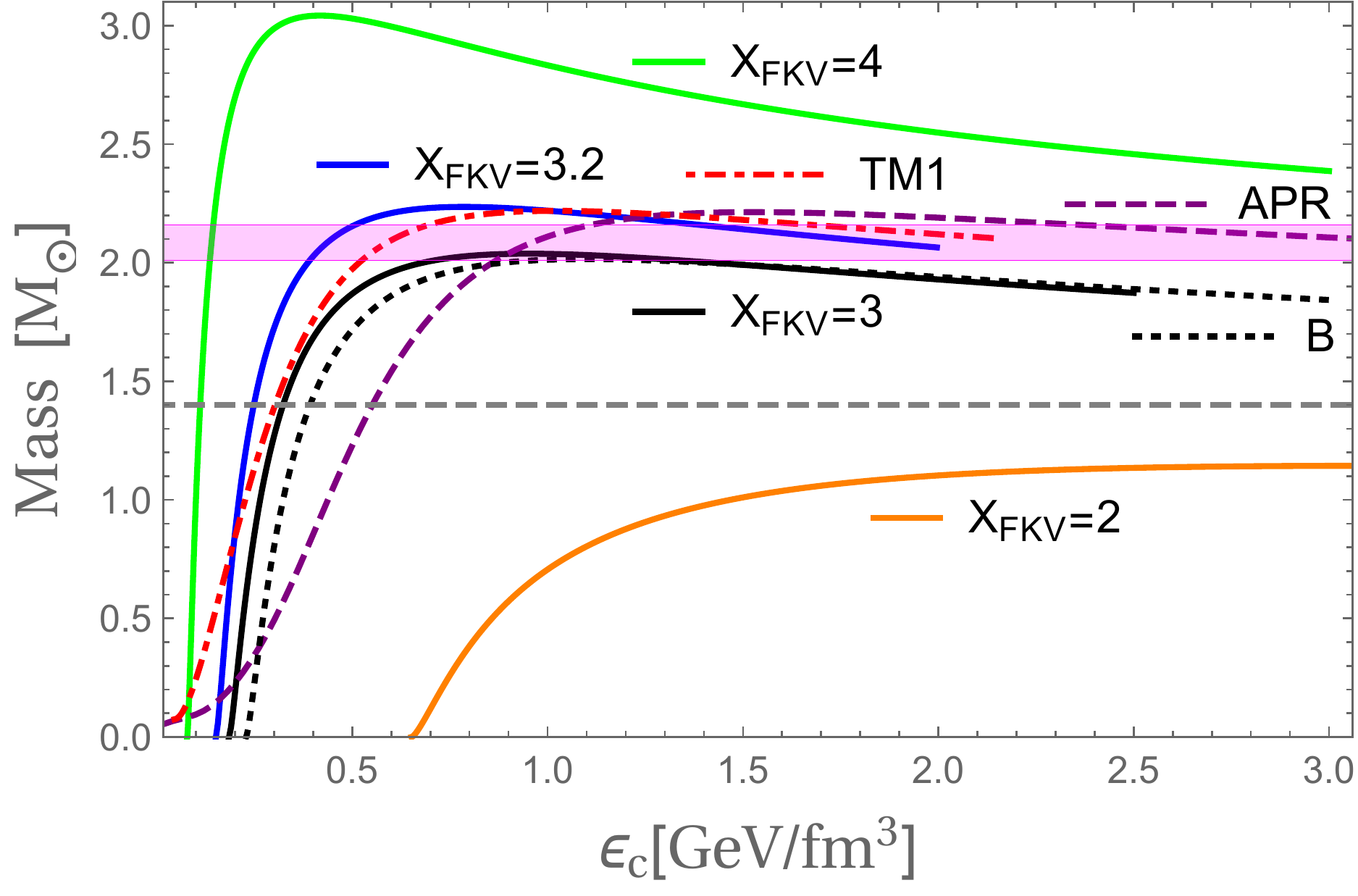}}
\end{center}
\vspace{5mm}
\caption{Total gravitational mass $M$ (in solar mass units) versus central energy density, $\epsilon_{c}$, for the same EoSs of Fig. \ref{fig:3EoSs}. The horizontal light-purple band represents the astrophysical constraint coming the gravitational wave signal GW170817 \cite{Rezzolla:2017aly}. The horizontal gray dashed line represents stellar configurations with $M=1.4M_\odot$. Taken from Ref. \cite{Jimenez:2019iuc}.}
\label{fig:3MassDens}
\end{figure}

\begin{figure}[h!]
\begin{center}
\resizebox*{!}{7.0cm}{\includegraphics{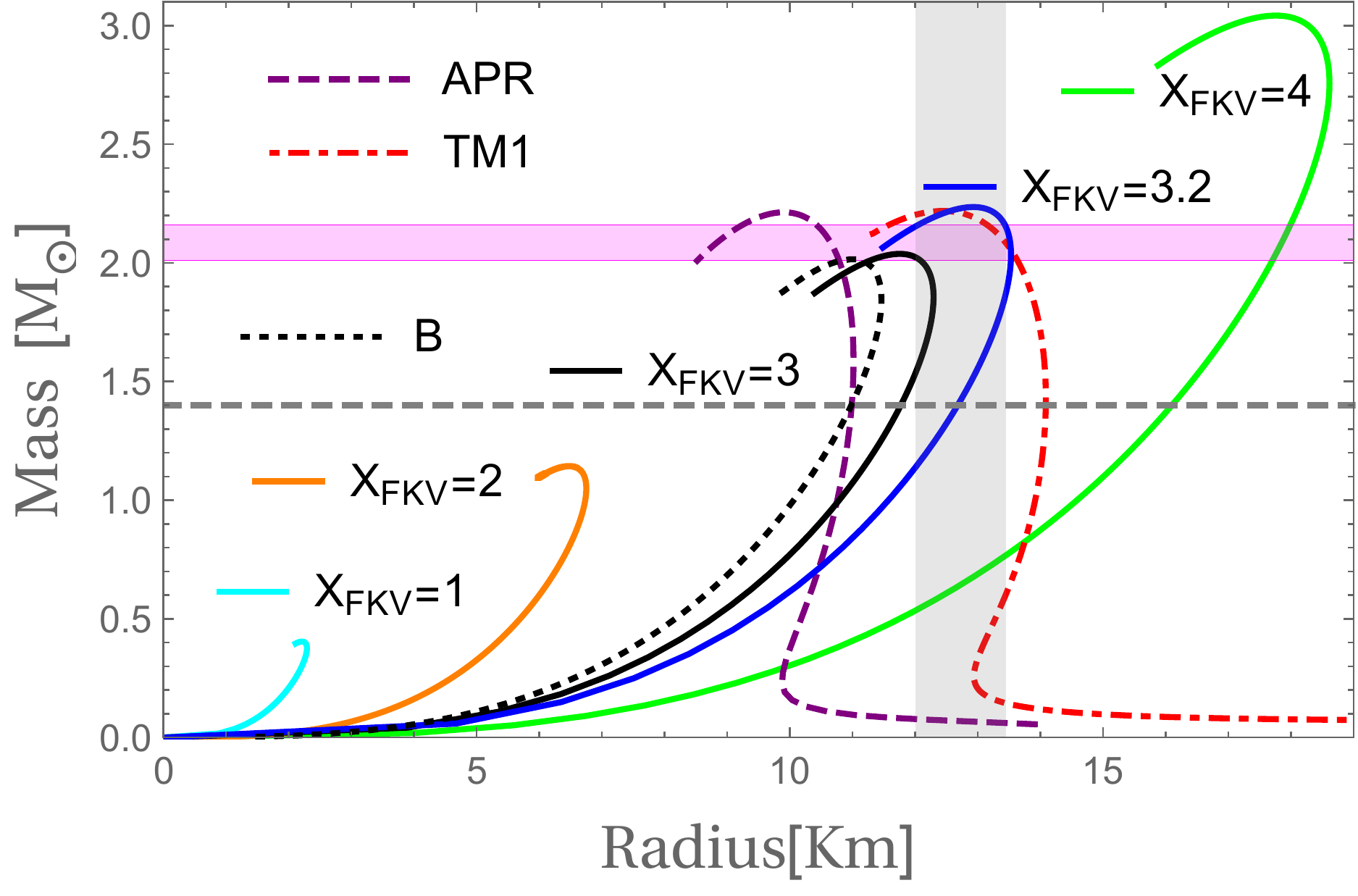}}
\end{center}
\vspace{5mm}
\caption{Mass-radius diagram for EoSs from pQCD for a few values of $X_{\rm FKV}$, the bag model $B=(145\rm MeV)^4$, and nuclear matter EoSs APR and TM1 (see text). The vertical light-gray band represents the maximum (13.45km) and minimal (12km) radii for a NS with $M=1.4M_\odot$ when using the gravitational wave constraint of Ref. \cite{Most:2018hfd}. Taken from Ref. \cite{Jimenez:2019iuc}.}
\label{fig:3MassRad}
\end{figure}

In Fig. \ref{fig:3MassDens} we show our results for the mass as a function of the central energy density and in Fig. \ref{fig:3MassRad} the mass-radius diagram. In both figures we indicate the astrophysical constraint on the maximum mass of NS obtained from the gravitational waves coming from the merger event GW170817, namely between $2.01$ and $2.16$ solar masses \cite{Rezzolla:2017aly} (horizontal light-purple band). This event additionally puts a constraint on the radius of a NS of $1.4 M_\odot$ to be between 12 and 13.45 km \cite{Most:2018hfd}, which we indicate in the right panel of this figure as a vertical gray band. From this panel it is straightforward to see that only values of $X_{\rm FKV}$ between 3 and 3.2 satisfy simultaneously the GW170817 constraints of mass and radius, whereas the APR and TM1 EoSs nearly satisfy the mass constraint but not the radius restriction.

To make our discussion more quantitative, we show a few illustrative tables. In Table \ref{tab:3table1} we list the associated values of minimal (at the quark star surface) and maximal (at quark star center) baryon chemical potentials, $\mu^{\rm (min,max)}_{B}$, corresponding to the star with maximum mass. We also present values of the associated central energy densities, $\epsilon^{\rm max}_{c}$, and radii, $R^{\rm max}$. Notice that the maximal values of $\mu_{B}$ for the APR and TM1 EoSs lie slightly above the quark analogues. However, a careful analysis of the APR EoS shows that it violates the causality limit (which demands that any EoS for strongly-interacting matter should satisfy $c\geq{c_{s}~\equiv~}\sqrt{(\partial{P}/\partial\epsilon)_{S}}$ , i.e. the adiabatic (isentropic) \textit{relativistic}\footnote{This $c_{s}$ appears when studying linear perturbations of the relativistic-hydrodynamic equations and it is not just a thermodynamic identity, but is measures the speed at which linear sound waves propagate \cite{Rezzolla:2013rel}. Its non-relativistic counterpart is defined as $(c_{s})_{\rm NR}{~\equiv~}\sqrt{(\partial{P}/\partial\rho)_{S}}$ \cite{Landau:1987flu}, where $\rho$ is the rest-mass density valid in a Newtonian regime. Notice that it is possible (after some careful physical manipulations) to relate both speeds of sound \cite{Rezzolla:2013rel}.} speed of sound $c_{s}$ cannot be larger that the speed of light $c$) before reaching the (central) energy density producing its maximum mass configuration listed in Table \ref{tab:3table1}. Moreover, it was estimated in Ref. \cite{Lattimer:2010uk} that the maximal value of baryon chemical potential at the center of NS would be $2.1$ GeV. 
\begin{table*}[t]
  \begin{center}
    \begin{tabular}{c|c|c|c|c|c} 
       ${\rm EoS}$ & $\mu^{\rm min}_{B}[{\rm GeV}]$ & $\mu^{\rm max}_{B}[{\rm GeV}]$ & $\epsilon^{\rm max}_{c} [{\rm GeV/fm}^3]$ & $M^{\rm max}[{\rm M_{\odot}}]$ & $R^{\rm max}[{\rm km}]$\\
      \hline
        $1$ 		&$2.01$ &$3.091$ &$26.29$ & $0.404087$ & $2.212$\\
        $2$ 		& $1.21$&$1.8376$ &$3.1$ & $1.14363$ & $6.47548$\\
        $3$ 		& $0.91248$ &$1.39$ &$0.982$ & $2.03809$ & $11.7532$\\
        $3.2$ 	& $0.87251$ &$1.323$ &$0.8$ & $2.23551$ & $12.9284$\\
        $4$ 		& $0.75267$ &$1.13$ &$0.416$ & $3.04224$ & $17.757$\\
        $B$ &$0.8285$&$1.2981$ &$1.0977$ & $2.02$ & $10.99$\\
        APR     & $0.9268$ &$2.269$ &$1.5337$ & $2.2$ & $10$\\
        TM1     & $0.932276$ &$1.628$ &$1.02$ & $2.2$ & $13.5$\\
        \end{tabular}
        \vspace{5mm}
        \caption{Equations of state from pQCD (for which we only show the value of $X_{\rm FKV}$), the bag model $B=(145\rm MeV)^4$, and nuclear matter (APR and TM1); minimal and maximal baryon chemical potentials; central energy densities for the maximum mass configurations; maximal masses; and corresponding radii. Taken from Ref. \cite{Jimenez:2019iuc}.}
      \label{tab:3table1}
  \end{center}
\end{table*}
\subsection*{Dynamically stable quark stars}
	
The stability criterion described above requires the non-trivial calculation of the eigenfrequency associated to the fundamental mode of the pulsation problem. However, for practical purposes, it can be reformulated in order to avoid such a long calculation and simplify the stability analysis to some simple rules to be followed. The simplest and standard ``practical'' criterion can be obtained assuming first that the adiabatic index in a pulsating star is the same as in slowly deformed matter, which is true for compact stars\footnote{Although not completely true for the most massive stars within a stellar family since non-linear general relativistic effects might change considerably the adiabatic index.}. This allow us to build stability criteria based on analyzing only the hydrostatic solutions of the TOV equations for a given EoS \cite{Haensel:2007yy}. We subdivide these criteria depending if we compare the gravitational mass against the compact star radii or the central energy densities. We pass to elaborate more on them.

\underline{Stability criteria on the mass--energy density diagram}

 In this case one can show that stellar matter is stable if its total gravitational mass $M$ increases with growing central energy density $\epsilon_{c}$, i.e.
\begin{equation*}
\partial{M}/\partial\epsilon_{c}>{0}.
\end{equation*}
This is called \textit{the static stability criterion} and it is widely used in the literature \cite{Harrison:1965gtg,Zeldovich:1971yak}. We note that this condition is \textit{necessary} but not sufficient. Besides, the opposite inequality $\partial{M}/\partial\epsilon_{c}<{0}$ always implies instability with respect to small deformations.

The behavior for the quark stars obtained from pQCD satisfying this condition is shown in Fig. \ref{fig:3MassDens}. However, the sufficient condition to verify stable configurations is by means of solving the pulsation equations for their radial perturbations, which allow us to study their associated eigenfrequencies.

\underline{Stability criteria on the mass--radius diagram}

Now, by considering a family of equilibrium stellar models parametrized by the central energy density, i.e. $M=M(\epsilon_{c})$ and $R=R(\epsilon_{c})$, one can construct the mass-radius diagram which depends on the EoS of stellar matter. The stability of stellar models with respect to radial oscillations is intimately related to the shape of the $M(R)$ curve \cite{Haensel:2007yy}.

In Fig. \ref{fig:3MRcurve} we show two examples of the $M(R)$ curves calculated for two different EoSs (the EoS on the right panel being overall stiffer). Each curve has three extrema, which will be called critical points ($C_{1}, C_{2}$ and $C_{3}$). These points divide the curves into four segments. A method which enables one to determine the precise number of unstable normal radial modes using the $M(R)$ curve was described by Ref. \cite{Bardeen:1966aa,Thorne:1967hea}. This stability criteria is formulated as follows:

\begin{itemize}

	\item \textit{Changing stability.} At each critical point of the $M(R)$ curve one and only one normal radial mode changes its stability (i.e. stable $\leftrightarrow$ unstable, or unstable $\leftrightarrow$ stable). There are no changes of stability associated with radial pulsations at other points of the $M(R)$ curves.
	
	\item \textit{Number of nodes for a mode which changes stability.} A mode with even number ``$n$'' of radial nodes changes its stability if and only if $dR/d\epsilon_{c}>0$ at the critical point. A mode with odd ``$n$'' changes its stability if and only if $dR/d\epsilon_{c}<0$.
	
	\item \textit{Bend at a critical point and the character of stability change.} One mode becomes unstable (stable) if and only if the $M(R)$ curve bends counterclockwise (clockwise) at the critical point. 

\end{itemize}
\begin{figure*}[h]
\begin{center}
\resizebox*{!}{5.5cm}{\includegraphics{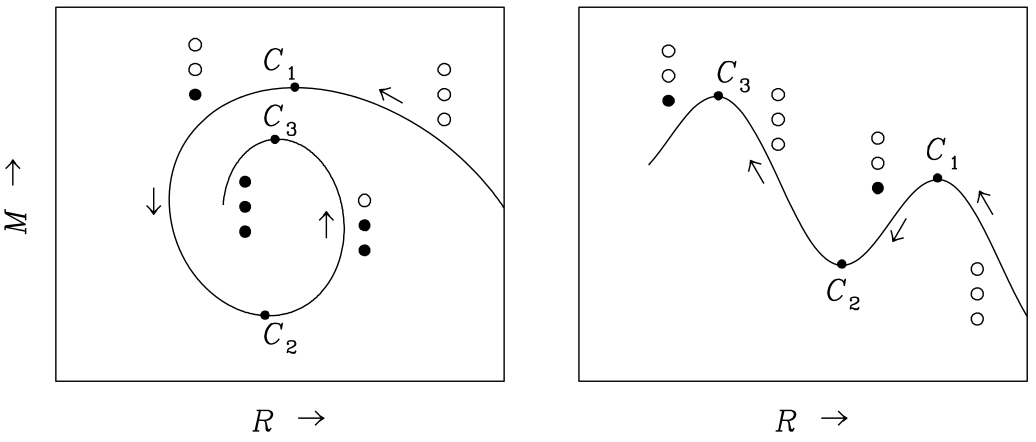}}
\end{center}
\caption{\label{fig:3MRcurve} Fragments of $M(R)$ curves for stellar models with different EoSs. Critical points are denoted by $C_{1}$, $C_{2}$, and $C_{3}$. Arrows near the curves indicate the direction of increasing $\epsilon_{c}$. Three lowest modes of radial pulsations on a given segment are represented by a column of three circles, with unstable modes filled in black. Taken from Ref. \cite{Haensel:2007yy}.}
\end{figure*} 
Let us apply these rules to the $M(R)$ curves in Fig. \ref{fig:3MRcurve}. Let the lowest density segment be stable for all radial modes. At $C_{1}$ the fundamental mode becomes unstable on both curves. It regains its stability at $C_{2}$ in the right panel, because $dR/d\epsilon_{c}<0$ there. However, on the left panel $(dR/d\epsilon_{c})_{C_{2}}>0$; therefore, the fundamental mode remains unstable. At $C_{3}$ the fundamental ($n=0$) mode becomes unstable in the right panel. In the left panel, the 2nd-excited ($n=2$) mode becomes unstable, so that beyond $C_{3}$ all three lowest radial modes are unstable. 

Let us now turn our attention to each particular dependence of the frequencies and periods on the central energy density, gravitational mass and redshift, respectively.

For simplicity, in the following we write only the eigenfrequencies, $\omega_{n}$, in terms of the linear frequency defined as $f_{n}~{\equiv}~{\omega_{n}/{2\pi}}$. In particular, the fundamental and first excited oscillation modes, i.e. $n=0,1$, are very relevant since they are the easiest to be excited by external (radial) perturbations. Besides, they will turn out to be very sensitive to the interactions encoded in the EoS from high-density perturbative QCD. Higher eigenfrequencies ($n=2,3,...$) can also be calculated and apparently they would be interesting for potential observations due to their larger numerical values. However, they are still very large to be detected by modern techniques in very convenient conditions, like gravitational waves coupled to radial modes which are around 1.6 kHz \cite{Passamonti:2007tm}, and thus we do not exhibit their values in this work. 

Generically, from Figs. \ref{fig:3f0f1dens} to \ref{fig:3t0t1Z}, we can see that the fundamental ($n=0$) and first-excited ($n=1$) mode frequencies, $f_{n=0}$ and $f_{n=1}$ respectively, behave differently for different values of $X_{\rm FKV}$, producing a large band of possibilities. As expected, their behavior is quite different from the nuclear matter EoSs (APR and TM1) which we plot for comparison. It is clear from these figures that different renormalization scales affect qualitative features for these radial oscillation frequencies. Moreover, the scaling law for the periods of the bag model (and some of its modified versions), $\bar{\tau}_{n}=(B/\bar{B})^{1/2}\times{\tau}_{n}$ 
\cite{Benvenuto:1991}, is not realized in the case of the equation of state coming from cold and dense perturbative QCD since conformal invariance is broken by interactions via the running of the strong coupling and quark masses.

\subsubsection*{Dependence on the energy density}

In Figs. \ref{fig:3f0f1dens} and \ref{fig:3t0t1dens} it is shown that, although the quantitative behavior of the fundamental and first excited ($n=0,1$) vibrational modes (as functions of $\epsilon_{c}$) of quark stars are very sensitive to $X_{\rm FKV}$, their qualitative behavior is similar to the bag model (${\rm B^{1/4}=145MeV}$) for all values of central energy density. As expected, nuclear matter stars (APR and TM1) behave quite differently. In Table \ref{tab:3table2} we list values of central energy densities, frequencies and periods for the modes $n=0,1$ for a canonical neutron star of $M=1.4M_{\odot}$ for the EoSs we use along this work. The choice of mass follows from the fact that most of the observed pulsars tend to have masses near this value. Notice that although the bag model surpasses the two-solar mass constraint only for high central energy densities, the EoS obtained from the FKV formula requires relatively low-energy central densities to produce heavy stars.

\begin{figure*}[h!]
\begin{center}
\hbox{\includegraphics[width=0.5\textwidth]{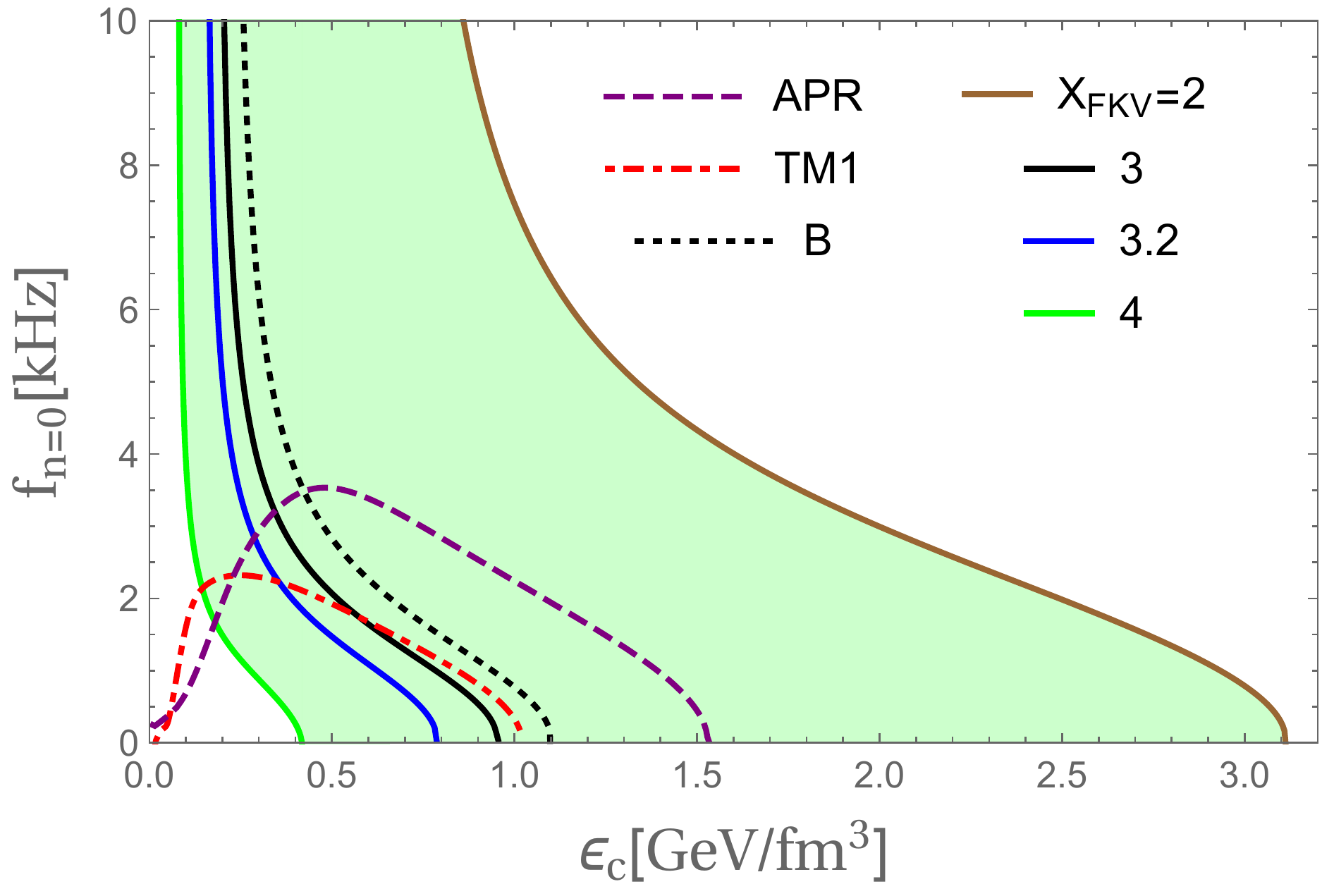}
	  \includegraphics[width=0.5\textwidth]{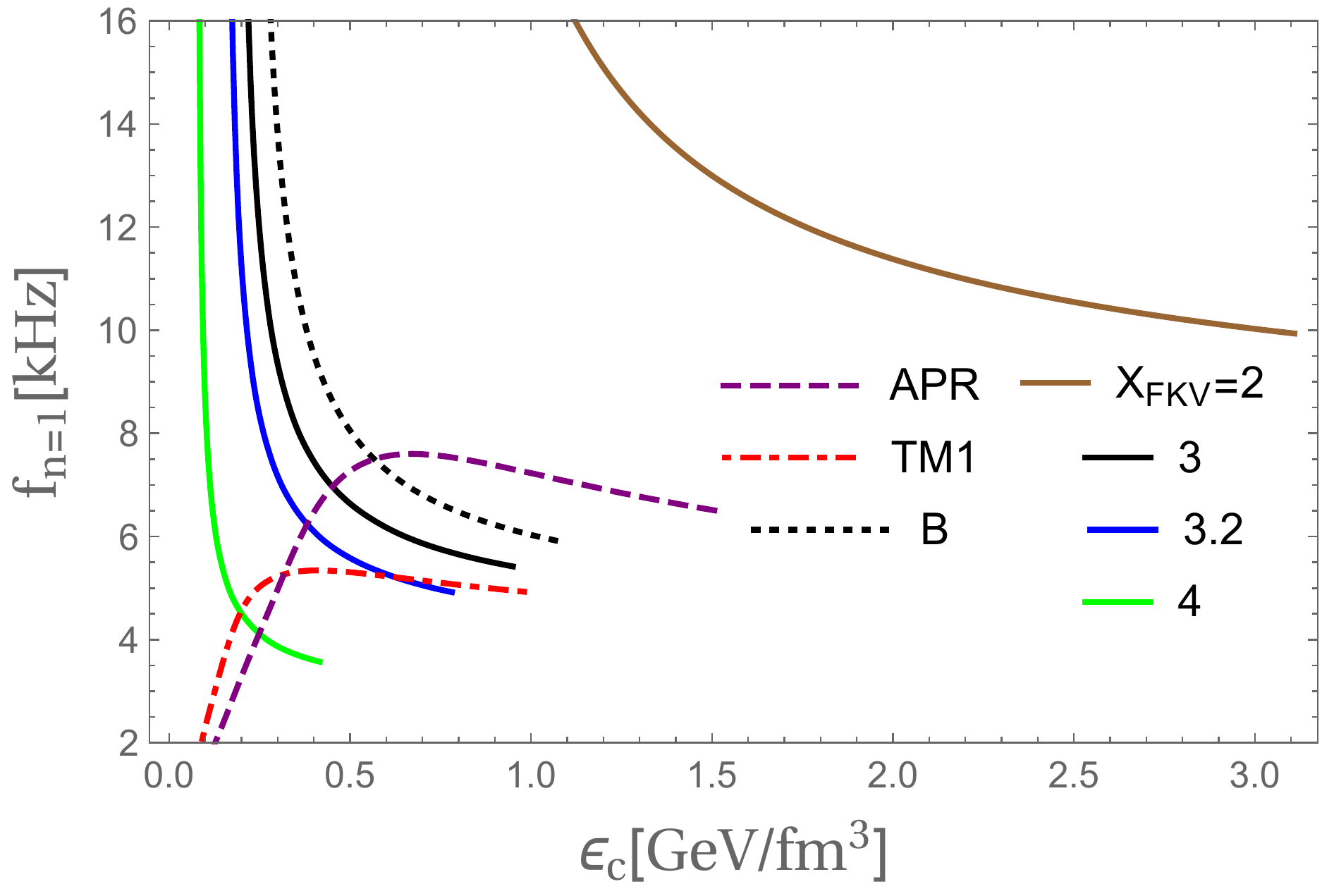}}
\vspace{5mm}
\caption{Fundamental mode, $f_{n=0}$, and first mode, $f_{n=1}$, frequencies as functions of the central energy density. We show results using EoSs from pQCD, the bag model ($B$) and nuclear matter (APR and TM1). Taken from Ref. \cite{Jimenez:2019iuc}.}
\label{fig:3f0f1dens}
\end{center}
\end{figure*}

\begin{figure*}[h!]
\begin{center}
\hbox{\includegraphics[width=0.5\textwidth]{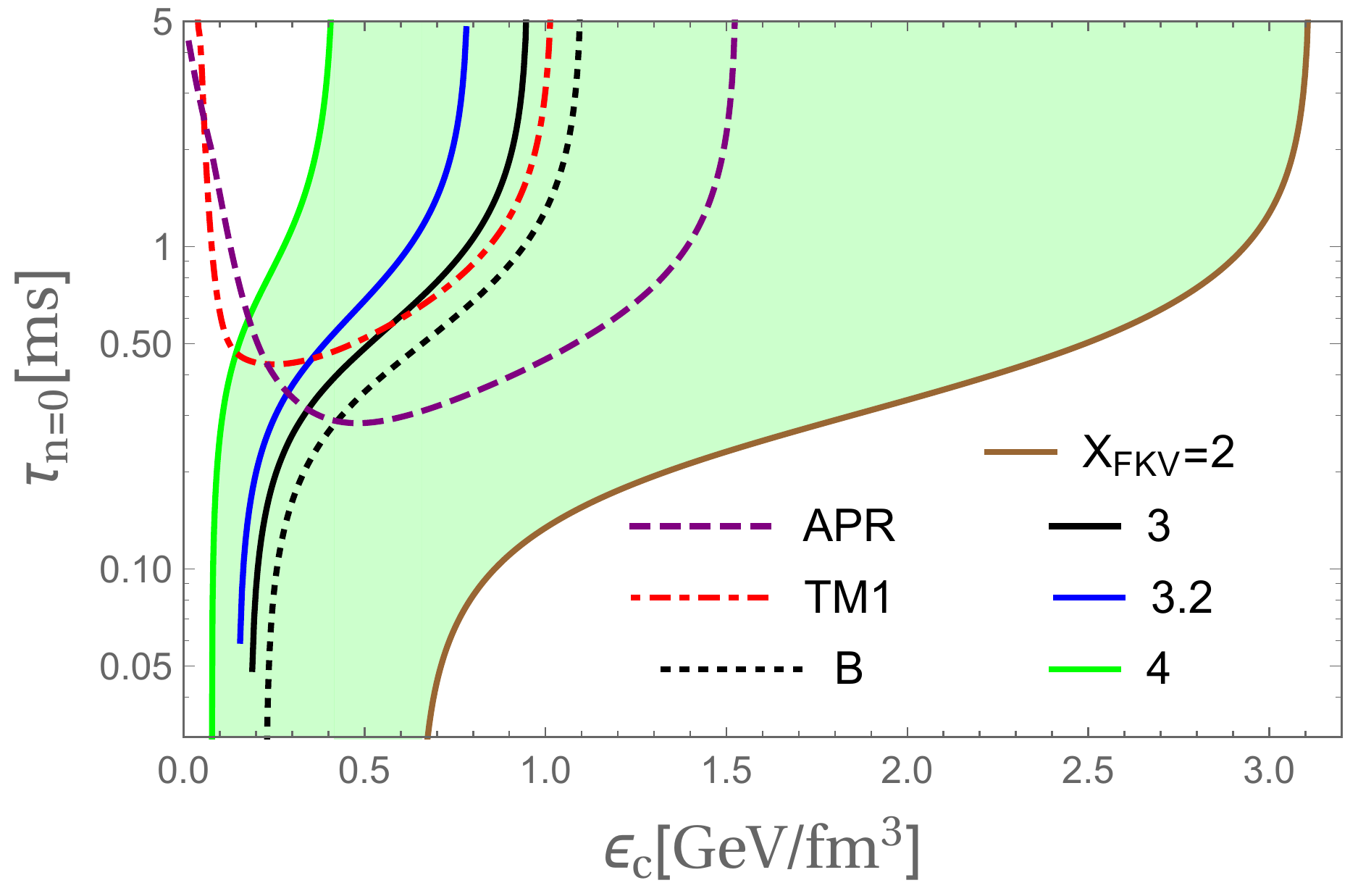}
	  \includegraphics[width=0.5\textwidth]{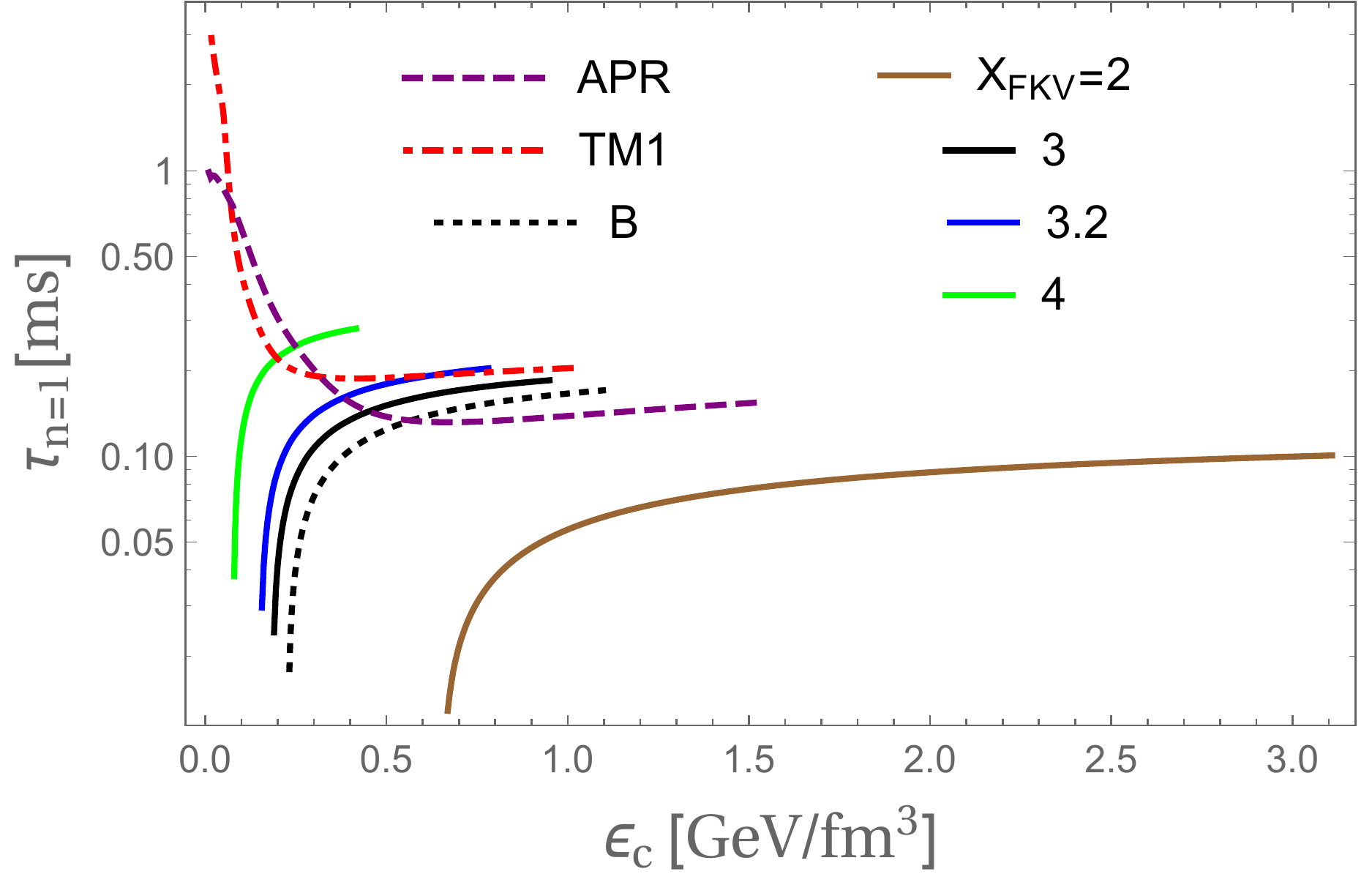}}
\vspace{5mm}
\caption{Fundamental mode, $\tau_{n=0}$, and first mode,  $\tau_{n=1}$, periods as functions of the central energy density. We show results using EoSs from pQCD, the bag model ($B$) and nuclear matter (APR and TM1). Taken from Ref. \cite{Jimenez:2019iuc}.}
\label{fig:3t0t1dens}
\end{center}
\end{figure*}

\begin{table*}[t]
  \begin{center}
    \begin{tabular}{c|c|c|c|c|c} 
      ${\rm EoS}$ & $\epsilon_{c}[{\rm GeV/fm^{3}}]$ & $f_{0}[{\rm kHz}]$ & $\tau_{0}[{\rm ms}]$ & $f_{1}[{\rm kHz}]$ & $\tau_{1}[{\rm ms}]$\\
      \hline
      $3$ & $ \approx 0.30$ &$ \approx 3.5 $ &$ \approx 0.29 $ & $\approx 9$ & $\approx 0.11$ \\
      $3.2$ & $ \approx 0.26$ &$ \approx 3.5 $ &$ \approx 0.29 $ & $\approx 9$ & $\approx 0.11$ \\
      $4$ & $ \approx 0.10$ &$ \approx 3.3 $ &$ \approx 0.30 $ & $\approx 8$ & $\approx 0.13$ \\
      ${\rm B}$ & $ \approx 0.40$ & $ \approx 4.0 $ &$ \approx 0.25 $ & $\approx 10$ & $\approx 0.10$ \\
      APR & $ \approx 0.56$ & $ \approx 3.5$ &$ \approx 0.29$ & $\approx 8$ & $\approx 0.13$ \\
      TM1 & $ \approx 0.29$ & $ \approx 2.0 $ &$ \approx 0.50 $ & $\approx 6$ & $\approx 0.17$ \\
    \end{tabular}
    \vspace{5mm}
        \caption{
Values of central energy densities ($\epsilon_{c}$), fundamental ($f_{n=0}$) and first-excited ($f_{n=1}$) mode frequencies, and their associated periods ($\tau_{n=0}$ and $\tau_{n=1}$, respectively) for stars with mass $M=1.4M_{\odot}$, obtained from equations of state from the FKV formula (for different values of $X_{\rm FKV}$), for nuclear matter and the bag model $B=(145\rm MeV)^4$ (see text). Taken from Ref. \cite{Jimenez:2019iuc}.}
    \label{tab:3table2}
  \end{center}
\end{table*}

\subsubsection*{Dependence on the gravitational mass}

It is clear from Figs. \ref{fig:3f0f1mass} and \ref{fig:3t0t1mass} that choosing $X_{\rm FKV}\gtrsim{1}$ yields very compact quark stars with higher (lower) values of frequencies (periods), in contrast to the ones provided by the bag model (the opposite happening for larger values of $X_{\rm FKV}$). For instance, the fundamental period of $X_{\rm FKV}=1$ takes a maximum value of approximately 0.1 milliseconds before it diverges at its maximum mass configuration. Notice also that for $X_{\rm FKV}$ approximately between $3$ and $4$, although producing heavy strange quark stars satisfying the two-solar mass constraint straightforwardly, their low-mass sector of the stellar sequence have lower values of frequency signalling that strong interactions play a role in making those stars less deformable (i.e. more compact) against external radial perturbations\footnote{The frequencies and periods were calculated for $X_{\rm FKV}\sim3(3.2)$ that generate maximum masses of $\sim2(2.2)M_{\odot}$, respectively.}.  

\begin{figure*}[h!]
\begin{center}
\hbox{\includegraphics[width=0.5\textwidth]{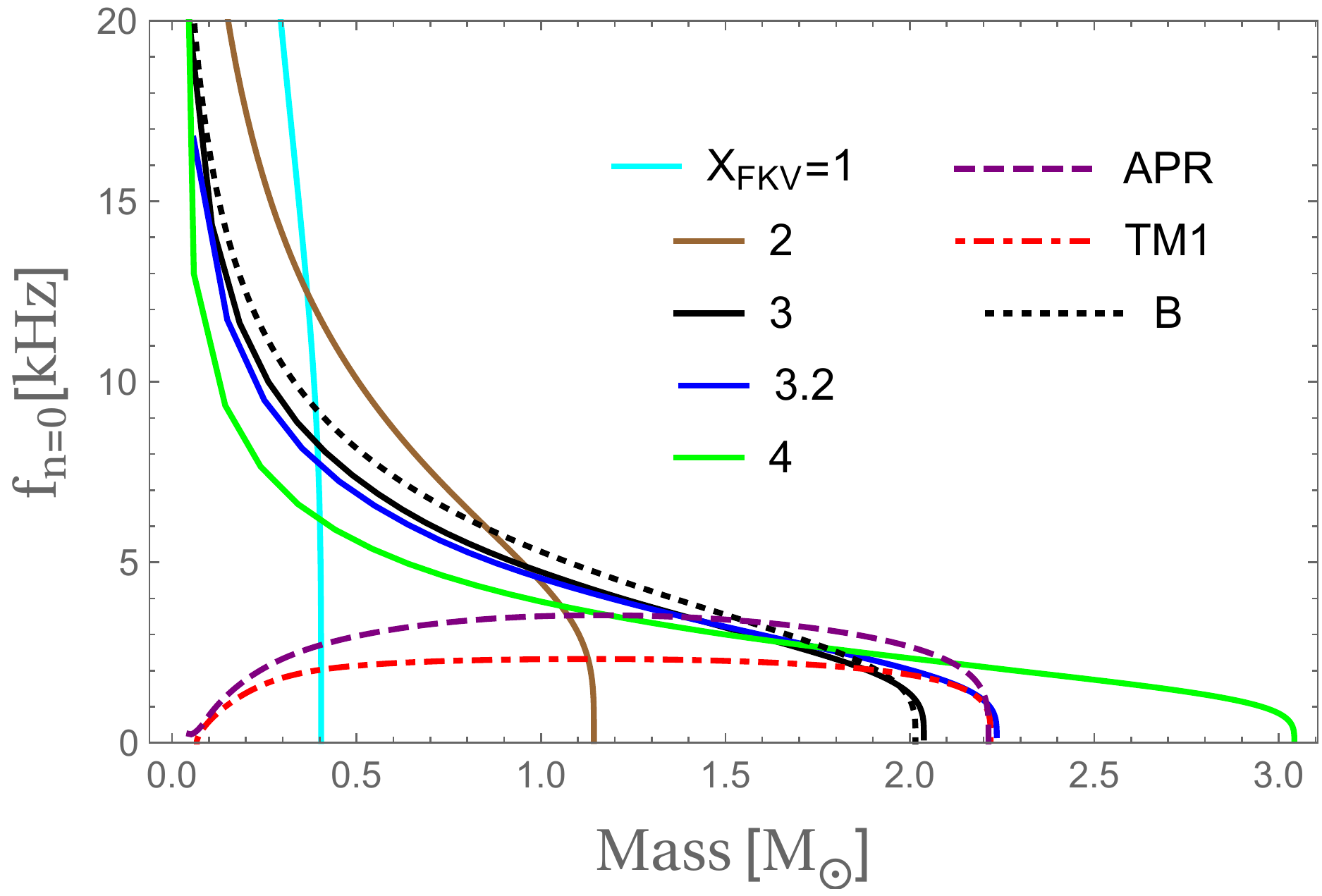}
	  \includegraphics[width=0.5\textwidth]{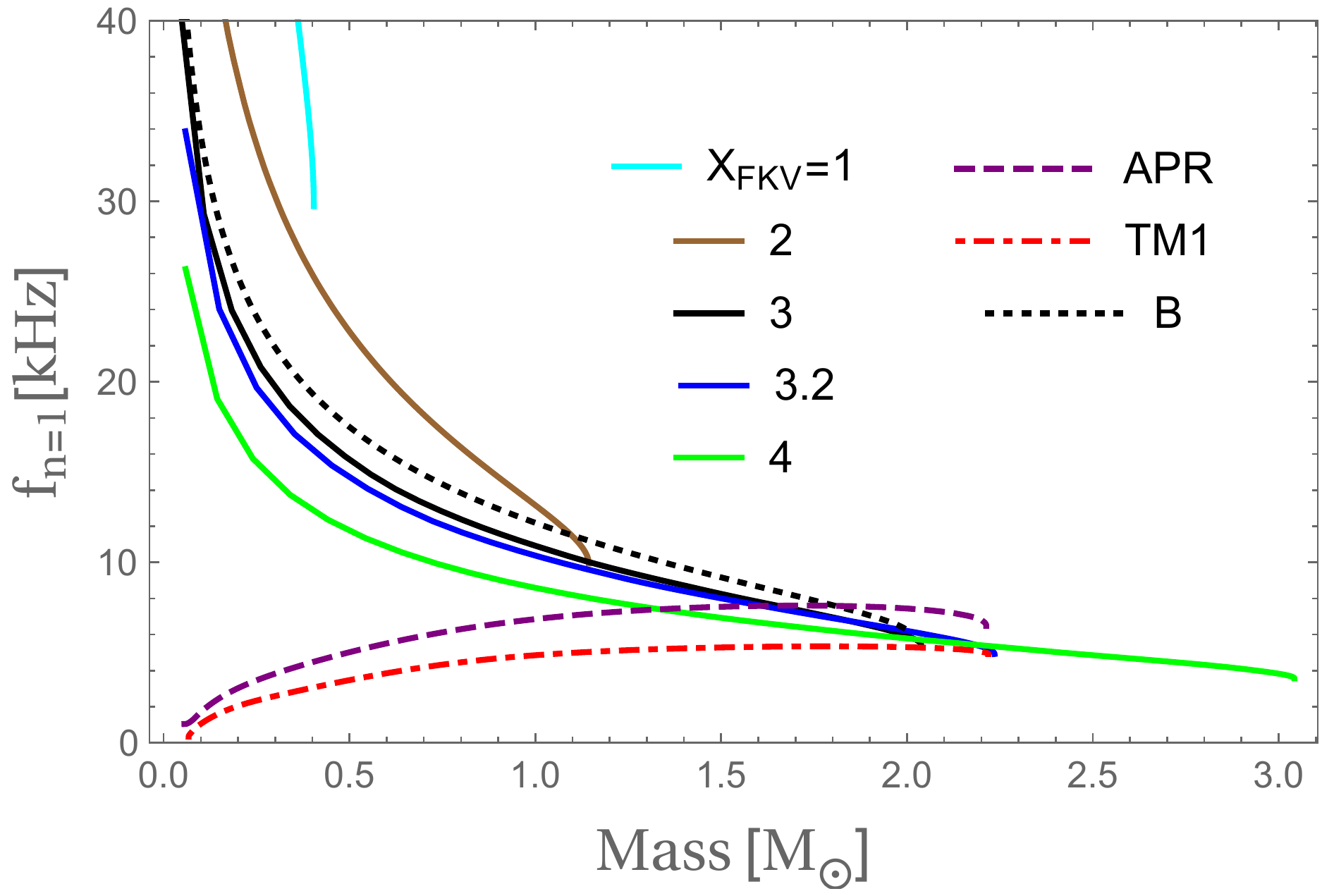}}
\vspace{5mm}
\caption{Same as in Fig. \ref{fig:3f0f1dens} but now the frequencies are functions of the total gravitational mass $M$. Taken from Ref. \cite{Jimenez:2019iuc}.}
\label{fig:3f0f1mass}
\end{center}
\end{figure*}

\begin{figure*}[h!]
\begin{center}
\hbox{\includegraphics[width=0.5\textwidth]{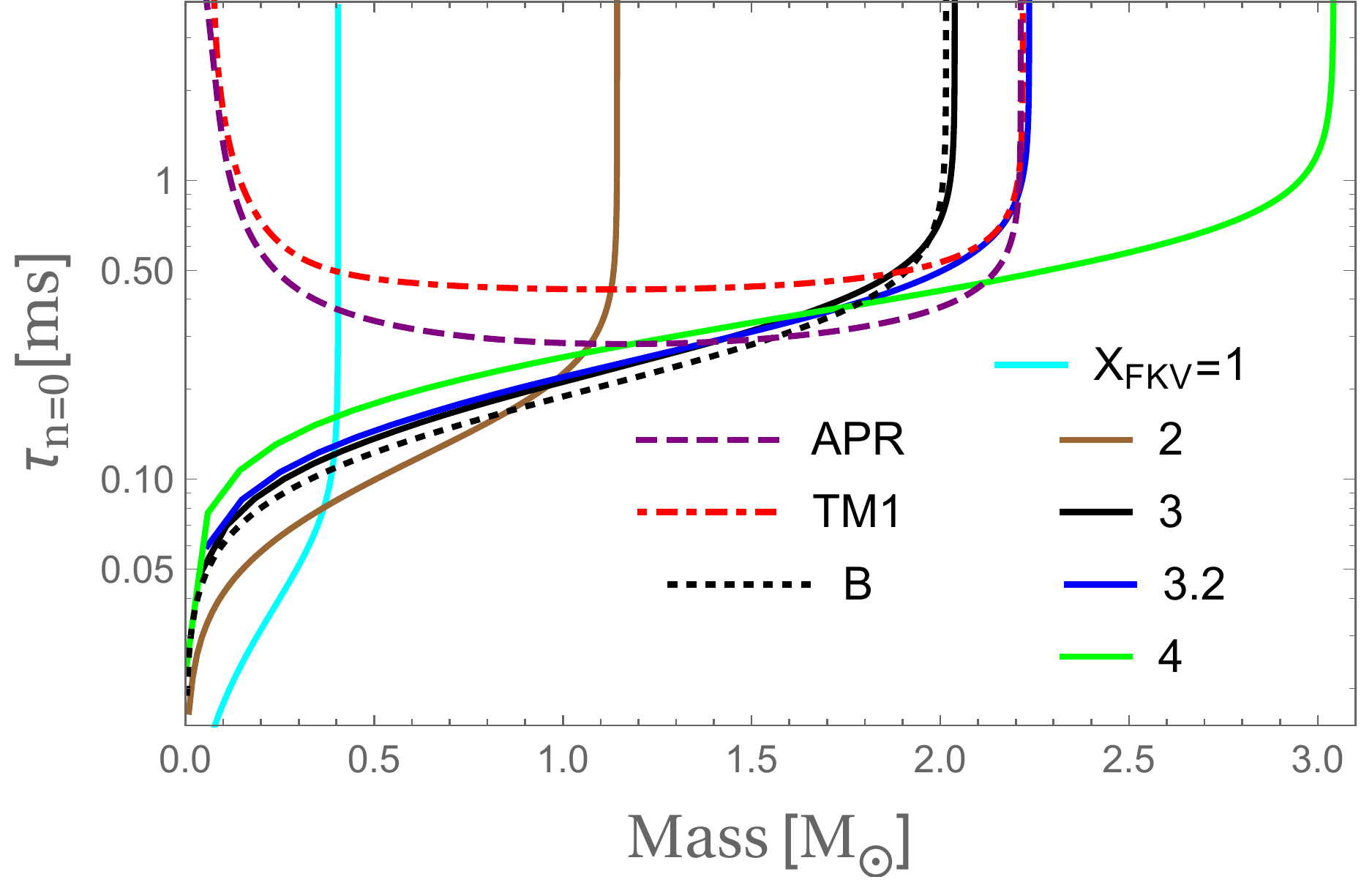}
	  \includegraphics[width=0.5\textwidth]{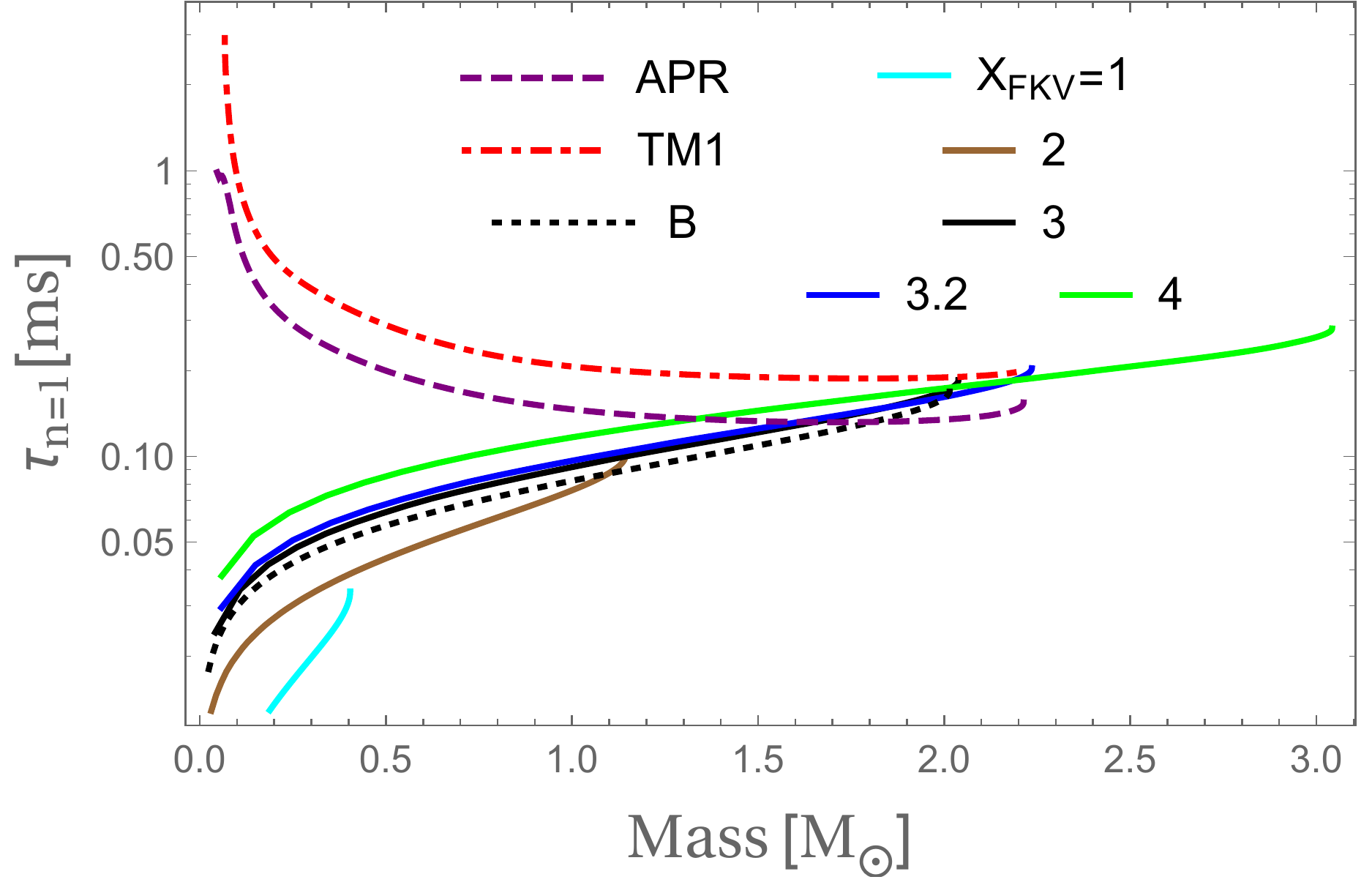}}
\vspace{5mm}
\caption{Same as in Fig. \ref{fig:3t0t1dens} but now the periods are functions of the total gravitational mass $M$. Taken from Ref. \cite{Jimenez:2019iuc}.}
\label{fig:3t0t1mass}
\end{center}
\end{figure*}

Taking into account the recent gravitational wave constraint from the GW170817 event on the maximum gravitational mass of neutron stars as being in the range $2.01M_{\odot} \leq M_{\rm max} \lesssim 2.16M_{\odot}$ \cite{Rezzolla:2017aly,Most:2018hfd}, we can extract additional limits on the values of oscillation frequencies and periods. We list the values of fundamental (first-excited) oscillation frequencies and corresponding periods in Table \ref{tab:3table3} (\ref{tab:3table4}) for stellar configurations within this range of maximum mass, indicating the values corresponding to the lower and upper limits in the previous inequality.

\begin{table*}[t]
  \begin{center}
    \begin{tabular}{c|c|c|c|c} 
       ${\rm EoS}$ & $f^{\rm lower}_{0}[{\rm kHz}]$ & $\tau^{\rm lower}_{0}[{\rm ms}]$ & $f^{\rm upper}_{0}[{\rm kHz}]$ & $\tau^{\rm upper}_{0}[{\rm ms}]$ \\
      \hline
        $3$ & $ \approx 1.0$ &$ \approx 1.0 $ & - & -\\
        $3.2$ & $ \approx 2.0$ &$ \approx 0.5 $ &$ \approx 1.0 $ & $\approx 1.0$\\
        $B$ & $ \approx 1.0$ &$ \approx 6.0^{(*)} $ & - & -\\
        APR & $ \approx 3$ &$ \approx 0.33 $ &$ \approx 2.0 $ & $\approx 0.5$\\
        TM1 & $ \approx 2.0$ &$ \approx 0.5 $ &$ \approx 1.0 $ & $\approx 1.0$\\
        \end{tabular}
        \vspace{5mm}
        \caption{Frequencies and periods of the fundamental mode for stellar configurations satisfying the gravitational wave event GW170817 on compact stars as having masses between $M^{\rm lower}_{\rm max}=2.01M_{\odot}$ and $M^{\rm upper}_{\rm max}=2.16M_{\odot}$. The period marked with (*) is notably different from the ones for other stellar configurations because it is very close to the maximum mass where the fundamental period diverges being $B=(145\rm MeV)^4$. Taken from Ref. \cite{Jimenez:2019iuc}.}
      \label{tab:3table3}
  \end{center}
\end{table*}
Strange stars with masses around the $2M_{\odot}$ limit have periods that tend to be higher than $1$ ms, whereas low-mass strange stars tend to have periods that are smaller and smaller, making them difficult to be detected by modern techniques including drifting subpulses and micropulses \cite{Benvenuto:1991jz}. The value of $X_{\rm FKV}$ is then constrained to be in the range of $\sim 3-3.2$. The period for the case of quark stars tend to be in the range of $\sim 0.4-2.9$ ms, which is something new from pQCD that the bag model $B$ cannot reproduce since although it can reach two solar masses, it cannot go above this limit without violating the Bodmer-Witten hypothesis unless effective interaction terms are added to the equation of state \cite{Weissenborn:2011qu}. 

\begin{table*}[t]
  \begin{center}
    \begin{tabular}{c|c|c|c|c} 
       ${\rm EoS}$ & $f^{\rm lower}_{1}[{\rm kHz}]$ & $\tau^{\rm lower}_{1}[{\rm ms}]$ & $f^{\rm upper}_{1}[{\rm kHz}]$ & $\tau^{\rm upper}_{1}[{\rm ms}]$ \\
      \hline
        $3$ & $ \approx 6.0$ &$ \approx 0.17 $ & - & -\\
        $3.2$ & $ \approx 6.0$ &$ \approx 0.17 $ &$ \approx 5.0 $ & $\approx 0.20$\\
        $B$ & $ \approx 6.0$ &$ \approx 0.17 $ & - & -\\
        APR & $ \approx 8.0$ &$ \approx 0.13 $ &$ \approx 8.0 $ & $\approx 0.13$\\
        TM1 & $ \approx 5.0$ &$ \approx 0.20 $ &$ \approx 5.0 $ & $\approx 0.20$\\
        \end{tabular}
        \vspace{5mm}
        \caption{Same as Table \ref{tab:3table3} but for frequencies and periods of the first-excited mode of oscillation for the mentioned EoSs with $B=(145\rm MeV)^4$. Taken from Ref. \cite{Jimenez:2019iuc}.}
      \label{tab:3table4}
  \end{center}
\end{table*}

\subsubsection*{Dependence on the redshift}

As it was mentioned some sections above, an \textit{in-principle} measurable astrophysical quantity associated to compact stars is the redshift parameter\footnote{It represents a way to compare the frequency of an \textit{observed} photon $\omega_{o}$ at infinity with that of the photon \textit{emitted} at the star's surface $\omega_{e}$. Strictly, this gravitational redshift is the fractional change between observed and emitted wavelengths compared to emitted wavelength i.e. $Z~\equiv~\Delta\lambda/\lambda_{e}=\omega_{e}/\omega_{o}-1$, which for a Schwarzschild star becomes our given definition.} defined by $Z=(1-2M/R)^{-1/2}-1$ \cite{Glendenning:2000}, which depends upon the ratio $M/R$ and is ``high'' for relativistic stars. Therefore, it is interesting to study the dependence of frequencies and periods on this parameter $Z$. This can be useful since it allows us to compare two observable quantities in astronomical measurements.

The dependence of the frequencies and periods of quark stars on the gravitational redshift parameter Z are displayed in Figs. \ref{fig:3f0f1Z} and \ref{fig:3t0t1Z}. From this figure, it becomes clear that, independently of the particular EoS used (for different values of $X_{\rm FKV}$ in pQCD, $B$ and TM1) and their maximum masses, the maximum gravitational redshift $Z$ tends to accumulate in the region between 0.42 and 0.48, which can be used to restrict\footnote{This serves as a consistency check since compact (neutron or quark) stars must have $Z~<~2$ as the maximum gravitational redshift in order to be stable, whereas black holes have $Z$ tending to infinity \cite{Glendenning:2000}.} the behavior of the EoS for dense matter when compared to current astronomical observations of $Z$. Although the APR case lies outside this region, one should recall that at high densities (before reaching its maximal mass configuration) it becomes superluminal, i.e. $c_{s}~>~c$. Notice that the first-excited mode displayed in Figs. \ref{fig:3f0f1Z} and \ref{fig:3t0t1Z} seems to distinguish low-mass quark stars from purely hadronic stars.

\begin{figure*}[h!]
\begin{center}
\hbox{\includegraphics[width=0.5\textwidth]{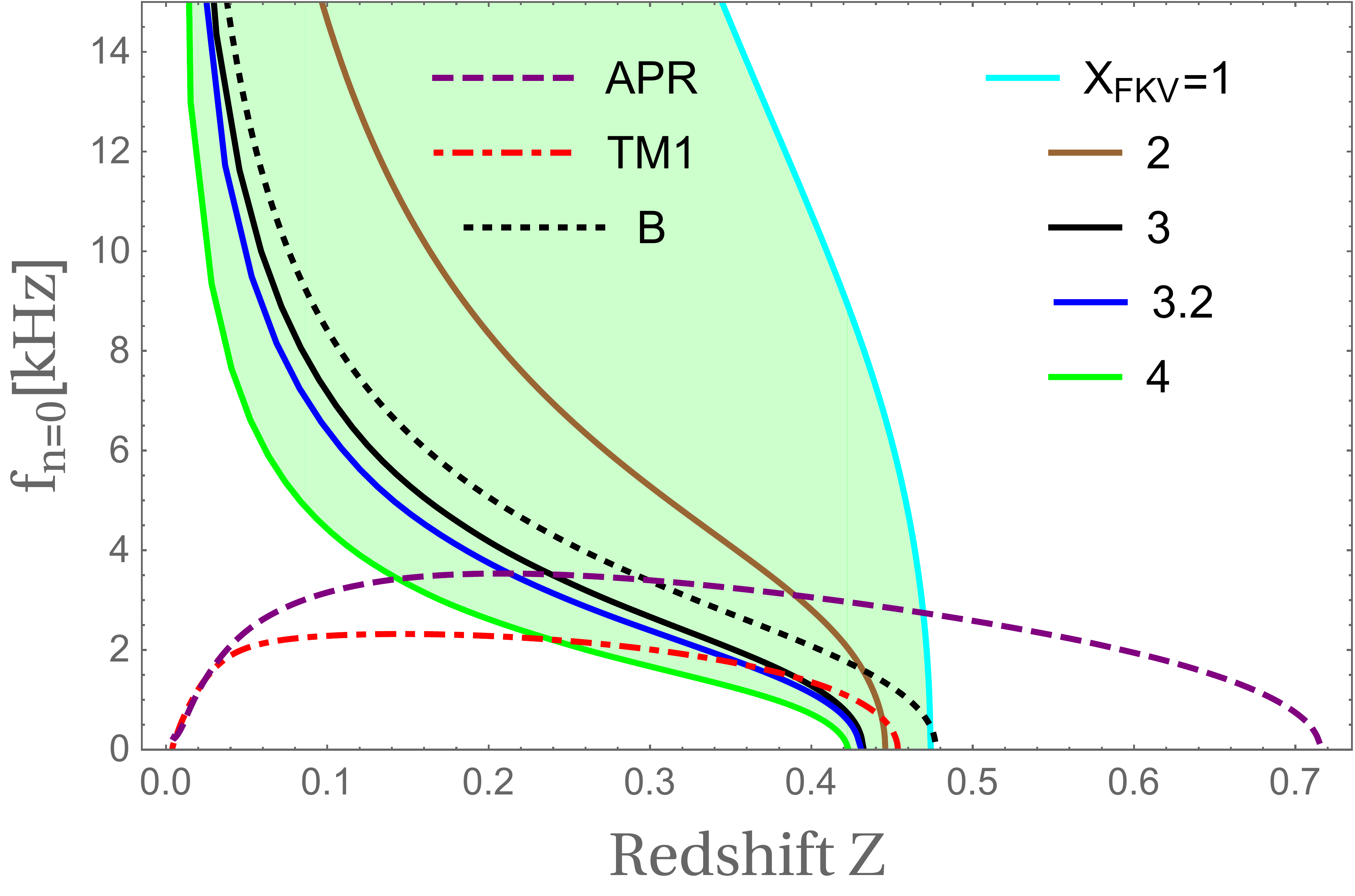}
	  \includegraphics[width=0.5\textwidth]{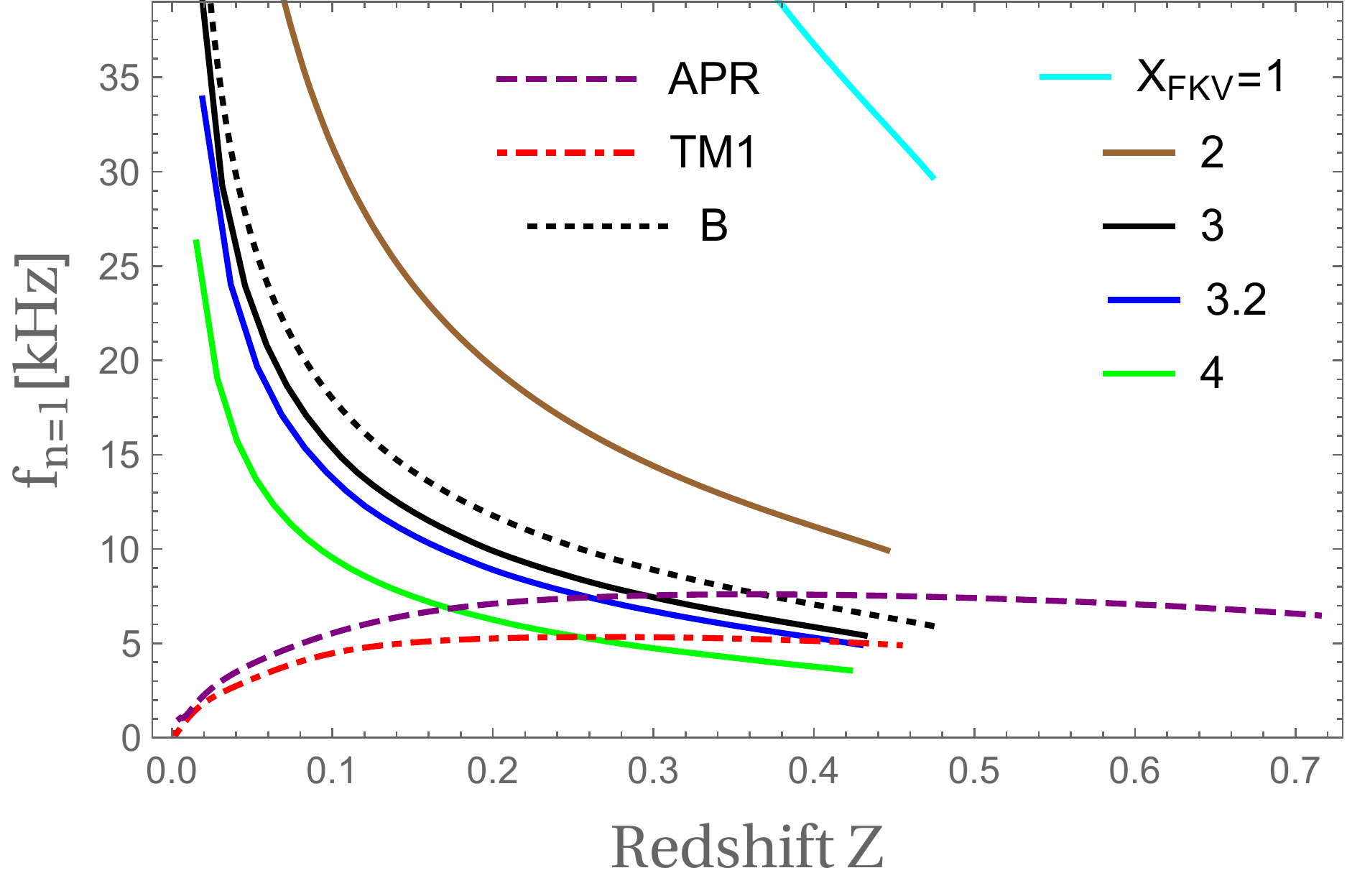}}
\vspace{5mm}
\caption{Same as in Fig. \ref{fig:3f0f1dens} but now the frequencies are functions of the redshift parameter Z. Taken from Ref. \cite{Jimenez:2019iuc}.}
\label{fig:3f0f1Z}
\end{center}
\end{figure*}

\begin{figure*}[h!]
\begin{center}
\hbox{\includegraphics[width=0.5\textwidth]{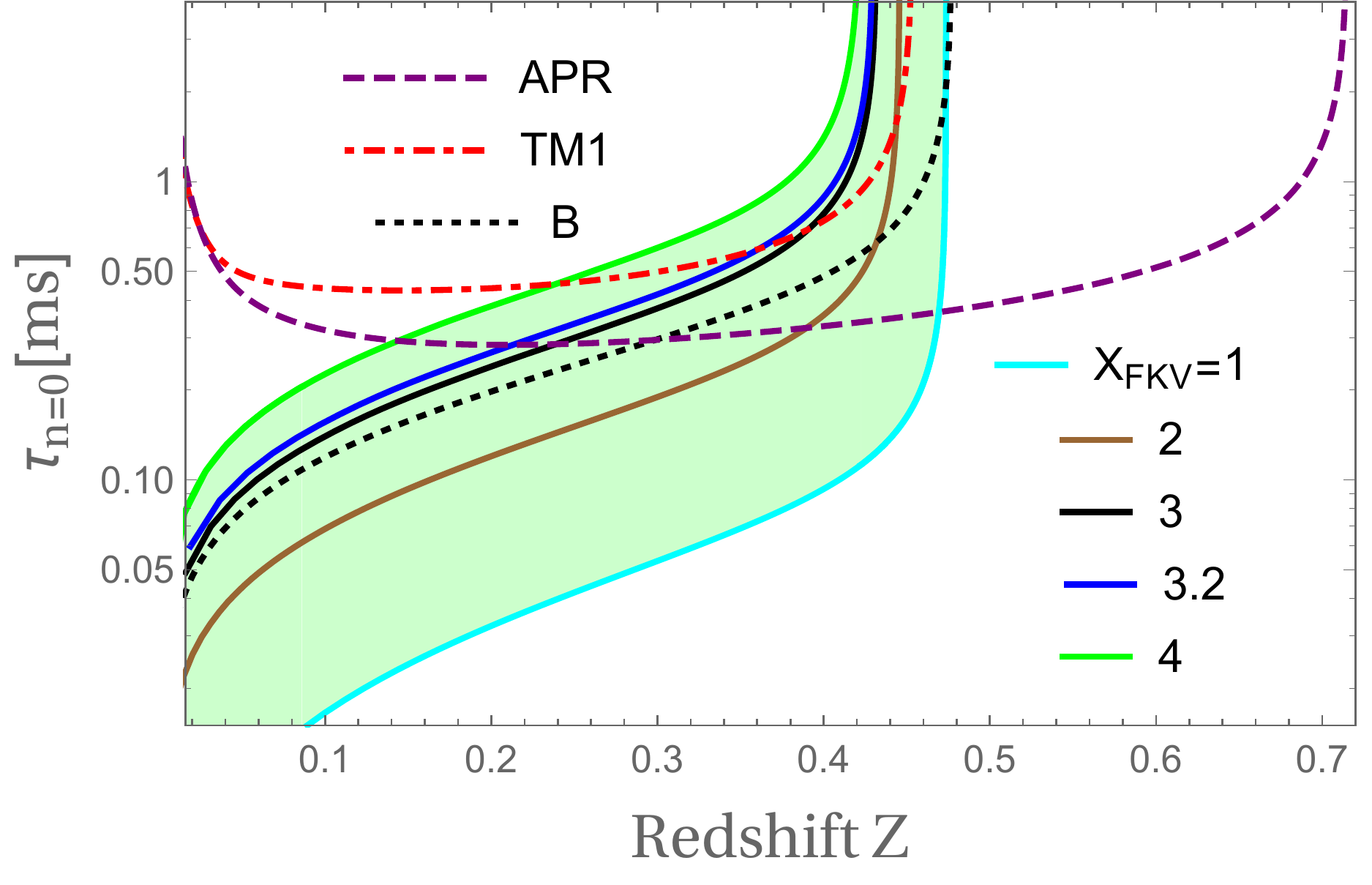}
	  \includegraphics[width=0.5\textwidth]{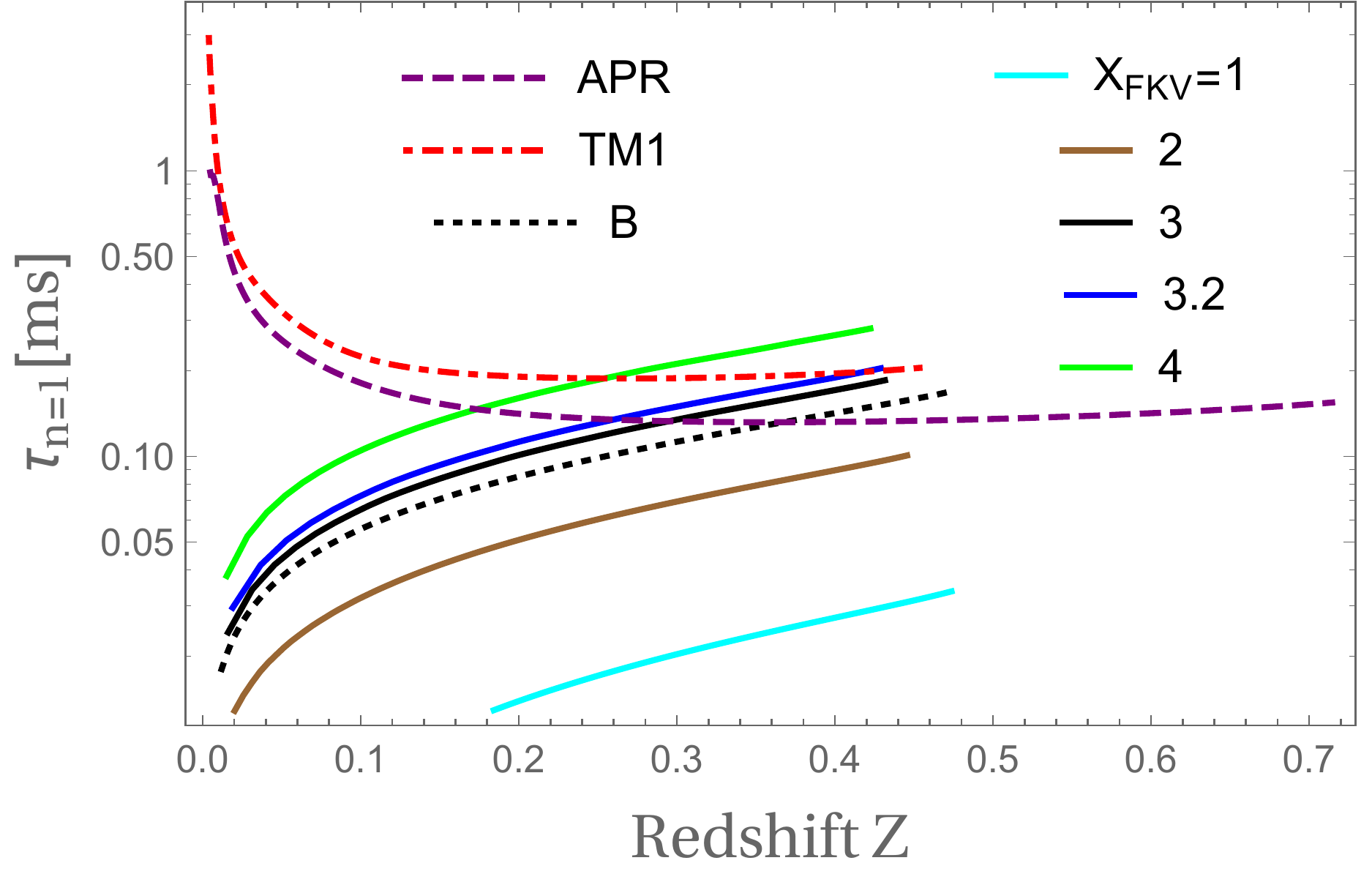}}
\vspace{5mm}
\caption{Same as in Fig. \ref{fig:3t0t1dens} but now the periods are functions of the redshift parameter Z. Taken from Ref. \cite{Jimenez:2019iuc}.}
\label{fig:3t0t1Z}
\end{center}
\end{figure*}

\section{Conclusions} 
	\label{sec:3conclusion}   
    
This chapter was aimed to investigate the relativistic radial oscillations of unpaired bare quark stars and strange stars using an equation of state from perturbative QCD, including up, down, and strange quarks in a cold, dense medium in $\beta$-equilibrium and electrically neutral. For the best of our knowledge, similar studies of the radial oscillation stability were only performed within the MIT bag model framework (occasionally including  minor modifications). Our results contains a natural estimate of the inherent systematic uncertainties in the evaluation of the equation of state, and therefore of all observables that follow, and might bring new insights into the phenomenology of quark stars and their possible observational searches.

Comparing the nucleonic and quark star results obtained in this work, one finds that their fundamental and first excited modes are quite distinguishable for low-mass stars. On the other hand, heavy stars become numerically indistinguishable in the region near the two-solar mass limit. Nevertheless, their curves are different and this could be important to map their vibrational behavior. In fact, this could be used to discriminate between hadronic and quark stars by comparing their non-radial pulsation modes (which are correlated to the radial modes in gravitational waves \cite{Passamonti:2005cz,Passamonti:2007tm}), especially in the case of heavy compact stars, close to the current constraint on their maximum mass \cite{Flores:2013yqa,Flores:2018pnn}.  
 
Our results represent an initial step towards the more realistic case of hybrid star pulsations (see, e.g., Ref. \cite{Sahu:2001iv}), where a hadronic mantle and crust effects should be included. They might also shed light onto the phenomenology of strange dwarfs, which seem to be unstable under radial  perturbations \cite{Glendenning:1994zb,Alford:2017vca}, or more exotic forms including the existence of condensed dark matter in neutron stars \cite{Deliyergiyev:2019vti} and strange stars \cite{Panotopoulos:2017eig}.	    
        
\end{chapter}

\begin{chapter}{Effects of Heavy Quarks on Neutron Star Matter} 
	\label{chap:charm}

\hspace{5 mm}

\section{Introduction}

Currently, the Standard Model of particle physics is considered an \textit{effective} theory since it is unable to describe all of physics up to infinitely high energies (or equivalently, down to infinitely short distances), i.e. all our fundamental theories are effective low-energy (or large-distance) ones. The key idea of these theories consists in setting a high-energy (mass) scale $M$, e.g. the Planck mass $M_{P}$, above which the effective theory breaks down.

In particular, the \textit{ab initio} theory of the strong interactions behaves as an effective theory due to the dependence of the strong coupling $\alpha_{s}$ on the number of active flavors $N_{f}$ in the system. This leads to an enhancement or diminishment of the decreasing nature of this coupling at high energies due to asymptotic freedom. From this feature one can understand the (usually considered obvious) reason of why QCD with all of its 6 flavors (up, down, strange, charm, bottom and top quarks) is rarely used. Besides, QCD theorems allow for a \textit{decoupling} of the heavy quarks (charm, bottom, and top) from the light ones (up, down, and strange) \cite{Shifman:1995dn}. So, the heavy physics does not need to be considered when only light quarks are present in the system. In fact, this QCD decoupling is omnipresent when a practical use of the theory is made at experimentally reachable energy scales. It would be a  mistake to use full 6-flavor QCD at characteristic energies of fractions of GeV. Thus, in most cases when doing calculations in QCD one usually uses an effective QCD version of the full theory, where the heaviest flavors have been \textit{eliminated} \cite{Shifman:1995dn}. For example, if we want to study a system composed only by light quarks (with masses $m_{i}\ll{M}$, where $M$ might be the heavy quark masses) at low energies, the effective QCD Lagrangian will only contain their associated fields, where processes with $\ll{M}$ can be described without introducing this heavy flavor \cite{Shifman:1995dn}. 

	The importance of these theoretical considerations lie at the heart of the physics of strongly interacting matter at extreme conditions. A full-flavor quark-gluon plasma EoS is undoubtedly important for cosmological studies, since this plasma may have existed microseconds after the big bang \cite{DeTar:2010xm}. Under these primordial conditions and longer time scales, heavy quarks probably participated in the thermal ensemble of light quarks as well, which implies that,  for the study of the early universe, the EoS with heavy flavors would be important \cite{Laine:2009ik}. For example, the scale factor of the early Universe is affected by the number of active quark flavors in the EoS used for its determination \cite{McGuigan:2008pz}. Additionally, the experiments at ion colliders create ``fireballs'' that thermalize within $\tau{~\approx~}10^{-24}$ seconds, where usually only the $u$, $d$, and $s$ quarks are considered in the thermal ensemble describing the state of the thermalized fireball. One can ask if the charm quark (which we mention without loosing generality) appears in this plasma under the current experimental conditions. However, it has been argued that probably it is not thermalized, and thus the 2+1 flavor (i.e., up and down massless quarks plus a massive strange quark) EoS is usually considered sufficient for hydrodynamic models \cite{DeTar:2010xm}. This issue is still under debate and intense studies \cite{Laine:2009ik}.
 	
Besides, lattice QCD calculations for large values of the strong coupling and heavy quark masses (although in these works \textit{all} the quark flavors are very heavy, even the up and down) have proved to be useful when studying large values of the baryon chemical potential\footnote{Although not completely well-established due to numerical technicalities which are still under research.}. These numerical calculations can be carried out since the Sign Problem does not appear to be too severe in these conditions and it can be non-trivially avoided \cite{Fromm:2011qi,Fromm:2012eb,
Langelage:2014vpa,Philipsen:2019qqm}. As an example, in Fig. \ref{fig:4heavyPhase} we show some results of these calculations which in turn allow us to draw the phase diagram of QCD with heavy quarks. For further details on the associated numerical uncertainties , see Ref. \cite{Glesaaen:2015vtp}.

    \begin{figure}[ht]
      \begin{center}
	\resizebox*{!}{7cm}{\includegraphics{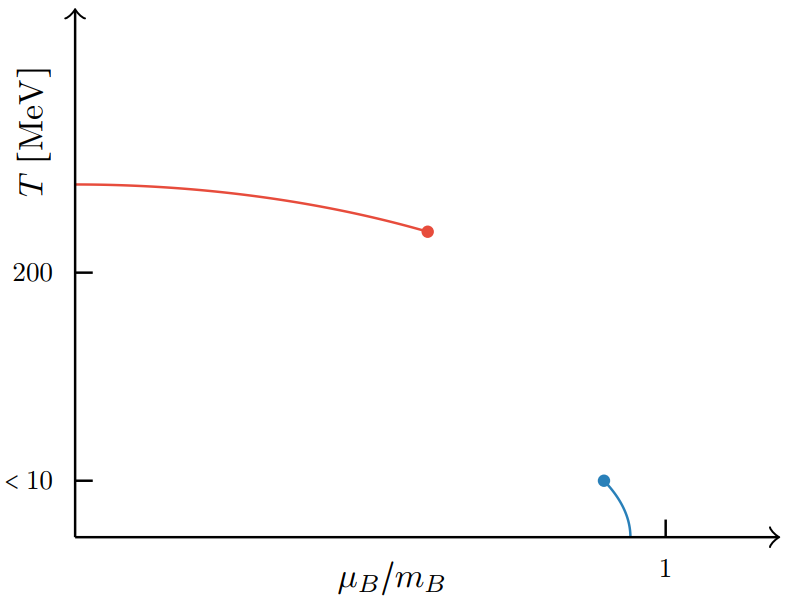}}
     \end{center}
     \vspace{5mm}
      \caption{The phase diagram of QCD with very heavy quarks. At zero density there is a first order deconfinement transition having a critical end point. At low temperatures and higher chemical potentials the nuclear liquid gas transition is obtained. In the infinite mass limit it moves to zero temperature \cite{Glesaaen:2015vtp,Glesaaen:2015aaw}. Figure taken from Ref. \cite{Glesaaen:2015vtp}.}
      \label{fig:4heavyPhase}
    \end{figure}

In addition to these heavy systems, it was pointed out in Ref. \cite{Yasui:2016svc} that the Kondo effect of condensed matter physics\footnote{Traditionally, the Kondo effect refers to an unusual scattering of conduction electrons in a metal with magnetic impurities, contributing to a term to the electrical resistivity that increases logarithmically with temperature as the temperature is lowered (as $\log(T)$).} can occur in a dense and cold light ($u$, $d$, and $s$) quark matter system with some heavy quarks being considered as impurities, known as the QCD Kondo phase. In compact stars, these heavy impurities can be charm quarks in low concentration. Their origin in strange quark stars can be related to neutrino interactions with quark matter since neutrinos emit a $W^{+}(\nu_{e}\rightarrow{W^{+}+e^{-}})$, which then is absorbed by an strange quark (or to a lesser extent by a down quark), which turns into a charm quark ($W^{+}+s\rightarrow{c}$). Thus, it is likely that a finite amount of charms will be present in these compact stars and which might generate the QCD Kondo effect \cite{Macias:2019vbl}.

From all these possibilities we will now focus on heavy quark matter, i.e. quark matter including heavy flavors, which could play a relevant role in extreme situations in the primordial quark-hadron transition \cite{Hajkarim:2019csy}.
Experimentally, it is expected that FAIR with its Compressed Baryonic Matter (CBM) experiment will be able to produce charm quarks immediately after heavy-ion collisions with energies close to or above the charm threshold \cite{Ablyazimov:2017guv}. Cold quark matter also brings about the possibility of charm stars. Since the critical density required for their appearance is far above the limit imposed from causality together with the existence of two-solar mass neutron stars, as discussed in Refs. \cite{Lattimer:2010uk,Lattimer:2019eez}, such stars might be realized in nature only as a new branch of ultradense hybrid compact stars.

As we go to higher values of quark mass, asymptotic
freedom makes the perturbative quantum chromodynamics
(pQCD) formalism more reliable \cite{Kapusta:2006pm}, so that this approach could be useful for heavy-ion collisions at low temperatures and high baryon chemical potentials as well as the physics of compact stars at ultrahigh densities, cases where charm quarks could play a role. One needs, then, to build the equation of state (EoS) for charm matter taking into account the constraints the system must respect below and above the charm threshold to generate matter configurations which are stable under electroweak interactions.

At high temperatures and zero quark chemical potentials,
perturbative QCD was employed by Laine and Schröder \cite{Laine:2006cp} to calculate the EoS including the charm quark contribution. These results were later compared to the ones provided by lattice QCD, including the charm and bottom contributions \cite{Borsanyi:2016ksw}, relevant for the study of the primordial Universe and its cosmological transitions \cite{Boeckel:2009ej,Boeckel:2011yj} (see also Ref. \cite{Mukherjee:2015mxc} for similar results around the transition temperature $T_{c}$). Considering nonzero chemical potentials for light and heavy quarks simultaneously implies some subtleties brought about by the heavy quarks at their mass thresholds. There, matching conditions should be imposed \cite{Rodrigo:1993hc}, having nontrivial effects on the possible values assumed by the renormalization scale $\bar{\Lambda}$. 

This chapter is concerned with the study of the effects of heavy quarks on the equation of state for cold and dense quark matter obtained from perturbative QCD (already studied in Chapter 2), yielding observables parametrized only by the renormalization scale. Additionally, in this chapter it is investigated the behavior of charm quark matter under the constraints of $\beta$-equilibrium and electric charge neutrality in a region of densities where in-medium perturbative QCD is in principle much more reliable. Given the equation of state, we discuss $\beta$ equilibrated and electrically neutral charm quark matter, and revisit the possibility of charm (quark) stars under the pQCD
perspective. Those were investigated in the past within the crudest version of the MIT bag model, being ruled out due to instabilities under radial pulsations \cite{Kettner:1994zs,Glendenning:2000,Haensel:2007yy} (see also Refs. \cite{SchaffnerBielich:1998ci,SchaffnerBielich:1997fx}). Quark stars with heavier quarks, i.e. bottom and top were considered also within the bag model in Ref. \cite{Prisznyak:1994ry}. This chapter follows the discussion found in Ref. \cite{Jimenez:2019kji} and it is organized as follows. In Sec. \ref{sec:4pQCDh} we summarize the main aspects of the pQCD formalism for $N_{f}=N_{l}+1$ flavors and present a systematic extension to include heavy quarks in the framework. In Sec. \ref{sec:4charmmat} we build the EoS for charm quark matter. Then, the structure equations for charm stars are solved, and their stability is studied under radial acoustic perturbations in Section \ref{sec:4charm-stars}. Section \ref{sec:4conclusion} presents our summary and conclusions.

\section{Perturbative QCD with $N_{m}$ heavy quarks}
	\label{sec:4pQCDh}
	
	It is expected that at high densities not only the light quarks will be present in a system of quark matter, but also some heavy flavors. In QCD, heavy quarks are meant to be the ones satisfying $m\gg\Lambda_{\rm \overline{MS}}$, i.e. quarks with masses that are very large compared to the QCD natural scale\footnote{This assumes implicitly that $m$ is also very large compared to the light quarks already present.}. Usually, their influence is neglected in most calculations by invoking the $\textit{heavy-quark QCD decoupling}$ theorem \cite{Symanzik:1973vg,Appelquist:1974tg,Bernreuther:1983zp}, which is essential when calculating any quantity in a wide range of energies. We now summarize this standard (although approximate) approach.

\subsection*{Effective Heavy QCD and QCD decoupling}

In principle, it would be correct to start the study of $N_{l}$ light massless quarks $\psi_{i}$ interacting with a single massive (later to be taken as heavy) flavor $\Psi$ having mass, $m$, by considering the (unrenormalized) QCD Lagrangian as (not including the gauge-fixing terms which are important only for the quantization process)
\begin{equation}
\mathcal{L}^{\rm QCD}=\sum^{N_{l}}_{i=1}\bar{\psi}_{i}i\slashed{D}\psi_{i}+
\bar{\Psi}(i\slashed{D}-m)\Psi-\frac{1}{4}F^{a}_{\mu\nu}F^{a\mu\nu}.
\end{equation}
Then, by taking the limit of $m$ being very large one obtains the low-energy effective theory containing only light fields \cite{Grozin:2012ec}:
\begin{equation}
\mathcal{L}^{\rm hQCD}=\sum^{N_{l}}_{i=1}{\bar{\psi}'}_{i}i
\slashed{D'}\psi{'}_{i}-\frac{1}{4}{F'}^{a}_{\mu\nu}{F'}^{a\mu\nu}+\mathcal{O}\left(\frac{1}{m^{2}}\right),
\end{equation}
where the primes indicate the redefinitions in the light and gauge fields carrying information of the heavy quark, plus corrections of the order $1/m^{2}$. When performing this approximation it is said that one is \textit{decoupling} heavy quarks. This reasoning in mostly employed in the community of high energy physics, i.e. neglecting heavy contributions in the theory to be compared with experiments. 

The most important effect of this elimination of the heavy field is carried by the strong coupling $\alpha_{s}(\mu)$ (at some energy scale $\mu$). For example, if one desires to relate the $\alpha_{s}(m_{\rm top}\approx{173}~{\rm GeV})$ to the $\alpha_{s}(m_{Z}\approx{80}~{\rm GeV})$, one needs to take into account both the renormalization group running (in each interval where it is controlled by the corresponding $\beta$ function) and consider the effects of decoupling\footnote{Parton distribution functions are also extracted from experiment at different values of $\mu$ where decoupling effects must be added.}. This same reasoning is applicable for the strange quark mass $m_{s}(\mu)$ through its dependence on $\alpha_{s}$, as already noted in Chapter 2. In this sense, QCD with one or more heavy flavors can be considered one of the simplest examples of an effective low-energy theory.

For completeness, in the following we show schematically how to calculate these effects within QCD. The full-QCD coupling $\alpha^{(N_{l}+1)}_{s}(\mu)$ (for $N_{l}$ massless and one massive quarks) is related to the effective coupling $\alpha^{(N_{l})}_{s}(\mu)$ (only for massless quarks) by the decoupling coefficient (see Ref. \cite{Grozin:2020bvd} for more details) as follows
\begin{equation}
\alpha^{(N_{l}+1)}_{s}(\mu)=\zeta^{-1}_{\alpha}(\mu)\alpha^{(N_{l})}_{s}(\mu),
\end{equation}
where the decoupling coefficient $\zeta^{-1}_{\alpha}(\mu)$ has to be determined by solving the renormalization group equation \cite{Grozin:2020bvd}
\begin{equation}
\frac{d\log\zeta_{\alpha}(\mu)}{d\log\mu}-2\beta^{(n_{l}+1)}(\alpha^{(n_{l}+1)}_{s}(\mu))+2\beta^{(n_{l})}(\alpha^{(n_{l})}_{s}(\mu))=0,
\end{equation}
with the initial condition being chosen by fixing its value at some fiducial scale $\mu=\bar{M}$ yielding \cite{Grozin:2020bvd}
\begin{equation}
\zeta^{-1}_{\alpha}(\bar{M})=1+\left(\frac{13}{3}C_{F}-\frac{32}{9}C_{A}\right)T_{F}\left(\frac{\alpha_{s}(\bar{M})}{4\pi}\right)^{2}+... , 
\end{equation}
where $C_{F}$, $C_{A}$, and $T_{F}$ are positive numbers (see Ref. \cite{Grozin:2020bvd} for more details). Then, after solving Eq. (4.4) one would realize that the QCD running coupling $\alpha_{s}(\mu)$ not only runs when $\mu$ varies but it also jumps when crossing heavy-flavor thresholds. For instance, the behavior of $\alpha_{s}(\mu)$ near the bottom mass $M_{b}$ is shown in Fig. \ref{fig:massthreshold}. For $\mu>M_{b}$, the correct theory is the full 5-flavor QCD ($\alpha^{(5)}_{s}(\mu)$, the solid line in this figure); at $\mu<M_{b}$, the correct theory is the effective low-energy QCD ($\alpha^{(4)}_{s}$, also the solid line), where the jump at the transition point $\mu=M_{b}$ is also shown. Of course, both curves can be continued across $M_{b}$ (dashed lines), and it is unimportant at which particular $\mu\approx{M}_{b}$ we switch from one theory to the other one since this is dependent on the fiducial scale to be used\footnote{These fiducial scales are usually chosen as $\mu=2M_{b}$ or $\mu=M_{b}/2$, which imply corrections of the order $\mathcal{O}(\alpha^{2}_{s})$ \cite{Grozin:2020bvd}.}. However, the on-shell mass $M^{os}_{b}$ is found conveniently useful because the jump is small, i.e. of the order of $\mathcal{O}(\alpha^{3}_{s})$ \cite{Grozin:2020bvd}.
 \begin{figure}[ht]
    \begin{center}
  \resizebox*{!}{7cm}{\includegraphics{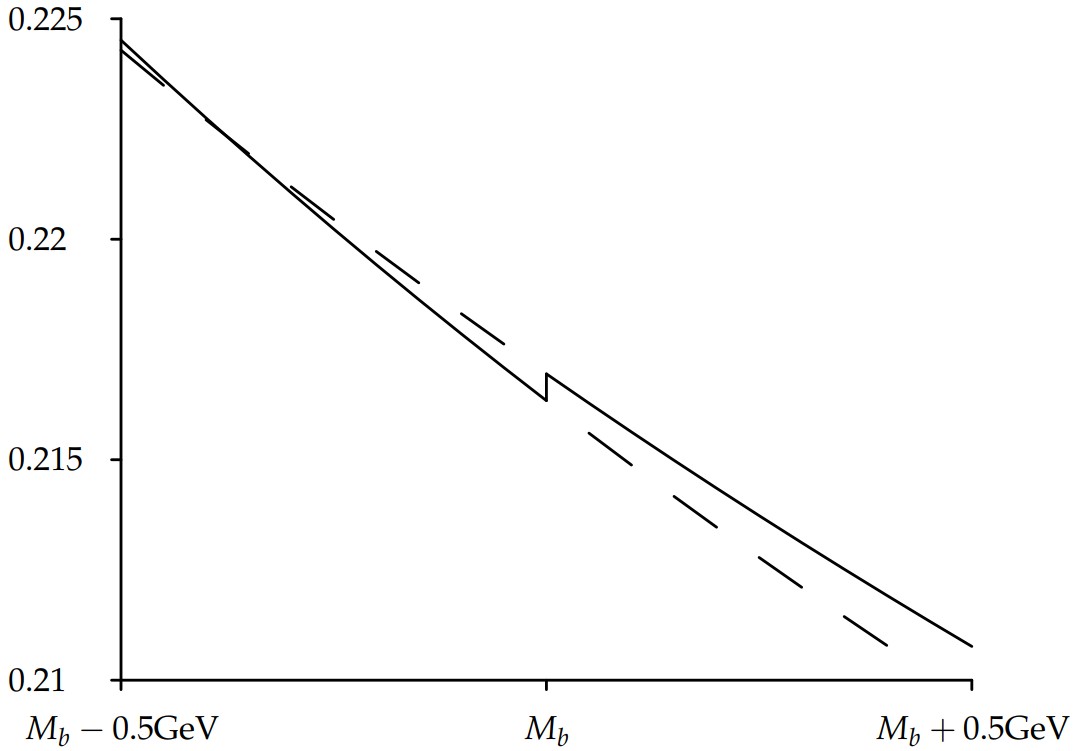}}
    \end{center}
   \vspace{5mm}
   \caption{ It is shown the behavior of $\alpha^{(5)}_{s}(\mu)$ and $\alpha^{(4)}_{s}(\mu)$ near $\mu=M_{b}$. Image taken from Ref. \cite{Grozin:2020bvd}.}
   \label{fig:massthreshold}
  \end{figure}
We note that the decoupling of heavy quarks also affects the light-quark masses since they also depend (perturbatively) on the strong coupling behavior. For this, a similar renormalization group equation must be solved (see Ref. \cite{Chetyrkin:1997un}), where the initial condition to solve this equation can be used at $\mu=\bar{M}$, with the massive decoupling coefficient taking the form \cite{Chetyrkin:1997un}
\begin{equation}
\zeta^{-1}_{m}(\bar{M})=1+\frac{89}{18}C_{F}T_{F}\left(\frac{\alpha_{s}(\bar{M})}{4\pi}\right)^{2}+...~. 
\end{equation}
%




\subsection{NNLO Heavy QCD Matter}

The formalism developed here for the inclusion of heavy quarks will rely fundamentally on the pQCD thermodynamic potential discussed in detail in Chapter 2. We begin by writing conveniently the total number of flavors in the form
	\begin{equation}
	 N_{f}=\sum^{N_{m}}_{i=1}(N_{l}+1)^{(i)},
	 \end{equation} 
where $N_{m}$ is the number of massive quarks present in the system, and respecting the constraint $N_l + N_m=N_f$. So, we add at least one massless quark for each massive flavor included. For example, for charm quark matter it will be convenient to write this sum over flavors as\footnote{In the literature for hot QCD matter this case is usually written also as $N_{f}=2+1+1$ for the charm inclusion.} 
\begin{equation}
N_{f}=(1+1)^{(1)}+(1+1)^{(2)}=2+2, 
\end{equation}
where $N^{(i)}_{l}=1$ and $N_{m}=2$. The usefulness of this way of writing will become clear when  summing the massless and massive contributions to the total free energy. Of course, this represents only a convenient way of writing the degrees of freedom at the level of the formalism. Additional physical conditions are needed in order to control when a heavy partner appears actively. Such conditions can be introduced by choosing appropriate values of the renormalization scale $\bar{\Lambda}$, depending on the chosen heavy flavor to be introduced in the system\footnote{Additional matching conditions on the renormalized QCD
parameters should be imposed at the quark thresholds, i.e., on $\alpha_{s}(\bar{\Lambda}_{\rm thr})$ and $m(\bar{\Lambda}_{\rm thr})$, in order to account for their behavior at different values of $N_{f}$, depending on the energy scale of the problem \cite{Rodrigo:1993hc}.}. We will see this in the next section for the particular case of \textit{charm} quark matter.
	
With this in mind, we write the QCD thermodynamic potential $\Omega_{\rm QCD}$ for $N_{l}$ massless and $N_{m}$ massive quarks as
	\begin{equation}
	\bar{\Omega}[{N_{f}}]=\sum^{N_{m}}_{i=1}\Omega_{\rm QCD}[{N_{f}}]=
	\sum^{N_{m}}_{i=1}\left\lbrace\Omega[{N^{(i)}_{l}}]
	+\Omega[{1^{(i)}}]\right\rbrace \; ,
	\label{eq:hQCDenergy}
	\end{equation}
where one must choose the number of massless flavors first when adding a massive one, so that
\begin{equation}
\Omega[{N^{(i)}_{l}}]~{\equiv}~(\Omega^{m=0}(\vec{\mu}))^{(i)} 
\end{equation}
is the massless contribution and 
	\begin{equation}
	\Omega[{1^{(i)}}]~{\equiv}~(\Omega^{m}+\Omega^{x}_{{\rm VM}}+\Omega_{{\rm ring}})^{(i)}
	\end{equation}
the mixed massive contribution, where $\vec{\mu}^{(i)}=(\mu_{1}, ...,\mu_{i})$ is the massless vector chemical potential, $\tilde{\mu}^{(i)}$ the massive (heavy) quark chemical potentials, and $m^{(i)}$ their corresponding masses. Here, $\Omega[...]$ indicates just the implicit parameter dependence (e.g. on $N_{f}$), whereas $\Omega(...)$ represents an explicit functional dependence.
 
In the following, we apply these results to the case of charm quark matter and charm stars. Interestingly, our calculations show that including heavy quarks makes the QCD thermodynamic potential less sensitive to the renormalization scale $\bar{\Lambda}$, i.e. its range of values chosen (before the heavy quark appearance) is reduced\footnote{This could has been anticipated on physical grounds since having only light quarks (as it is usually done also in hot QCD, where some improvements in the RG equations are under active study) does not impose restrictions on $\bar{\Lambda}$, what is not true when introducing heavy quarks that require further physical constraints at the particle threshold in order to have well-behaved thermodynamic observables.} in order to obtain a consistent thermodynamic transition between light/heavy quark flavors, similar to results obtained in hot QCD \cite{Laine:2006cp,Graf:2015tda}.

\section{Charm Quark Matter}
	\label{sec:4charmmat}	

Charm quarks were introduced into the theory of strong interactions in order satisfy the hadron phenomenology of many experiments carried out in the 1970's and they served to introduce the charm quantum number $C$, similar to the strangeness quantum number $S$ of strange quarks, being ``~+1~'' for particles (``~-1~'' for antiparticles) when conserved and zero when they are not conserved or for any other particles. In medium, they are important currently when calculating the equation of state for nuclear matter where charm baryons are created, therefore contributing to the EoS significantly. However, for cold and dense matter (like the one found in the interior of compact stars) these quarks are usually not taken into account since their chemical potential does not approach the magnitude of the charm quark mass (of approximately 1.3 GeV) in the usual density range of quarks stars having already strange quarks. Thus, it is mostly assumed that the charm quark states are not populated except at densities far above that found in neutron or analogous strange stars. 

In this section we consider the simplest case of heavy quark matter, charm quark matter\footnote{Strange matter is stable due to the Bodmer-Witten hypothesis and it may exist even with zero strangeness chemical potential as in the case of strange quark stars. Charm matter might have the same condition for stability with or without the charm quantum number being finite, being our case the latter.} in the star's bulk, which is composed of light quark matter plus charm quarks. Of course, it can only be realized above a given critical charm chemical potential. As we go to higher values of quark mass, asymptotic freedom makes the perturbative QCD formalism more reliable. To build the EoS for charm matter we first need to establish the constraints this system must respect below and above the charm threshold, in order to generate matter configurations stable under electroweak interactions.

\subsection{Heavy pQCD+ with $N_{f}=2+1+1$}

Using Eq. (\ref{eq:hQCDenergy}) with $N^{(1)}_{l}=1$ for the up, $N^{(2)}_{l}=1$ for the down, and $N_{m}=2$ for the strange and charm quarks, we have the following thermodynamic potential:
	 \begin{eqnarray}
 \bar{\Omega}[N_{f}=2+1+1]=\left\lbrace\Omega[N^{(1)}_{l}=1]+\Omega[1^{(1)}]
 \right\rbrace+
 \left\lbrace\Omega[N^{(2)}_{l}=1]+\Omega[1^{(2)}]
 \right\rbrace \;,
	 \end{eqnarray}
so that the flavors are counted as $N_{f}=(1+1)^{(1)}+(1+1)^{(2)}=(u+c)^{(1)}+(d+s)^{(2)}$. From this point, we follow the same strategy of the KRV prescription explained in detail in Chapter 2. 

Now we need to fix the running quark masses and strong coupling at some specific energy scale. For the strong coupling we use the result of Eq. (2.18), and its variation for different values of the renormalization scale is shown in Fig. \ref{fig:4runningCoupling}. Notice that from this figure one can realize that for energies below $\bar{\Lambda}\simeq{1}$ GeV, our pQCD results are less reliable since the strong coupling increases rapidly towards 1, being less rapidly when including the charm flavor, and where the uncertainties of $\Lambda_{\overline{\rm MS}}$ become more dominating. 

\begin{figure*}[h!]
\begin{center}
\hbox{\includegraphics[width=0.5\textwidth]{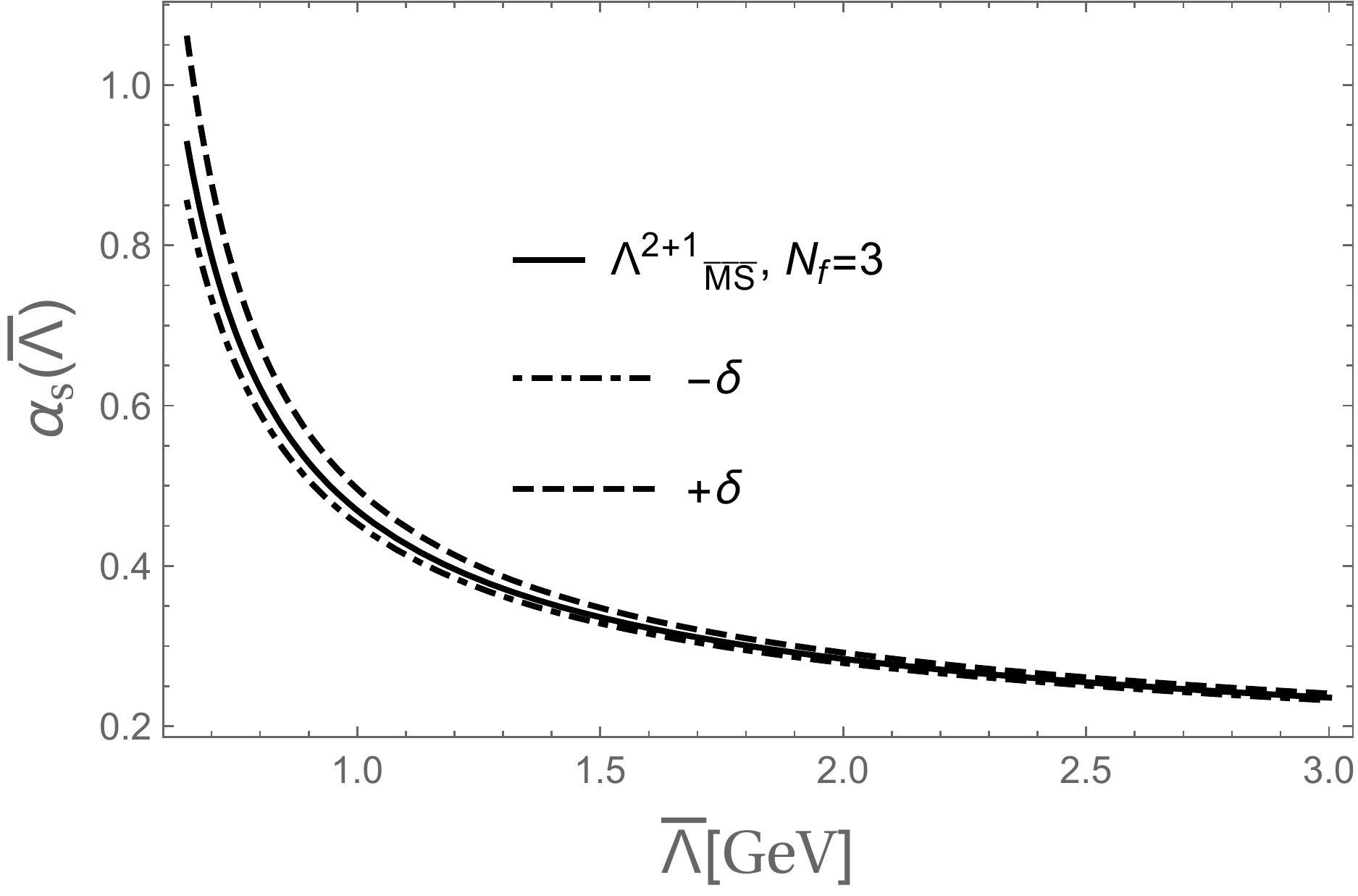}
	  \includegraphics[width=0.5\textwidth]{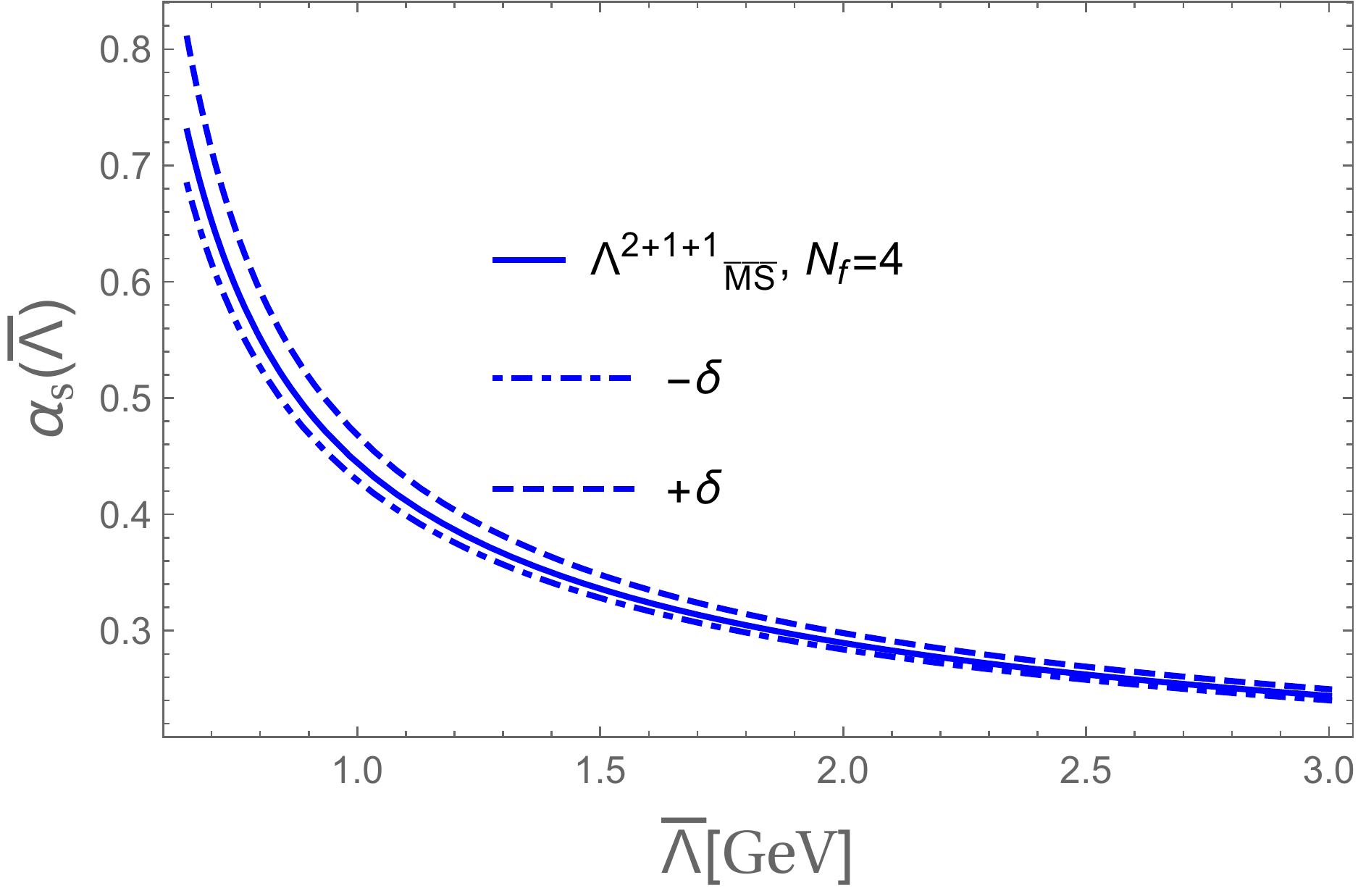}}
\vspace{5mm}
\caption{Strong coupling dependence on the number of flavors as well as on the renormalization scale $\bar{\Lambda}$, appropriate for the transition between strange and charm matter, where the $\delta$'s indicate the error variation in the values of $\bar{\Lambda}^{N_{f}}_{\overline{\rm MS}}$.}
\label{fig:4runningCoupling}
\end{center}
\end{figure*}

Additionally, by solving the renormalization group equations for the quark mass\footnote{Note that this is simply a QCD parameter in $\mathcal{L}_{\rm QCD}$ and does not correspond to the pole position of the propagator. Although quark confinement avoid physical poles in the full non-perturbative propagator, one can still define an on-shell mass perturbatively called the pole mass $m_{\rm pole}$ being useful in heavy-quark phenomenology \cite{Yagi:2005}.} parameters up to second order in the strong coupling $\alpha_{s}$, one obtains the following results for the strange and charm quarks \cite{Vermaseren:1997fq}
	\begin{eqnarray}
	\begin{aligned}
	m_{s}(\bar{\Lambda})=\hat{m}_{s}\times\left(\frac{\alpha_{s}}{\pi}\right)^{4/9}
	\times\left(1+0.895062\left(\frac{\alpha_{s}}{\pi}\right) +1.37143\left(\frac{\alpha_{s}}{\pi}\right)^{2}\right) \;,
	\end{aligned}
	\label{eq:smass}
	\end{eqnarray}
	\begin{eqnarray}
	\begin{aligned}
	m_{c}(\bar{\Lambda})=\hat{m}_{c}\times\left(\frac{\alpha_{s}}{\pi}\right)^{12/25}
	\times\left(1+1.01413\left(\frac{\alpha_{s}}{\pi}\right)+1.38921\left(\frac{\alpha_{s}}{\pi}\right)^{2}\right) \;,
	\end{aligned}
	\label{eq:cmass}
	\end{eqnarray}
with $\lbrace\hat{m}_{q}\rbrace$ being the renormalization group invariant quark masses, i.e. $\bar{\Lambda}$ independent\footnote{Expressing the quark masses in this way, one can see that their invariant masses can be fixed at independent energy scales, which is not obvious when using the quark mass function defined in Ref. \cite{Kurkela:2009gj}.}.
\begin{figure*}[h!]
\begin{center}
\hbox{\includegraphics[width=0.5\textwidth]{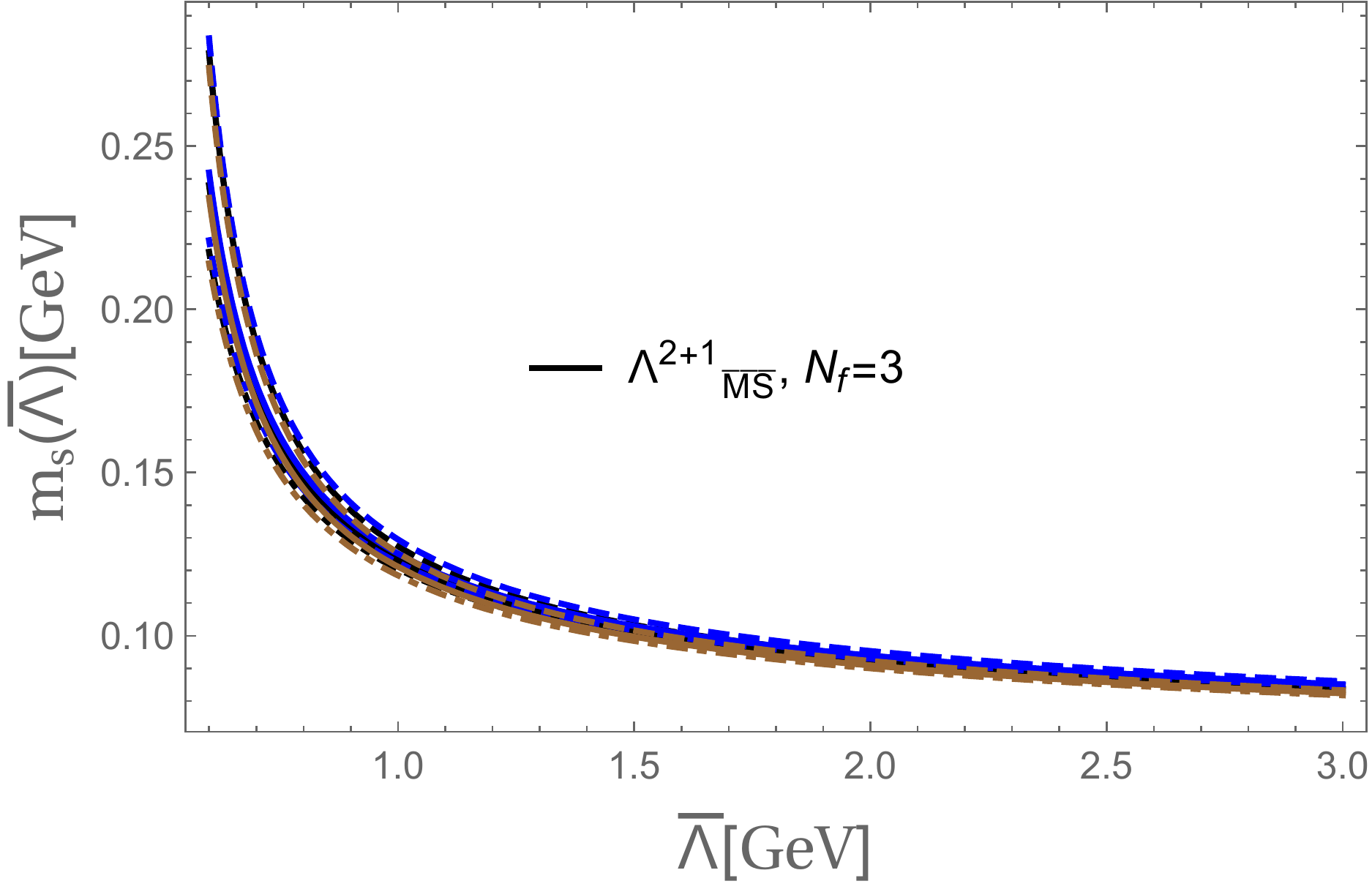}
	  \includegraphics[width=0.5\textwidth]{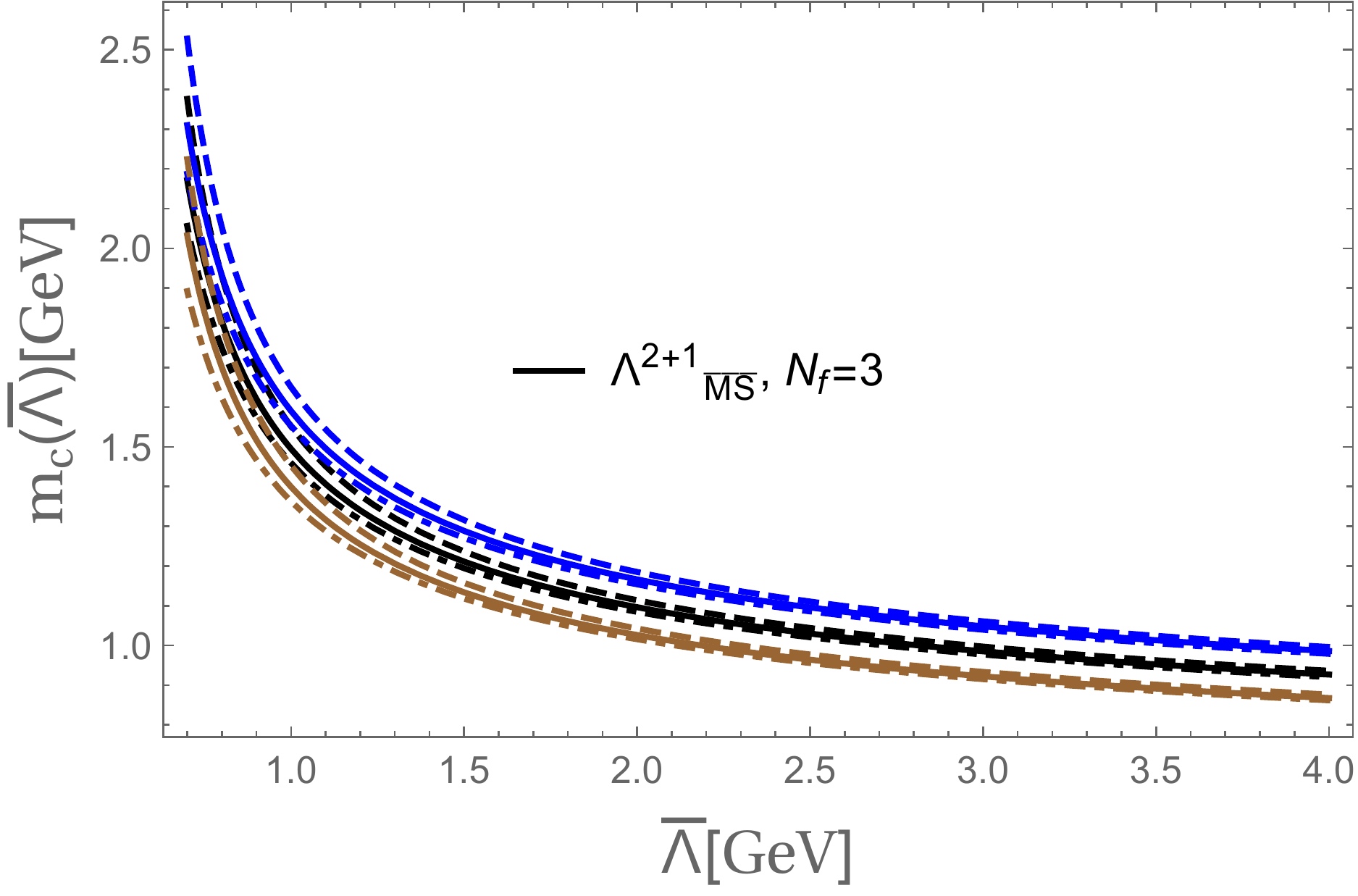}}
\vspace{5mm}
\caption{Running strange and charm quark masses dependence on the renormalization scale $\bar{\Lambda}$, appropriate for strange matter, where the error variations in the values of $\bar{\Lambda}^{N_{f}}_{\bar{\rm MS}}$ are considered for this plots.}
\label{fig:4runningMassNf3}
\end{center}
\end{figure*}
\begin{figure*}[h!]
\begin{center}
\hbox{\includegraphics[width=0.5\textwidth]{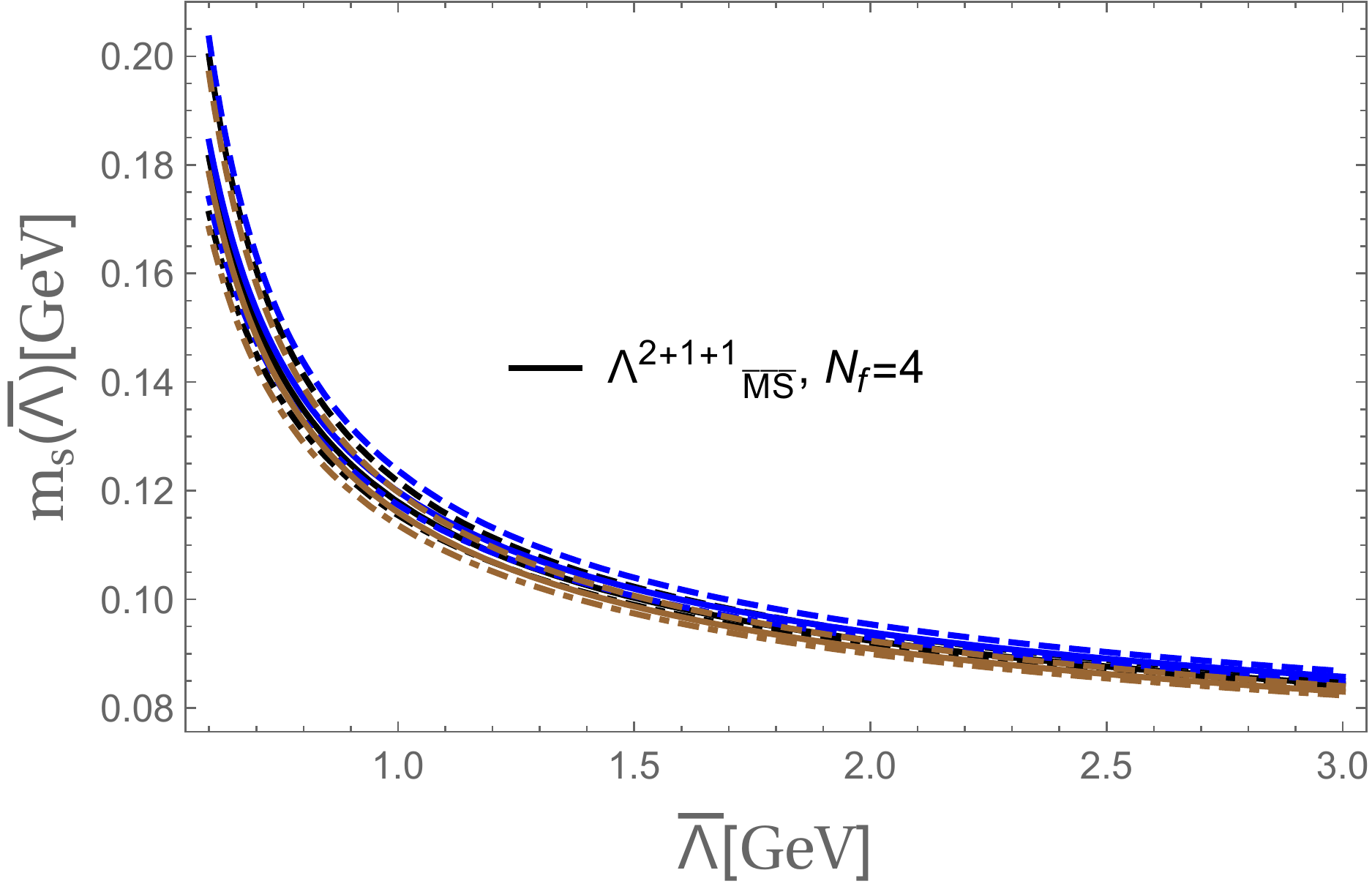}
	  \includegraphics[width=0.5\textwidth]{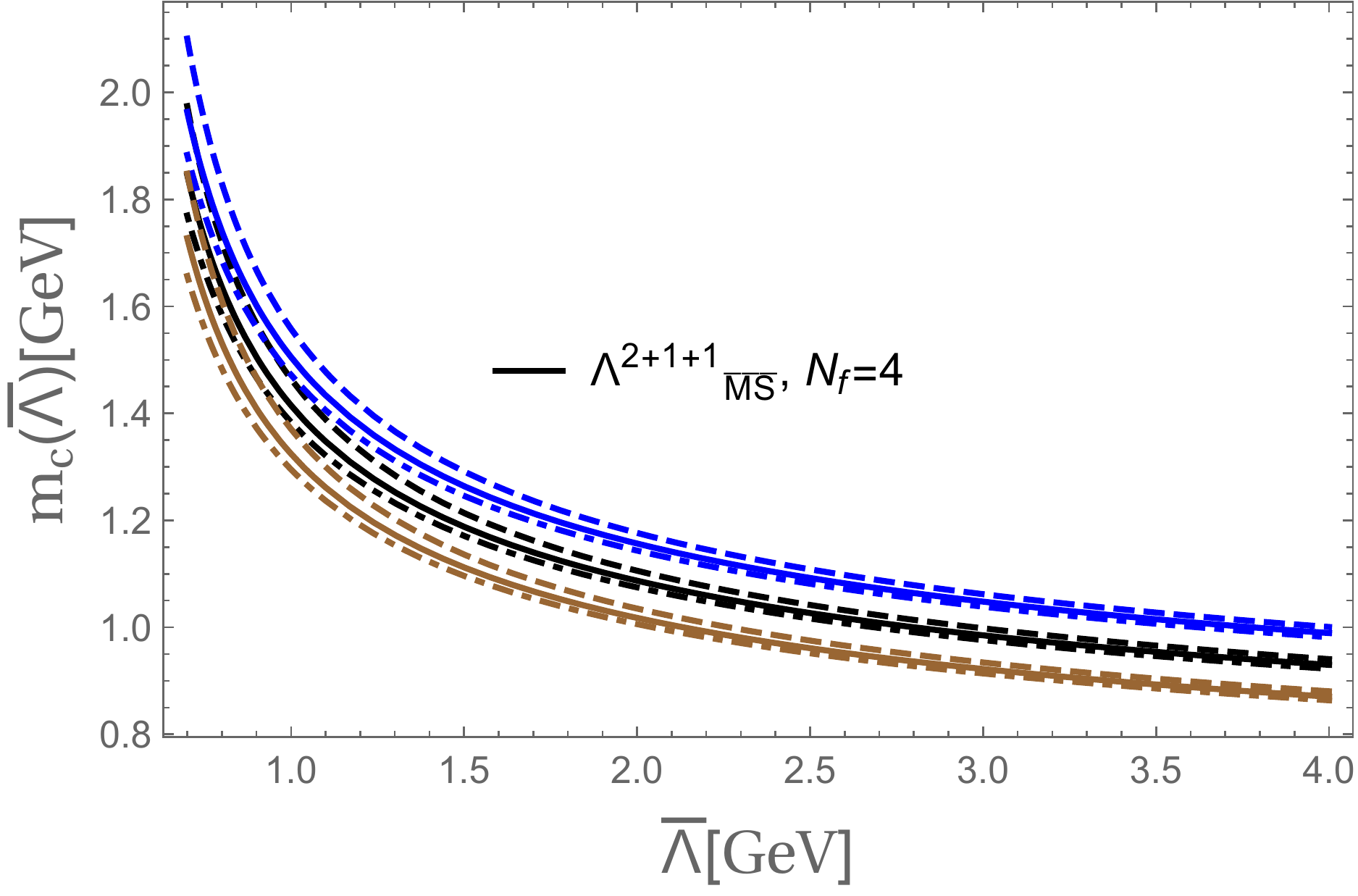}}
\vspace{5mm}
\caption{Running strange and charm quark masses dependence on the renormalization scale $\bar{\Lambda}$, appropriate for charm matter, where the error variations in the values of $\bar{\Lambda}^{N_{f}}_{\bar{\rm MS}}$ are considered for this plots.}
\label{fig:4runningMassNf4}
\end{center}
\end{figure*}
Since Eq. (\ref{eq:alphas}) for $\alpha_{s}$ tells us that different values of $N_{f}$ give different values of $\Lambda_{\overline{\rm MS}}$, by choosing  $\alpha_{s}(\bar{\Lambda}=1.5~{\rm GeV},~N_{f}=3,4)=0.336^{+0.012}_{-0.008}$ \cite{Bazavov:2014soa}, we obtain $\Lambda^{2+1}_{\overline{\rm MS}}=343^{+18}_{-12}~$MeV and $\Lambda^{2+1+1}_{\overline{\rm MS}}=290^{+18}_{-12}~$MeV, thus defining $\alpha^{2+1}_{s}(\bar{\Lambda})$ and $\alpha^{2+1+1}_{s}(\bar{\Lambda})$, respectively. In order to fix the quark masses, one should go to more-or-less UV energies where asymptotic freedom is reliable and then allow the RG equations to run towards lower energies in order to obtain their behavior for energies (temperatures and/or densities) of interest. Fixing the strange quark mass at $m_{s}(2~{\rm GeV}, N_{f}=3,4)=92.4(1.5)~$MeV \cite{Chakraborty:2014aca} gives $\hat{m}^{2+1}_{s}~{\approx}~246.2~$MeV when using $\alpha^{2+1}_{s}$ in Eq. (\ref{eq:smass}), and $\hat{m}^{2+1+1}_{s}~{\approx}~243.7~$MeV with $\alpha^{2+1+1}_{s}$ also in Eq. (\ref{eq:smass}). Additionally, fixing the charm quark mass at $m_{c}(3{\rm GeV}, N_{f}=4)=0.9851(63)~{\rm GeV}{~\equiv~}m^{0}_{c}$ \cite{Chakraborty:2014aca}, gives $\hat{m}^{2+1+1}_{c}~{\approx}~3.0895~$GeV when using $\alpha^{2+1+1}_{s}$ in Eq. (\ref{eq:cmass}). We define $m^{0}_{c}$ as the vacuum charm mass for convenience later. In Figs. \ref{fig:4runningMassNf3} and \ref{fig:4runningMassNf4} the running behavior of these masses is shown when using the parameters mentioned above. Notice from these figures that for $\Lambda$ below 0.9 GeV the strange and charm masses increase very rapidly to very large values. In this case one says that they are \textit{decoupled} from the system due to their very heavy nature becoming static, i.e. not appearing in the QCD Lagrangian.

\subsection*{Below the charm threshold: $N_{f}$=2+1}    
    
We now impose the compact-star conditions on this thermodynamic potential. The condition of electric charge neutrality for a system with $N_{f}=2+1$ quarks (plus electrons) is given by
	\begin{equation}
	\frac{2}{3}n_{u}-\frac{1}{3}n_{d}-\frac{1}{3}n_{s}-n_{e}=0 \; ,
	\label{eq:sneutral}
	\end{equation}
where $n_{i}(\mu_{i})$ are the associated particle number densities for quarks and electrons in the system. The electron number density is approximated, as usual, by that of a free Fermi gas, i.e. $n_{e}=\mu^{3}_{e}/(3\pi^{2})$. In Fig. \ref{fig:4chemical} we show the Fermi momenta for each particle added into the system in unpaired charm quark matter at their respective mass thresholds. 
\begin{figure}[ht]
    \begin{center}
  \resizebox*{!}{7cm}{\includegraphics{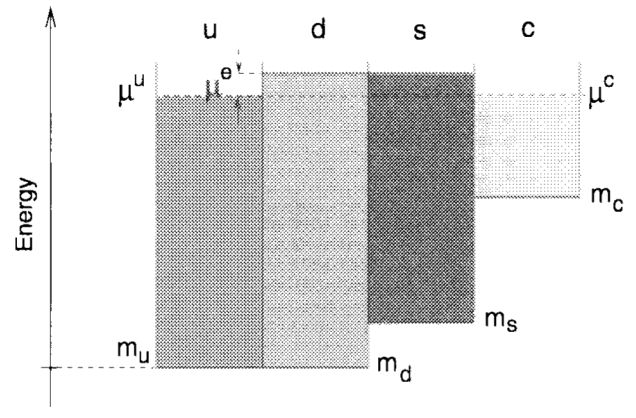}}
    \end{center}
   \vspace{5mm}
   \caption{Distribution of energies for the chemical equilibrium between $i=u,d,s,c$ quarks and electrons present in charm quark matter (no muons). Taken from Ref. \cite{Weber:1999}.}
   \label{fig:4chemical}
  \end{figure}

Then, weak reactions among light quark flavors are given by
    \begin{equation}
    d~{\rightarrow}~u+e^{-}+\bar{\nu}_{e^{-}},\hspace{0.5cm}s~{\rightarrow}~u+e^{-}+\bar{\nu}_{e^{-}}
    \; ,
    \label{eq:sweak1}
    \end{equation}
    \begin{equation}
 {s+u}\leftrightarrow{d+u} \;.
 	\label{eq:sweak2}
    \end{equation}
and yield the following relations between chemical potentials:
	\begin{equation}
\mu_{d}=\mu_{s},\hspace{1cm}\mu_{u}=\mu_{s}-\mu_{e} \;.
	\label{eq:sweak3}
	\end{equation}
We neglect the neutrino chemical potential since its mean free path is large compared to the size of a typical compact star. By solving simultaneously Eqs. (\ref{eq:sneutral}) and (\ref{eq:sweak3}) one is able to write all the quark and electron chemical potentials in terms only of the strange chemical potential, $\mu_{s}$.
	
\subsection*{Above the charm threshold: $N_{f}$=2+1+1} 
	
When $\mu_{s}$ crosses the charm quark threshold, the following weak equilibrium reaction is allowed to take place:
    \begin{equation}
   {u+d}\leftrightarrow{c+d} \;,
    \end{equation}
yielding the condition
	\begin{equation}
	\mu_{c}=\mu_{u} \;.
	\label{eq:cweak1}
	\end{equation}
Notice that at the threshold the charm quarks are essentially non-relativistic in the surrounding relativistic medium. Moreover, the electric charge neutrality condition turns into
	\begin{equation}
	\frac{2}{3}n_{u}+\frac{2}{3}n_{c}-\frac{1}{3}n_{d}-\frac{1}{3}n_{s}-n_{e}-n_{\mu}=0 \;,
	\label{eq:cneutral}
	\end{equation}
where we have included free muons, with $n_{\mu}=(\mu^{2}_{\mu}-m^{2}_{\mu})^{3/2}/(3\pi^{2})$, which appear when $\mu_{\mu}>m_{\mu}=105.7~$MeV and they are added to the system due to the reaction
\begin{equation}
e^{-}{~\leftrightarrow~}\mu^{-}+\nu_{e}+\bar{\nu}_{\mu}.
\end{equation}
Since neutrinos and antineutrinos escape from the star, lepton number conservation gives us $\mu_{\mu}=\mu_{e}$, which implies that for $\mu_{e}>m_{\mu}$ the muon states will be populated\footnote{In the case $N_{f}=2+1$ the muons were not considered since $\mu_{e}~<~50{\rm ~MeV}~<~{m}_{\mu}$.}. 

Again, by solving simultaneously Eqs. (\ref{eq:sweak3}), (\ref{eq:cweak1}) and (\ref{eq:cneutral}), we can express the quark and lepton chemical potentials only in terms of $\mu_{s}$. In the notation of Sec. \ref{sec:4pQCDh}, the charm matter free energy corresponds to the case 
\begin{equation}
N_{f}=(1+1)^{(1)}+(1+1)^{(2)}=(u+c)^{(1)}+(d+s)^{(2)},
\end{equation}
in Eq. \ref{eq:hQCDenergy}.	

We assume that charm quarks are allowed in the system when
	\begin{equation}
	\mu_{c}=\mu_{s}-\mu_{e}>m^{\rm medium}_{c}>m^{0}_{c},
	\label{eq:charmcondition}
	\end{equation}
where $m^{\rm medium}_{c}$ is the (unknown) in-medium charm mass\footnote{An exact value for the in-medium charm mass at finite density is still not known, whereas its vacuum mass at some fixed energy scale, $m^{0}_{c}$, can be extracted from lattice calculations.}. 
Then, the renormalization scale parameter below and above the charm threshold are given by\footnote{Alternatively, one could choose independent values of $\bar{\Lambda}$ when going from $N_{f}=3$ to $N_{f}=4$, the ``transition" point being found by a matching between strong couplings with different $N_{f}$ \cite{Rodrigo:1993hc,Graf:2016mzv}.}
	 \begin{numcases}
    {\bar{\Lambda}=}
      X\frac{(\mu_{u}+\mu_{d}+\mu_{s}+0)}{3} \;, &  $\mu_{s}~{\lesssim}~m^{0}_{c}$~,
      \label{A} \\
     X^{*}\frac{(\mu_{u}+\mu_{d}+\mu_{s}+\mu_{c})}{3} \;, &  
$\mu_{s}~{\gtrsim}~m^{0}_{c}$~,
     \label{B}
  \end{numcases}
where the approximations in the inequalities of Eqs. (\ref{A}) and (\ref{B}) represent that just before the threshold point the electron chemical potential takes its lowest value compared to the strange one, thus allowing us to make the approximation $\mu_{c}~{\approx}~\mu_{s}$. 

The only way to go from Eq. (\ref{A}) to Eq. (\ref{B}) continuously through the $N_{f}$ transition is by requiring the factors $X^{*}$ and $X$ to have the same range of possible values. To have them greater than $1$ with and without charm quarks, one needs values greater than $4/3=1.3333...$ for both, which implies a reduction in the renormalization scale band of the EoS when heavy quarks are included, something already found in thermal perturbative QCD with charm quarks \cite{Laine:2006cp}. For practical purposes, we will consider that $X$ only runs between 2 (since lower values produce low-mass quark stars being more difficult to produce charm quarks at their cores) and 5, where the latter value is chosen only to verify if the observables differ from $X=4$ when including heavy quarks.
	
We define the total quark number density for charm matter, for a given $X$, as
    \begin{equation}
    n_{q}(\left\lbrace{\mu_{f}}\right\rbrace, X){\hspace{0.1cm}}{\equiv}{\hspace{0.1cm}}n_{u}+n_{d}+n_{s}+n_{c} \;,
    \end{equation}
the total particle density as $n=n_{q}+n_{L}$, where $n_{L}=n_{e}+n_{\mu}$. We note that it is not clear if one can generalize these definitions for the baryon density as $n_{B}=n_{q}/3$ and baryochemical potential as $\mu_{B}=\mu_{u}+\mu_{d}+\mu_{s}+\mu_{c}$ since baryons only have three quarks.

In Fig. \ref{fig:4PartDensX4} we show the behavior of the relative particle populations for our $\beta$-equilibrated and electrically neutral charm quark matter system in the case of $X=3$. Only above the charm threshold, charm quarks begin to contribute to the total number density, $n$. The location of the threshold depends on the value we choose for $X$ and is within $\mu_{s}~{\approx}~$1.2--1.4 GeV for the band we consider\footnote{At first sight one would expect that these $\mu_{s}$--values might allow the addition of heavy baryons into the mixed and/or confined phase. By defining the baryon chemical potential as $\mu_{B}{~\equiv~}\mu_{u}+\mu_{d}+\mu_{s}+\mu_{c}$, one would obtain $\mu_{B}\approx{4\mu_{s}}{\approx}$ 4.8 GeV, which are the values of masses for the heaviest baryons within the modern baryon octet and decuplet. In this sense, as we will see, although bare charm quark stars cannot be realized in nature, it might be possible that a new branch of stable hybrid neutron stars might exist with charmed baryons in the nuclear sector and some fraction of charm quarks at their cores through a mixed phase built up using the Glendenning construction.}.

\begin{figure}[ht]
    \begin{center}
  \resizebox*{!}{7cm}{\includegraphics{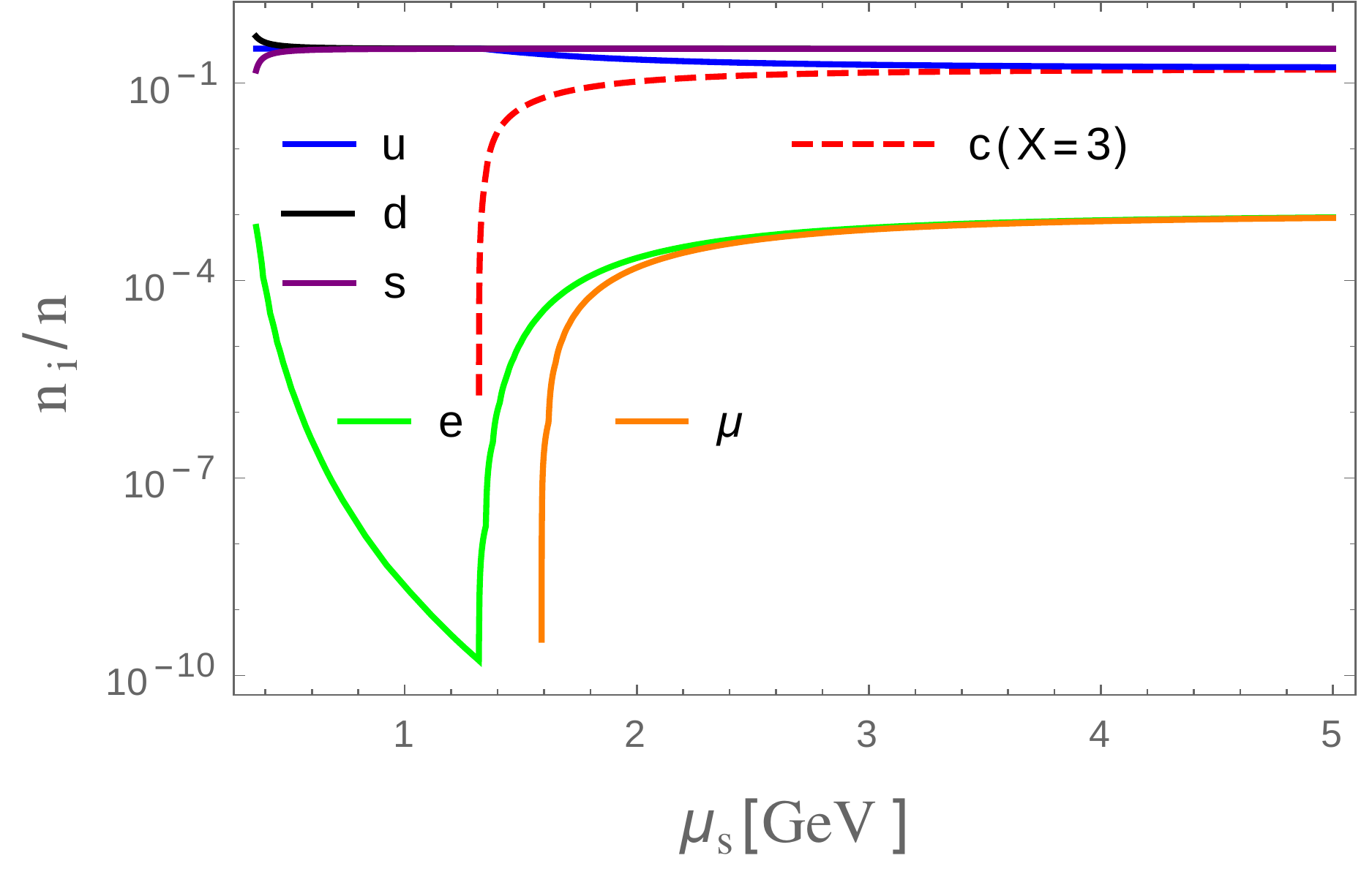}}
    \end{center}
   \vspace{5mm}
   \caption{Relative particle fractions for quarks and leptons, $n_{i}/n$, present in charm quark matter for $X=3$, where $i=u,d,s,c,e,\mu$. Above the charm quark threshold the lepton fractions increase to ensure electric charge neutrality at high densities. Below this threshold the electrons are necessary to balance the contribution from \textit{massive} strange quarks and it goes to zero when the strange becomes massless. Taken from Ref. \cite{Jimenez:2019kji}.}
   \label{fig:4PartDensX4}
  \end{figure}

To build the total pressure, one should be careful with the fact that the derivatives of the thermodynamic potential give rise to terms which have the form of $\partial\alpha_{s}(\bar{\Lambda})/\partial\mu_{f}$ since $\bar{\Lambda}\propto{\mu}_{f}$, as can be seen in Eqs. (\ref{A})--(\ref{B}) (see also Ref. \cite{Kurkela:2009gj}). To keep thermodynamic consistency, one can take the quark and lepton number densities as the fundamental ingredients and build the other thermodynamic observables (e.g., pressure and energy density) imposing consistency on the number densities.
	
Thus, we define the total pressure of the system as
\begin{equation}
P(\mu_{s},X)=\sum_{f=u,c}P^{(1)}_{f}+\sum_{f=d,s}P^{(2)}_{f}
+\sum_{L=e,~\mu}P_{L} \;,
	\label{eq:pressure}
\end{equation}
where we have separated the contributions coming from $N_{f}=(u+c)^{(1)}+(d+s)^{(2)}$, defining each term as
\begin{eqnarray}
P^{(1)}_{f}(\mu_{s},X)=\int^{\mu_{s}}_{\mu_{0}(X)}d\bar{\mu}_{s}
 \left[ n_{f}\left(1-\frac{d\mu_{f}}{d\bar{\mu}_{s}}\right)\right] \;,
\end{eqnarray}
\begin{eqnarray}
\begin{aligned}
P^{(2)}_{f}(\mu_{s},X)=\int^{\mu_{s}}_{\mu_{0}(X)}d\bar{\mu}_{s}~n_{f} \; ,
\end{aligned}
\end{eqnarray}
and the lepton contribution as

\begin{eqnarray}
\begin{aligned}
P_{L}(\mu_{s},X)=\int^{\mu_{s}}_{\mu_{0}(X)}d\bar{\mu}_{s}
~n_{L}\frac{d\mu_{L}}{d\bar{\mu}_{s}}\;.
\end{aligned}
\end{eqnarray}
We have chosen to include strange quarks even at zero pressure i.e. at $\mu_{0}(X)$. From these values we start the integration of the particle densities of the three light flavors together with the electrons. The charm and muons are included only when crossing their respective thresholds at $\mu_{0}=\mu^{\rm charm}_{\rm thr}(X)$ and $\mu_{0}=\mu^{\rm muon}_{\rm thr}(X)$, in their associated pressures. 

 \begin{figure}[ht]
    \begin{center}
  \resizebox*{!}{7cm}{\includegraphics{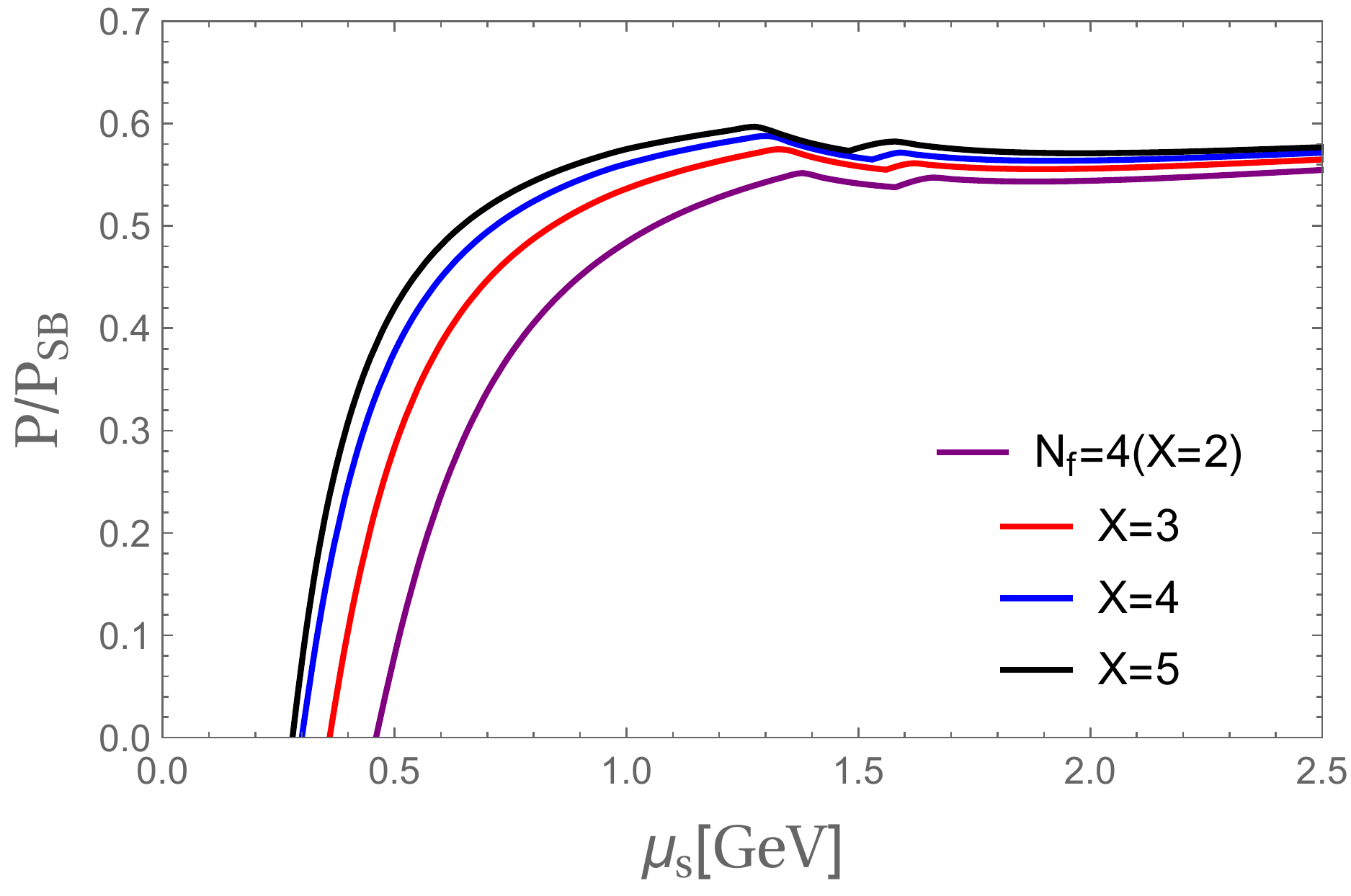}}
    \end{center}
   \vspace{5mm}
   \caption{Total pressure for a system of $N_{f}=2+1+1$ quarks plus leptons in $\beta$-equilibrium and electrically charge neutral normalized by the Stefan-Boltzmann massless free gas with $N_{f}=4$. We include the $X=5$ only to verify how the EoS depends on $X$ when including one additional massive flavor (see text for details in the kinks). Taken from Ref. \cite{Jimenez:2019kji}.}
   \label{fig:4ChPressure}
  \end{figure}

We define the energy density as
\begin{equation}
\epsilon(\mu_{s},X)=-P+\sum_{f=u,c}\epsilon^{(1)}_{f}
+\sum_{f=d,s}\epsilon^{(2)}_{f}
+\sum_{L=e,\mu}\epsilon_{L} \;,
	\label{eq:energydensity}
\end{equation}
where the quark and lepton contributions are
\begin{eqnarray}
\epsilon^{(1)}_{f}(\mu_{s},X)=\left[ \mu_{s}-\mu_{e}(\mu_{s}) \right]n_{f}(\mu_{s}) \;,
\end{eqnarray}
\begin{eqnarray}
\begin{aligned}
\epsilon^{(2)}_{f}(\mu_{s},X)=\mu_{s}n_{f}(\mu_{s}) \;,
\end{aligned}
\end{eqnarray}
\begin{eqnarray}
\begin{aligned}
\epsilon_{L}(\mu_{s},X)=\mu_{L}(\mu_{s})n_{L}(\mu_{s}) \;.
\end{aligned}
\end{eqnarray}
Following this recipe, we can build the EoS, $P=P(\epsilon)$, by combining Eqs. (\ref{eq:pressure}) and (\ref{eq:energydensity}) for a given $X$.

In Fig. \ref{fig:4ChPressure} we plot the total pressure for charm matter, normalized by a  Stefan-Boltzmann gas of quarks with $N_{f}=4$. From this one can see the usual behavior of the pressure for $N_{f}=2+1$ at intermediate densities, followed by a kink\footnote{These kinks are already visible in Ref. \cite{Kurkela:2009gj} for the critical chemical potentials at which the strange quark density drops to zero, and below which the quark matter is net strange quark free.} representing the charm threshold which softens the total (normalized) pressure. The charm quark
contribution reduces the renormalization-scale uncertainty band for X at high densities, which also affects the behavior of the EoS a lower densities, a feature which would be difficult extract from the pressure-density plane. An additional kink appears due to the muons. So, the charm EoS is largely softened, generating an apparent instability which could have astrophysical effects. In particular, it suggests
the possibility of another kind of ultradense compact star: charm stars.

For completeness, we note that it was verified numerically the matching between the pQCD EoSs with $N_{f}=2+1$ and $N_{f}=2+2$ at the charm threshold with the values mentioned above of $\mu_{s}$ around 1.3 GeV. This was done by using the original formalism of Kurkela \textit{et al.} which adds any number of massless flavors to a massive one, i.e. before the charm threshold one has a massive strange plus massless up and down quarks, being the charm quark decoupled from the system due its heavy nature. Above the threshold, the strange quark becomes effectively massless and only the charm quark is massive, so one adds massless up, down and strange quarks to a massive charm. This gives us confidence to our results which in a unified numerical code adds the charm quark only by manipulating appropriately the renormalization scale at the threshold. Additionally, with this reasoning it was verified that for values of $X$ below 4/3 the matching between pQCD with different number of flavors using the original result of Kurkela \textit{et al.} is impossible. Besides, it can be verified \cite{Graf:2015tda} that the pQCD thermodynamical potential only adds independently massive-flavor contributions, in other words, there does not exist massive cross-flavor interaction terms that may ruin our results for adding thermodynamic-potential terms corresponding to different massive quarks, at least perturbatively.

Besides, the charm quark contribution reduces the renormalization-scale uncertainty band for $X$ at high densities, which also affects the behavior of the EoS a lower densities. An additional kink appears due to the muons. So, the charm EoS is largely softened due to these pressure discontinuities, producing jumps in the number and energy densities that might be associated with instabilities having astrophysical effects. Specifically, this can be understood as a 1st-order phase transition whose control parameter is the flavor number $N_{f}$, like an external magnetic field, which might indicate a new stable branch for compact stars, in particular, it suggests the possibility of another kind of ultradense compact star: charm stars\footnote{Also the possibility of the charm presence in hybrid stars would be interesting to study.}.

\section{Can \textit{interacting} charm quark stars exist in Nature?}
	\label{sec:4charm-stars}

Although charm stars are excluded as two-solar mass neutron stars \cite{Lattimer:2010uk} given the high critical density for their appearance, they might be present as a new branch of hybrid compact stars. The first quantitative study of the possibility of the existence of charm quark stars\footnote{Quark stars might be called after the most massive quark flavor with which they are finitely populated, e.g., a strange star is composed of the three quark flavors $u,d,s$; a charm star of four flavors $u,d,s,c$.}, i.e. strange stars satisfying the Bodmer-Witten hypothesis and having a finite charm quark fraction at their cores, was carried out more than two decades ago Ref. \cite{Kettner:1994zs} (see also Ref. \cite{SchaffnerBielich:1998ci}). 
 \begin{figure}[ht]
    \begin{center}
  \resizebox*{!}{7cm}{\includegraphics{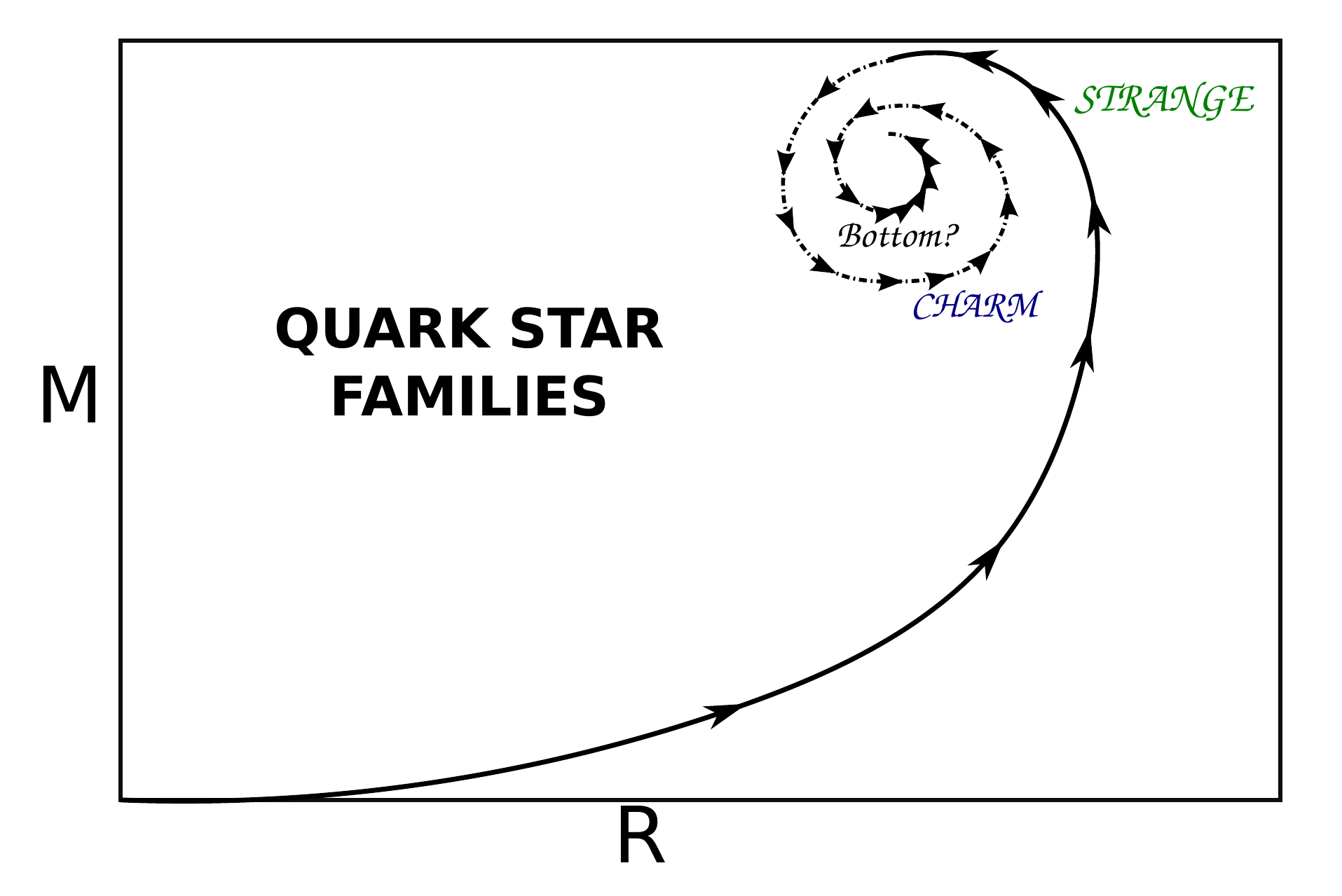}}
    \end{center}
   \vspace{5mm}
   \caption{Cartoon of mass-radius diagram for quark star families in hydrostatic equilibrium. The only stable branch seems to be the strange (continuous-arrowed line), whereas ultrahigh density stars are usually considered unstable against radial pulsations (dashed-arrowed line). Taken from Ref. \cite{Jimenez:2019kji}.}
   \label{fig:4MassRadCart}
  \end{figure}
The star bulk was described using the simplest version of the MIT bag model. For this, simple estimates were usually considered in the literature. For instance, the needed critical densities for the charm quark to appear, i.e. $n~{\geq}~n^{\rm crit}_{c}$ (for a charm mass of $m_{c}\simeq1.3$ GeV, $n^{\rm crit}_{c}=9m^{4}_{c}/(4\pi^{2})~{\geq}~1.4\times{10}^{17}{\rm g/cm^{3}}~{\geq}~5.2\times{10}^{2}n_{0}$ \cite{Haensel:2007yy}), where $n_{0}=0.16~{\rm fm^{-3}}$ is the nuclear saturation density of nuclear matter. This critical density is much higher than the maximal central density of strange stars, therefore charm quark stars would not exist. However, it would be better to solidify this estimate by means of more detailed simulations, i.e. a post-Newtonian calculation and a general relativistic stability analysis. Therefore, it was only after the stability analysis performed in Ref. \cite{Kettner:1994zs} that it was concluded more consistently that charm stars would be unstable (see Fig. \ref{fig:4MassRadCart} for an illustration of standard quark stars).
	
We revisit this question using our first-principle perturbative QCD for EoS for charm quark matter and also restrict our analysis to the simple case with no hadronic mantle. We choose the parameter space to be in the range $X\geq3$, which satisfies the Bodmer-Witten hypothesis, as shown in Ref. \cite{Kurkela:2009gj}, and perform the stability analysis as follows.
 \begin{figure}[ht]
    \begin{center}
  \resizebox*{!}{7cm}{\includegraphics{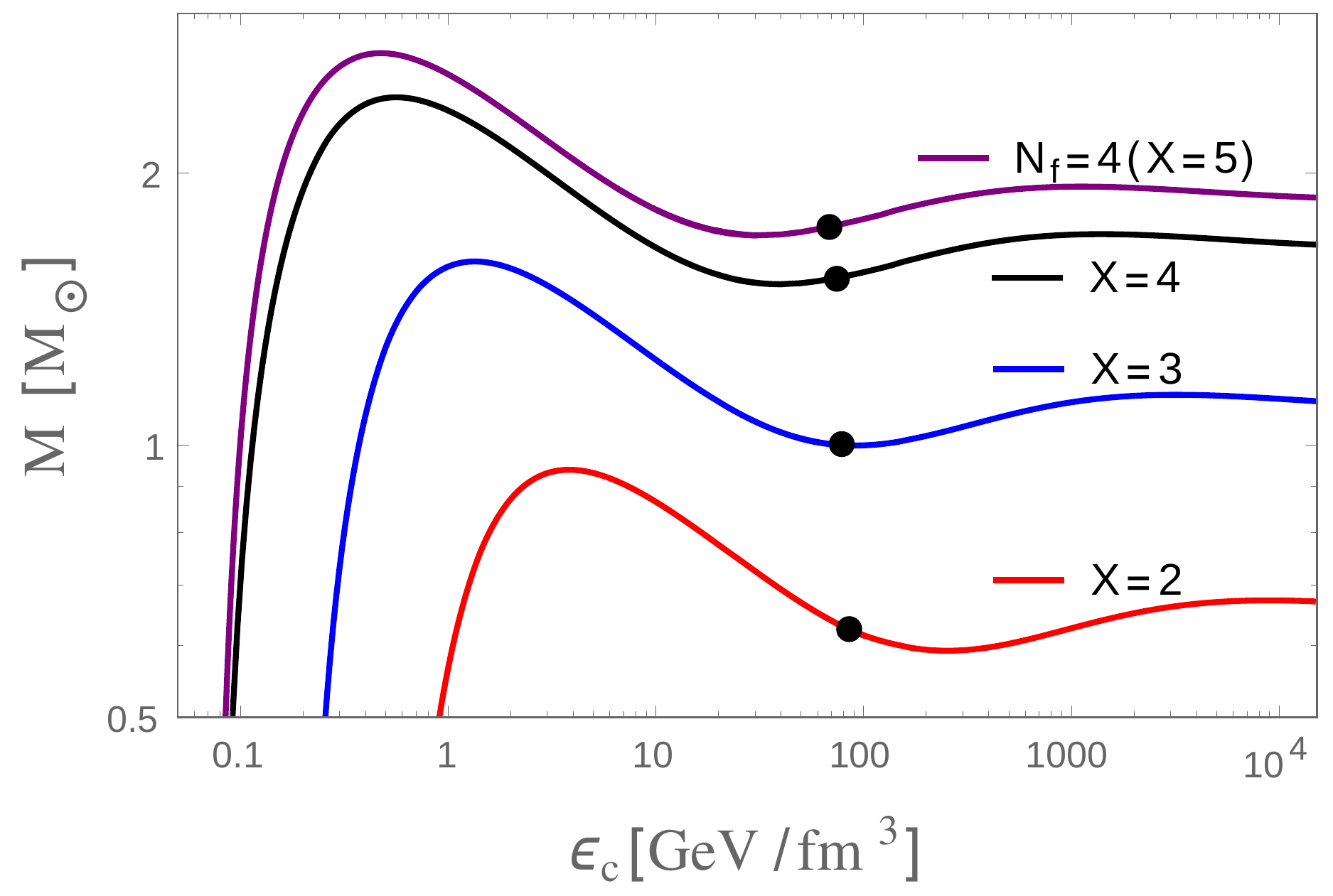}}
    \end{center}
   \vspace{5mm}
   \caption{Total gravitational mass vs. central energy density for a system with $N_{f}=2+1+1$ quarks plus leptons. The black dots indicate the minimal density required for the appearance of charm quarks in the system. Notice that it has been suggested in Ref. \cite{Misner:1964zz} that the second mass peak defines the sign for the next lowest eigenvalue $\omega^{2}_{1}$, for the neutron star sequence but with a polytropic EoS. Taken from Ref. \cite{Jimenez:2019kji}.}
   \label{fig:4GravMassDens}
  \end{figure}
 \begin{figure}[ht]
    \begin{center}
  \resizebox*{!}{7cm}{\includegraphics{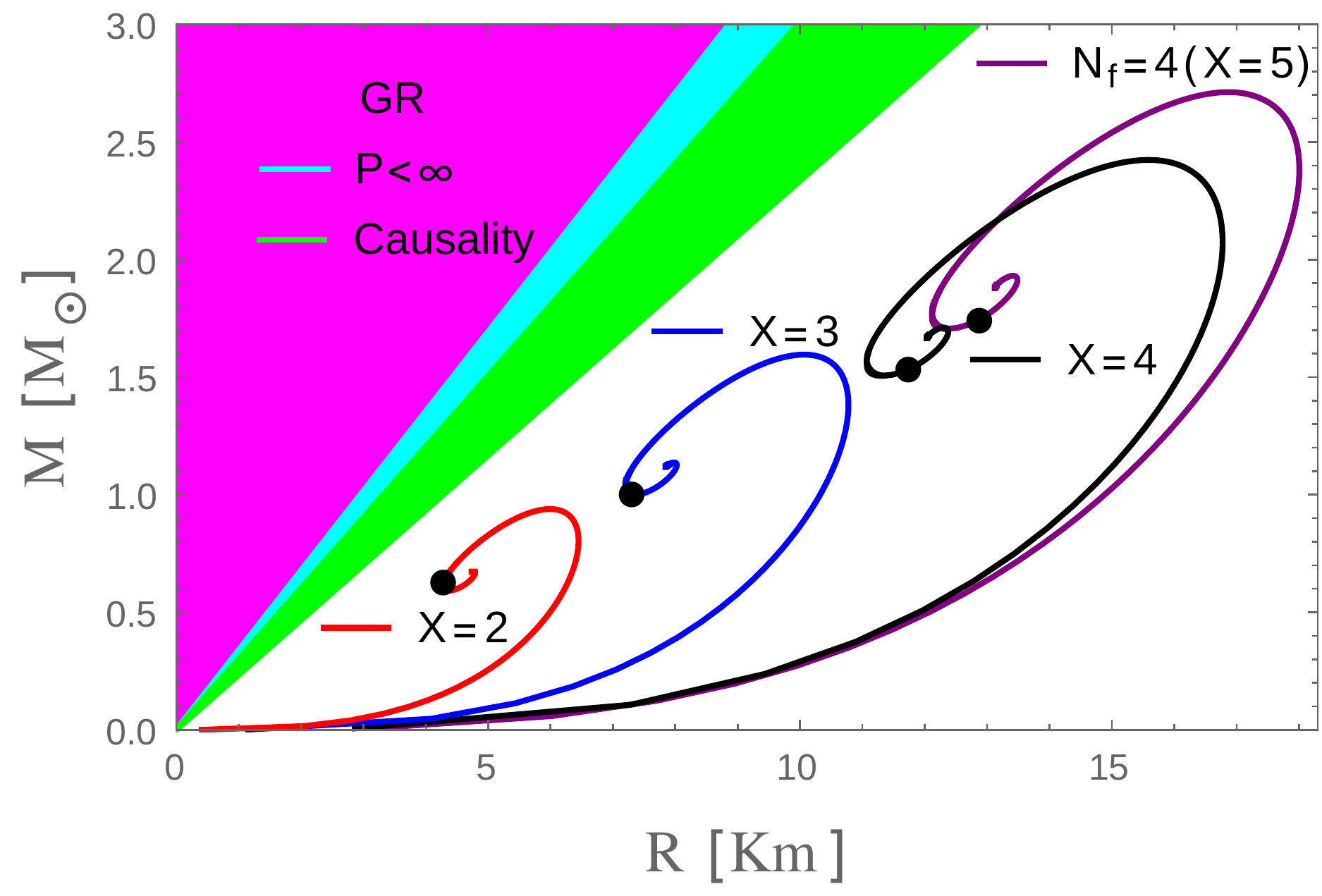}}
    \end{center}
   \vspace{5mm}
   \caption{Mass-radius diagram for quark stars made of $N_{f}=2+1+1$ quarks plus electrons and muons. The black dots signal the appearance of charm quarks in the system indicating possible charm star configurations. Stars not satisfying the general-relativistic (GR, magenta region), causality (green region), and finite pressure ($P<\infty$, light blue region) limits are excluded from this diagram \cite{Lattimer:2006xb}. Notice that this exclusion is less restrictive than the one presented in Ref. \cite{Lattimer:2010uk}, based on maximal star masses. Taken from Ref. \cite{Jimenez:2019kji}.}
   \label{fig:4MassRad}
  \end{figure}
As discussed in Chapter 2, the Tolman-Oppenheimer-Volkov (TOV) equations ensure the relativistic hydrostatic equilibrium of stellar configurations \cite{Glendenning:2000}. However, these configurations must also satisfy the thermodynamic condition $\partial{M}/\partial{\epsilon_{c}}~{\geq}~0$, where the maximum mass configuration for a given stellar family is identified with $\partial{M}/\partial{\epsilon_{c}}=0$ \cite{Glendenning:2000}. In Fig. \ref{fig:4GravMassDens} we show our results for the mass as a function of the central energy density. It can be seen that the necessary condition for thermodynamic stability is satisfied in the two branches, one at relative low and another at much higher energy densities\footnote{Such additional families of stars at ultrahigh nuclear densities are obtained not only for quark matter but were shown to exist for the neutron star sequence at ultrahigh densities too \cite{Misner:1964zz,Glendenning:2000}.}. However, we note that for the case $X=2$ this condition is not satisfied when charm quarks appear, which is indicated by the black dots in Fig. \ref{fig:4GravMassDens}. This is somewhat expected since it is difficult to have heavy quarks present in low-mass strange stars. In Table \ref{tab:4table1} we show the values of these observables at the charm threshold\footnote{In order to compare our results with the existing astrophysical literature on the subject, one can convert units and estimate energy density scales by using $ 60~{\rm GeV/fm^{3}}~{\approx}~{10^{17}}{\rm g/cm^{3}}$.}. On the other hand, for $X>3$ the thermodynamic condition is satisfied when charm quarks are present, which would correspond to charm stars. In Fig. \ref{fig:4MassRad} we show the mass-radius diagram for quark stars made of $N_{f}=2+1+1$ quarks plus electrons and muons for different values of $X$, where the usual instability of the radial modes is shown having the counter clockwise spiral behavior. Nevertheless, we pass to prove this qualitative indication in a precise quantitative manner within approximate and exact calculations performed within general relativity, respectively.

\begin{table}[h!]
  \begin{center}
   \begin{tabular}{c|c|c|c|c} 
     $X$ & $\mu^{\rm th}_{s}[{\rm GeV}]$ & $\epsilon^{\rm th}_{c}[{\rm GeV/fm^{3}}]$ & $M^{\rm th}[\rm M_{\odot}]$ &$R^{\rm th}[{\rm km}]$\\
     \hline
      $2$ & $1.377$ & $85.702$ &$0.625$ & 4.282\\
      $3$ & $1.340$ & $77.988$ &$0.999$ & 7.310\\
      $4$ & $1.290$ & $74.189$ &$1.531$  & 11.72\\
      $5$ & $1.295$ & $69.199$ &$1.745$  & 12.87\\
    \end{tabular}
    \vspace{5mm}
        \caption{Different values for threshold $\mu_{s}$, $\epsilon_{c}$, gravitational mass $M$ and its associated radii for different values of $X$. Taken from Ref. \cite{Jimenez:2019kji}.}
     \label{tab:4table1}
  \end{center}
\end{table}


\subsection*{Post-Newtonian Approximation Analysis}        
        
Before going into the full general-relativistic calculation, we first study the stability problem within the post-Newtonian (pN) approximation, where the effects of general relativity are treated as first-order (weak) corrections to Newtonian gravity \footnote{We think that the truncation at this order is reasonable as far as one desires to obtain additional physical insights beyond the insufficient Newtonian theory, instead of the numerical precision required for compact stars where the gravitational fields are strong.}. This approach requires the adiabatic index, $\Gamma~{\equiv}~(1+\epsilon/P)\partial{P}/\partial{\epsilon}$, to be larger than some characteristic index, $\Gamma_{0}~{\equiv}~{4}/{3}+{2M\kappa}/{R}$ (again up to 1st-order in the pN approximation) with $\kappa\sim{1}$ \cite{Shapiro:1983}, i.e. $\Gamma>\Gamma_{0}$ for a given $X$, in order to have stable charm quark configurations. In Table \ref{tab:4table2} we show our results for different values of $X$, from which it is concluded that they cannot exist in this 1st-order approximation of general relativity. This instability could also be understood in the inwardly spiralling behavior of the mass-radius curves in Fig. \ref{fig:4MassRad}, which is a typical behavior of unstable configurations when the underlying EoS does not exhibit any thermodynamical instability leading to a further family of compact stars at ultra-high densities. Notice that the same inwardly directed spiraling behavior is also obtained when extending the neutron star sequence to ultrahigh densities. Hence, this behavior is not particular of self-bound stars, but rather manifests the dominant role of gravity at ultrahigh densities.
\begin{table}[h!]
  \begin{center}
   \begin{tabular}{c|c|c|c|c} 
     $X$ & $\Gamma_{0}(M^{\rm Max})$ & $\Gamma(M^{\rm Max})$ & $\epsilon^{\rm Max}_{c}[{\rm GeV/fm^{3}}]$ &${\rm pN-Stable}$\\
     \hline
      $2$ & ${\approx}~1.75522$ & ${\approx}~1.338$ &${\approx}~14500$ & No\\
      $3$ & ${\approx}~1.77631$ & ${\approx}~1.341$ &${\approx}~4100$ & No\\
      $5$ & ${\approx}~1.75842$ & ${\approx}~1.334$ &${\approx}~1050$ & No\\
    \end{tabular}
    \vspace{5mm}
        \caption{Different values for the behavior of the polytropic index $\Gamma$ and $\Gamma_{0}$ are listed for different $X$'s. These indices are calculated for stellar configurations near the maximum charm mass configurations.}
     \label{tab:4table2}
  \end{center}
\end{table}

\subsection*{General-Relativistic Analysis}

The previous analysis provides a necessary but insufficient condition for stability of star configurations. One must still test the \textit{dynamical} stability under radial pulsations. For that we use the method of Gondek \textit{et al.} \cite{Gondek:1997fd} developed in detail in Chapter 3. Further details to deal with $\omega^{2}_{n}<{0}$, apart from the original treatment of Gondek \textit{et al.} for $\omega^{2}_{n}\geq{0}$, are in order\footnote{In this formalism, the maximum mass stellar configuration is characterized by having $\omega_{0}=0$ \cite{Glendenning:2000}.}. As it is well known \cite{Glendenning:2000}, additional (apparently) stable branches of stars appearing at high densities, e.g. the second branches of Fig. \ref{fig:4GravMassDens}, might change the sign of the squared frequency, so we must adapt our first-order formalism in order to deal with these situations. We do that by making $\omega^{2}_{n}\rightarrow{-\omega^{2}_{n}}$ in the pulsation equations, yielding a relevant change only in the equation for $\Delta{P}$
	\begin{multline}
	\frac{d\Delta{P}}{dr}=\xi\left\lbrace{-\omega^{2}e^{\lambda-\nu}(P+\epsilon)r-4\frac{dP}{dr}}\right\rbrace+ \\
	\xi\left\lbrace\left(\frac{dP}{dr}\right)^{2}\frac{r}{(P+\epsilon)}-8\pi{e^{\lambda}}(P+\epsilon)Pr\right\rbrace + \\
	\Delta{P}\left\lbrace{\frac{dP}{dr}\frac{1}{P+\epsilon}-4\pi(P+\epsilon)r{e}^{\lambda}}\right\rbrace \; ,
	\label{4Rad2}
	\end{multline}	
where the positiveness of $\omega^{2}$ represent a different pulsation equation Eq. (\ref{4Rad2}) probing unstable stars associated to the given EoS, and the boundary conditions are the same as already seen in Chapter 3. This is consistent with past results \cite{Glendenning:2000} where the ``negativeness'' was taken by the squared frequency and the equations to be solved being the same.

For purposes of the numerical calculations (as seen in the warm-up of Chapter 3), it is better to deal only the frequency $\omega_{n}={\rm Re}(\omega_{n})+i{\rm Im}(\omega_{n})$ and not its squared value. So, some comments are in order. In this first-order radial pulsation formalism, amplitudes oscillate harmonically when the frequencies are such that ${\rm Re}(\omega_{n})>0$ and ${\rm Im}(\omega_{n})=0$, or increase exponentially if ${\rm Im}(\omega_{n})>0$. Since our oscillation equations \cite{Gondek:1997fd} satisfy the ordering $\omega^{2}_{0}<\omega^{2}_{1}<\omega^{2}_{2}<\cdot\cdot\cdot< \omega^{2}_{n}$. So, if ${\rm Im}(\omega_{0})>0$ from some value of central energy density $\epsilon_{c}$, then all the higher modes will become complex too, representing the onset of the instability. So, if ${\rm Im(f_{0})>0}$ (being $f_{0}=\omega_{0}/2\pi$) continues finite and increasing for higher densities, all the configurations become unstable.

 \begin{figure}[h!]
    \begin{center}
  \resizebox*{!}{7cm}{\includegraphics{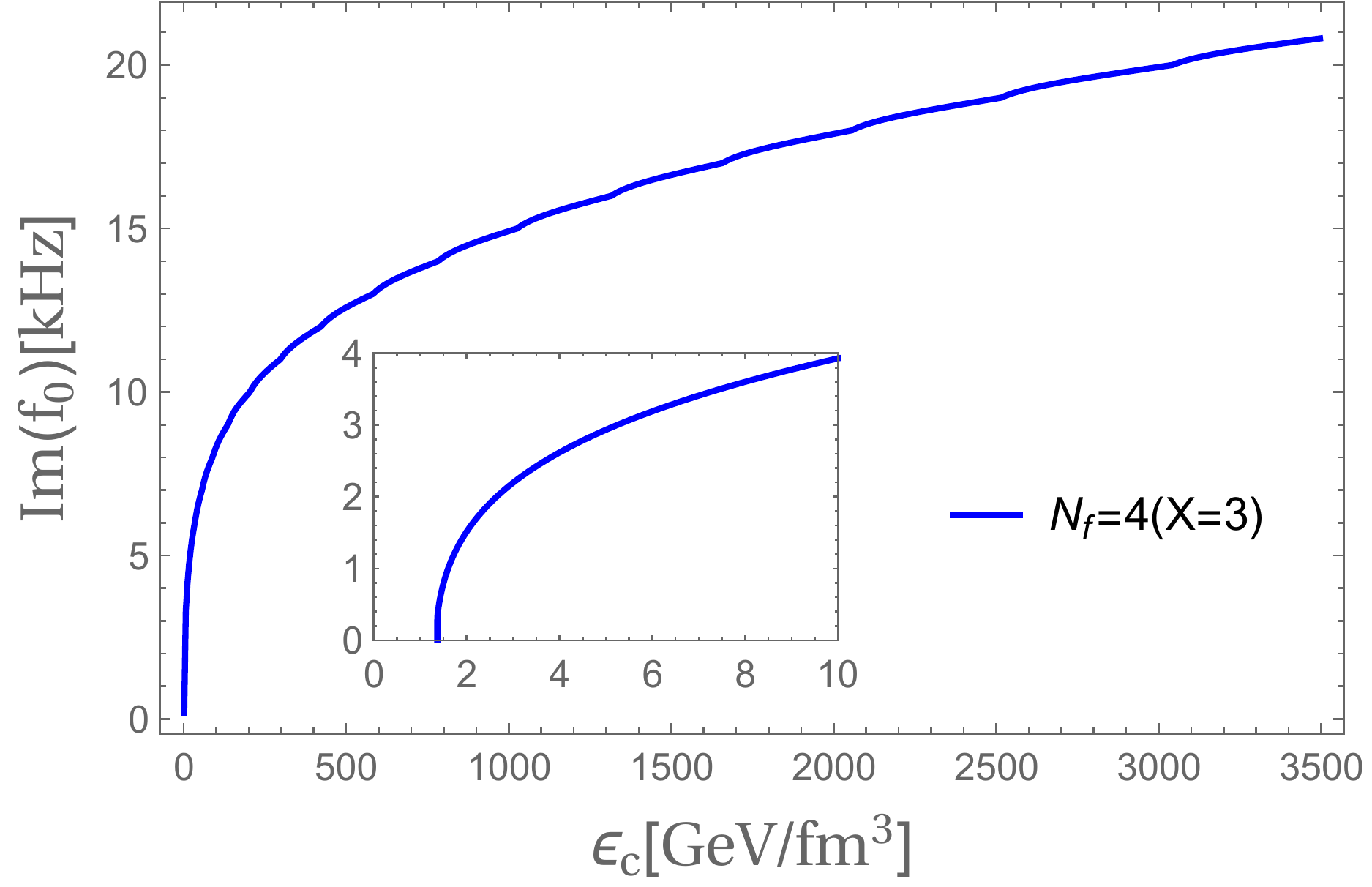}}
    \end{center}
   \vspace{5mm}
   \caption{Imaginary part of the fundamental mode frequency $f_0$ as a function of the central energy densities obtained after solving the radial pulsation equations for $X=3$. Notice that the minimal value of $\epsilon_{c}$ for ${\rm Im}(f_{0})=0$ coincides with the maximum-mass energy density of Fig. \ref{fig:4GravMassDens} at around 1.4 ${\rm GeV/{fm}^{3}}$. Taken from Ref. \cite{Jimenez:2019kji}.}
   \label{fig:4ImFreqDens_X3}
  \end{figure}
 \begin{figure}[h!]
    \begin{center}
  \resizebox*{!}{7cm}{\includegraphics{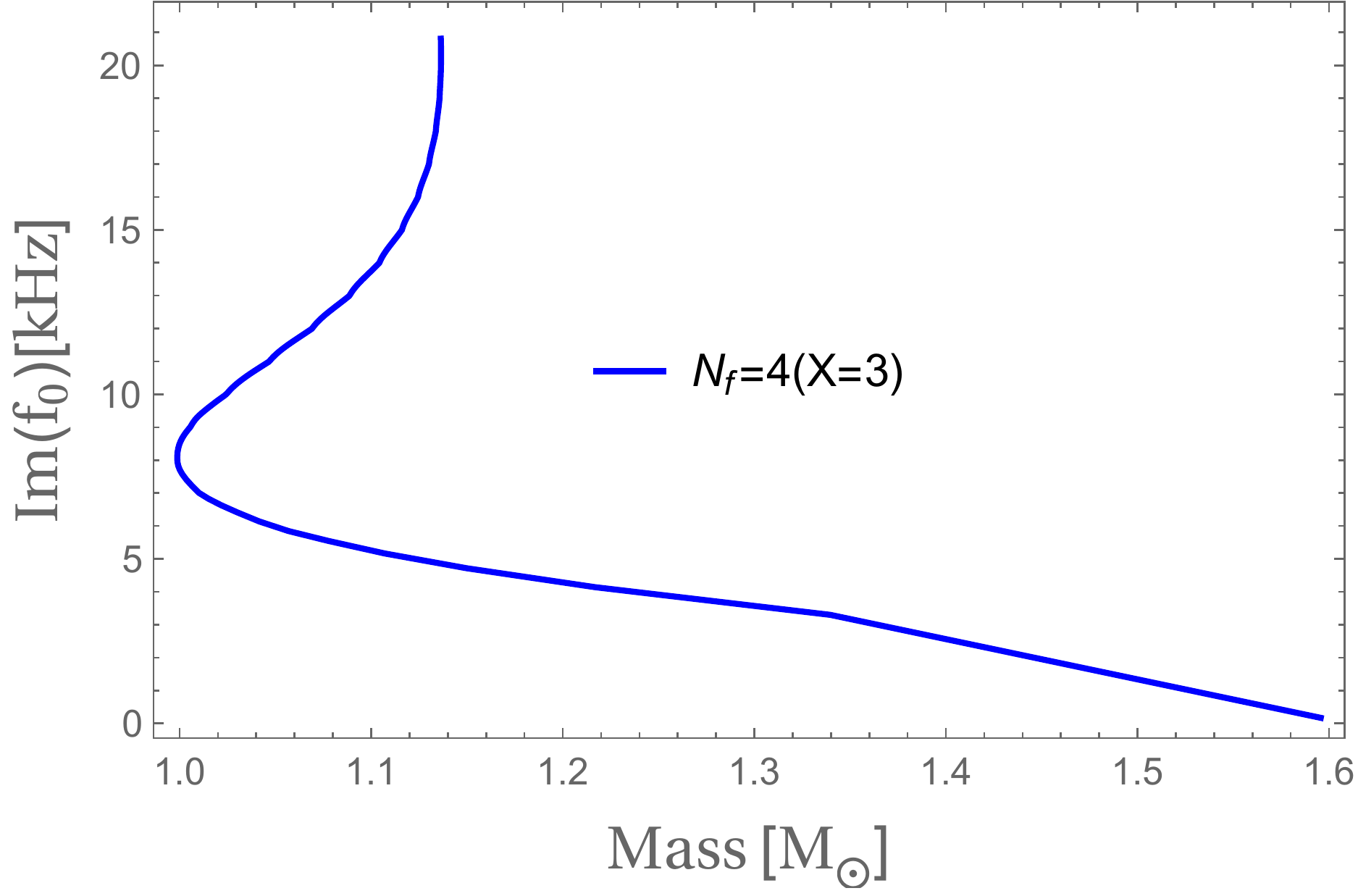}}
    \end{center}
   \vspace{5mm}
   \caption{Imaginary part of the fundamental mode frequency $f_0$ as a function of the total gravitational mass obtained after solving the radial pulsation equations for $X=3$.}
   \label{fig:4ImFreqMass_X3}
  \end{figure}
In Figs. \ref{fig:4ImFreqDens_X3} and \ref{fig:4ImFreqMass_X3} we show that for densities above the maximum-mass strange star configuration (for $X=3$), the stellar configurations increase their oscillation amplitudes exponentially even in the region where charm stars are expected, thus making them dynamically unstable\footnote{Although one could also build a figure with a complex plane of ${\rm (Re+Im)}(\omega_{0})$, i.e. frequency bands, it would not be useful since physically only one or the other are relevant. On the other hand, observables depending on ${\rm Re (Im)}(\omega_{n})$ might be calculated using the Kramers-Kronig relations.}. Since the same behavior was obtained for larger values of $X$, one can conclude from a perturbative QCD analysis that charm stars are unstable. One could ask if higher-order perturbative terms could in some way stabilize charm stars. However, a recent N$^{3}$LO weak coupling expansion, also including nonperturbative terms, yielded minor modifications to the EoS \cite{Gorda:2018gpy}.  

\section{Summary and Conclusions}
	\label{sec:4conclusion}

In this chapter we have extended the perturbative QCD $N_{f}=N_{l}+1$ formalism developed in Chapter 2 in order to allow for the inclusion of heavy quark flavors
in the EoS for cold and dense quark matter. In particular, we have investigated the effects of charm quarks in the equation of state in the case of $\beta$-equilibrium and electric charge neutrality, where a non-negligible range of the parameter space was discarded in order to go through the charm threshold in agreement with the EoS for light quarks. The, we have explored the possibility of charm stars within this model of heavy quark matter, spanning a range in quark chemical potentials where pQCD is in principle much more reliable and only estimates and conclusions within crude models were made previously. After performing a radial stability analysis, we concluded that these stars would be unstable. 
  
Although charm stars apparently are excluded by our analysis, and also due to the causality limits posed by the maximum mass constraints from neutron star observations \cite{Lattimer:2010uk,Lattimer:2019eez}, it is possible to have small amounts of charm quark matter in the core of the heaviest observed neutron stars (or, rather, hybrid stars), where a matching between a nuclear and a quark phase could be possible via a Glendenning construction for first-order phase transitions. Recently, a related possibility was investigated under the consideration of strange quark matter contaminated by charm quark impurities (in the sense of condensed matter physics) producing a QCD Kondo effect \cite{Yasui:2016svc,Macias:2019vbl}. Moreover, a non-negligible amount of charm quarks could contribute to the EoS at the early stage of neutron star mergers, when very high densities are reached \cite{Alford:2017rxf,Most:2018eaw}.
  
Our extended framework is appropriate to study the \textit{heavy} sector of the QCD phase diagram (see Ref. \cite{Maelger:2018vow} for related studies) which could exhibit new features, although it was shown in Refs. \cite{Fischer:2014ata,Burger:2018fvb} that heavy quarks affect negligibly the chiral and deconfinement transitions at finite temperature.   

\end{chapter}

\begin{chapter}{Conclusions and perspectives}
	\label{chap:conclusion}

\hspace{5 mm}


Strongly-interacting matter under extreme conditions has become a subject of great interest in the last years for theoretical and experimental reasons. Additionally, from the observational point of view, the measurement of the electromagnetic and gravitational waves coming from the merger of neutron stars in a binary system opened the multimessenger era of astrophysical observations. 

In this thesis we focused on the physics of compact stars, and addressed their interior by using the underlying theory of strong interactions. In particular, we have investigated cold and dense matter with perturbative QCD, only being reliable at high densities but giving us control of its unreliability at low densities through an energy scale inherited from QCD. This helped us to obtain results for thermodynamic observables and stellar properties of compact (quark or neutron) stars with a theoretical band parametrizing our ignorance of the nonperturbative sector of QCD.

In Chapter \ref{chap:PNS} it was assessed \cite{Jimenez:2017fax} the possibility of homogeneous nucleation in protoneutron star conditions for which the formation of quark matter droplets in a relatively hot and dense environment was studied. Usually in the literature this problem was tackled within the MIT bag model for the quark phase where only crude estimates were made. For the quark phase at high densities we built the lepton-rich pQCD EoS by including neutrinos into the framework of Kurkela \textit{et al.} \cite{Kurkela:2009gj} which furnish QCD interactions even at intermediate densities. A discontinuous matching process between the lepton-rich TM1 nuclear and pQCD EoSs was considered in order to mimic a first-order phase transition between phases. Having this framework available (valid at all densities), all the relevant nucleation parameters were extracted in the thin-wall approximation and, by providing a time scale estimate of the protoneutron star lifetime, we were able to compute a range of values for the surface tensions of the formed quark matter droplets. Our results show that the critical baryon densities between phases are increased due to the neutrino inclusion in the system (something already known when the bag model was used), but being of the order of 10$n_{0}$ for values of the renormalization scale producing hybrid stars having masses above the usual two-solar mass constraint.

Although the nucleation process was analyzed in the Chap. 2 in certain detail, it was only meant to be an application of the lepton-rich equation of state. Additional lines of research include using our lepton-rich pQCD EoS for the quark phase in detailed protoneutron star evolution simulations leading to a second-neutrino burst due to a hadron-quark transition (as discussed in Refs. \cite{Sagert:2008ka,Fischer:2010wp}), where short-ranged QCD interactions might change drastically the emission time and could offer a domain of values controlled by the renormalization scale. Besides, a Gibbs-Glendenning construction \cite{Glendenning:2000,Pagliara:2009dg} for a mixed phase can be used to impose global electric charge neutrality and global lepton fraction conservation with our EoS and some other standard nuclear matter EoS, which, if obtained, would be subject to a radial stability analysis (applying techniques of Chap. 3) of protoneutron star configurations, which was poorly investigated so far. 

Chapter \ref{chap:radialQS} (following Ref. \cite{Jimenez:2019iuc}) is devoted to the study of the dynamical stability of quark star configurations in hydrostatic equilibrium obtained after solving the TOV equations. This was done by studying the behavior of these quark star when perturbed radially and adiabatically within general relativity. As a preliminary step before performing this analysis, instead of the well-known Chandrasekhar second-order formalism written as a Sturm-Liouville problem we used a pair of first-order differential equations for appropriate Lagrangian variables for which boundary conditions could be imposed straightforwardly. An important ingredient for these studies was the quark matter model used along the calculations. For better numerical manipulations, it was better to use the pQCD result of Kurkela \textit{et al.} (already studied in Chap. 2) cast into a pocket formula obtained by Fraga \textit{et al.} \cite{Fraga:2013qra}. With this EoS our code produced the fundamental and first-excited mode frequencies and periods of quark stars as functions of the central energy density, total gravitational mass and gravitational red-shift parameter. Our results were given in terms of bands, again parametrized by the renormalization scale $X_{\rm FKV}$. 

It would be interesting to study the radial oscillations within the modified theory of gravity $f(R)$ but using the pocket pQCD FKV formula for quark star matter since up to now only polytropic equations of state for neutron star matter were considered in the literature \cite{Sbisa:2019mae}. On the other hand, the hybrid star case considering the FKV formula at high densities is of importance to understand the effects of QCD matter at the core of compact stars. In particular, we note that it is our aim to publish in the near future our results on the stability against radial oscillations of the constrained (hybrid) EoS for NS matter, i.e. study the stability of stellar configurations that satisfy the recent gravitational wave constraints from the merger of neutron stars for masses and radii on compact stars \cite{Annala:2017llu,Annala:2019puf}. 

Interpolated equations of state between well-known nuclear matter EoSs at low densities and the pQCD EoS were constructed almost a decade ago in Refs. \cite{Tews:2012fj,Hebeler:2013nza,Kurkela:2014vha}. However, no insight was obtained on the stability of these NS configurations. Only external constrains (maximal mass, minimal radii, tidal deformabilities, and so on) were imposed but not their stability within general relativity. This analysis is more complicated than the one elaborated in Chapter 3 since, for instance, it is not trivial to tame the numerical instabilities that appear at the phase boundaries where there are jumps in the energy density and a supplementary equation must be solved to get a continuous behavior of the normal modes. Our calculations are still in progress at this moment and we should report on this subject soon \cite{Jimenez:2020kk}.

Finally, a novel method \cite{Jimenez:2019kji} was presented in Chapter \ref{chap:charm} to deal with more than one heavy quark in the equation of state for cold quark matter. Initially we discussed the result of Kurkela \textit{et al.} \cite{Kurkela:2009gj}, technically called the $N_{f}=N_{l}+1$ EoS (developed in detail in Chap. 2). Then, we developed a formal method to reorganize it in order to include one by one $N_{m}$ massive quarks at the cost of adding at least one massless flavor to the quark matter system. For simplicity, we only focused on the charm quark matter case where many technical details must to be used when building up the pressure and energy density, mainly at the transition point between $N_{f}=2+1$ to $N_{f}=2+1+1$ flavors. After discussing these results, we considered the possible existence of charm quark stars in Nature. After performing the stability considerations developed in Chap. 3, we found that they cannot exist since all the stellar configurations after the maximal strange star configuration (for any $X$) have imaginary frequencies which means that their amplitudes increase in time, i.e. the star explodes. This is only one of the possible applications that can be made using our heavy EoS, e.g. in the primordial universe or charmlets in strange matter possibly existing in quark stars and heavy-ion collisions \cite{SchaffnerBielich:1998ci}.

\end{chapter}

	

\newpage
\phantomsection
\addcontentsline{toc}{chapter}{Bibliographic References}

\bibliography{Bibliography}


\end{document}